%% file: qmcourse.tex
\documentclass[a4paper,oneside,11pt]{book}

\usepackage{float} 
\usepackage{wrapfig} 

\usepackage{graphicx}

\usepackage{graphicx}
\usepackage{epsfig}

\usepackage{amsfonts}
\usepackage{amsmath}

\newcommand{\be}{\begin{equation}}
\newcommand{\ee}{\end{equation}}
\newcommand{\bea}{\begin{eqnarray}}
\newcommand{\eea}{\end{eqnarray}}
\newcommand{\ba}{\begin{array}}
\newcommand{\ea}{\end{array}}
\newcommand{\inft}{\int_{-\infty}^\infty}
\newcommand{\bit}{\begin{itemize}}
\newcommand{\eit}{\end{itemize}}

\newcommand{\f}{\frac}

\begin{document}
\input{Hmin1.tex}

\tableofcontents

\input{H0.tex} 
\part{The Laws of Quantum Mechanics}
\input{H1.tex} 
\input{H2.tex}
\input{H3.tex}
\input{H4.tex}
\part{Atoms and Wave Packets}
\input{H5.tex}

\input{H6.tex}
\input{H7.tex}
\input{H8.tex}

\part{Other Applications}
\input{H9.tex}
\input{H10.tex}
\input{H11.tex}
\input{H12.tex}
\input{H13.tex} 
\input{H14.tex} 
\input{H15.tex} 

\end{document}

%% file: Hmin1.tex
\chapter*{\begin{center}An introduction to Quantum Mechanics\end{center}}

%% file: H0.tex
\chapter*{Welcome!}
\addcontentsline{toc}{chapter}{Welcome!}
Welcome to this course on Quantum Mechanics. It is aimed at students who have not had any exposure to the subject before. Hopefully you enjoy this learning material. Extra material can be found on the course website:

\begin{center}
http://www.bramm.be/qmcourse
\end{center}
\subsection*{Background material}
The course assumes familiarity with basic calculus skills like differentiating, integrating, and complex numbers. Also, the reader should be familiar with the notion of vector spaces and matrices. On the physics side, it is probably useful to have knowledge of classical mechanics and electromagnetism. If you know these subjects, you should definitely be able to master this course.

\subsection*{Style}
This course is written in a rather loose style. The emphasis is mainly on physical understanding and not on mathematical rigor. Also, all chapters (with exception of the first one) have a fixed outline:
\begin{itemize} 
\item \includegraphics[width=0.04\textwidth]{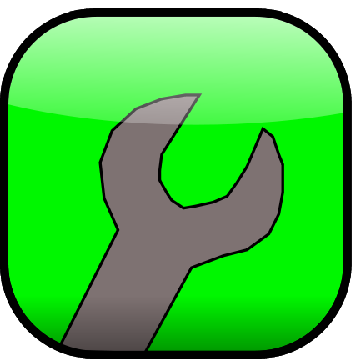} First, we introduce some necessary \textbf{tools}. Typically some math or a short physics review.  
\item \includegraphics[width=0.04\textwidth]{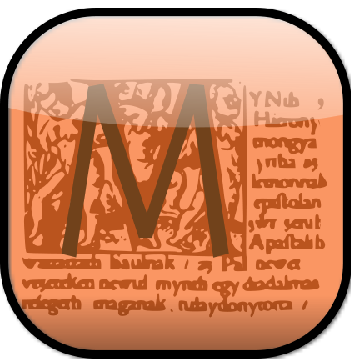} After that, a \textbf{story} is told. This is the theory/history part of the chapter.
\item \includegraphics[width=0.04\textwidth]{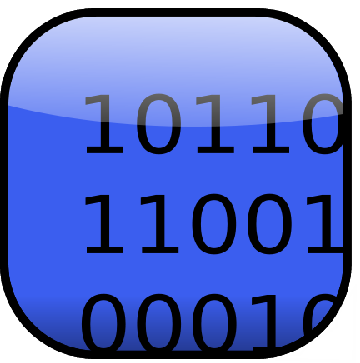} Finally, we combine the two previous ingredients (tools and story) to get to the physics of the chapter. Typically this involves a \textbf{computation} and some applications. 
\end{itemize}
Good luck!\\
\quad\quad\quad\quad\quad\quad\quad\quad\quad\textit{Bram Gaasbeek, juli 23, 2010}

%% file: H1.tex
\chapter{Intro: what is Quantum Mechanics?}

\subsection*{In this chapter...}
In this chapter you will learn the basic `quantum concepts'. You may find them a bit different from things you have learned before, but don't be afraid, they are far from incomprehensible. After reading the first few chapters, you may even see quantum mechanics as a quite natural description of the world on its smallest scale. 
\newpage 

\section{Unequal starts: reading guide}
As with any course, the biggest problem is the unequal start. Some students have learned more than others, some less. As a consequence, every course will always start off too fast for some and too slow for others. 

To solve this problem, there are several appendices to this chapter. They contain things you should know \textit{before} reading this course. However, you might know quite some bit of it already. The descriptions below should help you decide which appendices you should read, and (more important) which not. 

\begin{itemize}
\item Appendix 1: Complex numbers recap. \\
Only needed if your knowledge of complex numbers is poor. Review of complex conjugate, norm, inverse, $e^{i \theta}$, etcetera.
\item Appendix 2: Moving waves\\
Review of traveling waves like $y(x,t) = \sin (kx-\omega t)$ or $y(x,t) = e^{i(kx-\omega t)}$. Needed if you are not familiar with the concepts `wave amplitude', `wave number', `wave speed' or `complex wave'.
\item Appendix 3: Maxwell's equations. \\
Reviews how electromagnetism can be summarized in the four laws of Maxwell. If you don't know these laws by hard but understand them, you certainly do \textit{not} need to read this appendix.
\item Appendix 4: Light is an electromagnetic wave.\\
Needed if you never learned about this fact, or if you have never seen how the Maxwell equations imply the existence of waves.
\item Appendix 5: What is interference?\\
Needed if you do not know the
calculation of interference of (light)waves falling through two narrow slits.
\end{itemize}
If you know all of this, congratulations; forget about those appendices and read on. If there are some things you don't know yet, take your time to read them carefully - they will help making sense of what follows. Again: restrict your attention to what you don't know - ignore the rest.  After catching up, return to this point.

\section{Quantum Mechanics, what's up?}
As a warmup, some quick questions and answers about Quantum Mechanics...

\textbf{Why would I learn Quantum mechanics?}\\
Let's think a bit about the physics you have learned so far.  Throughout the past years, you got to know how bodies experience and exert forces, how gases behave, how electromagnetic forces act and all that. On top of that, you learned that molecules constitute all the matter we see in daily life. These molecules are arranged in gases, solids and liquids, mixed up in one big soup - the world around us. Better even, those molecules are made of atoms, and each single atom is a cloud of electrons flying around a nucleus of protons and neutrons. That's pretty impressive. Indeed, this insight explains almost the entire reality around us in terms of a very small set of particles: electrons, protons and neutrons. However impressive and unifying this fact may be, it does leave us with some very real questions. Questions that you -most likely- have not gotten an answer to yet. Why do those particles stick together in atoms? Why do they sit together very very close, but not on top of each other? And what \textit{are} these particles really? Just points? Or do they have some size? How do they behave? And so on. Precisely these questions are answered by quantum mechanics.

\textbf{What will I learn in this course?}\\
So, in this course, you will learn what those particles are, really. This means we will describe reality on the smallest scale. This description will be different from the physics you have learned so far. That does not mean one or the other is wrong. They are just descriptions of \textit{different} physical situations. If you want to describe the dynamics of individual particles (or just a few of them) then you have to use quantum mechanics. For gases, solids, and all that, we don't use this description: it would be both impractical, and impossible. Indeed: how could we describe a gas by mathematically keeping track of \textit{all} particles of such a system at the same time? We can't, that's why -for larger systems- we use gas laws, heat laws, Newton's laws, and so on. Those laws are very accurate for macroscopic systems, but one should always bear in mind that they actually \textit{follow} from the dynamics and physical laws that govern the microscopic world - that of particles. That is what makes Quantum Mechanics (short: `QM') - the fundamental theory of microscopic systems - so interesting. 

\textbf{Is QM difficult?}\\
QM may be more difficult than the physics you have learned up to this point, but definitely not `impossible to understand' or `comprehensible to only few'. You may wonder why it is more difficult. Why do we understand -for example- gas laws more easily then the quantum mechanical laws that describe fundamental particles? Actually, there is a simple answer to that question. Gas laws deal with systems of our scale, of our daily life. We have quite some feel for it, even before studying physics. You were not really \textit{shocked} when learning that the pressure of a gas rises when you compress it, right? This is just because you have experienced this law many times (f.e. when you inflate a bike tire) without really realizing it. However, none of us `feel' or `see' fundamental particles in ordinary daily life. If we did, the laws of quantum mechanics would feel very natural. You would just say `ah yes, thats indeed what it looks like' or `ah, yes, a particle indeed moves like that' - in the same way some people react on gas laws. Just keep in mind that our gas laws (or any other macroscopic physics) would look equally strange and funny for -say- some exotic \textit{submicroscopic} life form only used to the laws of QM. 

\textbf{How sure are we about QM?}\\
Or better even: how did we ever discover the laws of quantum mechanics? Like any other branch of science: by doing experiments and thinking about the results. You may know that the basic laws of quantum mechanics were discovered about a hundred years ago. Prehistory, that is, in terms of technology and science. Since their discovery, the laws of quantum mechanics have been confirmed by numerous experiments, often with a higher precision than anywhere else. So, it is no exaggeration to state that quantum mechanics is the best-tested theory of physics. On top of that, the number of inventions directly using quantum mechanics is significantly bigger than most people think, and very likely to keep on growing in the very near future. In conclusion, despite its beauty and exotic character, quantum mechanics is not at all `hypothetical' or `merely philosophical'. It is an accurate, tested and elegant description of the smallest and most fundamental part of the reality around us, that of elementary particles.

\section{How it all began... two slits}
Quantum mechanics was born together with a series of remarkable experiments. One of them was the so-called \textbf{two-slit experiment}\footnote{which you either learned about before, or have read about in Appendix 5}. The first version of this experiment was actually carried out by Thomas Young in 1803. Some people back then were arguing about the true nature of light. In our daily life, we see that light travels on straight lines -  you can't look around a street corner, right? This suggests light might be made up of tiny particles or `bullets' flying through space. However, some people -like Mr. Young- were convinced light has some wave aspect to it. Indeed, as you learned in your course on electromechanics\footnote{or else in Appendix 4}, you know that light is a wave in the electric and magnetic field. So to describe it, you need to give the distortion in the E and B field, depending on space and time. So you need to give $E(\vec{x},t)$ and $B(\vec{x},t)$ to describe a traveling piece of light. This contrasts with a particle, which can be described by only its position depending on time: $\vec{x}(t)$. Anyhow, Mr. Young did not know all the fancy formulas of Maxwell, so how \textit{did} he find out light is a traveling wave, not a traveling particle? 

\subsubsection{So, how did he find out? -- The two-slit experiment.}
\begin{figure}[h]
 \begin{center}
  \includegraphics[width=1.0\textwidth]{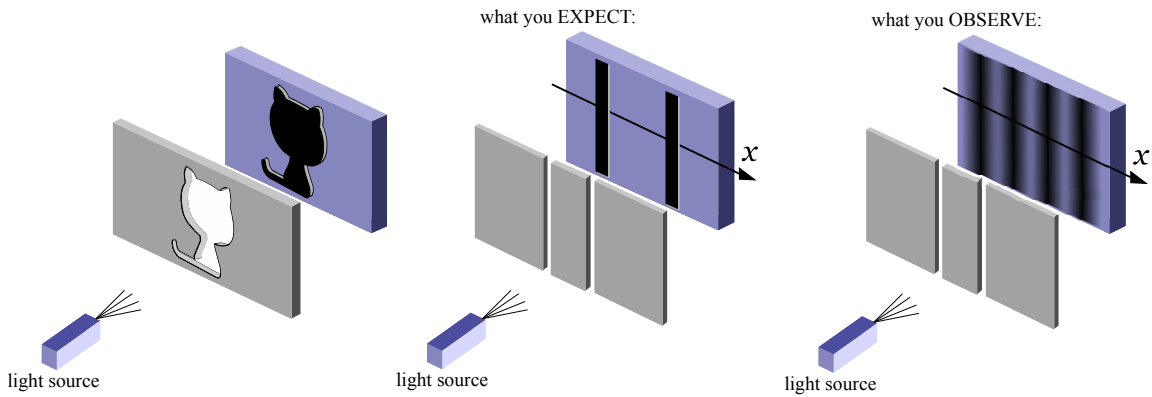}
  \caption{If you shine with a light source on a hole with the shape of a cat, light will impact along a cat-shaped region (left). Logically, you would guess that two very narrow slits lead to two vary narrow bands of light impacts (middle). Striking enough, what you observe is a whole series of bands (right). The conclusion is inevitable: light must be a wave.}  
  \label{cat}
  \end{center}
\end{figure}

He set up the following experiment: take a light source, a thin wall and a projection screen, as in Figure \ref{cat}. If you cut -for example- the shape of a cat away from the wall, and shine on it, you see a bright shape on the projection screen, in the shape of a cat. Now, what happens if instead two \textit{very} narrow slits are drilled, on a  small distance of each other? Naively, you expect two very narrow bright stripes, on a small distance of each other. Well, strange enough, that's not what you see. Instead, you see an alternating series of bright and dark bands. Those bands are precisely what you see when waves create an interference pattern. So Young could do nothing but conclude that indeed, light \textit{must} be a wave, even though he didn't know what kind of wave. Nice, huh? 

But what about that cat shaped hole then? Well, for larger holes, the interference pattern becomes much more complicated. It is still there, really, but you barely see it. Moreover, it turns out that for larger holes, the spot on the screen gets the same form as the hole. From this, it \textit{looks} like light just consists of `bullets' passing through, even though that's not really true. Got it? 

\subsubsection{The surprise}
Anyhow, nowadays we understand Young's experiment very well, since now we know precisely what kind of wave light is: an electromagnetic wave. But here comes the surprise. Let's do the two-slit experiment again, but now for electrons. More precise, tell your engineering friend to build a nice electron-source: an apparatus shooting electrons. He'll readily do that for you. Of course, as we can't see flying electrons with our eyes, we have to replace the projection screen by a set of detectors, which can measure the electrons arriving. So the setup is exactly that of Figure \ref{cat}, but with an electron source and a detecting screen instead of a light source and a projection screen. When sending electrons through the cat-shaped hole, the detectors will tell you that a cat-shaped region is under fire. Ah, ok, that's what you expect. Now do as Young did for light: cut out two narrow slits. What do the detectors tell you now? Let's think about it. Electrons really are particles (so imagine flying `bullets') so two slits should result in just two bands where electrons are detected. However, that is not what happens. \textit{Precisely like for light,} you see an alternating series of bands. On one band, there is much detection, then on the next there is few to none, and so on. Even if lower the intensity of the source (so that there is more time between the electrons, and you are sure they fly at the slits \textit{one by one}, not feeling the previous or next one) the result is the same. Bands, not just two stripes. 
\subsubsection{Interpretation}
There is only one conclusion we can draw. \textbf{Each electron has to be a wave.} Well, more precise, it is just a small piece of wave: a wave\textit{packet}. To a brutal and large observer like us, that small packet looks like a point particle in most experiments. However, it does have a non-zero size. It is a very small blob. And when each such wave packet goes through the two slits, it interferes with itself. (The part of the wave packet that went through one slit interferes with the part that went through the other slit.) So the wavelike nature of the electron is only made visible because of the specific setup of the two-slit experiment. In most other experiments, we don't notice, and the electron just looks like a point, even if, in fact, it is spread out over a small region. Now, if you repeat the experiment for neutrons and protons, you find precisely the same result. So they are all wave packets. People like to call this \textbf{the wave aspect of particles}.

\section{Another experiment: the photoelectric effect}
\begin{figure}[H]
 \begin{center}
  \includegraphics[width=0.35\textwidth]{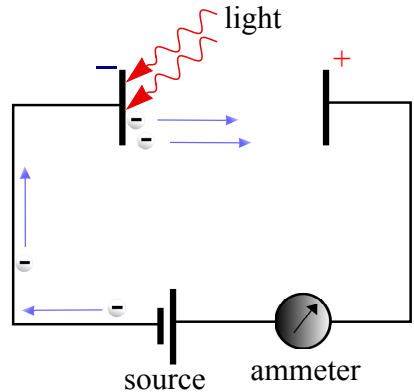}
  \caption{A setup showing the photoelectric effect. A source is connected to two pieces of metal. They get charged (electrons are displaced) but no net current flows through the circuit. When light shines on the negatively charged metal piece, things change. At least: if the light has a high enough frequency, it strips electrons out of the metal, so they can cross the gap, and reach the other side. This will cause a current to flow, as detected by the ammeter. On the other hand, if the frequency is too low, nothing happens, not even when one drastically increases the intensity of the light - the current stays exactly zero.}
 \label{PE}
 \end{center}
\end{figure}

Initially, the two-slit experiment completely stunned physicists. But the origin of QM lies at \textit{several} experiments and puzzles that (roughly) arose around the beginning of the 20th century. One of them was the \textbf{photoelectric effect}.  To see this effect, do the following. Take two pieces of metal. Take a direct current source, and connect it two the pieces of metal, as shown in Figure \ref{PE}. Since the metal pieces are not connected, no current can flow through the circuit. 
That is indeed what you see experimentally (your ammeter will indicate the current is zero). Well, wait: now do the following. Take a light source, of which you can adjust the color (the frequency). Since you can set it to shine at only one color, you call such a source \textbf{monochromatic}, which is Greek for one-color. Start with very red (low frequency) light, and shine on the negatively charged piece of metal (see figure). You still don't see anything special. Now if you increase the frequency of the light (go from red to yellow to green, towards blue light) there is suddenly some frequency/color at which you detect a current running. What? The only explanation is that the light strips off electrons from the metal; those fly through the air towards the (attracting) positive pole, making you detect a current. 
Now scientists partly understood this experiment. They knew light (or EM radiation in general) is a form of energy, so it is not so strange that it can strip off electrons from a metal. The strange thing, however, is that it only starts doing so \textit{beyond a certain frequency} (which depends only on the type of metal you are using). If the frequency is too low, there is zero current, \textit{regardless} of the intensity! So if you are below that critical frequency, no matter how much light you shine on the metal, you don't detect a current. Once above the critical frequency (which depends on the metal), the current is proportional to the applied intensity, so the amount of electrons ripped off grows with the amount of radiation you shine on the metal. 
So what is going on here?

\subsection*{Understanding the photoelectric effect}
This peculiar behavior remained unexplained for quite some while. Then, Albert Einstein proposed a simple explanation. Imagine that a ray of light consists of very many separate pieces. Let's call these fundamental bits of light \textbf{photons}. If you shine on the metal, all these photons bump into the atomic grid of the metal, one by one. Now on top of that, assume each photon has an energy proportional to its frequency. So 
\begin{equation}
E= h \nu
\end{equation}  
where E is the energy of the photon, and $\nu$ is its frequency (color). The factor of proportionality between the two, $h$ is called \textbf{Planck's constant}. Ok, suppose this is true. Einstein then reasoned further as follows. Say there is some energy $\Phi$ needed to strip off one electron. Then if the individual photons have an energy $E$ that is smaller than $\Phi$, a single photon cannot strip off one electron. And as they arrive one by one (however fast that may be) each of them has insufficient energy to kick an electron out. So nothing happens. Only if the individual photons have an energy $E$ \textit{greater} than $\Phi$, each of them can strip off one electron. In this case, the result is obvious: the more photons you send, the more electrons are released by the metal. Said differently: a higher intensity of light means more photons per second, so more electrons released, and a bigger current detected. 
A popular comparison is the `ball and the fence'-story. Imagine you want to kick a ball over a fence. If you kick too soft, it will not go over. Even if you repeat this, the result will be rather poor: the ball will just bounce back over and over again. However, if you kick hard enough, it will fly over the fence at once. 

As you see: the explanation of the photoelectric effect is not particularly difficult. To understand it, you don't need to be an Einstein at all. Note that the above provides more than just an explanation. It also gives a prediction. Since an electron needs an energy $\Phi$ to be kicked out (this energy is actually called the \textbf{work function}), and since the photons gives an energy $h\nu$, the kinetic energy $K$ of the electron after being kicked loose is given by:
\be
K = h\nu - \Phi
\ee
So the explanation of Einstein predicts a simple linear relation between the kinetic energy of the released electrons and the frequency of the light falling on the metal. With a slightly more careful setup one can measure the kinetic energy of the released electrons (by letting them cross a potential difference). And indeed, their kinetic energy precisely obeys the above relation. This result (and many other experiments) confirm that the photon-explanation has to be correct.

\subsubsection{Interpretation}

So what does the photoelectric effect tell us? Light must be made up of many individual particles, each with its own energy. They are really separate objects. How can we reconcile this with the knowledge that light is a wave? Each photon has to be a little piece of the electromagnetic wave. If you put many of these `elementary pieces of wave' together, it just looks like one big wave. So the particle-like behavior of light is only made visible because of the specific setup of the photoelectric effect. People like to call this \textbf{the particle aspect of light}.

\subsection*{Putting the pieces together}
The experiments we have just seen force us to think very differently about particles and radiation. Before the advent of quantum mechanics, people thought there was a very clear difference between radiation and particles. They thought particles didn't have any size and should be described by their position as a function of time. They thought radiation is a disturbance of the background, and should to be described by an amplitude (of the electric and magnetic fields) as a function of place and time. Quantum mechanics learns us that this is not correct. Instead, everything we see around us is built of wave \textit{packets}. If you see one such packet, as it tends to be very small, it may look like just a point of zero size. For radiation, if you see many of these wave packets at once, it looks like just one big wave, without obvious constituents - even if they really are still there. The idea that everything is built of wavepackets is really the core idea of quantum mechanics. We can actually summarize this entire chapter by one single drawing:
\\

\begin{figure}[H]
 \begin{center}
  \includegraphics[width=0.7\textwidth]{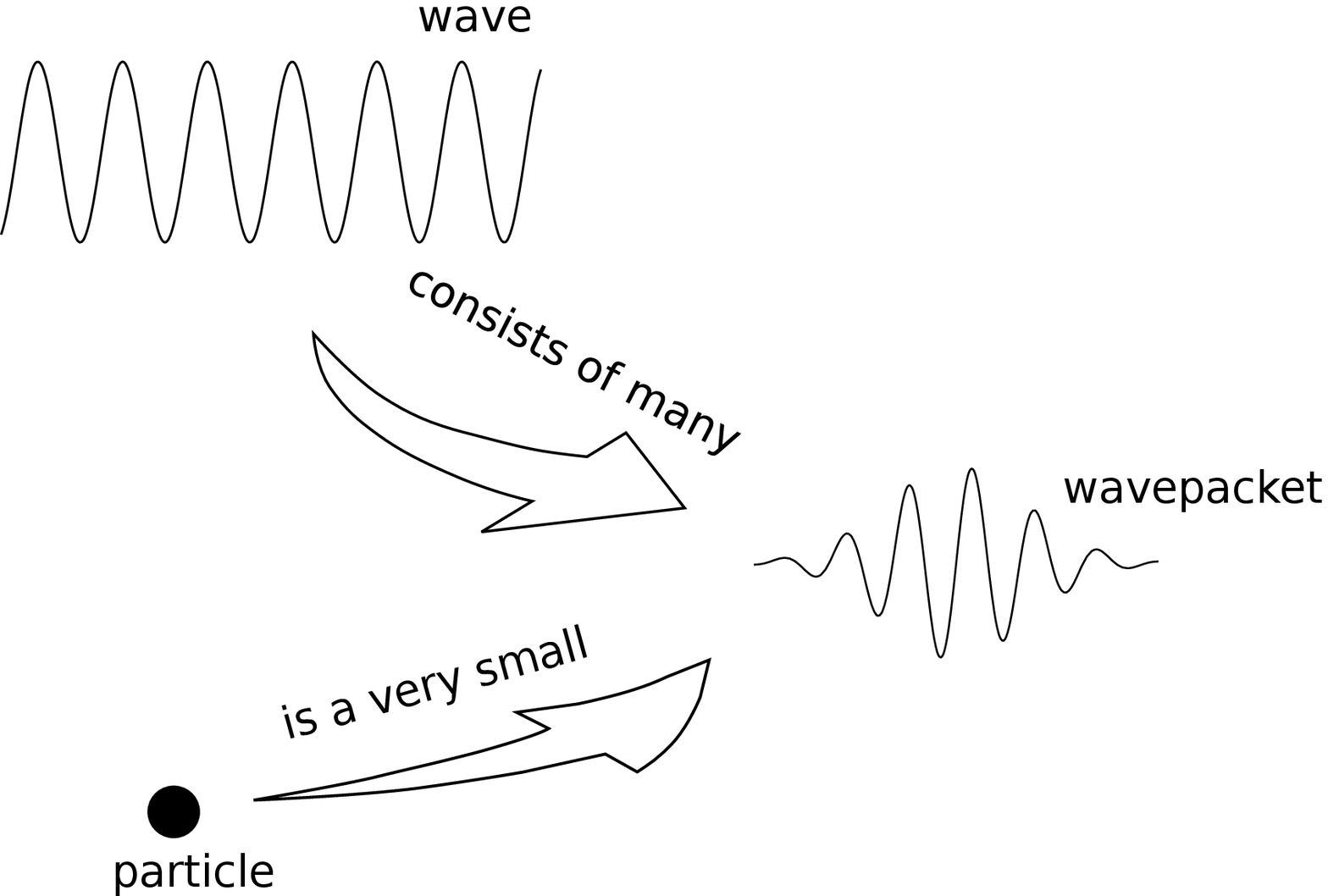}
   \end{center}
\end{figure}
This picture will be made more quantitative in the next section. 

\section{The de Broglie relations}

From the photoelectric effect, we know that photons have an energy proportional to the frequency. This is a quantitative expression of the particle aspect of light. There is also a formula expressing the wave aspect of particles. Here is a way to obtain it experimentally. You know (or have read in Appendix 5) that a two-slit interference experiment yields bright bands (a lot of detection) at positions $x$ on the screen satisfying
\be
x = \frac{n \lambda L}{d}
\ee
So from the a double-slit experiment, you can deduce the wavelength of the waves from the positions $x$ of constructive interference. (Given that you also know $L$ and $d$.) So if you do the double-slit experiment with particles, you can \textit{measure} the wavelength of these tiny wave packets. It turns out that the following holds:
\be
\lambda = \frac{h}{p}
\ee 
where $p$ is the momentum of the particles falling through the slits (their mass times their speed) and $h$ is again the Planck constant. In conclusion: the photoelectric effect and the double slit experiment show that
\be
\lambda = \frac{h}{p} \quad \textrm{and} \quad f = \frac{E}{h}
\ee
These equations are called the \textbf{de Broglie relations}.\footnote{The pronunciation of `de Broglie' is somewhere in between `duh Broy' and `duh Brey'.}
The first relation gives the wavelength associated to the wave packet of a particle. The second relation gives the energy of a photon in relation to its frequency. But actually, \textit{both} relations hold for photons and for particles. So you can read the de Broglie relations in two ways. First, for a particle with energy $E$ and momentum $p$, they give the wavelength and the frequency of the wave packet of the particle. Second, for a photon with frequency $f$ and wavelength $\lambda$, they give the energy and momentum of the photon. So a photon does not only have a certain amount of energy, it also has a tiny bit of momentum. Again, this puts particles and waves on a very symmetric footing.

We can also rewrite the above two equations a bit. Recall that the wavelength of a piece of wave is related to the wavenumber $k$ by the relation $\lambda = \frac{2 \pi}{k}$. Also, recall that the \textbf{angular frequency} $\omega$ of a wave is defined as $\omega = 2 \pi f$. Then the above two equations can be written as
\be
p = \hbar k \quad \textrm{and} \quad E = \hbar \omega
\ee
with $\hbar = \frac{h}{2\pi}$ is called the \textbf{reduced Planck constant}. The experimental values of $h$ and $\hbar$ are:\footnote{To see the conversion between units, recall that an electron volt (eV) is the energy you get from letting an electron cross a potential difference of one volt, so 
$1 eV = q_e (1V) = (1,6 \cdot 10^{-19} C) (1 V) = 1,6 \cdot 10^{-19} J $}
\bea
h\,\, =& 6.626\cdot 10^{-34}	\textrm{J$\cdot$s} = & 4.136\cdot10^{-15} \textrm{eV$\cdot$s}\\
\hbar = \frac{h}{2 \pi} =& 1.055\cdot10^{-34} \textrm{J$\cdot$s}  =& 6.582 \cdot 10^{-16} \textrm{eV$\cdot$s}
\eea
This constant is not just small, it's \textit{tiny} - that's one of the reasons why the quantum-behavior of particles is only manifest in very precise and specific experiments.
\newpage

\section{Conclusion}
The main conclusion of the chapter is the fact that the fundamental building block of nature is a wave\textit{packet} (a lump of wave). This is quite nice, since it rids us of the classical idea that nature consists of two separate building blocks (particles and waves). Whether we are talking about a photon, an electron, a proton, ... all these are a wave packet. Since and during the discovery of quantum mechanics, this fact has been proven again and again in experiments. Most textbooks mention not only the two experiments above, but several others. Many of those were equally important in the development of quantum mechanics. We have omitted those here, as they all lead to the same conclusion. If you are interested, go ahead and look them up. If you got the point though, maybe you don't really need to.

\section*{Excercises}
\begin{enumerate}
\item Explain to yourself (or better: to someone else) what the conclusion of this chapter was, and how it logically follows from the discussed experiments.
\item What is the momentum carried by a red photon? (Red light has a frequency $f= 450 \cdot 10^{12}$ Hz.) How many such photons do you need to obtain a momentum of $1$kg$\cdot$m/s (=a decently served tennis ball)?
\item For every wave, $\lambda f = v$ with $\lambda$ the wavelength, $f$ the frequency and $v$ the wave speed. Using that the speed of light is $c=3\cdot 10^8$ m/s, what is the wavelength of a red photon?
\item You are doing the double-slit experiment with electrons. The distance between the slits is 1 cm, and the impact screen is at 1 meter from the slits. What speed should the electrons have in order to obtain an interference pattern with the bright bands separated by $0.1$ m? (The mass of an electron is $m_e= 9.11\cdot 10^{-31}$ kg.) 
\item The work function $\Phi$ of silver is about 4.6 eV. What frequency do photons need to have in order to strip off electrons from a chunk of silver? 
Explain what happens at lower frequencies. What happens at higher frequencies? What is the kinetic energy of an electron stripped off by a photon of frequency $2 \cdot 10^{15}$ Hz?
\end{enumerate}

\newpage

\section*{Appendix 1: Complex numbers recap}
As you probably know, a complex number is a number of the form $a+bi$ where $a$ and $b$ are real numbers, and $i$ is the imaginary unit (sometimes denoted $I$ or $j$) which satisfies $i^2=-1$. In such a decomposition, $a$ is called the \textbf{real part} of $z$, and $b$ is the \textbf{imaginary part}:
\be
a = \textrm{Re}\, z \quad b = \textrm{Im} \, z 
\ee
The product and sum of two complex numbers $a+bi$ and $c+di$ are given by: 
\bea
(a+bi)+(c+di) &=& (a+b)+(c+d) i\\
(a+bi) \,\,.\,\,(c+di ) &=& ac + ad i + bc i + bd i^2 = (ac-bd) + (ad+bc) i
\eea
The \textbf{complex conjugate} of a complex number $z=a+bi$ is given by $\bar{z} = a-bi$. With this, the real and imaginary part of a complex number can be written as
\be
\textrm{Re} z = \frac{z + \bar{z}}{2} \quad \textrm{Im}  z = \frac{z - \bar{z}}{2 i}
\ee
The \textbf{norm} of a complex number is defined by $|z| = \sqrt{z\overline z} = \sqrt{a^2 + b^2}$. 
Also, each complex number can be written in the form $z = |z| e^{i\theta}$ with
\be
e^{i\theta} = \cos \theta + i \sin \theta
\ee
\begin{wrapfigure}{r}{0.30\textwidth}
  \begin{center}
    \includegraphics[width=0.30\textwidth]{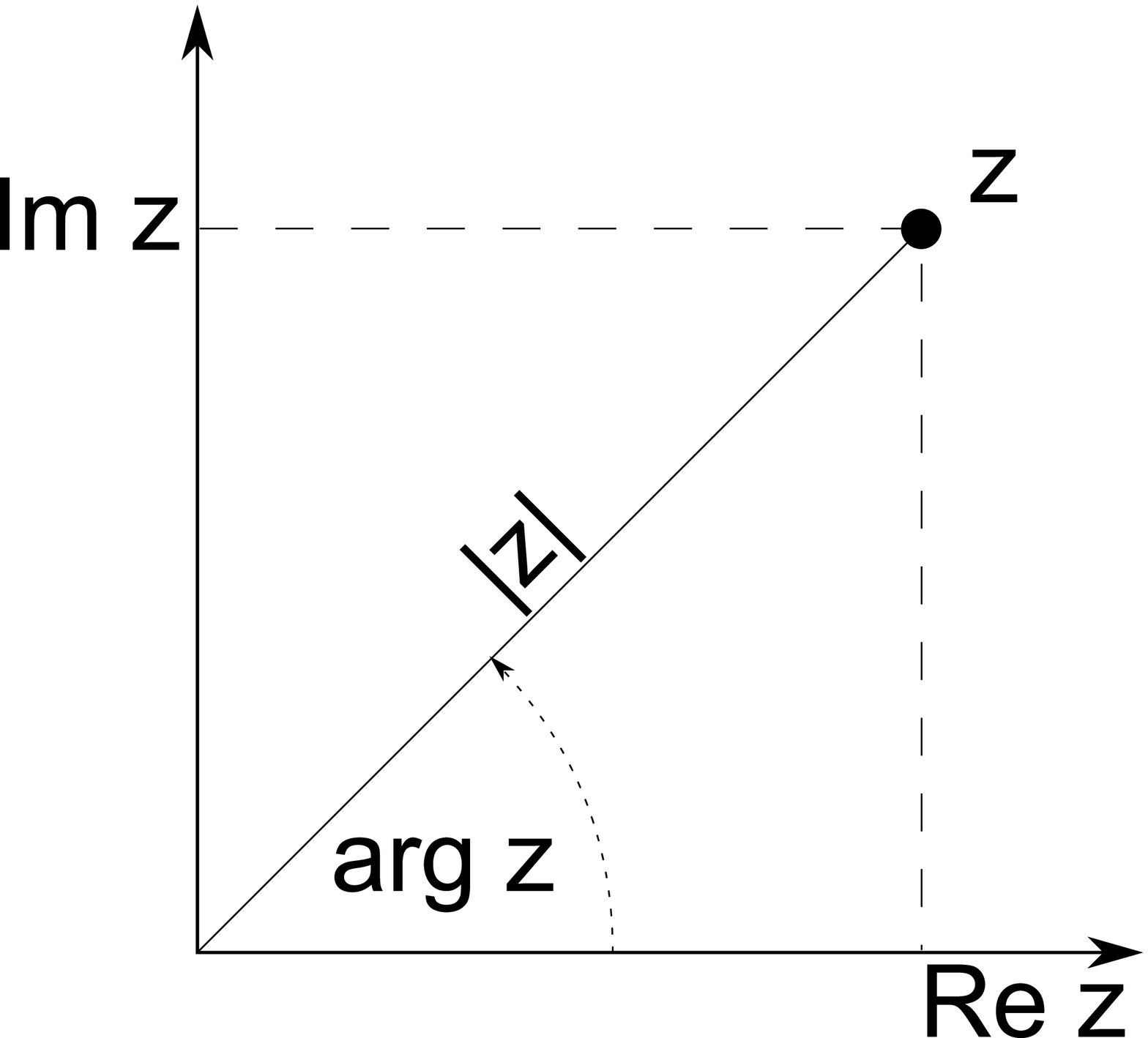}
    \label{complex_number}
  \end{center}
\end{wrapfigure}
The angle $\theta$ is called the \textbf{argument} of $z$.
\be
\theta = \textrm{arg} \, z
\ee
All these quantities can be shown very graphically, as shown in the figure on the side. 

For computations, there are some important properties that will show up again and again throughout this course. First, the inverse $z^{-1}$ of a complex number (defined by requiring $z^{-1} z =1$) is given by
\be
z^{-1} = \frac{\bar{z}}{|z|}
\label{eq:inv}
\ee
Also, the conjugate of the sum and the product of two complex numbers $z_1$ and $z_2$ are given by:
\be
\overline{z_1 + z_2} = \bar{z}_1 + \bar{z}_2 \quad \overline{z_1 z_2} = \bar{z}_1 \bar{z}_2.
\label{eq:property1}
\ee
Similarly, the norm of the product of two complex numbers is given by
\be
|z_1  z_2| = |z_1| |z_2|.
\label{eq:property2}
\ee
However, for the norm of the sum, one needs to be careful, since $|z_1 + z_2| \neq |z_1|+ |z_2|$. If you have forgotten some of these things, you could do the following short exercises: 
\begin{itemize}
\item Show that the norm of $ z = |z| e^{i\theta}$ really is $|z|$. (Write out real and imaginary parts explicitly, sum the squares.)
\item Check that the expression for the inverse $z^{-1}$ of a complex number indeed implies $z z^{-1}=1$.
\item Show the properties (\ref{eq:property1}) and (\ref{eq:property2}) listed above. 
\item Use these properties and (\ref{eq:inv}) to show that $(z_1 z_2)^{-1} = z_1^{-1} z_2^{-1}$.
\item Find an example of two complex numbers, such that $|z_1 + z_2| \neq |z_1|+ |z_2|$. (You can even find examples in the real numbers, which are just a subset of the complex numbers.)
\end{itemize}

\section*{Appendix 2: Moving waves}
Waves are all around us. Waves on water, sound, light, ... . In general , one can define a wave as a \textbf{disturbance, throughout some medium}. What this disturbance is, can differ significantly: for water waves, it is a disturbance in the height of the surface, for sound it is a disturbance in the pressure of the air, for light (see Appendix 4) it is a disturbance in the electric and magnetic fields. If we denote the quantity that is disturbed by $y$ then a wave is described by a function $y(x,t)$. Indeed, the disturbance typically depends on the position $x$, and changes throughout the time $t$. \footnote{Possibly, the position is described by more than just one coordinate. You need two coordinates for a wave on the 2 dimensional surface of water, 3 for a sound wave moving through a 3 dimensional room, etcetera. For simplicity, we consider only one coordinate $x$ here. To avoid confusion, we stress again that $y$ denotes the quantity being disturbed, it is not a spatial coordinate like $x$.} A nice feature is that for very, \textit{very} many systems, `small' waves (small disturbances) have the shape of a sine-function.\footnote{Or a cosine function, which is of course just the same shape, but shifted.} Superposing such sine-waves (=taking the sum, the total disturbance) then gives a large class of possible disturbances traveling through space. So our central question here is: \textbf{How can we describe such a sine-shaped wave?} Well, such a wave can always be put into the form
\be
y(x,t) = A \sin (k x - \omega t +\phi )
\ee
Let us dissect this expression. The right hand side indeed depends on the position $x$ and time $t$. The other variables $A$, $k$, $\omega$, $\phi$ are constants. Since $A$ multiplies the sine-function, it says how \textit{large} the disturbance is, so $A$ is called the \textbf{amplitude} of the wave. To see the meaning of the other constants, first note that at time $t=0$, the wave looks like
\be
y(x,t=0) = A \sin (kx+\phi)
\label{eq:tis0}
\ee 
This is a sine-function indeed. The distance between two peaks is called the \textbf{wavelength} $\lambda$, and is given by
\be
\lambda = \frac{2\pi}{k}.
\ee
To check this, just note that if there is a peak at $x_1$, there will also be a peak at $x_1 + \lambda$, since 
\be
y(x_1+\lambda, 0)=  A \sin (k x_1 + 2 \pi + \phi ) = A \sin (k x_1 +\phi) = y(x_1, 0).
\ee
The object $k$ itself is called the \textbf{wave number}. What about $\phi$? This constant determines how far the sine-function (\ref{eq:tis0}) is shifted to the left at $t=0$. It is called the \textbf{phase} of the wave. Now let us take a different point of view. What if we stay put at $x=0$, and watch how the disturbance at this specific position changes throughout time. Clearly, it is given by the function
\be
y(x=0,t) = A \sin (-\omega t +\phi)
\ee 
This looks very similar: again it is a harmonically oscillating behavior. Also, the disturbance is periodic: it repeats itself throughout time. Indeed, the disturbance at $t$ and $t+\frac{2\pi}{\omega}$ are the same:
\be
y(0,t+2\pi/\omega)= A \sin (-\omega t + 2\pi+\phi) = A \sin (-\omega t +\phi) = y(0,t).
\ee
The time $T=\frac{2\pi}{\omega}$ is called the \textbf{period} of the wave. The number $\omega$ is called the (angular) frequency. Here is what happens when we look at the entire wave, evolving through time. It starts of like the wave (\ref{eq:tis0}) at $t=0$, and then it runs along the x-axis (maintaining its shape) with a constant speed $v$. To see what the speed is, you can use the following trick. To follow the wave on its movement, you should sit at one point (a peak for instance) and ride along. To ride along means you always see the same disturbance along your way. So you want to see $y(x,t)$ to be constant, which is realized by $kx - \omega t = \textrm{constant}$. Reworking this expression, you get 
\be
x = \frac{\textrm{constant} + \omega t}{k } = \frac{\textrm{constant}}{k} + \frac{\omega}{k}t
\ee
This means you have to move at speed $\frac{\omega}{k}$. So the \textbf{wave speed} is given by $v = \frac{ \omega }{k}$.

\begin{figure}
 \begin{center}
  \includegraphics[width=0.8\textwidth]{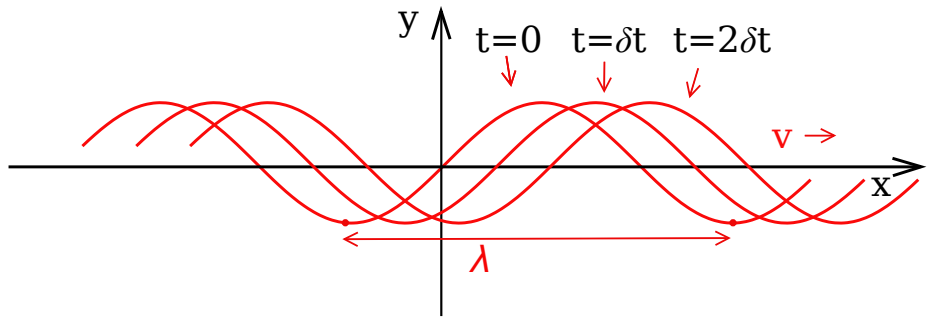}
  \caption{A moving wave. The displacement $y(x,t)$ is shown for times $t=0$, $t=\delta t$ and $t = 2 \delta t$. Clearly, the wave is moving to the right with increasing time. The wave speed is given by $v = \frac{\omega}{k}$}
  \label{lopende_golf_c}
  \end{center}
\end{figure}

\subsection*{Complex waves}
In quite some fields of physics, it turns out that the propagating waves in a system are most naturally described by a \textbf{complex wave}. This does not mean something wacky is going on, it just means this gives a convenient description, with simple equations etcetera. For example, in electrical circuits it can be handy to describe an alternating current by a complex wave. Maybe you have once seen this trick, and realized that such a description saves a lot of ugly trigoniometric manipulations that would be needed when working with ordinary (real) functions. However, if you really wanted to, you could do so, meaning that nothing strange is going on. It is just a matter of choice. Real functions are easier to understand, but complex functions save a lot of work. Anyway, how would you describe such a complex wave? Well, very similar to the above: by a function
\be
y(x,t) = A e^{i(k x - \omega t +\phi )}
\ee

\begin{figure}[h]
 \begin{center}
  \includegraphics[width=0.6\textwidth]{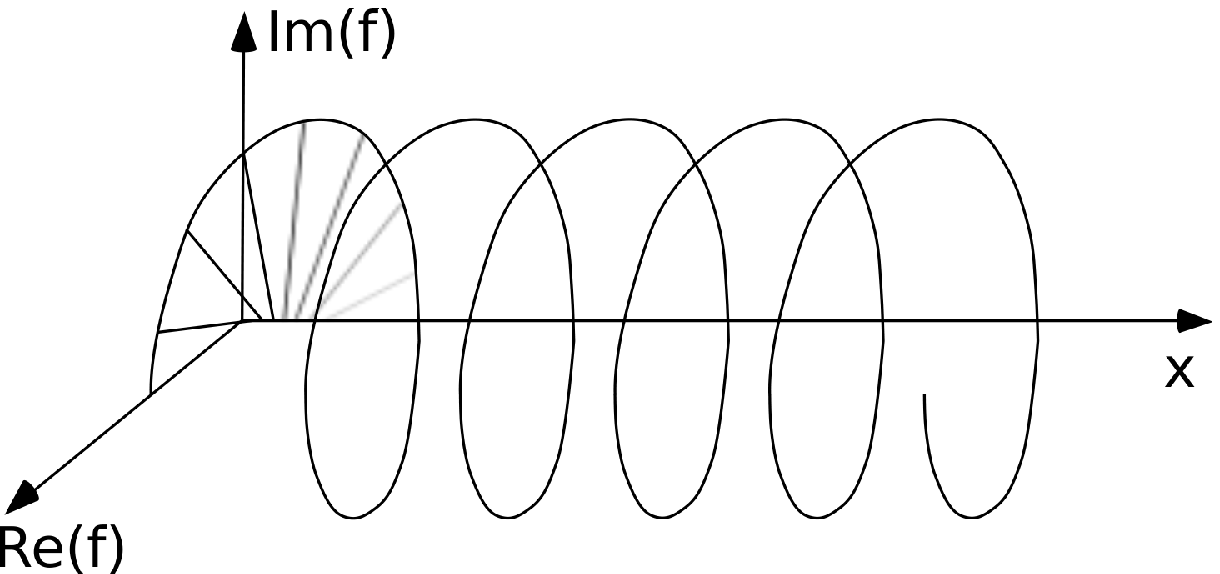}
  \caption{A graphical representation of a complex wave $z=e^{ix}$. The real and imaginary parts of $z$ are shown separately. The total figure has a helix shape.}
  \label{helix}
  \end{center}
\end{figure}
Actually, if we allow $A$ to be a complex number, we can absorb the phase $e^{i\phi}$ in it. Such a wave can be visualized as in Figure \ref{helix}. Note that the real and imaginary parts are just ordinary waves of the type we just discussed. Again, $k$ is the wavenumber, $\omega$ the (angular) frequency. Can we also describe a complex wave in three dimensions? Sure, this would be given by
\be
Y(x,y,z,t) = A e^{i(k_x x + k_y y + k_z z - \omega t +\phi )} = A e^{i(\vec{k}\vec{x} - \omega t +\phi )} 
\ee
(We capitalized the name of the wave to avoid confusion.) In this case, the vector $\vec{k}=(k_x,k_y,k_z)$ is called the \textbf{wave vector}. It can be shown that such a wave travels in the direction of $\vec{k}$, with speed $v = \frac{|\vec{k}|}{\omega} = \frac{\sqrt{k_x^2+k_y^2 + k_z^2}}{\omega}$.

To conclude, a remark about units. It is often stated that functions like the sine and exponential function can only have dimensionless arguments. Since $x$ and $t$ have units of length and time (meters and seconds), $k$ and $\omega$ necessarily have units $m^{-1}$ and $s^{-1}$. With this, you can verify that the period $T$ and wavelength $\lambda$ of a wave (with the above expressions) then have units time and length, as you would expect.

\section*{Appendix 3: Maxwell's equations}

You have probably seen the laws of electromagnetism several times throughout your life. They just happen to be very important to a lot of interesting physical phenomena. Quite some physics books phrase these laws in terms of physical situations. Like: `the magnetic field around a conductor carrying a current, is given by ...' or `a changing magnetic flux will create an electric field inside the conductor, given by...'. These phrases are very important, but are not the most elegant description, since they require a lot of words and explanation. A quite nice fact is that \textit{all} these laws can be summarized in a very compact way: only four equations. They are called Maxwell's equations, although they are only a different version of the laws discovered by others before him. They look as follows:
\begin{itemize}
\item $\vec{\nabla} \cdot \vec{E} = \frac {\rho} {\varepsilon_0}$ (Gauss's law)
\item $\vec{\nabla} \cdot \vec{B} = 0$ (Gauss's law for magnetism) 
\item $\vec{\nabla} \times \vec{E} = -\frac{\partial \vec{B}} {\partial t}$ (Faraday's law)
\item $\vec{\nabla} \times \vec{B} = \mu_0\vec{J} + \mu_0 \varepsilon_0 \frac{\partial \vec{E}} {\partial t} $ (Amp\`{e}re's law)
\end{itemize}
Here, $\vec{E}$ and $\vec{B}$ are the electric and magnetic fields, $\rho$ is the charge density, and $\vec{J}$ the electrical current. The constants $\varepsilon_0$ and $\mu_0$ are the electrical permittivity, and the magnetic permeability. The object $\vec{\nabla}$ is a vector of derivatives: $\vec{\nabla} = (\frac{d}{dx}, \frac{d}{dy},\frac{d}{dz} )$. So for example $\vec{\nabla}\vec{E} = \frac{d E_x}{dx} + \frac{d E_y} {dy} + \frac{d E_z}{dz}$. So what do these equations say? First of all, note that the relevant variables ($\vec{E}$, $\vec{B}$, $\rho$, $\vec{J}$) depend on the position $\vec{x}$ and time. So they are \textit{local} equations: they say how all involved quantities are related at \textit{every} position and time. Also, the equations involve derivatives of $\vec{E}$, $\vec{B}$: the equations of Maxwell are four \textbf{differential equations}. Sometimes, people say the above expressions are the `differential' form of the laws of electromagnetism. It might not be immediately clear how to extract real physics (say, the force between two conductors carrying a current) from them, but sure they are very compact and elegant! That's why people like them so much. A very important consequence that can be extracted from them, is described in the following appendix.

\section*{Appendix 4: Light is an electromagnetic wave}
A question Maxwell asked himself, was the following. What happens if there are no charges or currents present? Could something interesting happen then? In that case ($\rho=0$, $\vec{j}=0$) the four equations become:

\bea
 \nabla \cdot \vec{E}  \;&=&\; 0\\
 \nabla \times \vec{E} \;&=&\; -\frac{\partial \vec{B}} {\partial t}\label{eq:curl1}\\
 \nabla \cdot \vec{B}  \;&=&\; 0\\
 \nabla \times \vec{B} \;&=&\; \mu_0 \varepsilon_0 \frac{ \partial \vec{E}} {\partial t}\label{eq:curl2}
\eea
Taking the curl ($\nabla \times$) of the second and fourth equation and then using them again gives:
\bea
\nabla \times \nabla \times \vec{E} \;&=\; -\frac{\partial } {\partial t} \nabla \times \vec{B} = -\mu_0 \varepsilon_0 \frac{\partial^2 \vec{E} }  {\partial t^2}\\
\nabla \times \nabla \times \vec{B} \;&=\; \mu_0 \varepsilon_0 \frac{\partial } {\partial t} \nabla \times \vec{E} = -\mu_o \varepsilon_o \frac{\partial^2 \vec{B}}{\partial t^2}
\eea
After this, using the vector identity
\be
\nabla \times \left( \nabla \times \vec{V} \right) = \nabla \left( \nabla \cdot \vec{V} \right) - \nabla^2 \vec{V}
\ee
(valid for every vector function $\vec{V}(\vec{x})$) and the first and third equation ($\nabla \vec{E} = \nabla \vec{B} =0$) we get
\bea
{\partial^2 \vec{E} \over \partial t^2} - {c}^2 \cdot \nabla^2 \vec{E} \;&=\; 0\label{eq:elwav}\\
{\partial^2 \vec{B} \over \partial t^2} - {c}^2 \cdot \nabla^2 \vec{B} \;&=\; 0\label{eq:elwav2}
\eea
where $ c= { \frac{1}{ \sqrt{ \mu_0 \varepsilon_0 } }} $. 
You may recognize that these equations are standard \textbf{wave equations}. That means they have solutions with the shape of a sine-function, like $\sin(kx-\omega t)$. For example, if one takes the electric field to be
\be
\vec{E} = (0,E_0\sin(kx - \omega t),0 ).
\ee
Then the x- and z- component of (\ref{eq:elwav}) are trivially solved, and the y-component becomes:
\be
\omega^2 - c^2 k^2 = 0
\ee
which can be satisfied by taking $\omega = c k$. From (\ref{eq:curl1}) and (\ref{eq:curl2}) we see that we should take 
\be
\vec{B} = (0, 0, B_0 \sin(kx-\omega t))
\ee
\begin{figure}[H]
 \begin{center}
  \includegraphics[width=1.0\textwidth]{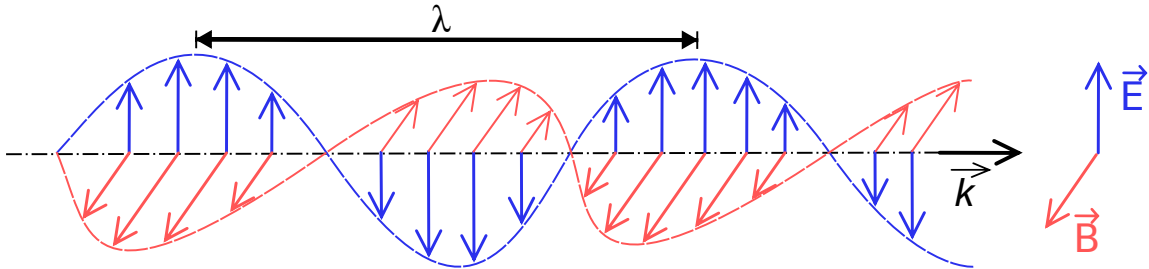}
  \caption{This figure shows an electromagnetic wave, with the electric and magnetic fields along the x-axis shown at a specific instant in time. If we let time run, the wave moves (`shifts') to the right.}
    \label{max}
  \end{center}
\end{figure}
with $B_0 =\frac{E_0}{c}$. (Just check by plugging in that this works.) Clearly, the above also satisfies the wave equation (\ref{eq:elwav}). So what does this look like? This solution to the Maxwell equations is drawn in Figure \ref{max}. It is wave of oscillating electric and magnetic fields, traveling through space in the x-direction. Note that $E$ and $B$ are perpendicular to the direction of motion. We can intuitively understand why such a wave is possible. From electromagnetism, we know that a changing magnetic field induces a changing electric field, and vice versa. So the above wave precisely embodies this principle - even without any charges or currents being present. So the equations of Maxwell imply the existence of electromagnetic waves. That's nice, but there is more. The speed of this wave is given by
\be
v = \frac{\omega}{ k} =c
\ee
and the numerical value of $c$ is given by plugging in the values of the permittivity and permeability:
\be
c= { \frac{1}{ \sqrt{ \mu_0 \varepsilon_0 } }} = 2,998 \times 10^8 \,\,\textrm{m/s}
\ee
That's precisely the numerical value of the speed of light! From this, Maxwell concluded that light has to be a form of electromagnetic radiation. A very neat and surprising conclusion, and - better even - very right. Today we know that light is only one form of electromagnetic radiation. Microwaves, radio waves, X-rays,... are all electromagnetic waves, with the single difference of comprising waves of different frequency regions. 

\section*{Appendix 5: What is interference?}
\begin{figure}[h]
 \begin{center}
  \includegraphics[width=1.0\textwidth]{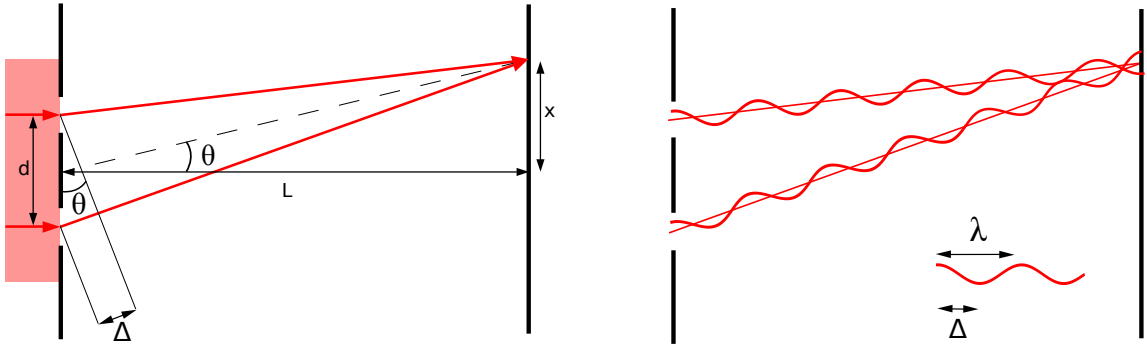}
  \caption{An interference experiment: two narrow slits/holes, through which waves fall. The distance between the two holes is $d$. The distance between the holes and the wall on the right is $L$. We are interested at the total (superposed) wave arriving at position $x$, corresponding to an angle $\theta$ (shown at two places). The path difference is clearly $\Delta = d \sin \theta$. On the right, the waves emerging from the two holes and traveling towards one specific $x$ are shown, at a particular instant in time. Since the waves emerging from the holes necessarily have the same phase. Indeed, here, they both show a crest at the left most position. However, at the height of the wall, the two waves are out of phase. The upper wave shows a trough, whereas the lower wave shows a crest. This means destructive interference occurs at this particular x. This is not so surprising, since the path difference is precisely half of the wavelength. (Shown in the right corner.)}
  \label{doble}
  \end{center}
\end{figure}

An important property of waves is that they can be put `on top of each other'. Such a sum of two waves is called a superposition. This can be seen in water waves for example. If a duck on water creates waves $y_1(x,t)$ and another duck makes waves $y_2(x,t)$ then the total disturbance pattern they create is given by $y_1(x,t) + y_2(x,t)$.  A very interesting thing that occurs when two disturbances are superposed is \textbf{interference}. It could be that at some position and time, $y_1$ and $y_2$ are exactly opposite. This means the total disturbance is zero, even if $y_1$ and $y_2$ individually would have been nonzero. Such a situation is called destructive interference. Similarly, if $y_1$ and $y_2$ have the same sign, the total wave will be larger than the individual disturbances. This is called constructive interference. You can easily reproduce these patterns yourself  -  by throwing two stones into an undisturbed pond for example. An experiment that makes constructive and destructive interference particularly clear is the \textbf{double-slit experiment}, shown in Figure \ref{doble}. In this case, incoming waves fall on two narrow slits/holes. At the other side of the slits, the waves come out. In that region, the waves emerging from the two slits interfere. A question worth asking is what the total wave pattern will look like on a wall at some distance of the slits. For ease, we suppose the waves are of a single wavelength $\lambda$. Also, suppose the wall is at a distance $L>>\lambda$ of the slits. Since the waves start out the same way at the slits, they must have the same phase there. This means that the disturbances at the two slits show maximum positive displacement (a \textbf{crest}) and maximal negative displacement (a \textbf{trough}) at the same moments in time. More precise, the wave amplitude $y(x,t)$ satisfies:
\be
y(\textrm{slit} \,1,t) = y(\textrm{slit}\, 2,t) \quad \forall t
\ee
Consider the pieces of wave traveling upwards, towards a position $x$ on the screen. Note that the distance the two pieces of wave have to travel is not the same. This is called the \textbf{path difference}. From \ref{doble} it is clear that the path difference $\Delta$ is given by 
\be
\Delta = d \sin \theta
\label{eq:pathdiff}
\ee
Here is the crux of the story: if the path difference is a multiple of $\lambda$, the two waves will have the same phase at $x$, so they will interfere constructively. If the path difference is $\lambda/2$ plus a multiple of $\lambda$, the wave will be exactly opposites when arriving at $x$. So in that case, destructive interference will occur. So
\bea
\Delta =& n \lambda \quad \textrm{(n integer)} \quad &\Rightarrow \textrm{constructive interference at $x$}\label{eq:constri}\\
\Delta =& \left(n+\frac{1}{2}\right) \lambda \quad \textrm{(n integer)} \quad &\Rightarrow \textrm{destructive interference at $x$}
\eea
So combining (\ref{eq:pathdiff}) with (\ref{eq:constri}), we see that constructive interference occurs if
\be
 d \sin \theta = n \lambda  
\ee
Note that the position $x$ on the wall is hidden in $\theta$. If $L$ is large, $\theta$ is small and $\sin \theta \approx \tan \theta = x/L$. In this approximation, the above condition can be rewritten as
\be
x = \frac{n \lambda L}{d} 
\ee
At points $x$ satisfying the above relation, constructive interference occurs, so the combined waves give rise to a wave with large amplitude. In between such points (at $x = \left(n+\frac{1}{2}\right) \lambda L/d$) destructive interference occurs - the waves of the lower and upper slit exactly cancel each other. This means there is no disturbance at all at these points. In between the two extremes, there is a smooth transition.  This pattern shows alternating bands (regions) of constructive and destructive interference, and is called the \textbf{interference pattern}.

%% file: H2.tex
\chapter{Schr\"{o}dinger's equation}
\subsection*{In this chapter...}
The introductory chapter gave a quick qualitative view on one of the central theses of quantum mechanics, namely that the fundamental building blocks of all matter and radiation are tiny wave packets. As a consequence, we know that we should actually describe particles as waves, not just a moving points. The question that now arises is obvious: how can do so? How do we describe such a wave? And what is the dynamics of such a wave; how does it evolve in time? These questions will be addressed in this chapter. We will move on from a \textit{qualitative} to a \textit{quantitative} description.

As mentioned in the foreword, all chapters will from now on have a fixed structure. The tool part of this chapter will be an introduction on the mathematics of operators. Then we tell the story of the Schr\"{o}dinger equation. Finally, in the computation part we will do our first real quantum mechanical description of a particle. Good luck!
\newpage

\section[Operators and functions]{\includegraphics[width=0.04\textwidth]{tool_c} Operators and functions}
In this section, we will learn about \textbf{operators}. It will be a very central tool for the following chapters. First we explain what an operator is, then we give some examples. We end with the notion of an eigenvector and eigenvalue of an operator. 

\subsection{An operator, what's that?}

About the most ubiquitous object in mathematics is the `function'. Take for example the real function $f(x)=x^2$. It maps each number to another number. More pictorially, we could say that such a function is a \textit{machine}: if you feed it a number $x$, it will return to you a different number, $x^2$. Now what is an \textbf{operator}? Its a machine too, but now the in- and output aren't numbers, but functions. Lets give a simple example. Given a nice smooth function $f$, we can compute its derivative $f'$. We can see the process of differentiation as a map:
\begin{equation}
\textrm{differentiation}: f\rightarrow f'
\end{equation}
We can do this for every nice smooth function, so `differentiation' is an operator: if you feed it a function, it will return a new function, $f'$. If we denote this differentiation operator with $D$, we have
\begin{equation}
D: f\rightarrow D(f)=f'
\end{equation}
So $D(x^2)=2x$, $D(e^x)=e^x$, etcetera. We can give another example of an operator: take a number $n$, and define the operator $X^n$ as follows
\begin{equation}
X^n:f \rightarrow X^n(f)= x^n \cdot f
\end{equation}
Since $X^n$ sends every function to a new function ($x^n$ times the original one) it's an operator indeed. Take for example $X^2$. We have $X^2(\sin x)=x^2 \sin x$, $X^2(e^x)=x^2 e^x$, etcetera.
\footnote{For notation naggers: there is actually some abuse of notation here: we use $x^n$ and $\sin x$ as shorthand for the \textit{functions} $x\rightarrow x^n$ and $x\rightarrow \sin x$. 
In principle we could write out things more carefully, but this would make all formulas look ugly.}
Now both the above examples are of a special type: they are \textbf{linear operators}. We call an operator $L$ linear if for all functions $f$ and $g$ and for all numbers $a$ and $b$
\begin{equation}
L(af+bg)= a L(f) + b L(g). 
\end{equation}
This indeed holds for the differentiation operator $D$, and also for the operators $X^n$. In fact, \textit{all} operators we will meet from here on will be linear operators. They are the ones that are relevant for quantum mechanics. So, we will drop the adjective `linear' from now on, and just talk about `operators'. Also, we will try to stick to the following notation: a capital letter is an operator, while a minuscule denotes a function. (Numbers will be denoted by either a capital or a minuscule.) Besides that, we will also be more careful with the differentiation operator, and denote with a subscript the variable with respect to which the derivative is taken:
\be
D_x :f\rightarrow\f{\partial f}{\partial x}
\ee
and analogous for any other variable.
\begin{figure}
 \begin{center}
  \includegraphics[width=0.7\textwidth]{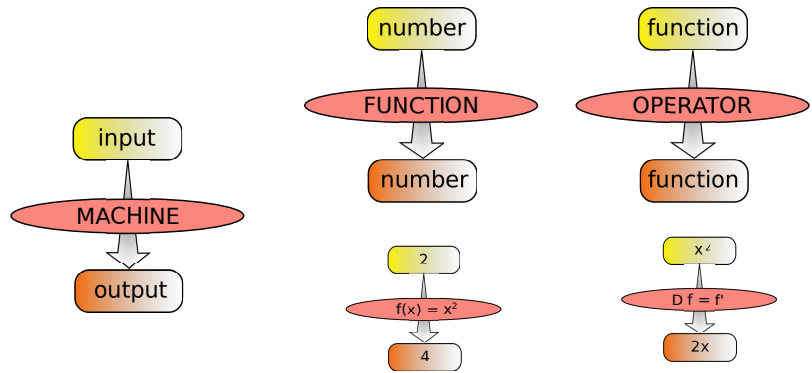}
  \caption{A pictorial way to think about functions and operators. Both are machines. A function has a number as in- and output. An operator takes a function as input and gives another function in return.}
  \end{center}
\end{figure}

\subsection{Eigenfunctions and eigenvalues.}

If you had any course on linear algebra, you are probably familiar with the notion of an eigenvector. Given a matrix $A$,  we say $v$ is an eigenvector with eigenvalue $\lambda$ if
\begin{equation}
 A v=\lambda v.
\end{equation}
(Being lazy, we didn't put an arrow on the vector $v$.) For example:
\begin{equation}
\left( \begin{array}{cc}
2 & 0  \\
-2 & 1 \end{array} \right) 
\left( \begin{array}{c}
1  \\
-2   \end{array} \right) 
=
2\left( \begin{array}{c}
1  \\
-2   \end{array} \right) 
\end{equation}
So the vector $\left( \begin{array}{c} 1  \\
-2   \end{array} \right)$ is an eigenvector of the matrix $\left( \begin{array}{cc}2 & 0\\
-2 & 1 \end{array} \right)$ with eigenvalue $2$. However simple the idea of eigenvector is, it is the basic concept for a lot of interesting math. Not surprising, it will turn out to be a very useful step to generalize it to the context of operators. There, the input and output are functions, not vectors. So it could be the case that for a certain function $f$ and an operator $A$:
\begin{equation}
 A(f)= \lambda f.
\end{equation}
with $\lambda$ a number. In that case, we say that the function $f$ is \textbf{eigenfunction} of the operator $A$, with eigenvalue $\lambda$. If $A$ is a \textit{linear} operator (and all operators we will meet from here on will be so) then every multiple of an eigenfunction is an eigenfunction again. Indeed, in the case of a linear operator $A$ and an eigenfunction $f$, we have 
\be
A(kf)=kA(f)=k \lambda f= \lambda (kf)
\ee
for every number $k$.
\begin{figure}[h]
 \begin{center}
  \includegraphics[width=0.6\textwidth]{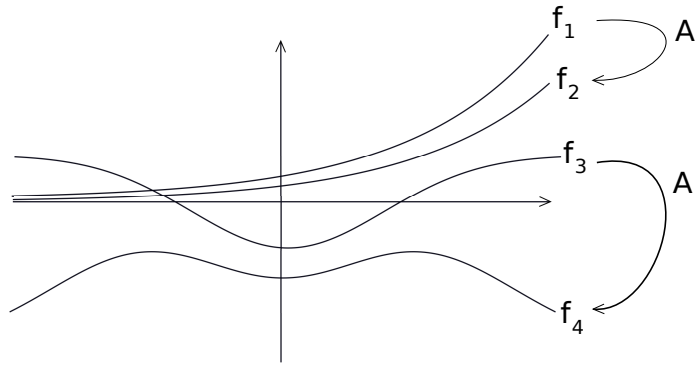}
  \caption{An example of two functions $f_1$ and $f_3$, acted upon by an operator $A$. Function $f_1$ is sent to a multiple of itself: it is an eigenfunction of the operator $A$. Function $f_3$ is sent to something completely different, and clearly is \textit{not} an eigenfunction of $A$.}
  \end{center}
\end{figure}
The parallel with ordinary eigenvectors in linear algebra is very strong. To make the resemblance even more explicit, we can rewrite the condition to be an eigenfunction in a slightly shorter way:
\begin{equation}
A f= \lambda f.
\end{equation}
Here it is understood that the operator A works on the object to its right - just like a matrix working on a vector. From now on we will always use this notation. So if we don't put brackets around its argument explicitly, it is understood that \textbf{an operator works on the object standing to its right}. So for example we may simply write $A B f$ instead of $A(B(f))$. At first sight, it may look like this could give rise to ambiguous expressions, but just like with matrix products, the notation works quite well. For example, given two operators $A$ and $B$ we can define a new operator, denoted by $A B$, which acts on any function with $B$ first, and then with $A$:
\begin{equation}
 A B: f \rightarrow (A B)(f)\equiv A(B(f))
\end{equation}
Let's do some random examples. The function $e^{7x}$ is an eigenfunction of $D_x$, with eigenvalue $7$, because
\begin{equation}
 D_x\,e^{7x}=7 e^{7x}.
\end{equation}
Another one: $\sin(x)$ is an eigenfunction of the operator $D_x D_x$, with eigenvalue $-1$,  because
\begin{equation}
 D_x D_x \sin(x)= D_x(D_x(\sin (x)))=D_x(\cos(x))=-\sin(x).
\end{equation}
In the first step we have added brackets, to make connect with the previous notation.
Another example. Define the unity operator `$1$' as the operator that sends each function to itself:
\begin{equation}
1: f\rightarrow f
\end{equation}
(In fact, this is just the operator $X^0$.) It is obvious that \textit{all} functions are eigenfunctions of the unity operator, with eigenvalue 1. Last example: take a function depending on some variables, but not on $x$. So, for example a function $g(y,z)$. If we take the derivative with respect to $x$, we get
\be
D_x g(y,z)=\frac{\partial g(y,z)}{\partial x} =0
\ee 
So any such function $g(y,z)$ is an eigenfunction of $D_x$, with eigenvalue $0$.

\subsubsection*{Sums of operators}
A last remark: we can also add operators, and multiply them by numbers, in a quite trivial way. The operator $A+B$ is defined as follows: it sends every function to $A f+B f$. In the same way, for any number $c$, the operator $cA$ is defined as follows: it sends every function $f$ to $c A f$ (=$c$ times the function $Af$). As an application of these rules, you can check that the operator $X^2 - 2 X^3$ acts as:
\be
X^2 - 2 X^3: f \rightarrow (x^2 - 2 x^3) \cdot f
\ee 
just like you would guess naively. In general, to any function $g(x)$ we can associate the operator $g(X)$, defined as follows:
\be
g(X): f \rightarrow g(x)\cdot f 
\ee
So for example $\exp(X) x= e^x x$. That's the math for this chapter, let's move on to the story part!

\section[Wavefunctions and Schr\"{o}dinger]{\includegraphics[width=0.04\textwidth]{once_c} Wavefunctions and Schr\"{o}dinger}
\subsection{The  wave function}
In the previous chapter, we saw how the two-slit experiment and the photoelectric effect force us to describe both radiation and matter as wave packets. Especially for particles this requires a new framework. From here on, our main goal will be to answer the question: \textbf{how can we describe the wave aspect of particles}? Here particle can mean: electron, proton, neutron, or any other of the (more rare) particles that make up our universe. Let's take for example one single particle on a specific moment time $t$. The pioneers of quantum mechanics did the following (very reasonable) step: let us describe this particle by a function:
\begin{equation}
\psi(x)
\end{equation}
The symbol $\psi$ is the Greek letter Psi, pronounced as `\textit{sigh}'. Like for a wave, it gives the \textit{disturbance} as a function of space. As we want to describe a wave \textit{packet}, we expect $\psi$ to slope of fast outside some region which is the location of the wave packet. Schematically, you might see something like this
\begin{figure}[H]
 \begin{center}
  \includegraphics[width=0.8\textwidth]{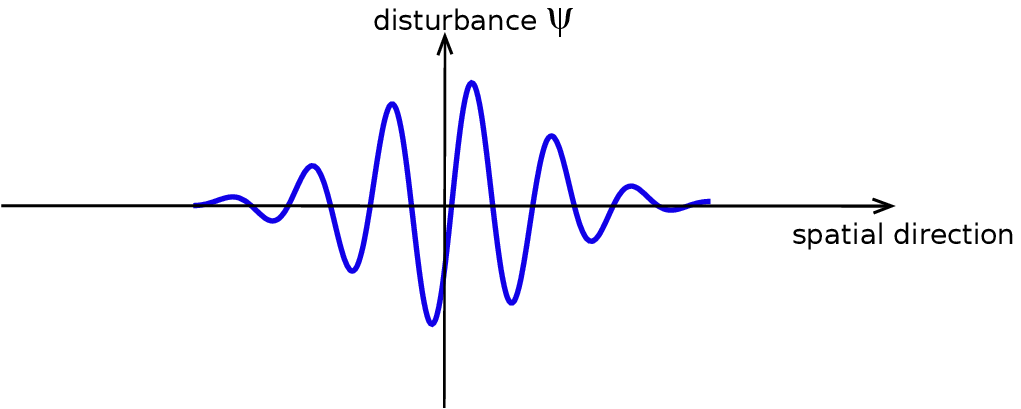}
  \end{center}
\end{figure} 
For simplicity we have drawn only one coordinate axis. In principle the disturbance $\psi$  will depend on the tree different coordinates, but for now we will describe disturbances depending on one single coordinate. So a wave packet is just a function $\psi(x)$. But then what \textit{is} that disturbance $\psi$? For radiation, there are two oscillating fields: E and B. For a sound wave, there is a space-(and time) dependent perturbation of the pressure and the density. Here however, we are dealing with a fundamentally new concept: for sure $\psi$ is not a type of disturbance that you have met before. So, since a better name is lacking, $\psi$ is simply called the \textbf{wave function} of the particle. After all, it is a \textit{function} describing the particle's \textit{wave} packet. A peculiar thing is that the wave function is taken to be a complex function. \\
\subsection{Why complex?}
Notice that in the examples just given (radiation and sound waves) there were several oscillating quantities. This is very general: most waves consist of \textit{two} types of disturbance, which are constantly being transformed into each other, at each point in space. For electromagnetism, a changing electric field creates the magnetic field, and the changing magnetic field creates the electric field again, and so on - and this happens at each point in space. For a wave of water, there is the movement (velocity) of water on one hand and the vertical displacement on the other hand. Again, these two types of disturbance are constantly transformed into each other. 
So many waves are described by two real functions (disturbances). This already suggests it is possible to group these together into one single complex function. 

There is another argument to do this `grouping'. As we have stressed, for each position in space, the wave exhibits a cyclic behavior. The specific stage of the cycle can be expressed by a \textit{phase} (an angle). On the other hand, a wave often has a notion of \textit{total amplitude} (a total of the oscillating quantities). As you know, a phase and an amplitude can be economically summarized in a complex number, via the \textit{argument} and the \textit{norm} of the complex number.  

So this suggests one can economically describe a wave by a single complex function. Indeed, in -for example- electromechanics alternating currents are often described by complex waves. This is not for fun, or to increase the complexity of the situation. There, such a description really is quite natural \textit{and} simplifies a lot of computations. Indeed: wave phenomena naturally involve trigoniometric functions ($\cos x$, $\sin x$,...) and instead of using tedious trigoniometric identities, one can see these functions as the real/imaginary part of a single imaginary function like $e^{i x}$. (See Appendix 2 of the previous chapter.) For water waves and pure electromagnetism, a complex description of waves is less usual. One reason is that there, the constituents of these waves have a very clear physical meaning (electric and magnetic fields, displacement of water, ...) and people like to hold on to those concepts. Of the many scientific fields in which people deal with waves, some use complex descriptions, some don't. It's a bit of a trade-off. A complex description can lead to easier expressions (sometimes, not always) but is of course slightly less intuitive. 

Back to quantum mechanics and the wave function of a single particle. There, the disturbance is not of a kind we know already. So there is no point at trying to give a description in terms of real fields - this would just make the mathematical expressions look more ugly. This suggests one should just go ahead and use a complex function to describe these wave packets. And indeed you will see this works pretty well. Just recall that there is nothing fancy about using complex functions: it does not means some intrinsic strangeness is involved. Complex functions are just the most natural way to deal with waves - that is why we like to use them.


\subsection{Time evolution}
We now know we should describe a particle by a complex function, the wave function. The next question is obvious: how does such a wave function \textit{evolve} in time? Otherwise stated, say you have a wave function at a given time $t_0$, call this $\psi(x,t_0)$,  how will it move/change its shape as time runs? So what is $\psi(x,t)$ for later times $t$? A solution to this problem would be to find a relation of the form
\be
\partial_t \psi(x,t) \quad \textrm{in terms of}\quad \psi(x,t)
\ee 
In other words, we want to find a differential equation relating the state $\psi(x,t)$ of the wave function to its change $\partial_t \psi(x,t)$ in time.

\subsubsection*{To get a hint}
To get a hint what such a relation might look like, lets first try to generalize an (infinite) real wave to the complex description we will use. An infinite wave with amplitude $A$, frequency $\omega$ and wave number $k$ is typically described by
\begin{equation}
A \sin(k x - \omega t).
\end{equation}
Indeed, this is an oscillation, moving to the right with speed $v=\omega/k$. Now the \textit{complex} function
\begin{equation}
A e^{i(k x - \omega t)}
\label{eq:complwav}
\end{equation}
is an oscillation as well. Take some time to try to visualize this wave. At each point in space the value of the function is a complex number, with modulus $A$, and with a phase that evolves with angular frequency $\omega$. So the above function describes a complex wave with angular frequency $\omega$, wavenumber $k$, and amplitude $A$. If you draw the real and imaginary parts separately, you get a helix shape, like in the following figure:
\begin{figure}[H]
 \centering
  \includegraphics[width=0.5\textwidth]{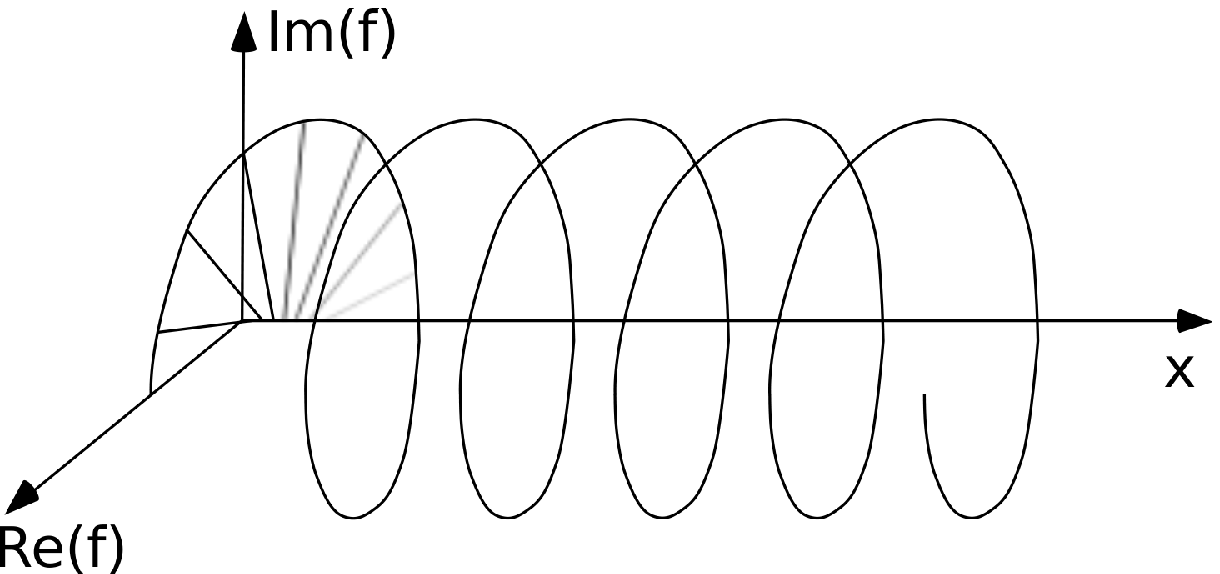}
  \caption{The complex wave $f(x) =e^{ix}$, showing its real and imaginary parts separately.}
\end{figure}
Throughout time, the wave (\ref{eq:complwav}) will move: the above helix will run to the left or to the right. Also, note that (\ref{eq:complwav}) is an \textit{eigenfunction} of the operators
\be
D_t\equiv \frac{\partial}{\partial t} \quad \textrm{with eigenvalue}\,\,\, -i \omega
\ee
and of 
\be
D_x \equiv \frac{\partial}{\partial x} \quad \textrm{with eigenvalue}\,\,\, i k.
\ee
Loosely speaking, the values of $\omega$ and $k$ can be obtained by acting with the time- and spatial derivative on the wave. Moreover, we know from the first chapter that the frequency and wave number of a particle are related to the energy and momentum,
\begin{equation}
E=\hbar \omega \quad p=\hbar k
\end{equation}
Combining with the above, we expect `some' relation between the quantities $E$ and $p$ of the particle and \textit{operators} acting on its wave function:
\be
D_t \leftrightarrow - i \omega = - i \frac{E}{\hbar} \quad \textrm{and}\quad D_x \leftrightarrow  ik =i \frac{p}{\hbar}
\label{correspondentie 1}
\ee
On the other hand, we know from classical mechanics that the total energy $E$ and momentum $p$\,($=mv$) of a particle are related by the following expression:
\be
E=V(x)+\f{mv^2}{2} = V(x)+\f{p^2}{2m}.
\label{correspondentie 2}
\ee
where $V(x)$ is the potential energy of the particle at position $x$. 
\subsubsection*{A guess}
Now, using (\ref{correspondentie 1}) to replace $E$ and $p$ in \ref{correspondentie 2}, we get
\be
i \hbar D_t \,\,``=  " \,\, V(x) + \frac{1}{2m} (-i \hbar)^2 D_x^2.
\ee
Of course, this equation looks a bit strange. However, if we let both sides act on $\psi$, we get the following
\be
i \hbar D_t \psi(x,t) = V(x)\psi(x,t)+ \frac{(-i \hbar)^2}{2m} D_x^2 \psi(x,t)
\ee 
which can be rewritten as:
\be
i \hbar\f{\partial \psi(x,t)}{\partial t} = V(x)\psi(x,t) - \f{\hbar^2}{2m} \frac{\partial^2 \psi(x,t)}{\partial x^2}
\ee 
This expression is called \textbf{the Sch\"{o}dinger equation}, after the physicist Erwin Schr\"{o}dinger who wrote it down first.\footnote{Pronounced `SHROEding-uh'}
It has the shape of an evolution equation: it tells you what the time evolution ($\f{\partial}{\partial t}$) of a wave function is, given the wave function on that moment. That is precisely what we were looking for! It turns out that \textit{almost all} particles obey the above time evolution equation: electrons, protons, neutrons, ... . Hence, the above equation is really the core of quantum mechanics - much of the rest of the course will be using this equation. 
\subsubsection{Disclaimer}
If you have vaguely understood the motivation and arguments leading to the Schr\"{o}dinger equation, do congratulate yourself. Do not think that the arguments present are even close to being solid though, we did \textit{not} derive the equation. Such a derivation doesn't even exist. To see why, just think back of how the laws of Newton were introduced to you a long time ago. People can try to convince you that those formulas are more or less reasonable, but their exact form can not derived in a mathematical way: they are only found by doing lots of experiments, by many people thinking long and hard about how to put the experimental results into a small set of formulas. Newton did this for gravity and classical mechanics (building on the work of many others), and Schr\"{o}dinger (equally dependent on others' ideas and results) did so for quantum mechanics.

Of course, things will get more clear after \textit{using} the Schr\"{o}dinger equation. Good news: in the next section we will do our first real quantum mechanics computation.

\section[Using the Schr\"{o}dinger equation]{\includegraphics[width=0.04\textwidth]{comput_c} Using the Schr\"{o}dinger equation}
\subsection{The Hamiltonian}
Lets look at the Schr\"{o}dinger equation more carefully. On the right hand side there is  
\be
V(x)\psi(x,t) - \f{\hbar^2}{2m} \frac{\partial^2 \psi(x,t)}{\partial x^2}
\ee
This whole object is a function, depending on $x$ and $t$. If you know $V(x)$, and someone gives you some input $\psi(x,t)$, you can readily compute the above object for him. Wait wait wait, we have heard that before: this sound like an operator! Indeed, using notation of the first section \includegraphics[width=0.04\textwidth]{tool_c}, we can define the operator 
\be
H\equiv V(X) -  \f{\hbar^2}{2m} D_x^2
\ee 
which sends any function to $V(x)$ time itself, plus $-  \f{\hbar^2}{2m} $ times its second derivative. This operator is called the \textbf{Hamiltonian operator}. Using this operator, we can then rewrite the Schr\"{o}dinger equation in the following concise form:
\be
i \hbar\f{\partial \psi(x,t)}{\partial t}  = H \psi(x,t)
\ee
Some people like to write partial derivatives like $\f{\partial}{\partial t} $ and $\f{\partial}{\partial x} $ simply as $\partial_t$ and $\partial_x$. They can stylize even further:
\be
i \hbar \partial_t \psi = H \psi.
\ee
In this form the Schr\"{o}dinger equation looks less scary - although the notation is more abstract.
\subsection{A real computation: a particle in a box}
We are now ready for our first real computation. Imagine a particle enclosed in a small space, or a ``box" - for instance an electron trapped in a small region. What does its wave function look like, and how does it evolve in time?
Let's see if we can answer these questions.
First let's try to see if we can make the word ``box" more precise. Consider the following potential
\be
V(x)=
\left\{ 
\begin{array}{l l}
0 & \quad \textrm{for} \,\,0< x <L \quad\\
+\infty& \quad \textrm{everywhere else}
 \end{array} \right.
\ee 
\begin{figure}[h]
 \begin{center}
  \includegraphics[width=0.5\textwidth]{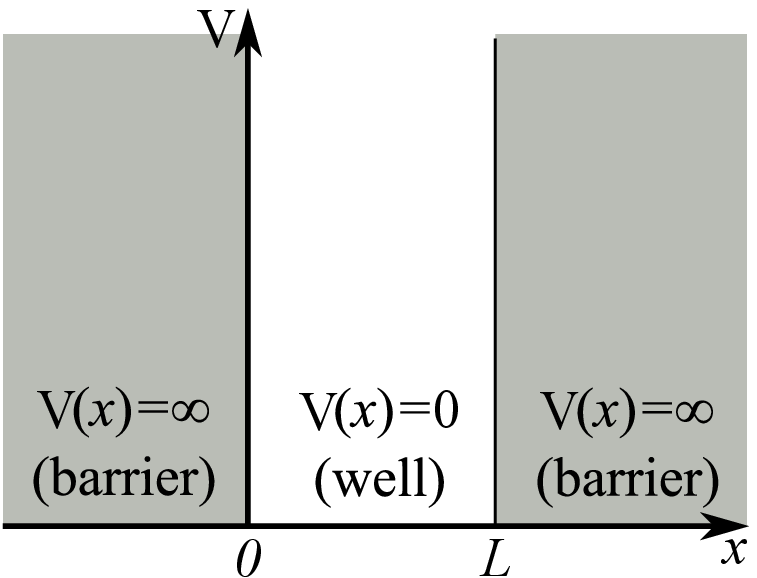}
  \label{well_well}
  \end{center}
\end{figure}
Between $x=0$ and $x=L$, the potential is constant, so the particle moves freely. Outside that region however, the potential energy is infinite, meaning the particle can not go there. In terms of the wave function: we impose that $\psi$ has to be zero outside $[0,L]$. If we also demand the wave function to be continuous, it also has to vanish on $x=0$ and $x=L$. Since outside the box $\psi=0$, the Schr\"{o}dinger equation is trivially satisfied there, so we can focus on solving the equation inside the box. For a general initial wave function $\psi(x)\equiv\psi(x,t=0)$ it is not so clear how to explicitly solve the Schr\"{o}dinger equation. So it is hard to tell what the wave function will look like at a later moment. However, something special happens if the initial wave function is an \textit{eigenfunction} of the Hamiltonian operator. In that case,
\be
H \psi(x) = E \psi(x)
\label{eq:tise}
\ee
for some \textit{number} $E$. 
In that case, the time evolution of the wave packet is rather easy: it is given by. 
\be
\psi(x,t)=e^{- i E t/\hbar}\psi(x)
\label{time}
\ee 
You can easily check that the above $\psi(x,t)$ indeed satisfies the Schr\"{o}dinger equation $i \hbar \partial_t \psi(x,t) = H \psi(x,t)$. So for each  wave function that is an eigenfunction of the Hamiltonian, we can write down a fully time-dependent solution to the Schr\"{o}dinger equation. For this reason, (\ref{eq:tise}) is called the \textbf{time-independent Schr\"{o}dinger equation}: it's an equation not involving $t$, and the time evolution of its solutions is very easy. Let's try to solve that equation for a particle in a box. Here $V(x)=0$ in the relevant region, so that
\be
H=-\f{\hbar^2}{2m} D_x^2= -\f{\hbar^2}{2m} \f{\partial^2}{\partial x^2}
\ee
Can we find eigenfunctions of this operator? Well, you know that the functions $\sin x$ and $\cos x$ are proportional so their second derivative. This suggests we should look in that direction. Putting in some extra constants, we find that all functions of the form: 
\be
\psi(x) = A \sin(kx) + B \cos(kx)\quad
\ee
satisfy $E \psi = -\f{\hbar^2}{2m} \f{\partial^2 \psi}{\partial x^2}$, given that
\be
k^2 =\f{2 E m}{\hbar^2}.
\ee
Only the function $\sin (kx)$ vanishes at $x=0$ and demanding that also $\psi(L)=0$ imposes 
\be
k=\f{n\pi}{L} \quad \textrm{with n a positive natural number}.
\ee
So the eigenfunctions of $H$ (satisfying the boundary conditions $\psi(0)=\psi(L)=0$) are given by
\be
\psi_n(x)=A \sin \f{n \pi x}{L}
\ee
and they have corresponding eigenvalues 
\be
E_n = \frac{\hbar^2 k^2}{2m}=\frac{n^2\hbar^2 \pi ^2}{2 m L^2}
\ee
A priori, the prefactors $A$ have arbitrary values. For reasons that will be clear later, we will put the $A$'s equal to $\sqrt{\f{2}{L}}$.
If we now use (\ref{time}), we see that the full time-dependent solutions to the Schr\"{o}dinger equation are
\be
\psi_n(x,t)=e^{ - i E_n t/\hbar} \sqrt{\f{2}{L}}  \sin \f{n \pi x}{L}
\ee
The first few $\psi_n$ are shown in figure below.
\begin{figure}[H]
 \begin{center}
  \includegraphics[width=0.6\textwidth]{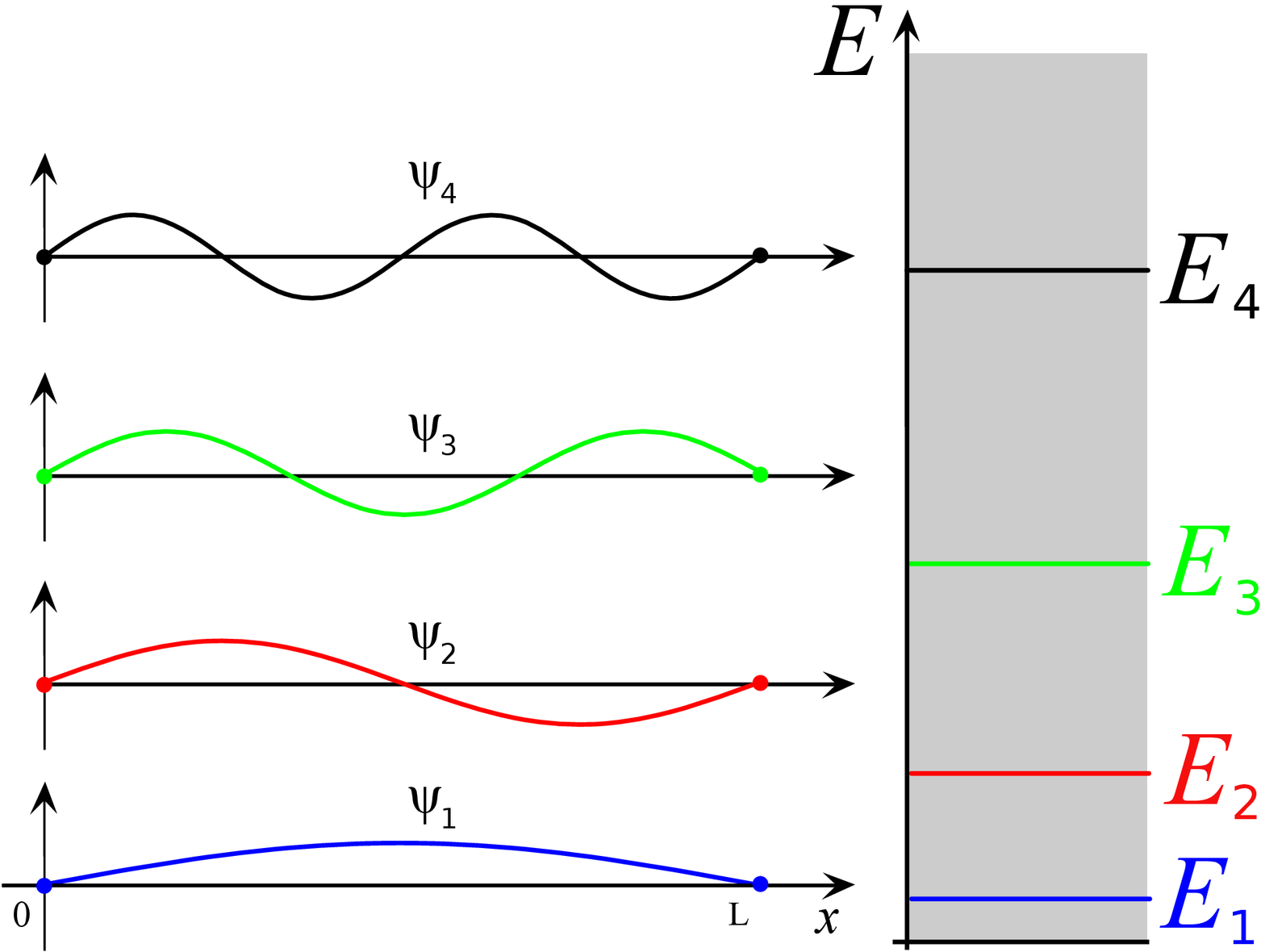}
  \label{fig:pib}
  \end{center}
\end{figure}
The time-dependent solutions are just the above graphs, but `rotating' in time: multiplying by $e^{ - i E_n t/\hbar}$ keeps the norm of the  wave function the same throughout time, but the argument runs with angular velocity $\omega=\frac{E_n}{\hbar}$.
This is a pretty nice result. We find an infinite series of allowed  wave functions for a particle in a box. The solutions are labeled by the integer $n$, and each such solution is an eigenfunction of the hamiltonian, with eigenvalue $E_n$. Let's try to understand this result better.

\subsubsection{Interpretation}
It is very striking that only a \textbf{discrete set} of  wave functions (obeying the Schr\"{o}dinger equation and the matching conditions) can be found. By \textit{discrete} we mean the set of solutions is labeled by an integer, not by a continuous parameter. Each such solution is called a \textbf{state}. 
So a particle in a box can only be in a discrete series of states. As we saw before, the Hamiltonian operator is closely related to the concept of energy, so it is natural to see the eigenvalue of a state under $H$ as its \textbf{energy}. So according to quantum mechanics, the energy of a particle in a box can only take on a discrete set of values $E_n$.

Let's compare this to the classical (incorrect) description. If a fundamental particle really was a \textit{point}, it would just look like a bullet bouncing from side to side in the box, with speed $v=\f{p}{m}$. The states would then just be labeled by the momentum $p$ of the particle. So in such a situation, any value of $p$ would be allowed, and the energy $E=\f{p^2}{2m}$ of the particle could take on any positive value. 

This means the classical (point particle) description is drastically different from the (correct) quantum description. How can we check the quantum description is really true? Imagine a particle jumping from one energy level to a lower one. In such a process, the energy difference has to be released - for example in the form of radiation: the emission of a photon. The energy of such a released photon can be measured. This way, we can measure the energy \textit{differences} between the different states of a system. Sure enough, for particles enclosed in a small region these energy differences take on only very special values, confirming the spectrum is actually discrete, just like QM predicts. Of course, the details of that spectrum depend on the exact shape of the potential in which the particle is locked up. The potential discussed here is a bit simplistic/idealized. There is more than one dimension (here we considered only an $x$-dependent wave function) and it is technically not possible to build such a perfect box potential. However, very similar systems \textit{can} be built. A \textbf{quantum dot} f.e. is a configuration in which an electron is locked up in a very small zone of a material. These dots can have f.e. a nice round shape. The calculation to solve the Schr\"{o}dinger equation is a bit more difficult there, but the result is very similar: one finds a discrete series of states, each with its own energy. And the computed energy levels indeed match experimental results, backing up the quantum description, once again. 

\newpage
\section*{Exercises}
\begin{enumerate}
\item In the beginning of the chapter, we promised that all operators relevant to QM are linear operators. Check that the Hamiltonian operator is linear, by explicitly writing out its action on a linear combination of functions.
\item What is the eigenvalue of the function $f(x)=1$ under the operator $D_x$? And of the function $f(x)=x$ under the operator $XD_x$? 
Give an example of a function which is \textit{not} an eigenfunction of these operators (should be easy). 
\item Under what condition (on $n$ and $m$) is the function $x^n$ an eigenfunction of $X^m D_x^2$? What's the corresponding eigenvalue?
\item Consider an electron locked up in a box, with length $L=10$ nm (1 nm =$10^{-9}$ m, the electron mass is $m_e = 9,11 10^{-31}$ kg). What is the energy $E_2$ of the second level?\footnote{Be careful with units here! Unless you are using the de relation $E=hf$ (where you can express both the energy and Planck constant in eV-units) you \textbf{have to} express Planck's constant in standard units: J$\cdot$s and not eV$\cdot$s, since all other quantities (lengths, masses, ...) are expressed in standard units. Of course, at the end of your calculation you may choose to re-expess the energy you found in eV again.} And what's the energy of the first (lowest) level? If electron drops down from the second to the first level, the corresponding energy difference has to be released, for example in the form of a photon. What would be the frequency of this photon? 
\item The nucleus is a very small region where protons and neutrons are squeezed together (or `locked up') in a region of about 1 fm $=10^{-15}$ m. Be very crude and pretend that a nucleus really \textit{is} a `particle in a box' potential trapping the nucleons. If a neutron ($m_n = 1,67 \cdot 10^{-27}$ kg) drops from the third to the first level, what energy is released? To convince you that the `particle in a box' approximation is not completely ridiculous here: it is indeed possible for a nucleus to be in an excited state - with one or more nuclei occupying a higher energy state (such states are typically produced in nuclear decay processes). When decaying, typically a gamma ray escapes, with an energy around 1 MeV ($=10^9$ eV). Is this comparable to what you found with your `toy calculation'? (If not, you may have done something wrong with units - check the footnote.)
\item In classical mechanics, the potential (and all other energy levels) can often be shifted without changing the physics. To see how things work out here: solve the problem of a particle in a box, but now with $V=$(some constant) instead of $V=0$ in the allowed zone. What happens to the energy levels? Try to see what happens in general if you send $H\rightarrow H+$constant: do the eigenstates change, and their eigenvalues? What about time evolution?
\end{enumerate}

%% file: H3.tex
\chapter{The measurement}

\subsection*{In this chapter...}
We are now ready to introduce another central aspect of quantum mechanics: performing measurements on a particle's  wave function. The tool we need to introduce is the notion of inner product between functions. We then tell the story of the measurement in QM. In the last section, make all this more concrete with the study of the particle in a harmonic potential.


\newpage

\section[Inner products and bra(c)kets]{\includegraphics[width=0.04\textwidth]{tool_c} Inner products and bra(c)kets}

\subsection{Inner product of functions}
In quantum mechanics, a very central role is played by complex functions. A simple but important property is that the sum of two complex functions is again a function, and a complex multiple of a complex function is a complex function as well. This means the space of complex functions forms a (complex) \textbf{vector space}. Many vector spaces have notion of distance, angles and all that. All these useful concepts are usually derived from one single notion: the \textbf{inner product}. Can you define an inner product for complex functions as well? The answer is: yes, you can! It is usually denoted by $\langle f,g\rangle  $ and is a map from any two complex functions $f$ and $g$ to a complex number, as follows:
\be
\langle f,g\rangle  \equiv \int_{-\infty}^{+\infty} \overline{f(x)} g(x) dx 
\label{inner product}
\ee
Here the bar on $f$ denotes complex conjugation. Some people denote complex conjugation with an asterisk, so they write $f(x)^*$ instead of $\overline{f(x)}$. The result of the above integral is indeed a single complex number. You may worry that the above integral is not always finite. Here is a relief: we will use inner products like the above specifically for \textit{wave functions} $f$ and $g$. Since wave functions fall off fast outside some region (they are wave \textit{packets})  they have finite inner products $\langle f,g\rangle  $ amongst themselves. So in the case of our interest, the above object will be well-defined. Roughly speaking, the inner product is large if the wave packets $f$ and $g$ have a similar shape and location, otherwise it is small. The inner product has a special property, called \textbf{bilinearity}:
\be
\langle f,a g_1+b g_2\rangle  =a \langle f,g_1\rangle  +b\langle f, g_2\rangle   
\ee
and
\be
 \langle a f_1+b f_2, g\rangle  = \bar{a} \langle f_1,g\rangle  +\bar{b}\langle f_2,g\rangle  .
\ee
This is sometimes put as follows: the inner product is \textbf{linear} in its second argument and \textbf{anti-linear} in its first argument. Lets check that for example the second of these properties is true:
\begin{eqnarray}
\langle a f_1+b f_2, g\rangle  &=& \int_{-\infty}^{+\infty} \overline{a f_1(x)+b f_2(x)} g(x) dx \\ 
&=& \int_{-\infty}^{+\infty} \overline{a} \overline{f_1(x)}+\overline{b} \overline{f_2(x)} g(x) dx \\
&=& \bar{a} \langle f_1,g\rangle  +\bar{b}\langle f_2,g\rangle  
\end{eqnarray}
Here we first used the properties of complex conjugation (see Appendix 1 on complex numbers) and then the linearity of an integral. Another essential property of the inner product (which you can easily check) is the so-called \textbf{conjugate symmetry}:
\be
\langle f, g \rangle = \overline{\langle  g,f \rangle }
\ee
Of course, we can also take the inner product of a function with itself:
\be
\langle f,f\rangle  =\int_{-\infty}^{+\infty} \overline{f(x)} f(x) dx =\int_{-\infty}^{+\infty} |f(x)|^2 dx 
\ee
The last expression, being the integral of a real and positive function $|f(x)|^2$, has to be real and positive. 
Hence,
\be
\langle f,f\rangle  \,\,\,\geq\,\,0.
\ee
This means we can always take the square root $\sqrt{\langle f,f\rangle  }$. This gives a real number, and says how ``big'' the (complex) function $f$ is. (Something like the total size of the complex function $f$.) It is called the \textbf{norm} of the function $f$, and is denoted by $\|f\|$:
\be
\|f\| = \sqrt{\langle f,f\rangle  }
\ee
Using our fresh and shiny inner product tool, we can introduce yet another new concept (besides the norm): the \textbf{Hermitian conjugate} of any operator $A$. It is denoted by $A^\dag$ and defined as following: $A^\dag$ is the operator such that for every pair of functions $f$ and $g$,
\be
\langle A^\dag f,g\rangle  =\langle f,A g\rangle  .
\ee
One can show that 
\be
(A^\dag)^\dag=A
\label{eq:Adagdag}
\ee 
and 
\be
(aA + bB)^\dag = \bar{a} A^\dag + \bar{b} B^\dag
\label{eq:LCHermitian}
\ee 
for all operators $A$ and $B$, and all complex numbers $a$ and $b$.  Side remark: as mentioned in Appendix 1 of the first chapter, we denote complex conjugation either by a bar on top (like in the above expression) or by an asterisk, at will. So both $\bar{c}$ and $c^*$ denote the conjugate of $c$. Another property of Hermitian conjugation that one can show, is 
\be
(AB)^\dag = B^\dag A^\dag
\label{eq:reverse}
\ee
So the order of operators switches upon taking the Hermitian conjugate. 
\subsection{Column- and row vectors}

Maybe you recognized several terms of the previous section. Indeed, you might be familiar with the \textit{inner product of (complex) vectors} from the context of linear algebra. Given two $n-$dimensional complex vectors $a=(a_1,...,a_n)$ and $b=(b_1,...,b_n)$ it is defined as 
\be
\langle a,b\rangle  = \sum_{i=1}^{n} \bar{a}_i b_i
\ee 
You can check that this, just like the inner product of functions, is a bilinear operation. The only difference is that here it maps two vectors, not two functions, to a complex number. In linear algebra, one usually denotes vectors in a \textit{column version}. For example the vector $\vec{v}$ with components $1$ and $i$ is written as:
\be
v=\left( \begin{array}{c}
1  \\
i  \end{array} \right) 
\ee
Next to that, vectors also have a \textit{row version}, called their \textbf{Hermitian conjugate} and denoted by $\vec{v}^\dag$. It is obtained by taking the complex conjugate of all components, and then writing the numbers in a row. For the above vector, this gives:
\be
v^\dag=\left( \begin{array}{cc}
1 & - i \end{array} \right)
\ee
Now what is the point of making two different versions of the \textit{same} object? Well, if you write a column vector after a row vector, the ordinary matrix multiplication of the two objects gives you precisely the inner product of the two vectors. So, by using the right version of the vectors, people know where to take the inner product: it just happens \textit{automatically} if you perform the matrix multiplications. For example: if someone wants to calculate the inner product of the above vector $v$ with the matrix product of $A=  \left( \begin{array}{cc}  1 & 0 \\     1 & 1   \end{array} \right)$ times the vector $w=\left(\begin{array}{c} 1 \\ 0 \end{array} \right)$, he should write down
\be
\langle \left( \begin{array}{c}
1  \\
i  \end{array} \right) ,  \left( \begin{array}{cc}
    1 & 0 \\ 
    1 & 1 \\ \end{array} \right)  \cdot \left( \begin{array}{c}
1  \\
0 \end{array} \right) \rangle  
\ee 
This looks horrible. However, with the row-column trick he can just write the above expression as 
\be
v=\left( \begin{array}{cc}
1  &
-i  \\ \end{array} \right)   \left( \begin{array}{cc}
    1 & 0 \\ 
    1 & 1 \\ \end{array}\right)\left( \begin{array}{c}
1  \\
0  \end{array} \right) 
\ee
This looks more appealing, and is completely clear: just take the matrix product twice, and the result you get is indeed the required inner product.
It is important to remember that the difference between row and column vectors is not fundamental: it is just a trick to indicate: ah, if the two are next to each other, take the (inner- or matrix-) product.

\subsection{Bras and kets}

Let's try to repeat that convenient trick for the inner product of functions. Doing so, we will also upgrade our notation. Up to this point, we were a bit sloppy with the difference between a state of a particle and a function. For sure, the state of a particle is \textit{described} by a complex function (the particle's wave function) but very strictly speaking it is not the same thing. A similar example: a temperature can be \textit{described} by a number, but a temperature is more than `just a number': it has a special meaning. To be more careful with this difference: if a particle has  wave function $f$, we will say that it is in the state $| f \rangle  $. So using the vertical bar and a right bracket, we indicate that we are talking about a \textit{state}, not just any function. (In the same way that an $^{\circ}F$, $^{\circ}C$ or $K$ indicates that the number in front is a temperature, not just any number.) Now, analogous to linear algebra, for a state $|f\rangle  $, we define its \textbf{hermitian conjugate}, and denote it as $\langle f|$. In some sense, it really is the same object (it specifies the state of a particle) but in a different version - just like row versus column. Here the visual difference is not row or column, but the shape of the brackets. Just like for ordinary vector spaces, the only use is the following: if you write a conjugate $\langle g|$ next to $|f\rangle  $, it is understood that you should take the inner product. A state written in the form $|f\rangle  $, is called a \textbf{ket}, and if you write it in the (conjugate) form $\langle f|$, you call it a \textbf{bra}. The reason for these funny terms is as follows: if you write a \textit{`bra'} and then a \textit{`ket'}, you have to take the \textit{bracket} (which is slang for `inner product'). Yes, this proves that physicists pull bad jokes.
\be
\ba{ccccc}
\langle  f | \,\,\,\, | g \rangle   &=&  \langle f|g\rangle  &=& \langle f,g\rangle   \nonumber\\
\textrm{(bra and then ket)} &=& \textrm{(short notation)} &=&\textrm{(bracket)} \nonumber
\ea
\ee
This notation is a bit strange when you first meet it. Don't worry: its use will become clear later on. Hopefully you understood that there is not something deep behind it: it is just a convenient bookkeeping on where to take the inner product, that's all.

\subsubsection*{When operator meets (brac)ket}
Now what about operators? When acting with an operator $A$ on a function $f$ you get a new function $A f$. In the context of states: $|f\rangle  $ describes a state with  wave function f, and $|Af\rangle  $ describes a state with  wave function $Af$. So this means we can see every operator not only as a map between functions, but also as a \textbf{map between states}, as follows:
\be
A: |f\rangle  \rightarrow A|f\rangle   = |Af\rangle  
\ee
In this last line, $A|f\rangle  $ means: `act with operator $A$ on state $|f\rangle  $', and $|Af\rangle  $ means again: `the state with  wave function $Af$'. This is a bit of a trivial remark - mostly a nuance in notation and interpretation. 
Time for the story part!

\section[The measurement]{\includegraphics[width=0.04\textwidth]{once_c} The measurement}
In the first chapter, we learned about the two slit experiment for electrons. Since the result of the experiment was very clearly pointing at interference (even if single particles went through) we concluded that each particle has to be a wave packet. This conclusion has been confirmed by many other experiments since - particles \textit{really are} extended objects, described by a  wave function. Yet, there is a subtlety that we have not touched upon so far. 
How does detection occur when one of the electrons arrives at the wall? In principle, the detecting wall \textit{measures the position} of the incoming particle. But how does this work, given that the particle arriving is -in fact- an extended object? What outcome does a measuring device give? Of course, if a measuring device could speak and gesture, it would answer: ``Well, the particle is spread out a bit, and is more or less here", waving its hand at the location of the wave packet. But that is not what it is made to do, it is made to give you an exact answer. It turns out that a measuring device \textit{will} give you a very precise answer for the position, one particular $x$. But what will it say? \textit{Which} $x$ will it give you? Well, the \textbf{chance for a device to give you $x$ as the particle's position is proportional to the amplitude $|\psi(x)|^2$ of the  wave function there}. So in some sense, the measuring device does a good job: the position it gives you is usually a place where the  wave function is big, which is indeed an indication for where the  wave function is located. Since the function $|\psi(x)|^2$ is proportional to the chance to detect the particle's position to be $x$, we call this function the \textbf{probability amplitude}. 
\begin{figure}[H]
 \begin{center}
  \includegraphics[width=0.8\textwidth]{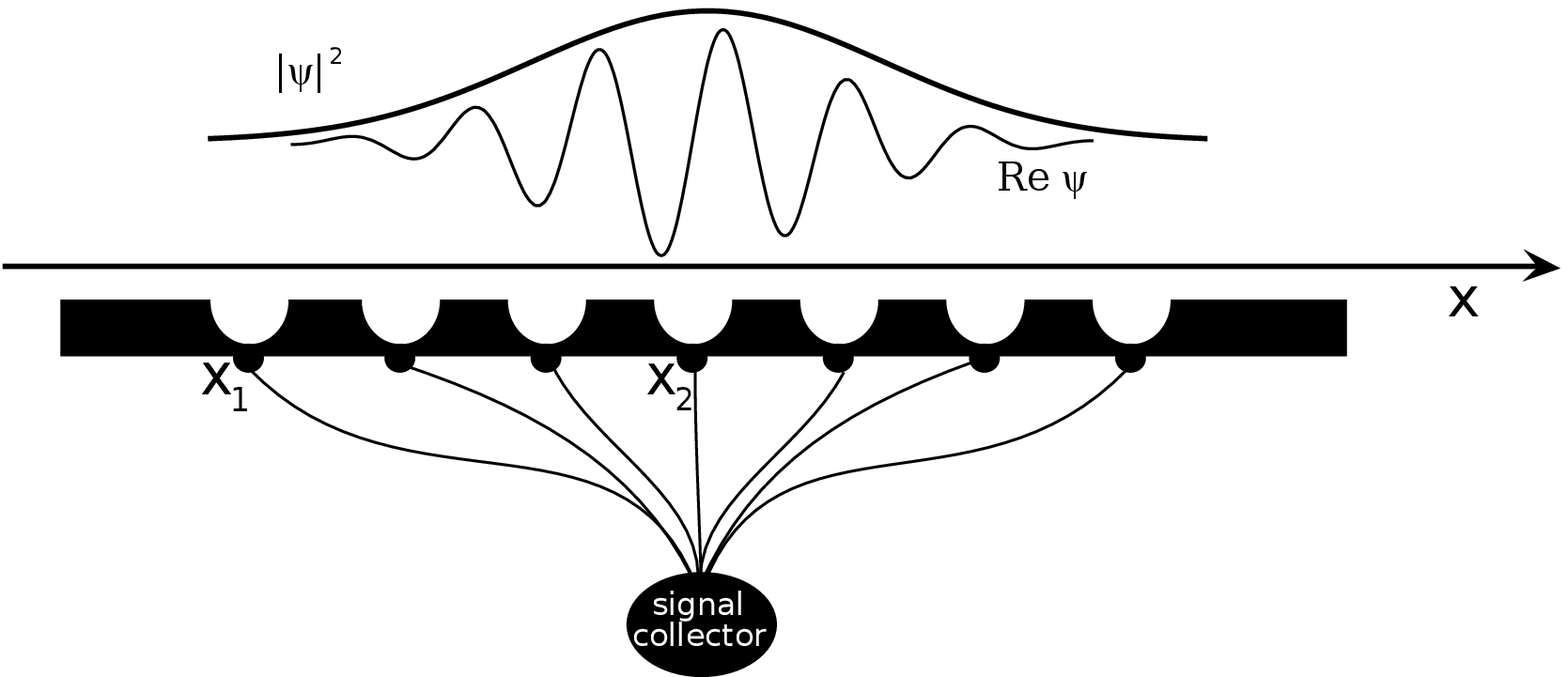}
  \end{center}
\end{figure}
On this figure, several things are shown: a wave packet (consisting of a single lump here) arrives at a detector wall. The detector wall consists of many individual detectors (the holes) connected to a data collector. To indicate the shape of the  wave function, its real part $\textrm{Re} \,\psi$ is shown. The total probability amplitude $|\psi|^2 = (\textrm{Re} \,\psi)^2+(\textrm{Im} \,\psi)^2$ is also shown. The odds for a detector to give you $x_1$ as an answer is really small as the amplitude $|\psi(x_1)|^2$ is very small there. An answer like $x_2$ is much more likely to be given. Vaguely speaking: a measuring device will give a \textbf{random, but reasonable} anwer. Now if you like to experiment, the above explanation probably brings the following question to your mind. What if, right after measuring a first time, you measure again? This could in principle be done by making a `transparent' detecting wall that does not stop the particle, and placing a second detecting wall behind it. Is there again some statistical chance for the possible outcomes? The answer is no! It turns out that if you do so, you will get the \textit{same} answer. So no statistical chances anymore, nothing like that, just a $100\%$ sure that you get the same outcome as the first measurement.
If this confuses you, don't worry, many great minds in physics of the last century were very puzzled by this fact, and the only answer they could come up with is the following. Apparently, \textbf{performing a measurement on a  wave function drastically alters it}: it seems to collapse onto the state corresponding to your measurement outcome. So if you measure its position, and get as answer $x_1$, the wave packet simultaneously collapses into a  wave function that is very sharply localized around $x_1$, even if it wasn't so before. 
\begin{figure}[H]
 \begin{center}
  \includegraphics[width=0.6\textwidth]{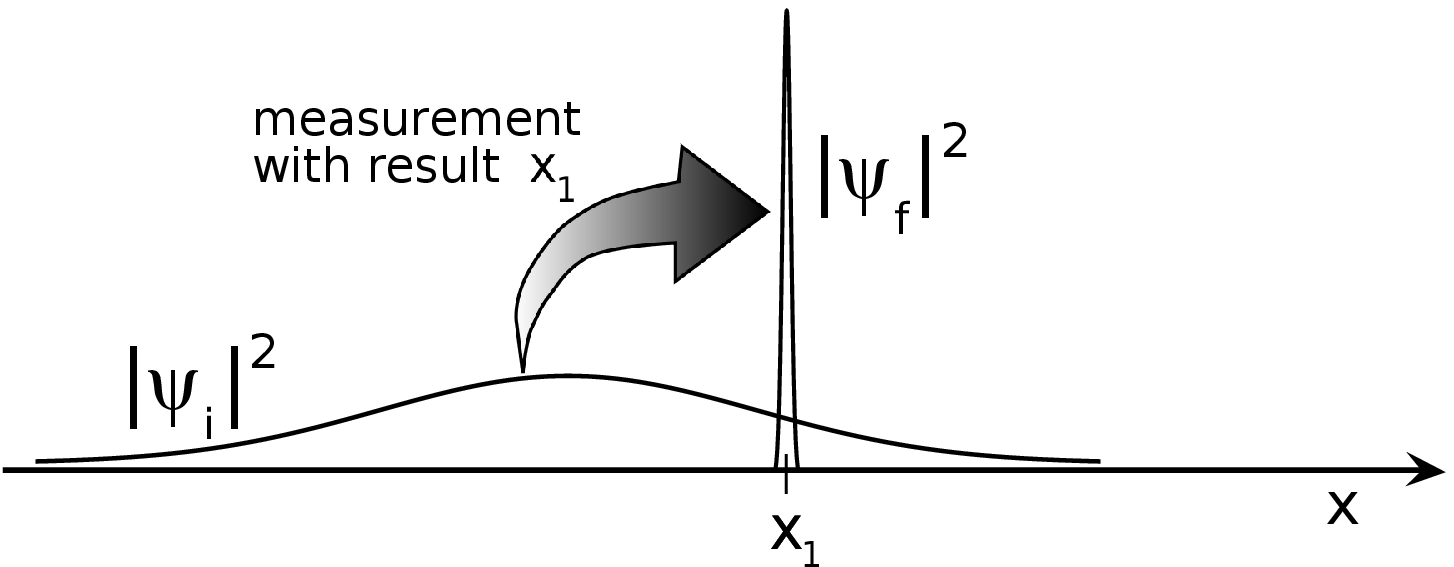}
  \end{center}
\end{figure}
This phenomenon is called the \textbf{collapse of the  wave function}. There are some theories on how and why it happens, but despite all the debate, there is no consensus on which one is true. The great difficulty is of course talking about a particle (microscopic object) and a detector (macroscopic object) at the same time. To be consistent you would need to describe the detector in the quantum mechanical language as well. That is of course a very hard task, and can only be modeled poorly. Maybe the next generation physicists will be able to solve this problem - lets hope so. Here is the good news: even though we don't know \textit{why} it happens, it is very sure \textit{that} this happens - to very high accuracy - so this gives a partial relief. 

\subsubsection*{The two slit experiment, revisited}
To illustrate the above behaviour, let's go back to the two slit experiment. Say we use an electron source and put it at a very low intensity, to make sure it shoots out the electrons one by one. As a particle arrives at the slit, its  wave function is still a nice blob. Then, as it passes through the holes, the two parts of the  wave functions (passing through the left and right slit) interfere with each other. The result is a ``striped"  wave function: only on places of constructive interference, the amplitude $|\psi|^2$ is big. Now the striped  wave function arrives at the detecting wall. The chance for the detecting grid to answer to you that the particle arrived at position $x$ is large at the stripes only, since the probability amplitude is large there only.  So very probably, it will answer you a position -say some $x_1$- lying within one of the stripe's positions. Now a second electron comes through. The shape of its  wave function after passing through the slits is \textit{identical} to that of the first electron. However, the result of the measurement is not necessarily the same. The measurement may yield any outcome $x_2$, with probability $|\psi(x_2)|^2$. Obviously, after a lot of particles have gone through the slits, the \textit{total} detection rates will be nicely proportional to the probability amplitude $|\psi(x)|^2$ of the individual particles. So even though individual measurements give very few information about the  wave functions, the total detection count reveals this information very accurately. This way, the wave property of the individual particles inevitably becomes apparent. So we conclude the first interpretation of the experiment was correct, although the physics involved has a statistical subtlety. Got it? 

\subsubsection*{Wave packets look like point particles}
Also, the above explanation should make clear why particles appear as points to us in many experiments. This is because the position measurement gives very sharp results, and simultaneously collapses the wave function on a very localized one. In a more lyrical phrasing: the measurement forces the wave function to act like a particle, or better: you may easily get the impression they are strict points indeed, although this is more an artifact of the measuring process.

\section[More precise]{\includegraphics[width=0.04\textwidth]{comput_c} More precise}
So far the story part, let's now make all this more concrete and precise. We said that the chance to measure a particle with  wave function $\psi$ to be at position $x$ is proportional to $|\psi(x)|^2$. By this we actually mean: the chance to measure the particle to be in a very small interval $[x,x+\Delta x]$ is given by $\Delta x \cdot |\psi(x)|^2$. For a larger interval the probability is given by the \textit{integral} of 
$|\psi(x)|^2$:
\be
P([a,b])=\int_{a}^{b} |\psi(x)|^2 dx.
\label{eq:problaw}
\ee
where by $P([a,b])$ we mean the chance to detect the particle in the position-interval $[a,b]$. For this reason, people say that $|\psi(x)|^2$ is a \textbf{probability density}. 
\subsubsection*{Consequence 1}
The \textit{total chance} getting a measurement result between $x=-\infty$ and $x=+\infty$ has to be $1$ of course. This means we need
\be
\int_{-\infty}^{+\infty} |\psi(x)|^2 dx =1
\ee
In the terminology of the first section: the square of the norm of $\psi$ should be $1$: 
\be
\langle \psi|\psi\rangle  =1
\ee
If this is not the case for some  wave function, you should multiply it by a number so that it \textit{does}. Concretely, you can do this by dividing $\psi$ by its norm:
\be
\psi\rightarrow\frac{\psi}{\sqrt{\langle \psi|\psi\rangle  }}
\ee
This is called \textbf{normalization of the  wave function}. If a wave function (or state) is not normalized, that isn't a deep problem, but in that case you can't immediately use the probability law (\ref{eq:problaw}). You can now understand why we put the $A$'s involved for the particle in a box to a specific value: we wanted the  wave functions to be properly normalized. More precise: the norm of the particle-in-a-box wave functions can be shown to be
\be
\langle \psi_n,\psi_n\rangle   = |A|^2 \frac{L}{2}
\ee
So to achieve normalization, we have to put $A = \sqrt{\frac{2}{L}}$

\subsubsection*{Consequence 2}
Another thing. Lets try to calculate the \textit{expectation value} of the position. So given a  wave function, what is the \textit{average} position you would measure? This is the number you would get if you would measure the position again and again and take the average - at least if after each measurement you are able to put the particle back in its original state, undoing the effect of the collapse of the  wave function. More practical: it is the average of all the consecutive position measurements of a large collection of particles, which you manage to put in the same state. (Just like in the two slit experiment for example.) In words, this quantity is given by summing for all intervals $[x,x+\Delta x]$ the chance of finding the particle in that interval \textit{times} the corresponding position $x$. So
\be
\langle X\rangle  _\psi=\int_{-\infty}^{+\infty} x \cdot |\psi(x)|^2 \,dx
\ee
Here we have denoted the expectation value of the position for a  wave function $\psi$ by $\langle X\rangle  _\psi$. Now the above can be written in a very compact way using the bra-ket notation:
\be
\langle X\rangle  _\psi=\langle \psi|X|\psi\rangle  
\ee
Indeed, acting with the operator $X$ on the function $\psi(x)$ gives $x \cdot \psi(x)$, so the state $X |\psi\rangle  $ (\,=$|X\psi\rangle  $) has  wave function $x \psi(x)$ and
\bea
\langle \psi|X| \psi\rangle   &=& \langle \psi| X \psi\rangle   \\
&=& \inft \overline{\psi(x)}\, (x\cdot \psi(x)) \,\, dx \\
&=& \inft x |\psi(x)|^2\,\, dx
\eea
The object $\langle \psi|X|\psi\rangle  $ is sometimes called the \textbf{sandwich} of the operator $X$ inside the state $\psi$. This origin of this term is probably clear. In general, we call the object $\langle \psi|A|\phi\rangle  $ the sandwich of operator $A$ between states $\psi$ and $\phi$:
\begin{figure}[H]
 \begin{center}
  \includegraphics[width=0.4\textwidth]{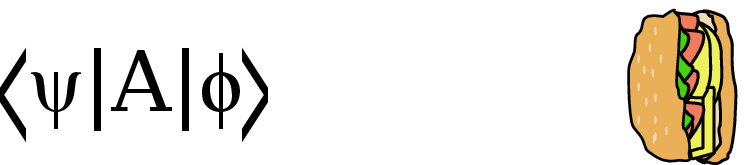}
  \caption{Two sandwiches}
  \end{center}
\end{figure}
We can cryptically summarize the above paragraph by stating that the expectation value of the position for a state is given by the sandwich of the position operator $X$ inside that state. If you understand this sentence, you have clearly gotten the above. We are now ready for another important example: a particle in a harmonic potential. There, we will be able to apply all the bra-ket machinery.

\subsection{Example: the harmonic oscillator}
Let's consider a particle with mass $m$ in a quadratic potential:
\be
V(x)= \f{1}{2}m\omega^2 x^2
\ee
Classically, an object in such a potential swings perpetually around its equilibrium, with frequency $\omega$. That is indeed what a marble would do in such a potential (ignoring friction). But to see what a elementary \textit{particle} in such a potential looks like, we need to use the quantum mechanical description. The key is of course the SE (short for Schr\"{o}dinger Equation):
\be
i \hbar \partial_t \psi(x,t)= H\psi(x,t)
\ee
For this potential, the Hamiltonian is given by
\be
H=-\f{\hbar^2}{2 m}D_x^2+V(X)=\f{P^2}{2m}+\f{m\omega^2}{2}X^2
\label{eq:sumsq}
\ee
In the last step we have also defined the \textbf{momentum operator}
\be
P = -i\hbar D_x.
\ee
\begin{figure}[h]
 \begin{center}
  \includegraphics[width=0.3\textwidth]{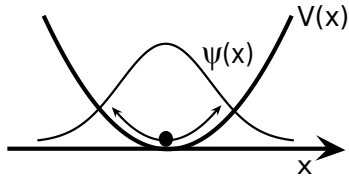}
  \caption{The (classical) movement of a macroscopic object in a harmonic potential is shown by the oscillating point. For a microscopic particle things will look differently: the particle will take on some specific  wave function $\psi(x)$ inside the potential.}
  \end{center}
\end{figure}

\subsection*{General recipe}
The strategy is exactly the same as for the particle in a box. Assume we can solve the time-independent SE. That is, assume we can find energy eigenstates $|\psi_n\rangle  $:
\be
H|\psi_n\rangle  =E_n|\psi_n\rangle  
\ee
Then their fully time dependent counterparts are given by 
\be
\psi_n(x,t)=e^{- i E_n t/\hbar} \psi_n(x) 
\label{eq:timedep}
\ee
and these automatically satisfy the SE. So the problem is reduced to one real task: finding all the  wave functions $\psi_n(x)$ that solve to the equation
\be
-\f{\hbar^2}{2 m}\f{d^2}{dx^2}\psi_n(x)+\f{1}{2}m\omega^2 X^2\psi_n(x)=E_n \psi_n(x).
\ee
Now there is bad news and good news. The bad news: unlike for the particle in a box, we now have a nasty potential term in the Hamiltonian. This makes the above equation is not easy to solve. The good news: there is a neat trick to solve the problem in an elegant fashion. Here goes.

\subsection{A neat trick}
Let us define the following operators:
\bea
a &=& \sqrt{m\omega \over 2\hbar} \left(X + {i \over m \omega} P \right) \nonumber\\
a^{\dagger} &=& \sqrt{m \omega \over 2\hbar} \left( X - {i \over m \omega} P \right)
\eea
\be
(a)^\dag =  \sqrt{m\omega \over 2\hbar} \left(X^\dag + \left({i \over m \omega}\right)^* P^\dag \right) = \sqrt{m \omega \over 2\hbar} \left( X - {i \over m \omega} P \right) = a^\dag
\ee
In the first step, we used rule (\ref{eq:LCHermitian}) for Hermitian conjugation, namely 
\be
(a A+ b B)^\dag= \bar{a}A^\dag+\bar{b}B^\dag \quad\textrm{for all numbers $a$, $b$ and operators $A$, $B$.}
\ee
and in the second step, we used the fact that $X^\dag=X$ and $P^\dag=P$, a property you can easily verify (see exercises) - but just buy it for now.
If we now compose those operators into a new operator $a^\dag a$, we get:
\bea
a^\dag a&=& {m\omega \over 2\hbar}  \left(X + {i \over m \omega} P \right) \left( X - {i \over m \omega} P \right)\nonumber\\
&=&{m\omega \over 2\hbar} \left(X^2 +{i \over m \omega}(XP-PX)+P^2\right)\nonumber\\
&=&{m\omega \over 2\hbar} \left(X^2 +{\hbar \over m \omega}(X D - D X)-\f{\hbar^2}{m^2\omega^2} D^2\right)
\eea
In the last line we did not write the index $x$ under the differential operator $D$ - since it is clear we mean differentiation with respect to $x$.
If $X$ and $D$ were ordinary numbers, the object $XD-DX$ would be zero. But they are \textit{operators}, and hence $XD$ is not necessarily the same as $DX$! Lets think about this more careful. The operator $XD$ takes the derivative of a function, and then multiplies by the identity function $x$. 
\be
XD: f\rightarrow X D f  =  x f'
\ee 
But operator $DX$ does something else: \textit{first} it multiplies by $x$, \textit{then} it takes the derivative of that object. Using the product rule of differentiation, we get
\be
DX: f\rightarrow DXf = D (x \cdot f)=f + x f'
\ee
So $XD$ and $DX$ are really different operators. In general, for operators $A$ and $B$, the object $AB-BA$ is called the \textbf{commutator}, and is denoted by $[A,B]$:
\be
[A,B]=AB-BA
\ee
If the commutator of two operators is zero, then $AB=BA$, and we say the two operators \textbf{commute}. In words: for commuting operators their order doesn't matter. If the commutator of two operators is not zero, we say they \textbf{don't commute}, and then $AB\neq BA$. Ok, back to the above. As $XD\neq DX$, we see that $D$ and $X$ don't commute. In fact, for any $f$ we have
\be
(XD-DX) f= x f' -(f + x f')=-f = -1\cdot f
\ee
So here the commutator is given by:
\be
[X,D]=-1
\ee
where $1$ is the \textbf{unity operator} we met before. Using this, we conclude that 
 \be
a^\dag a= {m\omega \over 2\hbar} \left(X^2 - {\hbar \over m \omega}+\f{\hbar^2}{m^2\omega^2} D^2\right)
\ee
So the Hamiltonian can be written as
\be
H= \hbar \omega\left(a^\dag a+\f{1}{2}\right)
\label{eq:Haa}
\ee
Woah, what a coincidence! In terms of the a's, the Hamiltonian looks quite simple. With a similar calculation to the above, you can check that 
\be
[a,a^\dag]=1.
\label{eq:aa}
\ee
And from (\ref{eq:Haa}) and (\ref{eq:aa}) one can show
\be
\left[H , a \right]  = - \hbar \omega a \quad \textrm{and} \quad \left[H , a ^\dagger\right] =   \hbar \omega a^\dagger
\label{eq:Ha}
\ee
Giving us even more remarkable results. But what do we get from all these formulas? Hold on for a moment, we will extract some important consequences ater the next section. First, we need one last sidestep.

\subsection{Sidestep: getting familiar with bras} 
Suppose you have a state $|\psi\rangle  $. Then $a|\psi\rangle  $ is another state. Here is a question: what is the \textit{bra} state corresponding to this ket state $a|\psi\rangle  $? Recall that the bra and ket are related by Hermitian conjugation:
\be
\langle \psi| = |\psi\rangle  ^\dag
\ee
So the bra version of $a|\psi_E\rangle  $ is given by $(a|\psi\rangle  )^\dag$. From formula (\ref{eq:reverse}) we know that for two operators $(AB)^\dag = B^\dag A^\dag$. It turns out that for the product of a state and an operator, the \textit{same} property holds. We will not prove this, and use it as a given rule. In particular this means that
\be
(a|\psi\rangle  )^\dag = (|\psi\rangle  )^\dag a^\dag = \langle \psi | a^\dag
\ee
So the object on the right hand side is the bra version of the ket $a|\psi\rangle  $. Now, recall that the norm squared of a state is obtained by putting the bra and the ket together: $\| \,\,|\psi\rangle  \,\, \|^2 = \langle \psi|\psi\rangle  $. This means that the squared norm of the state $a |\psi\rangle  $ is given by
\be
\| \, a |\psi\rangle   \|^2 = \langle \psi| a^\dag a |\psi\rangle  
\label{eq:adagnorm}
\ee
You can read the object on the right hand side in different ways. Either you see it as the bra $\langle \psi| a^\dag$ followed by the ket $a |\psi\rangle  $, or you see it as the operator $a^\dag a$ sandwiched inside the $\langle \psi|$ and $|\psi\rangle  $. What viewpoint you take does not really matter - all really describe the same object. In mathematical terms: building objects like in the above line is an associative operation: it does not matter which objects you want to group or put around brackets, so
\be
(\langle \psi| a^\dag)  ( a |\psi\rangle  ) = \langle \psi| (a^\dag a) |\psi\rangle   = \langle \psi| (a^\dag a |\psi\rangle  ) =...
\ee
and so on. This might seem a bit suspicious at first, so a comparison with ordinary vectors and matrices might be in place. Imagine you want to compute the norm of the vector $v$, with 
\be
v= \left(\begin{array}{cc} 1 & 0 \\ i & 1 \\ \end{array} \right) \left( \begin{array}{c} 1 \\ 0\\ \end{array}\right) 
\label{eq:vee}
\ee
You can do this in two ways. Of course, you can compute that $v =  \left( \begin{array}{c} 1 \\ i \\ \end{array}\right)$ and hence the norm is given by
\be
\|v\|^2 = v^\dag v =  \left( \begin{array}{cc} 1 & -i \\ \end{array}\right)  \left( \begin{array}{c} 1 \\ i \\ \end{array}\right) = 2
\ee
However, you can also directly take the conjugate of (\ref{eq:vee}), so that
\be
\|v\|^2 = v^\dag v = \left( \begin{array}{cc} 1 & -i \\ \end{array}\right)   \left(\begin{array}{cc} 1 & -i \\ 0 & 1 \\ \end{array} \right) \left(\begin{array}{cc} 1 & 0 \\ i & 1 \\ \end{array} \right) \left( \begin{array}{c} 1 \\ 0\\ \end{array}\right) 
\ee
If you compute the above line, you will again find 2 as an answer. But what we want to stress, is that it does not matter in what order you perform the matrix products. You can start out left, or you can multiply the matrices in the middle first, whatever you like. So here too, you can view the object in different (equivalent) ways, thanks to associativity. The situation with (\ref{eq:adagnorm}) is precisely the same.

\subsection{The fling}
We can now start using all the above. Suppose there is a (normalized) state $|\psi_E\rangle $ that is an energy eigenstate (=eigenstate of the Hamiltonian) with energy $E$. Then $a|\psi_E\rangle  $ is another state, and from the above we know that its norm is given by $\langle \psi_E|a^\dag a|\psi_E\rangle  $. Because the norm of that state is positive (this is always the case), we must have
\be
0\,\,\leq\,\,\langle \psi_E |a^\dag a|\psi_E\rangle  
\ee
Expressing $a^\dag a$ from the above expression in terms of the Hamiltonian, we get
\bea
0\,\,&\leq&\,\, \left\langle \psi_E \left| \left({H \over \hbar \omega} - {1 \over 2}\right) \right|\psi_E\right\rangle \\ \label{eq:may}
&=&\left\langle \psi_E \left| \left({E \over \hbar \omega} - {1 \over 2} \right) \right|\psi_E\right\rangle \\
&=&\left({E \over \hbar \omega} - {1 \over 2} \right)  \langle \psi_E|\psi_E\rangle    \\
&=&\left({E \over \hbar \omega} - {1 \over 2} \right) \label{eq:look}
\eea
In the first step we have used that $|\psi_E\rangle  $ is an eigenstate of $H$ with eigenvalue $E$ (by assumption). Then, we dragged some things out of the bracket to the left: this is OK because they are all just numbers.
The last step then uses the normalization of $|\psi_E\rangle$. In fact, we have also used the linearity property of the bracket implicitly in the first two steps. If you now compare the first and last part, you see that $E \ge \hbar \omega / 2$. So every state has an energy of \textit{at least} $\f{1}{2}\hbar \omega$. That is already a nice result. But there is more. What if we act with the Hamiltonian on this state $a |\psi_E\rangle  $?
\bea
H (a \left| \psi_E \right\rangle)
 &=& (\left[H,a\right] + a H) \left|\psi_E\right\rangle \nonumber\\
 &=& (- \hbar\omega a + a E) \left|\psi_E\right\rangle \nonumber\\
 &=& (E - \hbar\omega) (a\left|\psi_E\right\rangle)\nonumber
\eea
The second step uses (\ref{eq:Ha}). In the last step, we dragged $E$ to the left. This is OK because it is just a number, so there is no problem when you pass it to the other side of an operator like $a$.  Comparing the first and last part, you see the state $a \left| \psi_E \right\rangle$ is an eigenstate of the Hamiltonian with energy $E-\hbar \omega$. So it is a state with one `step'($=\hbar \omega$) energy less than $|\psi_E\rangle  $. Similarly, you can show that
\be
H (a^\dagger \left| \psi_E \right\rangle) = (E + \hbar\omega) (a^\dagger \left| \psi_E \right\rangle)
\ee
Meaning that $a^\dag \left| \psi_E \right\rangle$ is an energy eigenstate as well, but with one `step' more energy. This means that (given one energy eigenstate) we can build many more energy eigenstates, with lower or higher energy, by acting with the operators $a^\dag$ and $a$. That's why $a^\dag$ and $a$ are called the \textbf{raising and lowering operator}, as they raise and lower the energy of a state by one energy step $\hbar \omega$. 

Now there might seem to be a contradiction here: by acting repeatedly with $a$, we can get a series of states with lower and lower energy, but we just found that the minimum energy is $E_{\textrm{min}}=\f{1}{2}\hbar \omega$. This can only be reconciled if there is some state $|0\rangle  $, for which 
\be 
a \left| 0 \right\rangle = 0 
\ee
so that the lowering process ends there. The right hand side here is the \textbf{zero ket}, a state with  wave function equal to zero, meaning there is no particle at all. Any  subsequent application of the lowering operator will just give the zero ket again, instead of additional energy eigenstates. So the state $|0\rangle  $ necessarily is the state with lowest energy:
\be
H\left|0\right\rangle = \frac{1}{2}\hbar\omega \left|0\right\rangle
\ee
Starting from this lowest energy state we can now build up a series of states (using $a^\dag$), which have higher and higher energy. (We may have to normalize them properly, but that does not affect their energy.) We denote this tower of states by
\be
\left\{\left| 0 \right \rangle, \left| 1 \right \rangle, \left| 2 \right \rangle, ... , \left| n \right \rangle, ...\right\}
\ee
Their energies are given by
\be
H \left|n\right\rangle = \hbar\omega \left(n + \frac{1}{2}\right) \left|n\right\rangle .
\ee
So, in a very elegant way we have solved the spectrum of the harmonic oscillator. We have constructed a series of energy eigenstates, and know that these are the only ones around. The time-dependent solutions are given by multiplying each state $|n\rangle  $ by $e^{- i E_n t /\hbar}$, like in (\ref{eq:timedep}). So this solves the problem of describing a particle in a harmonic potential.

Of course, you may have some difficulties imagining what these states \textit{look like}. If so, you will find relief in the next section.

\subsection*{Their shape}
Maybe we should start with the lowest energy state. The lowest energy state satisfies $ a \left| 0 \right\rangle = 0 $ so by definition of $a$ the wave function $\psi_{n=0}(x)$ corresponding to $|0\rangle$ satisfies
\be
x\psi_0(x) + \frac{\hbar}{m \omega} \frac{d \psi_0}{dx}(x) = 0
\ee
which can be shown to be solved by the  wave function
\be
\psi_0(x)\propto e^{-{m\omega \over 2\hbar}x^2}
\ee
or (after normalization)
\be
\psi_0(x)= \left({m\omega \over \pi\hbar}\right)^{1 \over 4}e^{-{m\omega \over 2\hbar}x^2}.
\ee
and its time evolution is given by
\be
\psi_0(x,t)= \psi_0(x) e^{-iE_0t/\hbar}=  \psi_0(x) e^{-i\omega t/2}
\ee
In a similar fashion, one can explicitly find the  wave function of the other energy eigenstates $|2\rangle  $, $|3\rangle $, etcetera. This is a bit tedious, and we will just show the result pictorially: 
\begin{figure}[H]
 \begin{center}
  \includegraphics[width=0.6\textwidth]{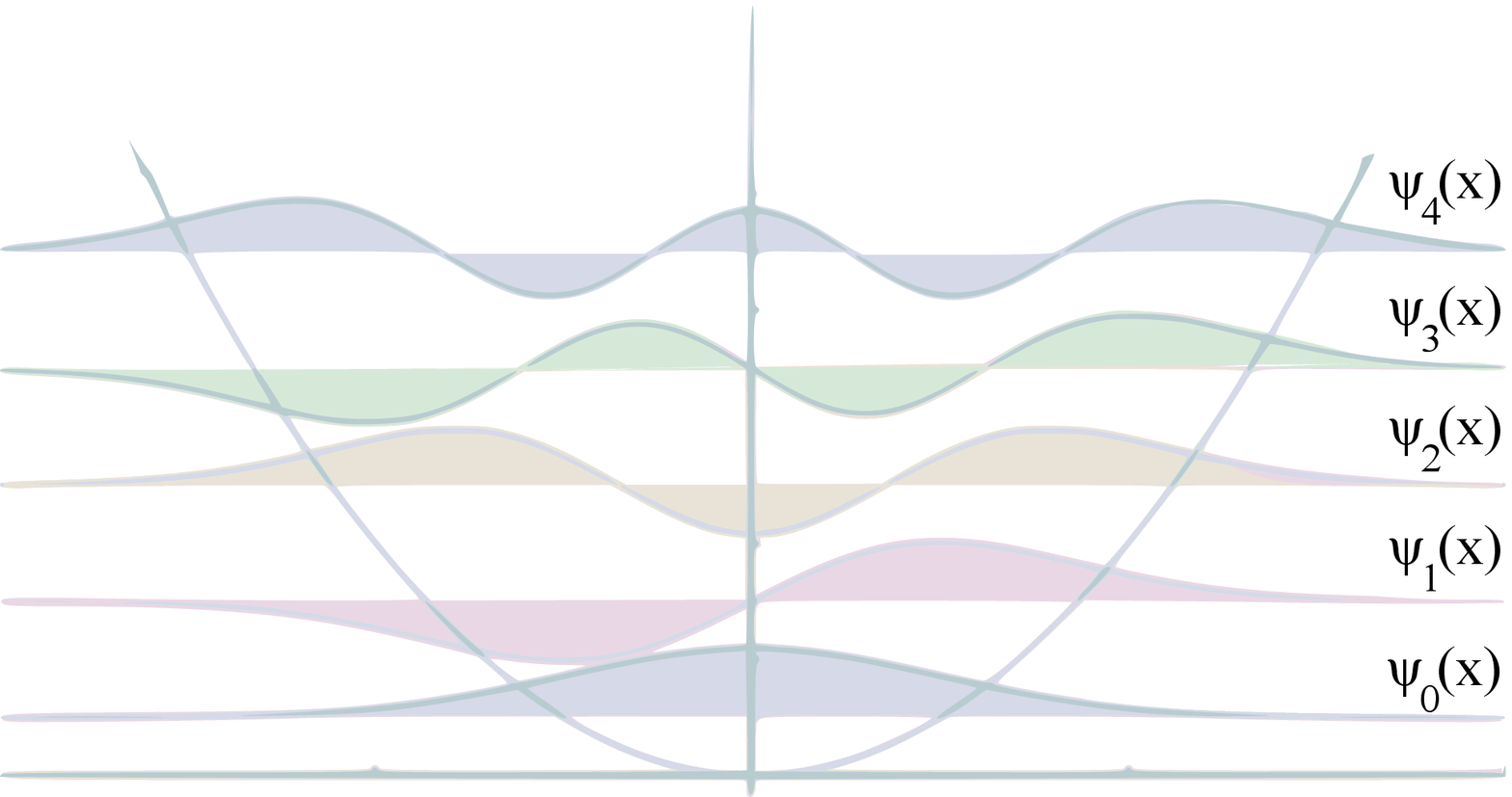}
  \label{fig:SHO}
  \end{center}
\end{figure}
What about measuring the position of a particle in one of these states? As you know, the chance to get an outcome $x$ from a measuring device is proportional to 
$|\psi(x)|^2$. From the figure you see $\psi_n(x)$ becomes more and more spead out with increasing $n$. So for higher energy states, it becomes more and more likely to measure the particle to be far away from the origin. Intuitively, the particle spreads more throughout the potential as it gains energy, somewhat like a classical oscillator gaining amplitude with increasing energy. What about the \textit{expectation value} of the position? This is given by
\be
\langle X\rangle  _{\psi_n} = \inft x |\psi_n(x)|^2 dx
\ee
As you see from the figure, the  wave functions $\psi_n(x)$ are all symmetric or anti-symmetric: $\psi_n(x) = \pm \psi_n(-x)$. Hence the squared amplitude has $|\psi_n(x)|^2 = |\psi_n(-x)|^2$. So the integrand $x |\psi_n(x)|^2$ above is an anti-symmetric function: the value at $x$ and $-x$ are exactly opposite. Such functions have integral zero, since the contributions from the positive and negative integration domain exactly cancel. So
\be
\langle X\rangle  _{\psi_n} = 0  \quad \forall n
\ee
So we find that the expectation value of the position measurement is zero, for all states. This is not so surprising: after all the system is left- right symmetric. Of course, this does not mean you will always get $0$ as an outcome of the measurement: the above is just a `weighted average' - which happens to be zero. 

\subsection{Conclusion}
Summarising, there are several special aspects to this system. First, it is remarkable that the lowest energy is not zero. Classically, a particle at rest in the potential has zero kinetic and potential energy, and hence zero total energy. For a  wave function, this turns out not to be the case: the lowest energy it can reach is $\f{1}{2} \hbar \omega$. Some people express this (in a somewhat obscure, but poetic phrasing) as: ``Quantum mechanically, a harmonic oscillator can never be at rest." By this they just mean: ``A particle in a harmonic potential can not have zero energy." Besides that, we also see some similarities with the particle in a box. Again, only certain energy levels are admitted. There is no state with energy $3.1 \hbar \omega$ for example. This means the transitions between different energy levels of the particle can release only specific amounts of energy: only multiples of $\hbar \omega$. 
In the classical theory of (point-like) electrons and electromagnetism, an electron that is harmonically oscillating at frequency $\omega$ emits radiation with frequency $\omega$, hereby lowering its energy. However, this loss of energy can occur in arbitrary small steps. According to the correct (quantum mechanical) description, the situation is different. Here the particle can emit energy too, but it is clear that this radiation should always have energy $n \hbar \omega$ for some natural number $n$. Do you smell $n$ photons of frequency $\omega$ coming out of the particle? Nice, huh?

You may now understand better where the name `Quantum Mechanics' comes from. In the situation here, for the particle in a box and in many more examples we will see that discrete amounts of energy are typically involved. We meet discrete energy levels, a photon is a discrete package of energy, and so on. The word `quantum' (derived from Latin) precisely means `discrete amount' or `small package'. So Quantum Mechanics is in essence the description of systems involving discrete amounts of energy: the world of particles.
\newpage
\section*{Exercises}
\begin{enumerate}
\item Prove (\ref{eq:Adagdag}), (\ref{eq:LCHermitian}) and (\ref{eq:reverse}). (Hint: you'll need the conjugate symmetry and bilinearity.)
\item Compute $[X^2,D_x]$.
\item Prove that if an energy eigenstate is normalized, then it will still be normalized at any later moment in time. (Just write down its time evolution and take the norm.)
\item Verify $\left[H , a \right]  = - \hbar \omega a \quad \textrm{and} \quad \left[H , a ^\dagger\right] =   \hbar \omega a^\dagger$.
\item Compute $\psi_1(x)$ from the expression for $\psi_0(x)$. (You don't need to normalize it.) Try to draw this function. Does it match the graph shown in section \ref{fig:SHO}? What is the time-evolution of this state?
\item Imagine you can create an electric field of the form $E= -  \kappa x$ with $\kappa = 100 V/m^2$. (Check that the units are right.) Put a proton ($m_p=1,67\cdot 10^{-27}$ kg, $q_p=+1,6\cdot10^{-19}$ C) in that field - it will stay trapped around $x=0$. What frequency should a photon have to be able to kick up the proton from the lowest to the next-to-lowest energy level? 
\item Prove that the operators $X$ and $P$ are Hermitian. (For the second: think of partial integration. Why can you drop the boundary terms for a wave packet?)
\end{enumerate}

%% file: H4.tex
\chapter{Observables}
\subsection*{In this chapter...}
In the previous chapter, we learned how a position measurement works out in quantum mechanics. In this chapter, we will extend this further, and explain how general measurements (such as the energy or velocity of a particle) work. Doing so, we will learn more about the space of states of a particle. We first introduce the concepts `Hermitian operator' and `state basis'. Then follows the story part (which generalizes the previous chapter to other measurements) and we conclude by a simple example.

\newpage

\section[Hermitian operators and the Hilbert space]{\includegraphics[width=0.04\textwidth]{tool_c} Hermitian operators and the Hilbert space}

\subsection{Hermitian operators}

You know by now that to each operator $A$, we can associate another one, its Hermitian conjugate $A^\dag$. It turns out that for some special operators, the following holds:
\be
A=A^\dag
\ee
By this we just mean that for every function $f$, the outputs $A f$ and $A^\dag f$ are equal. Such operators are called \textbf{Hermitian operators}. As we will learn in this chapter, such operators are very important in quantum mechanics. One special property they exhibit, is that they only have \textbf{real eigenvalues}. Indeed, suppose a Hermitian operator $L$ has an eigenvalue $\lambda$. Then there exists a function $f$ such that $L f = \lambda f$, and 
\be
\overline{\lambda} \langle f,f\rangle  = \langle \lambda f, f\rangle  = \langle L f, f\rangle  = \langle f,L f\rangle  = \lambda \langle f,f\rangle .
\ee
In these steps we used bilinearity of the inner product, Hermiticity of $L$, and the fact that $f$ is an eigenfunction.
Since $\langle f,f\rangle $ is nonzero (true for every nonzero $f$) the left and right side of the above equation gives $\overline{\lambda}=\lambda$, so $\lambda$ must be real. This shows that every eigenvalue $\lambda$ of a Hermitian operator indeed has to be real. 
Besides that, an important property of Hermitian operators is that \textbf{their eigenfunctions are orthogonal}. By this we mean that eigenfunctions with different eigenvalues have an inner product equal to zero. Indeed, imagine functions $f_1$ and $f_2$ with eigenvalues $\lambda_1$ and $\lambda_2(\neq\lambda_1)$ under a Hermitian operator $L$. Then
\be
\lambda_2 \langle f_1,f_2\rangle  =\langle f_1,L f_2\rangle  = \langle  L f_1,f_2\rangle = \overline{\lambda_1}\langle f_1,f_2\rangle =\lambda_1\langle f_1,f_2\rangle 
\ee
In the last step we used the fact that an eigenvalue $\lambda_1$ has to be real. Once again combining the first and last side of the equation, we now get
\be
(\lambda_2-\lambda_1)\langle f_1,f_2\rangle  = 0
\ee
Which (as $\lambda_2\neq\lambda_1$) means that $\langle f_1,f_2\rangle  = 0$, so indeed $f_1$ and $f_2$ are orthogonal.

\subsection{A space of states}
In the previous chapters, we solved two systems: a particle in a box and a particle in a harmonic potential. In either case, we found an infinite tower of states. Each state describes a specific wave function the particle can be in. You may wonder what the meaning is of \textit{sums} of these wave functions - or in general: any linear combination. These are complex functions as well. And they satisfy the same good properties: they are smooth and have a decent fall-off (so they can be normalized). On these grounds, we could be tempted include these combinations as valid wave functions as well. Doing so, the space of states becomes a complex vector space! That space is called the \textbf{Hilbert space}. So if $\psi_1(x)$ and $\psi_2(x)$ are the two wave functions (i.e.: elements of the Hilbert space) then
\be
\chi(x) =a \psi_1(x) + b \psi_2(x)
\ee 
is a wave function as well, for any value of the complex numbers $a$ and $b$. You may still need to normalize this wave function $\chi(x)$, but that is just a matter of multiplying by a constant - not really a deep issue. In the language of kets: 
\be
|\chi\rangle  = a |\psi_1\rangle  + b |\psi_2\rangle 
\ee
gives a new state $|\chi\rangle $ built from the states $|\psi_1\rangle $ and $|\psi_2\rangle $. So if you take the sum of two kets, it is understood that you mean the state whose wave function is the sum of the wave functions:
\be
 a |\psi_1\rangle  + b |\psi_2\rangle  = |a \psi_1 + b \psi_2\rangle 
\ee
Of course, we can also take linear combinations of more than two elements. This way, all linear combinations of the original set of states become elements of the Hilbert space. So the tower of states we found for the harmonic oscillator and the particle in a box form a \textbf{basis} of the Hilbert space. 

From a `the more the merrier' point of view, the above is not really an unpleasant fact. However, something might bother you. We were very proud that we could explicitly write down the time-evolution of the states we found. But now we have a whole new zoo of states that we have to consider. How do all these evolve in time? This turns out to work out nicely. Let us look at the Schr\"{o}dinger equation once again:
\be
i \hbar \partial_t \psi(x,t) = H \psi(x,t)
\ee
This differential equation is of a special kind: it is a \textbf{linear} differential equation. That means that if you have two solutions satisfying it, every linear combination satisfies it too. Let us make this more concrete. Say at $t=0$ we consider two wave functions $\psi_1(x,0)$ and $\psi_2(x,0)$ that are energy eigenfunctions with values $E_1$ and $E_2$. We know that their time-evolution (dictated by the SE) is just given by
\be
\psi_1(x,t) = \psi_1(x,0) e^{-iE_1 t /\hbar} \quad \psi_2(x,t) = \psi_2(x,0) e^{-iE_2 t /\hbar} 
\ee
But what if at $t=0$ we start out with the state $\chi(x,0) = a \psi_1(x,0) + b \psi(x,0)$. How does this evolve in time: what is $\chi(x,t)$? A priori this could be everything, but because of linearity, the function
\be
\chi(x,t) = a \psi_1(x,t) + b\psi_2(x,t) = a  \psi_1(x,0) e^{-iE_1 t /\hbar} + b \psi_2(x,0) e^{-iE_1 t /\hbar} 
\ee
\textit{automatically} satisfies the SE since $\psi_1(x,t)$ and $\psi_2(x,t)$ do. In conclusion: the time evolution of a linear combination is just the linear combination of the time evolutions. This means we do not need to do any extra work when describing the entire Hilbert space. Time for the story...

\section[Observables]{\includegraphics[width=0.04\textwidth]{once_c} Observables}

In the previous chapter, we learned about the probability amplitude and how measuring a particle's position deforms its wave function. Of course, besides position there are many other properties of a particle that you may measure: energy, velocity, angular momentum, or any combination like `position times momentum' or `position squared' and so on. Every quantity of a particle that you can measure (or `observe') is called an \textbf{observable}. It turns out that every observable is associated to a unique operator. So there is a strict correspondence between the numerical quantities we can observe in experiments with particles, and operators acting on the space of wave functions (Hilbert space). Here is a list of pairs we have met so far:
\bea
\textrm{energy (E)}&\leftrightarrow& \textrm{Hamiltonian operator}\,\,\, H\nonumber\\
\textrm{position (x)}&\leftrightarrow& \textrm{position operator}\,\,\, X\nonumber\\
\textrm{momentum (p)}&\leftrightarrow& \textrm{momentum operator}  \,\,\,P\,(\,=-i\hbar D_x)\nonumber\\
\eea
Now comes a very nice surprise: it turns out that \textbf{all observables -including the above- are Hermitian}. That is a very nice property: we just learned that the eigenfunctions of a Hermitian operator are orthogonal. If you normalize those states, you get an ortho\textit{normal} basis. 

The value of such an orthonormal basis can not be stressed enough. To see why, think back of an ordinary complex vector space for a little moment. If you have a basis $\{e_1,...e_n\}$ then every element $v$ in the vector space can be written as
\be
v = v_1 e_1 + ... + v_n e_n
\ee
for precisely one combination $\{v_1,...v_n\}$ of complex numbers. (We don't write any vector arrows, out of laziness.) So the advantage of a basis is clear: it allows specifying every element of the vector space uniquely, by a set of numbers. The problem is that (for a given vector) it is not immediately clear how to find the above decomposition. But if a vector space is endowed with an inner product, things are different. One can try to find a \textit{special} basis, namely an \textbf{orthonormal} one. By orthonormal we mean that
\be
\langle e_i,e_j\rangle =\delta_{ij}
\ee
Here $\delta_{ij}$ is the Kronecker delta function, which equals one if $i=j$ and equals zero if $i\neq j$. For such a basis, and any vector $v$ in the vector space,
\be
\langle e_1,v\rangle =\langle e_1,v_1 e_1 + ... + v_n e_n\rangle  = v_1 \cdot 1 + 0 + ...+ 0 = v_1.
\ee
Where we used linearity of the inner product and orthonormality of the basis. So the number $v_1$ (and analogously any of the $v_i$) can be obtained just by taking the inner product of $v$ with $e_1$ (or any of the $e_i$). This is a very strong fact: not only is every vector uniquely specified by a set of numbers, this decomposition can be obtained easily by taking the inner product. The quantities $v_i$ are called the \textbf{projection} of $v$ on $e_i$. They tell you how much $v$ is directed along the vector $e_i$. An example in two dimensions, the vector $\vec{v} = 2 \vec{e}_1+ \vec{e}_2$ and its projections:
\begin{figure}[H]
 \begin{center}
  \includegraphics[width=0.5\textwidth]{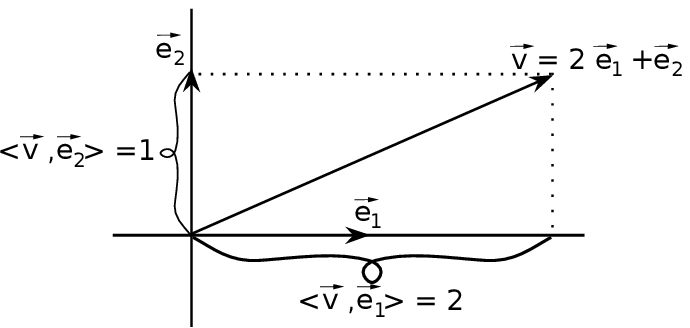}
  \end{center}
\end{figure}

Things are very similar for a Hilbert space. There too, we have a basis. There too, we want to be able to write every element in terms of this basis. But in the context of the Hilbert space, the advantage of an \textit{orthonormal} basis is even more drastic. In that case the set of basis elements is typically infinite. If someone gives you a function, and asks to decompose it as a combination of basis elements, you will have a very hard time. For an orthonormal basis, things are not that harsh: you can obtain that decomposition in a straightforward manner. Indeed, the projections on the basis elements are just given by the inner product, which requires just performing a definite integral. Later on, we will perform some concrete decompositions. What you should remember from this, is that the set of eigenstates of an observable provides a very good basis of the Hilbert space: an orthonormal one. So every observable is a bit like a fairy, it can help making your deepest wish come true.

Think back of -say- the particle in a box for a moment. There we found eigenstates of the Hamiltonian. You now know that this operator is Hermitian, so you conclude that the different energy states we found were orthonormal. Indeed, if you like performing integrals, feel very welcome to try to compute the inner product between different energy states. It will always give you zero. 

Now imagine that -for a given problem- you have found the eigenstates/ eigenfunctions of some observable $O$. What is the physical meaning of these eigenstates? Say f.e. that you have a eigenstate under the operator $O$ with eigenvalue $\lambda$. It is natural to interpret this as \textbf{a state for which the observable corresponding to $O$ has value $\lambda$}. Actually, we used a specific case of this interpretation already: we interpreted eigenstates of the Hamiltonian with eigenvalue $E$ as states with \textit{energy} $E$. Similarly, an eigenstate of the momentum operator $P$ with eigenvalue $p$ will just mean a particle with momentum $p$, etcetera. 

\subsection{Non-eigenstates}
Also the negation of the above interpretation makes sense. A state that is \textit{not} an eigenstate of $O$ is a particle for which the measurable quantity corresponding to $O$ does \textit{not} have a specific value. Let's be more concrete. Suppose the eigenstates of an observable $O$ are given by states $|\psi_n\rangle $, with corresponding eigenvalues $\lambda_n$. If we have also normalized the $|\psi_n\rangle $, then they form an orthonormal basis:
\be
\langle \psi_n|\psi_m\rangle =\delta_{mn}.
\ee
Now a state of the form
\be
|\psi\rangle =\sum_{n} c_n |\psi_n\rangle 
\ee
(with the $c_n$'s complex numbers) is clearly \textit{not} an eigenstate of $O$. (Unless only one of the $c_i$ is nonzero.) If you measure the observably quantity corresponding to $O$ of a particle in the above state, your apparatus will give you any of the values $\lambda_i$ \textbf{at random}, but the probability of getting a specific $\lambda_i$ as an answer is given by
\be
P_\psi(\lambda_i)= \textrm{probability to measure } \lambda_i = \f{|c_i|^2}{\sum_n |c_n|^2}
\label{eq:probability}
\ee 
Just as with measuring the position, the measurement drastically changes the state. If your measurement results in some $\lambda_i$, the state will \textbf{collapse} onto the corresponding eigenstate, so
\be
\textrm{if measurement gives outcome $\lambda_i$ then}\quad |\psi\rangle \rightarrow|\psi_i\rangle 
\ee
Let us look at the denominator in (\ref{eq:probability}) for a moment. This object is just the norm of the state $|\psi\rangle $.
\bea
\langle \psi|\psi\rangle &=&\left(\sum_n \bar{c}_n \langle \psi_n|\right) \left(\sum_m c_m |\psi_m\rangle \right)\\
&=& \sum_n \sum_m \bar{c}_n c_m \delta_{nm}\\
&=& \sum_i |c_i|^2
\eea
From this, it follows that expression (\ref{eq:probability}) does not change if one replaces the state $|\psi\rangle $ by any complex multiple $a|\psi\rangle $ of itself: both the numerator and denominator will scale up with a factor $|a|^2$. This implies that the seemingly different states $|\psi\rangle $ and $a |\psi\rangle $ give the same measurement probabilities, so they represent the \textbf{same physical object}. Stated more roughly: they are just different names for the same thing. This is of course related to the issue of normalization. States $|\psi\rangle $ and $a |\psi\rangle $ may look differently, after normalization, they are the same object.\footnote{To be overly precise: even if one asks $|\psi\rangle $ to be normalized, there is still some freedom of renaming: all states $a|\psi\rangle $ with $a=e^{i\theta}$ (a complex number of unit modulus) are still normalized, and all represent the very same physical state.} 
If we demand $|\psi\rangle $ to be normalized from the beginning, (\ref{eq:probability}) simplifies a bit. In that case the denominator on the right equals one, so that 
\be
\quad \quad \quad P_\psi(\lambda_i)= |c_i|^2 \quad\quad\textrm{(for \textit{normalized} $|\psi\rangle $)}
\ee 
Note that the \textit{total chance} of measuring any of the $\lambda_i$ is one, as should be. The expectation value of the observable $O$ is then given by
\be
\langle O\rangle _\psi=\sum_i  \lambda_i  P_\psi (\lambda_i)= \f{\sum_i \lambda_i |c_n|^2}{\sum_n |c_n|^2} 
\label{eq:expval1}
\ee
You can check that the above can be rewritten in a compact fashion using the \textit{sandwich} of $O$:
\be
\langle O\rangle _\psi=\frac{\langle \psi|O|\psi\rangle }{\langle \psi|\psi\rangle }
\label{eq:expval2}
\ee
which is very similar to the expression for the position expectation value. For a normalized state, the denominator is again one. So this is how the measurement works for observables other than the position. Time for an example!

\section[Working with the Hilbert space]{\includegraphics[width=0.04\textwidth]{comput_c} Working with the Hilbert space}
\subsection{Example}
To illustrate all the above, let's go back to -say- the harmonic oscillator. Recall that we found 
\be
H |n\rangle  = \hbar\omega \left(n + \frac{1}{2}\right) |n\rangle 
\ee
or by using  $H = \hbar \omega \left(a^\dag a+ \frac{1}{2}\right)$, 
\be
a^\dag a |n\rangle  = n |n\rangle 
\ee 
For this reason, the object $a^\dag a$ is called the \textbf{number operator} - it just gives you the state number. (More precise: the $n$-th energy eigenstate is an eigenstate of the number operator as well, with eigenvalue $n$.) If we require that $|n\rangle $ is a normalized state, we have
\be
\langle n| a^\dag a |n\rangle  = n \langle n|n\rangle  = n
\ee
We also found that $a|n\rangle $ is a state with one quantum less energy. So necessarily $a |n\rangle  \sim |n-1\rangle $. There could be a proportionality constant involved here: $|n-1\rangle $ is normalized but $a |n\rangle $ may not be so. Indeed, the above equation precisely shows that $a |\psi\rangle $ has norm squared equal to $n$. So we conclude
\be
a |n\rangle  = \sqrt{n} |n-1\rangle 
\label{eq:lowermat}
\ee
Similarly, acting with $a^\dag$ on this equation and renaming $n\rightarrow n-1$ shows that
\be
a^\dag |n\rangle   = \sqrt{n+1} |n+1\rangle .
\label{eq:othermat}
\ee
Actually, there is a nice way to make all these things more concrete. Let us denote the state $|n\rangle $ by the n-th basis vector:
\be
|n\rangle  \leftrightarrow \left(\begin{array}{c}0 \\ \vdots \\0\\1\\0\\\vdots \end{array}\right)
\ee
where the $1$ is on the n-th position. Obviously, to include all states we need to take the vector on the right hand side to be infinitely long. This is a bit  funny, but the notation works fine. A general state in the Hilbert space is then denoted in vector form as follows:
\be
|\psi\rangle  = \sum_n c_n |n\rangle  \leftrightarrow  \left(\begin{array}{c} c_0\\c_1 \\ c_2 \\c_3 \\ \vdots  \end{array}\right)
\label{eq:genstat}
\ee
We start labeling by $0$ because the state labels start from zero on. In a similar way, we can denote the bra version of states by a row vector. For the state $|\psi\rangle $ above, we have
\be
\langle \psi| \leftrightarrow \left(\begin{array}{cccc} \bar{c}_0 & \bar{c}_1 & \bar{c}_2 & \ldots  \end{array}\right)
\ee 
In this way, the norm squared of a state is just the matrix product:
\be
\langle \psi|\psi\rangle  = \left(\begin{array}{cccc} \bar{c}_0 &\bar{c}_1 & \bar{c}_2 & \ldots  \end{array}\right)   \left(\begin{array}{c} c_0 \\ c_1 \\ c_2 \\ \vdots  \end{array}\right)= \sum_n |c_n|^2
\ee
With this notation, we can represent operators as matrices acting on the vectors. For example, looking at (\ref{eq:lowermat}) it follows that the lowering operator is given by
\be
a\leftrightarrow \left(\begin{array}{cccccc}
0 & \sqrt{1} & 0 & 0 & 0 & \ldots \\
0 & 0 & \sqrt{2} & 0 & 0 & \ldots \\
0 &0  & 0 & \sqrt{3} & 0 & \ldots \\
0 & 0 & 0 & 0 & \sqrt{4} & \ldots \\
\vdots & \vdots &  & \ddots & \ddots & \ddots \\
\end{array}\right)
\ee 
The raising operator is a similar object:
\be
a^\dag\leftrightarrow \left(\begin{array}{cccccc}
0 & 0 & 0 & 0 & 0 & \ldots \\
\sqrt{1} & 0 & 0 & 0 & 0 & \ldots \\
0 &\sqrt{2}  & 0 & 0 & 0 & \ldots \\
0 & 0 & \sqrt{3} & 0 & 0 & \ldots \\
\vdots & \vdots &  & \ddots & \ddots & \ddots \\
\end{array}\right)
\ee
In fact, this is just the Hermitian conjugate of the previous matrix - this just stems from the fact that operators $a$ and $a^\dag$ are Hermitian conjugate operators. If you perform the (infinite) matrix product, you see that $H = \hbar \omega \left(a^\dag a + \frac{1}{2} \right)$ should be written as
\be
H \leftrightarrow \hbar \omega \left(\begin{array}{cccccc}
0 + \frac{1}{2} & 0 & 0 & 0 & 0 & \ldots \\
0 & 1 + \frac{1}{2} & 0 & 0 & 0 & \ldots \\
0 &0  & 2 + \frac{1}{2} & 0 & 0 & \ldots \\
0 & 0 & 0 & 3 + \frac{1}{2} & 0 & \ldots \\
\vdots & \vdots &  & \ddots & \ddots & \ddots \\
\end{array}\right)
\ee
There are several observations here. First, this is a Hermitian matrix. This is actually due to the fact that $H$ is a Hermitian operator. Second, $H$ is a diagonal matrix. This expresses that the basis we use consists of eigenstates of $H$. More than ever, the parallel with linear algebra is clearly present - although here we work with infinitely big matrices. 

Let us now consider a particular state $|\psi\rangle  = |0\rangle  + |1\rangle $: 
\be
|\psi\rangle   =  \left(\begin{array}{c} 1 \\ 1 \\0 \\ \vdots  \end{array}\right)
\ee
This state is not normalized however: $\langle \psi|\psi\rangle  = 2$. So to normalize, we need to divide by a constant:
\be
|\psi\rangle   \rightarrow \frac{|\psi\rangle }{\sqrt{2}} = \left(\begin{array}{c} 1/\sqrt{2} \\ 1/\sqrt{2} \\0 \\ \vdots  \end{array}\right)
\ee
The state equally combines both first energy levels, so if you measure the energy, you expect that there is $\f{1}{2}$ chance to detect an energy $E_0$, and $\f{1}{2}$ for $E_1$. Indeed:
\be
P(E\rightarrow E_0) = |c_0|^2 = \frac{1}{2} = |c_1|^2 = P (E\rightarrow E_1)
\ee
Now say you measured the energy of the particle to equal $E_0$. The state then collapses onto that level: $|\psi\rangle \rightarrow |\psi'\rangle =|0\rangle $ after the measurement. If you measure the energy again, you are sure that you get $E_0$ as an answer again, because now
\be
P(E\rightarrow E_0)=|c'_0|^2=1
\ee
and all the other are zero. Indeed, the wave function collapsed.
Now what was the \textit{original} expectation value of the energy? One-half chance on either of the energies means that
\be
\langle E\rangle _\psi=\f{E_0+E_1}{2} =\frac{\frac{1}{2} \hbar\omega +\frac{3}{2}\hbar\omega }{2} = \hbar \omega
\ee
just their average. Is this equal to the sandwich of $H$ inside $|\psi\rangle $? Yes it is, 
\bea
\langle \psi|H|\psi\rangle &=&\f{1}{2}(\langle 0|+\langle 1|)H(|0\rangle +|1\rangle )\nonumber\\
&=&\f{1}{2}(\langle 0|+\langle 1|) (E_1|0\rangle +E_2|1\rangle ) \nonumber\\
&=& \f{1}{2} (E_0+E_1).
\eea
For the time evolution, we note that the time dependent version of $|0\rangle $ and $|1\rangle $ are given by
\be
e^{-i E_0 t/\hbar} |0\rangle  \quad \textrm{and }\quad e^{-i E_1 t/\hbar} |1\rangle  
\ee
such that
\be
|\psi(t)\rangle  = \frac{1}{\sqrt{2}} e^{-i E_0 t/\hbar} |0\rangle  + \frac{1}{\sqrt{2}} e^{-i E_1 t/\hbar} |1\rangle  
\ee
is the time evolution of the initial state $|\psi\rangle $.

A last remark: the above implies that the energy (or any other observable) of a particle needs not be sharply defined: a particle may be `smeared out' over several values of this observable. This is very similar to the fact that a wave function does not have a sharply defined position: simply because it is spread out over some region in space. This does not mean the quantum description is fuzzy or so: \textit{states} of particles are well defined; it's just the naive point particle idea (where position, energy, ... are all unambiguously defined) which you need to get rid of.

\subsection{Recap: the postulates of quantum mechanics.}

Some people like to summarize. Some people like to highlight. For those souls, get your sharpened markers out, we are about to summarize this and all previous chapters in just one section. Why will we do so? Well, the summary below is actually due to the pioneers of quantum mechanics, who had been trying to understand the theory so hard that -on the moment they got it all straight- they couldn't resist listing the set of rules they met on the way. 
This list bears the pompous name \textbf{`the postulates of quantum mechanics'}. These postulates are true not only for the examples we have done so far, but still hold for all other examples and situations we will describe in this course. So, OK, maybe they did have some reason to pick such a solemn name. Here goes.\\
\\
When we wish to describe a particle quantum mechanically, we use the following rules:
\subsubsection*{Postulate 1} All possible states form a complex vector space, the Hilbert space. Complex multiples of a state $|\psi\rangle $ correspond to the \textit{same} physical state, and we prefer working with the normalized version: $\langle \psi|\psi\rangle =1$. 

\subsubsection*{Postulate 2}
The time evolution of a state is given by the Schr\"{o}dinger equation
\be
i \hbar \partial_t |\psi\rangle  =H |\psi\rangle 
\ee
where $H$ is the Hamiltonian. 

\subsubsection*{Postulate 3}
Every observable (=measurable quantity of the system) corresponds to a Hermitian operator.

\subsubsection*{Postulate 4}
If the system is in a state that is an eigenstate of an observable $O$, with eigenvalue $\lambda$, then measurement of the quantity corresponding to $O$ will yield $\lambda$ as an outcome. The eigenstates $|\psi_i\rangle $ of an observable $O$ form an orthonormal basis of the Hilbert space, so every state in the Hilbert space can be written as
\be
|\psi\rangle =\sum_i c_i |\psi_i\rangle .
\ee
For such a (general) state, the outcome of measuring $O$ is probabilistic. The chance that the measuring device gives $\lambda_i$ as an answer is given by
\be
P_\psi(\lambda_i)=\frac{|c_i|^2}{\sum_n |c_n|^2}
\ee
(If the state is properly normalized, the denominator falls away (it's one) and the above expression looks somewhat simpler.) After measuring $O$, the state of the system collapses into the eigenstate corresponding to the measured eigenvalue.
\be
|\psi\rangle \rightarrow \textrm{measuring O results in some $\lambda_a$} \rightarrow \textrm{state collapses into } |\psi_a\rangle 
\ee

\subsubsection*{Postulate 5} 
The probabilistic nature of the measurement means we can associate to a general state not an exact value of an observable $O$, but merely an expectation value. This is the average result one would obtain when measuring $O$ again and again, each time making sure to put back the system in its original state. The expectation value is given by the sandwich of the operator $O$ inside the state:
\be
\langle O\rangle _\psi=\frac{\langle \psi|O|\psi\rangle }{\langle \psi|\psi\rangle }
\ee
(Again, if the state is normalized, the denominator can be left out.)

\subsection{Conclusion}
That's it! These are the laws of quantum mechanics. If you recognized all the above postulates (they all showed up earlier), and understood their meaning, congratulations. To get here was the hardest task of this course. From now on, we will use the developed framework to get more insight into the quantum-world. So we will make the things we have learned more concrete, and extend to more general situations. What does this last thing mean? Well, so far we have been considering a particle living in one dimension $x$. A real particle (as described in the next chapter f.e.) lives in 3 dimensions, $x$, $y$, $z$. So to be more precise, particle states are wave functions $\psi(x,y,z)$. Also, we have not yet lived up to our promise to explain the structure of atoms. These and other things will be done in the next chapters. So some other things will be added, we will do more examples, but the framework (the postulates above) stays exactly the same. Even the most involved computations in particle physics are, in fact, a mere application of the above rules. 
Nice, no?

\newpage
\section*{Exercises}
\begin{enumerate}
\item Without looking at the text, try to recall as much from the postulates of quantum mechanics as you can.
\item The number operator for a particle in a harmonic potential is a Hermitian operator. What can you conclude about the eigenvalues of this operator. Is this indeed the case?
\item For the harmonic oscillator, write out the operators $X$ and $P$ in (infinite) matrix form. Compute the expectation values of $\langle n|X|n\rangle $ and $\langle n|P|n\rangle $. Did you expect this result? (Note that we `computed' $\langle n|X|n\rangle $ already in the previous chapter. Your computation here was a bit more precise.) By taking matrix products, find $\langle|XP|\rangle$.
\item Write down your favorite quantum state of a particle in a harmonic potential that is \textit{not} and energy eigenstate. What is the expectation value of the energy for this state? 
\item Consider an observable $O$ (which does not involve time explicitly) and a state $|\psi(t)\rangle $ (which is not necessarily an eigenstate of $O$ or the Hamiltonian). Obviously, $\langle \psi|O|\psi\rangle $ may depend on time, since the state depends on time. Show that $$\partial_t\langle \psi(t)|O|\psi(t)\rangle =\frac{i}{\hbar}\langle \psi(t)|[H,O]|\psi(t)\rangle. $$ (Hint: you'll need the Schr\"{o}dinger equation.) By taking $O$ to be the unity operator, prove that the norm of a state does not change in time.
\end{enumerate}

%% file: H5.tex
\chapter{The hydrogen atom}

\subsection*{In this chapter...}
One of the main motivations of the development of quantum mechanics, also mentioned in the introduction, was getting a better understanding of atoms and molecules. How can we explain the behavior of these crucial building blocks of nature? Nice enough, we have already collected the necessary pieces to start answering this puzzle: in this chapter, we will describe in detail the most elementary atom: hydrogen. 

We start with some more this-and-thats on Hilbert spaces, then comes the usual bite of (hi)story, and in the last part we get down to business - describing the hydrogen atom quantum mechanically.
\\
\\
\\

\begin{figure}[H]
 \begin{center}
  \includegraphics[width=0.3\textwidth]{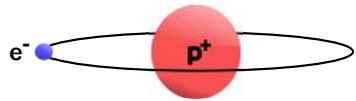}
  \caption{A naive depiction of the atom. This chapter, we will learn better.}
  \end{center}
\end{figure}

\newpage

\section[Operators in three dimensions]{\includegraphics[width=0.04\textwidth]{tool_c} Operators in three dimensions}
\subsection{3D sensation}
In the previous chapter, we used only one spatial variable, $x$, on which wave functions could depend. That describes particles living in one dimension only, so on a line. Doing so makes life easy, and such a description can even be quite good for systems where the $y$ and $z$ dependence is trivial. But what if one really wants to describe a wave function propagating in three dimensions? In such case, we need to let the wave function depend on three coordinates
\be
\psi=\psi(x,y,z).
\ee
So that is just a minor change. As there are more variables now, we need to be a bit more careful with derivatives. To avoid confusion, we will write an index for each differential operator:
\be
D_x=\f{\partial}{\partial x} \quad D_y=\f{\partial}{\partial y} \quad D_z=\f{\partial}{\partial z}.
\ee
As the order of taking derivatives never matters, these differential operators all commute:
\be
\f{\partial}{\partial x} \f{\partial}{\partial y} f= \f{\partial}{\partial y}\f{\partial}{\partial x} f\Rightarrow [D_x,D_y]=0 
\ee 
and similarly $ [D_x,D_z]=[D_y,D_z]=0$. Now to each derivative, we can associate a corresponding momentum operator:
\be
P_x= - i \hbar D_x \quad P_y= - i \hbar D_y \quad P_z= - i \hbar D_z 
\ee
This makes sense, as for a particle in 3d you can measure three different components of the momentum. Using the \textbf{grad(ient) operator} $\vec{\nabla}=(\partial_x,\partial_y,\partial_z)$, the above formula can be written in vector form:
\be
\vec{P} = - i \hbar \vec{\nabla}
\ee
So here left- and right hand side are \textbf{vectors of operators}. Some people are a bit lazy, and drop the vector arrows on top. We will do so too, sometimes. 

Experimentally, one can also measure the \textit{total size} of the momentum of a particle. (Regardless its direction.) Which operator corresponds to this observable? Classically, the square of the total momentum is given by
\be
|p|^2= p_x^2+p_y^2+p_z^2.
\ee
This suggests that the corresponding quantum operator is given by
\be
P^2 = P_x^2+P_y^2+P_z^2 = - \hbar^2 \nabla^2 = - \hbar^2 (\partial_x^2  + \partial_y^2  +\partial_z^2  )
\ee
Note that this is a single operator again - not a vector. Ah, so when going from one to three dimensions, the operator for the momentum squared just becomes a grad(ient) squared. This immediately suggests the form of the Schr\"{o}dinger equation in 3d:
\be
i \hbar \partial_t \psi(\vec{x},t)= \f{- \hbar^2}{2m} \nabla^2 \psi(\vec{x},t)+ V(\vec{x}) \psi(\vec{x},t) 
\ee
The changes are: $\psi$ now depends on three coordinates $\vec{x}$, the same for the potential, and the derivative squared is now a grad squared. Not all too different, right?
\subsection{Angular momentum}
When passing from 1D to 3D, there is another thing that changes: in 3 dimensions, objects can rotate. In the same way that linear momentum is the `amount of motion', angular momentum measures an object's `amount of rotation'. The angular momentum is a vector $\vec{l}$, and for a point particle with position $\vec{x}$ and momentum $\vec{p}$ it is given by:
\be
\vec{l}=\vec{x}\times \vec{p}
\ee
or writing out the cross product in components:
\be
\vec{l}=(y p_z - z p_y, z p_x- x p_z, x p_y - y p_z).
\ee
Again the corresponding quantum operator is a straightforward guess:
\be
\vec{L} = \vec{X} \times \vec{P}
\ee
where by $\vec{X}$ we mean the vector of the position operators $X$, $Y$ and $Z$. Writing out the momentum operator, we get
\be
\vec{L} = - i \hbar \vec{X} \times \vec{\nabla}.
\ee
So for example the $x$ component of the angular momentum operator is given by:
\be
L_x=  Y P_z - Z P_y = - i \hbar ( Y \partial_z - Z \partial_y )
\ee
\subsubsection*{Mini-example}
The function $f(x,y,z)=y+ i z$ is an eigenfunction of $L_x$, with eigenvalue $\hbar$, since
\be
L_x f(x,y,z) = - i \hbar (y i - z ) = \hbar (y + iz) = \hbar f
\ee
Of course, such a function is not a decent wave function: it is nowhere near normalizable. But well - it's just an example.

\section[Story: cracking the mystery of atoms]{\includegraphics[width=0.04\textwidth]{once_c} Story: cracking the mystery of atoms}

The idea of atoms making up the basic constituents of nature is not so new. Even some greek philosophers thought of the possibility of  fundamental, indivisible pieces of matter\footnote{Actually, `atom' comes from the Greek word for indivisible.}. Almost two millennia later, the first scientifically motivated atomic theories came to life. Through a long and fascinating history of discoveries, it became clear that there really have to be basic constituents of all matter: atoms. First a mere model, more and more proofs of their existence accumulated, up to their classification, and finally even the possibility of `looking' inside those strange building blocks. At the end of a long debate, scientists got convinced that atoms have to be made of a small positive core, with a cloud of electrons around it. For a moment, one could have thought that the mystery of atoms was unraveled. Electric attraction indeed explains that such a cloud of electrons would stay close to a positive core, orbiting around it. It could not be more beautiful, the smallest known scale in almost perfect similarity to our very own solar system. However prefect this may seem, there were some serious problems indicating that the mystery was not completely solved yet.
 
\subsection{The end of the world}
From your basic electromagnetic course, you know that opposite charges attract. A bit similar to gravitation keeps a planet circling around a star, a light negative charge would neatly orbit around a heavy positive one. But there is a problem. The circular movement means that the orbiting charge experiences an acceleration: the centripetal force realized by the electrostatic attraction. Now a more advanced course on electromagnetism (whether you had one or not) will readily tell you that an accelerated charge necessarily \textbf{radiates away energy}.  Roughly speaking, the circular movement of the charge gives rise to a changing electric field, in such a way that electromagnetic waves are created. This is similar to a moving object in water: it creates waves that carry away energy from the moving object. So an orbiting charge will gradually lose its energy, and is deemed to (after some time) end up at the positive core. One could think this process just takes long, and therefore does not occur on visible timescales, but it turns out that such a collapse would go very quickly. So if the classical view on the electron and nucleus was correct, they were doomed to collapse quickly. Sounds very much like the end of the world, which we luckily don't see, so something else has to be going on inside the atoms around us.
\begin{figure}[H]
 \begin{center}
  \includegraphics[width=0.5\textwidth]{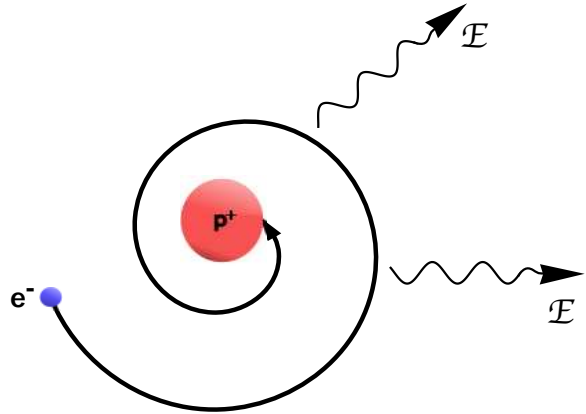}
  \caption{If electrons were point particles, they would lose all their energy by the emitted radiation, and spiral down to the nucleus. In no time, all atoms would collapse, and the universe would end in a drastic collapse. Obviously, this is not what happens.}
  \end{center}
\end{figure}

\subsection{Balmer and friends}
Nevertheless, there are some things that do make sense if you think of an atom in the `planetary' way. If electrons are orbiting around some core, you expect that they can change their kinetic energy. So if one invests energy in an electron, it will just move to a larger orbit. If it falls back to the smaller orbit, the absorbed energy will be emitted again. This suggests atoms can absorb and emit energy. And such was indeed observed. At the end of the 19th century, the pioneers in spectroscopy had found clear indications that some gases absorb light (so, energy) and emit it again. To the scientists working on the atomic model, it was clear that this absorption of energy had to do something with the electrons of an atom absorbing it, and spitting it out again later. But there was a very strange peculiarity. Gases did not seem to absorb \textit{any} radiation, but only at very specific wavelengths. For \textrm{hydrogen} gas, this behavior was put into an empiric formula by Johann Balmer. He found that absorbed wavelengths $\lambda_{\textrm{abs}}$ were all of the form
\be
\f{1}{\lambda_{\textrm{abs}}}\propto \left(\f{1}{2^2} - \f{1}{n^2}\right) \quad \textrm{with $n$ some integer $>2$.}
\ee
Other scientists later found some other absorbed wavelengths, and these are were all of the form
\be
\f{1}{\lambda_{\textrm{abs}}}= R_H \left(\f{1}{n_1^2} - \f{1}{n_2^2}\right) \quad \textrm{with $n_1<n_2$ integers.}
\ee
The above equation is called the \textbf{Rydberg formula}, after the discoverer of this generalization of Balmer's formula. The number $R_H$ is the \textbf{Rydberg constant}.
This formula very strongly suggests that the an electron in a hydrogen atom can only `sit' on special energy levels. To see that, first note that for light, the inverse wavelength is proportional to the frequency. Indeed, the product of wavelength and frequency is the wave speed (light speed): 
\be
f \lambda=c
\ee
So because the speed of light $c$ is just a constant, we have $\frac{1}{\lambda} \sim f$. Next the frequency of a light quantum (a photon) is proportional to energy: $E=h f$. So $\lambda_{\textrm{abs}}^{-1}$ is really a measure for an energy difference: the energy of a photon emitted in a transition between two electron orbits. So the Rydberg formula implies that the allowed energy differences are given by
\be
\Delta E \propto \left(\f{1}{n_1^2} - \f{1}{n_2^2}\right) \quad \textrm{with $n_1<n_2$ positive integers.}
\ee
This then suggests the only energy levels occupied by the (hydrogen) electron are given by
\be
E=E_0 \f{1}{n^2} \quad \textrm{with $n$ positive and integer.}
\ee
With $E_0$ a constant. If you track down the constants involved in all proportionalities, you can check that $E_0= h c R_H$.  

\subsection{Bohr}
This  observed discreteness of the atomic spectrum of hydrogen (and other atoms) is of course not so easily explained by classical mechanics. The only situation in which physicists had been confronted with discrete amounts of energy was the photon-hypothesis. As explained in the introduction, the photoelectric effect (amongst other experimental results) showed that light has to consist of discrete packets, of which the energy is determined completely by the frequency of the light:
\be
E= h f (=\hbar \omega).
\ee
In part inspired by this result, the physicist Niels Bohr cooked up the following explanation for the energy spectrum of hydrogen. He suggested that electrons around an atomic nucleus can only move on circular orbits, whose angular momentum is an \textit{integer multiple} of Planck's constant $\hbar$. This assumption was never derived or even slightly motivated. Professor Bohr just pulled it out of the hat. Any motivation (or lack of it) aside, let us try to see what the \textit{consequence} is of such an assumption. If we take the origin of the system at the nucleus, the length of the position vector $\vec{r}$ of the electron is equal to the orbital radius $r$. The momentum is of course given by $mv$ with $m$ the electron mass, and $v$ its velocity. (Although the direction of the velocity changes throughout the circular motion, its size remains the same.) Since the velocity is perpendicular to the position vector we have $|\vec{r}\times \vec{p}| = |\vec{r}| |\vec{p}|$ and the size of the angular momentum is given by
\be
l = |\vec{l}| = |\vec{r} \times \vec{p}| = r m v.
\ee
So Bohr's hypothesis amounts to 
\be
r mv = n \hbar \quad \textrm{with $n$ a positive integer}.
\label{bohr}
\ee
\begin{figure}[t]
 \begin{center}
  \includegraphics[width=0.5\textwidth]{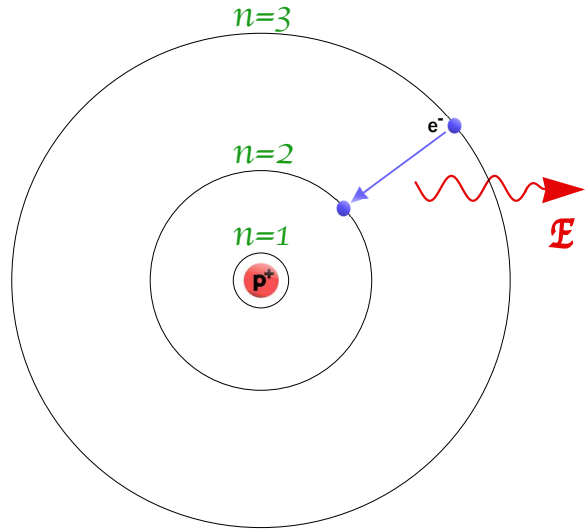}
  \caption{According to the Bohr model, electrons can move on specific orbits only. As a consequence, an electron falling from one energy level to another (here from $n=3$ to $n=2$) can only emit a specific amount of energy. For the same reason, only specific amounts of energy can be \textit{absorbed} by an atom. This explains the findings of spectroscopic experiments and the formula of Rydberg.}
  \label{fig:bohr_zelf}
  \end{center}
\end{figure}
Which orbits satisfy this relation? First note that the centripetal force is given by the attraction between the nucleus and the electron. Putting in the charge of the proton ($+e$) and the electron ($-e$): 
\be
F_\textit{centr} =  F_{\textit{coul}} \quad \Rightarrow \quad \f{m v^2}{r}= k \f{e^2}{r}
\label{forces}
\ee
So that
\be
r=\f{k e^2}{m v^2}
\label{eq:bohrradeq1}
\ee
for every circular orbit. This allows us to rewrite Bohr's hypothesis (\ref{bohr}) as
\be
\f{k e^2}{v}=n \hbar.
\label{eq:bohrradeq2}
\ee
We are now getting close. The energy of the system is given by
\be
E= E_{\textit{kin}} + E_{\textit{pot}} = \f{1}{2} m v^2 - k \f{e^2}{r} = -\f{1}{2} m v^2
\ee
where in the last step we used (\ref{forces}). Plugging in Bohr's hypothesis (\ref{eq:bohrradeq2}), we find that the energy of the electron is given by
\be
E = -\f{1}{2} m\left( \f{k e^2}{n \hbar} \right)^2 = - E_0 \f{1}{n^2} 
\ee
with $n$ a positive integer and $E_0=\f{1}{2} m\left( \f{k e^2}{ \hbar} \right)^2$ a numerical constant with units of energy. Woah, ye pirate, Bohr! Using his seemingly random hypothesis, we find a discrete set of energies. And not just some set, precisely the one matching Rydberg's formula. Indeed, it has the right $\f{1}{n^2}$ form, and moreover the numerical value of $E_0$ turns out to coincide sharply with the Rydberg constant $R_H$. 

\subsection{A better explanation?}
That's all very nice of Bohr. The great trick he pulls is just assuming a discrete behavior of the angular momentum, which then in the end of course leads to a discrete behavior of the energy. However perfect the final result matches with experiment, it just shifts the problem to another one: why is the angular momentum discrete, then? It also does not answer the problem raised in the `end of the world' section. The model still describes an electron moving on circles, which should radiate away all its energy. 

What can we do about that? 
Well, first of all we know that the electron should be described by a smeared out wave function, not a circular orbit of a point particle. In the examples we have met so far, solving the SE of such a wave function leads to an eigenfunction problem. Moreover, in the examples we have seen, this eigenfunction problem only had a special set of solutions. So a wave function `locked up' in a attractive potential naturally has a discrete set of energy states. This is a much more satisfying answer than the magic Bohr pulled out of his hat. But can we also reproduce this result here? It's to say, if we solve the SE for the hydrogen atom, do we get the right spectrum? We will see in the next section...

\section[The quantum approach]{\includegraphics[width=0.04\textwidth]{comput_c} The quantum approach}
Let us try to apply what we have learned in the previous chapters, and understand the hydrogen atom using quantum mechanics.
We know that the nucleus is very small, and that it creates a potential for the bound electron:
\be
V = - k  \f{e^2}{r}.
\ee
With $r=\sqrt{x^2+y^2+z^2}$ the distance to the origin. This means the Schr\"{o}dinger equation for the electron wave function reads
\be
i \hbar \partial_t \psi = \left(V-\frac{\hbar^2}{2m}\Delta\right)\psi
\ee
Here $V= -k e^2/ \sqrt{x^2+y^2+z^2}$ is the potential, $\psi=\psi(x,y,z)$ the three-dimensional wave function, and $\Delta=\partial_x^2+\partial_y^2+\partial_z^2$ the Laplacian. This gives a first situation where we have to deal with a realistic (and fully 3D) problem. In principle, the only thing one need to do to understand the quantum behavior of an electron in a hydrogen atom is to solve the above equation. Again, we first look for energy-eigenstates.
The time-evolution of these states is trivial ($e^{-iEt/\hbar}$), and the most general state of the Hilbert space is just given by taking linear combinations. So the only real problem is solving:
\be
E \psi = \left(V-\frac{\hbar^2}{2m}\Delta\right)\psi.
\label{eq:energy eigenstate}
\ee
Phrased differently: we want to find the eigenfunctions of the hamiltonian $H=V-\frac{\hbar^2}{2m} \Delta$.
\subsection{A better coordinate system}
The above equation is not so easy to solve. The naughty element is the potential, which involves a square root. Actually, since the potential (and hence the entire problem) is spherically symmetric, it is natural to switch to an adapted coordinate frame. Spherical coordinates $(r,\theta,\varphi)$ are related to Cartesian ones by
\bea
x&=&r \, \cos\varphi \, \sin\theta \\
y&=&r \, \sin\varphi \, \sin\theta\\
z&=&r \, \cos\theta \quad
\eea
or
\bea
r&=&\sqrt{x^2 + y^2 + z^2}\\
{\varphi}&=&\textrm{arctan}(y/x)\\
{\theta}&=&\arccos \left( {\frac{z}{\sqrt{x^2 + y^2 + z^2}}} \right)
\eea
The ranges are $r \in [0,\infty[$, $\theta \in [0,\pi]$ and $\varphi \in [0,2\pi[$. 
If we want to rewrite the energy-eigenstate equation (\ref{eq:energy eigenstate}) in terms of $(r,\theta,\varphi)$, three things change. First, we need to replace $\psi(x,y,z)$ by $\psi(r,\theta,\varphi)$. Next, the potential just becomes $- k  e^2/r$ instead of $ -k e^2/ \sqrt{x^2+y^2+z^2}$. The last thing is to rewrite the Laplacian in terms of derivatives $\partial_r$, $\partial_\theta$ and $\partial_\varphi$. 
Doing so is unusually nasty (not even fun) and we only sketch the procedure here.
For example, to express $\partial_z^2$, we remark that
\be
\partial_z=\frac{\partial}{\partial z} = \frac{\partial r}{\partial z} \partial_r +\frac{\partial \theta}{\partial z} \partial_\theta+\frac{\partial \varphi}{\partial z} \partial_\varphi
\ee
This is just the chain rule for derivatives. (Both sides are operators, so one should imagine them acting on a function. Writing a function on the right of both sides, you indeed see nothing but the chain rule.) You can then compute the three partial derivatives occurring on the right-hand side, using the expressions for $r(x,y,z)$, $\theta(x,y,z)$ and $\varphi(x,y,z)$ and next express the results back in spherical coordinates. This way you can obtain 
\be
\partial_z = \cos\theta\, \partial_r - \frac{\sin \theta}{r} \partial_\theta
\ee
Now to get $\partial_z^2$, one must be a bit careful. Once again, recall that we are dealing with operators. So for example the square of an operator $(r\partial_r)$ would be 
\be
(r\partial_r)^2 = (r\partial_r)(r\partial_r) = r (\partial_r + r \partial_r^2 ) \neq r^2 \partial_r^2
\ee 
If you are confused by these steps, just write a general function $f$ on the right of every term. In the same fashion, one can obtain an expression for $\partial_z^2$. (We omit it here.) Doing the same for $\partial_x^2$ and $\partial_y^2$ and taking the sum, one obtains the Laplacian in spherical coordinates:
\be
\Delta  = {1 \over r^2} {\partial \over \partial r}  \left( r^2 {\partial  \over \partial r} \right) + {1 \over r^2 \sin \theta} {\partial \over \partial \theta}
  \left( \sin \theta {\partial  \over \partial \theta} \right)  + {1 \over r^2 \sin^2 \theta} {\partial^2  \over \partial \varphi^2}.
\label{eq:spherical laplacian1}
\ee

\subsection*{Back to the equation}
After this sidestep, let us move on to what we are here for, solving 
\be
E \psi(r,\theta,\varphi) = \left(V(r)-\frac{\hbar^2}{2m} \Delta\right) \psi(r,\theta,\varphi)
\label{eq:spher SE}
\ee
With $\Delta$ given by (\ref{eq:spherical laplacian1}). Let us suppose that the wave function $\psi(r,\theta,\varphi)$ can be written as 
\be
\psi(r,\theta,\varphi)= R(r)F(\theta,\varphi)
\ee
This is called \textbf{separation of variables}. In terms of $R$ and $F$, the energy eigenstate equation (\ref{eq:spher SE}) becomes (after multiplying by $\frac{-2m}{\hbar^2}r^2$)
\bea
&F& \left(  r^2 \frac{\partial^2 R}{\partial r^2}+ 2 r \frac{\partial R}{\partial r} - \frac{2m}{\hbar^2} r^2 (V(r)-E) R \right)\\
+&R& \left( \frac{1}{\sin \theta} \frac{\partial}{\partial \theta}(\sin \theta \frac{\partial F}{\partial \theta})+ \frac{1}{\sin^2 \theta}\frac{\partial^2 F}{\partial \varphi^2}\right)=0
\eea
This looks pretty disastrous, but there is an important simplification. Formally, the above says
\be
F (\textrm{some function of r}) + R (\textrm{some function of $\theta$ and $\varphi$})=0
\ee
So assuming R and F are not zero - a typical kind of physics sloppiness: 
\be
\frac{(\textrm{some function of r})}{R} =- \frac{(\textrm{some function of $\theta$ and $\varphi$})}{F}.
\ee
The last equation is very special. On the left hand side things only depend on $r$, whereas on the right hand side only on $\theta$ and $\varphi$. This can only be true if both sides are equal to some constant, say $C$. This then reduces our problem to solving
\be
\frac{1}{\sin \theta} \frac{\partial}{\partial \theta}(\sin \theta \frac{\partial F}{\partial \theta})+ \frac{1}{\sin^2 \theta}\frac{\partial^2 F}{\partial \varphi^2}=-CF
\label{eq:angular equation}
\ee
\be
r^2 \frac{\partial^2 R}{\partial r^2}+ 2 r \frac{\partial R}{\partial r} - \frac{2m}{\hbar^2} r^2 (V(r)-E) R= C R
\ee
Ah, we have cracked down the original equation in two parts! That's good news. The first part is called \textbf{the angular equation}, the second is \textbf{the radial equation}. We will solve them separately.
\subsection{The angular equation}
For those who were masochistic enough to have done the Laplacian calculation, another goodie. Verify that the angular momentum operator, defined in the beginning of the chapter, has the following form in spherical coordinates:
\be
L^2=L_x^2 + L_y^2 + L_z^2 = - \hbar^2 \left[\frac{1}{\sin \theta} \frac{\partial}{\partial \theta}\left(\sin \theta \frac{\partial }{\partial \theta}\right)+ \frac{1}{\sin^2 \theta}\frac{\partial^2 }{\partial \varphi^2}\right]
\label{eq:spherical laplacian}
\ee
If you have better things to do in your life than performing tedious algebra, just believe the above statement. It will leave with more energy to discover something special: the angular equation (\ref{eq:angular equation}) is just an eigenvalue equation of the angular momentum operator: 
\be
L^2 F = \hbar^2 C F
\ee
Now this equation turns out to be well-studied in physics. It shows up in virtually any problem with spherical symmetry (quantum or not). It only has solutions for the following values of $C$:
\be
C=\ell (\ell+1) \quad\textrm{where} \quad \ell=0,1,2,...
\ee
Striking enough, for each such $C$ (except for $C=0$), there are \textit{several} solutions. So to be clear on which solution we are talking about, not only do we need to specify $\ell$, but also an extra label distinguishing the different solutions corresponding to that $\ell$. It turns out that the different solutions corresponding to the same $\ell$ are all eigenfunctions of the operator $L_z$, but with different eigenvalues. This means we can use the eigenvalue with respect to $L_z$ as the extra label needed to specify the solution uniquely. This label is conventionally called $m$, although it has nothing to do with mass. The corresponding eigenvalues of $L_z$ are $\hbar m$. In conclusion: the angular equation $L^2 F = \hbar^2 C F$ has an infinite set of solutions, denoted $Y^m_\ell(\theta,\varphi)$, where the labels $\ell$ and $m$ are related to the eigenvalues of the solution with respect to $L$ and $L_z$:
\be
L^2 \, Y^m_\ell= \hbar^2 \ell (\ell+1)\,Y^m_\ell
\label{eq:angpart1}
\ee
and
\be
L_z \,Y^m_\ell= \hbar m \,Y^m_\ell
\label{eq:angpart2}
\ee
It turns out that the values $m$ can take are bound by $\pm\ell$: so $m=0, \pm1,\pm2,...\pm \ell$. The functions $Y^m_\ell(\theta,\varphi)$ are called \textbf{spherical harmonics}. To see where this name comes from: a function that behaves `well' under the Laplacian is called \textit{harmonic}, and $L^2$ is just the angular (=\textit{spherical}) part of the Laplacian.

The general form of the spherical harmonics $Y^m_l(\theta,\varphi)$ can be written down explicitly. This expression is not so illuminating, so we just restrict to giving a table of the first few solutions (smallest $\ell$ and $m$):

\begin{center}
    \begin{tabular}{ | l | l | l |}
    \hline
    $\ell$ & $m$ & $Y^m_\ell$ \\ \hline
    0 & 0 & $ \sqrt{\frac{1}{4\pi}}$\\ \hline \hline
    1 & 0 & $\sqrt{\frac{3}{4\pi}} \, \cos \theta$ \\ \hline 
    1 & $\pm 1$ & $\mp \sqrt{\frac{3}{8\pi}} \sin \theta e^{\pm i \varphi}$ \\ \hline \hline
    2 & 0 & $\sqrt{\frac{5}{16\pi}} (3 \cos^2 \theta -1)$\\ \hline 
    2 & $\pm1$ & $\sqrt{\frac{15}{8\pi}} \cos \theta \sin \theta e^{\pm i\varphi}$ \\ \hline
    2 & $\pm 2$ & $ \sqrt{\frac{15}{32\pi}}\sin^2 \theta e^{\pm 2 i \varphi}$\\
    \hline
    \end{tabular}
\end{center}
You also see that $m$ runs from $-\ell$ to $+\ell$ indeed. Next, all the spherical harmonics have a simple $\varphi$-dependence, of the form $e^{i m \varphi}$. This guarantees they are eigenfunctions of $L_z =- i \hbar \partial_\varphi$ with eigenvalue $\hbar m$.

\subsection{The radial equation}
Having solved the angular equation, let us move on to the radial part. Using $C = \ell (\ell +1)$, we can write the radial equation as
\be
\frac{\partial^2 R}{\partial r^2} + \frac{2}{r}\frac{\partial R}{\partial r} - \frac{\ell(\ell +1)}{r^2} R + \frac{2}{a r}R - b^2R =0
\label{eq:radial}
\ee
with $a=\frac{\hbar^2}{m k e^2}$ and $-b^2=\frac{2 m E}{\hbar^2}$. Note that the $-b^2$ is proportional to the energy $E$. Classically, an elektron is bound to an attractive potential is its total energy (kinetic+potential) is negative. This is also the case in quantum mechanics. So we really are interested in solutions where $E$ is negative, meaning $b$ is a \textit{real} constant. 
For large $r$ the middle terms are small and the equation becomes
\be
\frac{\partial^2 R}{\partial r^2} -b^2 R =0
\ee
So $R\propto e^{\pm b r}$. The solution with a plus sign is a bit pathetic: the wave function would then blow up for large $r$. We conclude there is only one healthy asymptotic behavior: $R\propto e^{- b r}$. This gives a hint on what solutions to the full equation (\ref{eq:radial}) might look like: 
\be
R(r)=\textrm{(some function of r)} \cdot e^{- b r}.
\ee
Concretely, by defining 
\be
R(r)=\frac{u(r)}{r}e^{- br}
\ee
the radial equation becomes
\be
\frac{\partial^2 u}{\partial r^2} - 2 b \frac{\partial u}{\partial r} + \left(\frac{2}{a r} - \frac{\ell(\ell+1)}{r^2} \right) u=0.
\label{eq:ueq}
\ee
Now what kind of solutions do we want? We need $u(r)$ to be a function of $r$, but it should be decent enough, so that the asymptotic behavior is still dominated by the exponential. Studying the asymptotic behavior more carefully, one can show that $u$ has to be a polynomial. (Otherwise, $R(r)$ will blow up for large $r$.) So the problem of solving the radial equation has been reduced to: `what polynomials satisfy the equation above?' Ah, again this is a well-understood problem for mathematicians. Don't you just love them? The problem only has solutions when
\be
\frac{1}{a b} = n \quad \textrm{with $n$ a positive integer}
\ee
Also, we need $\ell+1\leq n$ for a solution to exist. For these special values of $a$ and $b$, the solution of (\ref{eq:ueq}) is given by
\be
u(r)=r^{\ell+1} L^{2 \ell+1}_{n-\ell-1} (2 b r)
\ee
with $L^\alpha_\beta$ the so-called \textbf{generalised Laguerre polynomials}. They are defined by
\be
L_\beta^{\alpha}(x)=
{x^{-\alpha} e^x \over \beta!}{d^\beta \over dx^\beta} \left(e^{-x} x^{\alpha+\beta}\right)
\ee
and the first relevant combinations become:
\begin{center}
    \begin{tabular}{ | l | l | c |}
    \hline
    $n$ & $\ell$ & $r^{\ell+1} L^{2 \ell+1}_{n-\ell-1}(2 b r)$ \\ \hline
    1 & 0 & $r$\\ \hline \hline
    2 & 0 & $\left.r(-2br+2)\right.$ \\ \hline 
    2 & $ 1$ & $r^2$ \\ \hline \hline
    3 & 0 & $r (2b^2r^2-6br+3)$\\ \hline 
    3 & 1 & $r^2(-2br+4)$ \\ \hline
    3 & 2 & $r^3$\\
    \hline
    \end{tabular}
\end{center}
Again, you can check for your favorite element of the above list that it satisfies the corresponding differential equation. Returning to the original function $R$, the solution corresponding to a particular value of $n$ and $\ell$ is
\be
R(r) = r^\ell L^{2 \ell+1}_{n-\ell-1} (2 b r) e^{- b r}  =r^\ell L^{2 \ell+1}_{n-\ell-1} (2 b r) e^{-b r}  
\ee 

\subsection{Conclusion}
Congratulations! You have gotten through solving the radial and angular part of the Schr\"{o}dinger equation of the hydrogen atom. The conclusion is: there is an infinite set of solutions, labeled by the numbers $n$, $\ell$ and $m$. They are given by
\bea
\psi_{n,\ell,m}(r,\theta,\varphi) & = & R(r) F(\theta,\varphi) \\
&=& N r^\ell L^{2 \ell+1}_{n-\ell-1} (2b r) e^{-b r} \, Y^m_\ell(\theta,\varphi). \label{eq:wavcomplete}
\eea
Where 
\bea
n&=&1, 2, 3,...\\
\ell &=& 0, 1, ... n-1 \\
 m & = & -\ell, -\ell +1 , ... \ell-1, \ell
\eea
A constant $N$ has been included to take account of the normalization of these solutions. You may wonder what those numbers $n$, $\ell$ and $m$ mean physically. Because the operators $L^2$ and $L_z$ involve only radial coordinates, equations (\ref{eq:angpart1}) and (\ref{eq:angpart2}) also imply that
\bea
L^2 \psi_{n,\ell,m} &=& \hbar^2 \ell(\ell+1) \psi_{n,\ell,m}\\
L_z \psi_{n,\ell,m} &=& \hbar m \psi_{n,\ell,m}
\eea
Hence the state $|\psi_{n,\ell,m}\rangle$ has (squared) angular momentum $\hbar^2 \ell(\ell+1)$, and the z-component of the angular momentum is $\hbar m$. The label $\ell$ is called the \textbf{orbital quantum number} and $m$ is the \textbf{magnetic quantum number}. Furthermore, using the expressions for $b$ and $a$, the energy is given by 
\be
E = -\frac{\hbar^2 b^2}{2m} = - \frac{\hbar^2 (1/a n)^2}{2m} = -\frac{m k^2 e^4}{2  \hbar^2 n^2}
\ee
Defining again $E_0 = \frac{m k^2 e^4}{2  \hbar^2} \,\,(\approx 13,6 eV)$, the energy is given by
\be
E = -E_0 \frac{1}{n^2}.
\ee
So the label $n$ determines the energy. It is called the \textbf{principal quantum number}. 

Although it took us quite a bit of work to get here, this is a very beautiful result. Not only have we found the wave functions that can be taken by the electron of the hydrogen atom, we have also found the corresponding values of the energy. On top of that, these energies \textit{precisely} coincide with the measured energy levels. (Since the result is in accordance with the empirical formula of Rydberg.) In contrast to Bohr, we did not have to make some funny assumptions. Also, these wave functions just \textit{sit} there: they are not an orbiting point particle as Bohr assumed, which had the problem of inevitably radiating away its energy and ending up in the nucleus. So in some sense the quantum-approach also solves the `end of the world'-problem. More precise: there is a state ($n=1$ and $\ell=m=0$) that has the lowest energy: this is the \textrm{ground state}. When the electron is in this state, there is no possible process taking it to a lower energy state. The system thus is stable, and cannot decay or collapse. This contrasts with the numerous plagues one encounters when trying to describe the atomic structure using classical mechanics.

\subsection*{Closing remarks}

We have obtained our desired solutions, but it might be clarifying to spend some more time on them. First, as you may know, the different possible states of the electron in the atom are called \textbf{orbitals}. There is an alternative notation to indicate the principal and orbital quantum number of each orbital: write the value of $n$ and then the letter s, p, d, f, ... for $\ell = 0, 1, 2, 3, ...$. So for example
\bea
1s &\leftrightarrow& n=1, \,\ell = 0\\
2s &\leftrightarrow& n=2, \,\ell = 0\\
2p &\leftrightarrow& n= 2 , \,\ell = 1\\
&\vdots&
\eea
etcetera. The strange letters indicating the orbital quantum number derive from some purely visual aspects in spectroscopic experiments: they stand for s(harp), p(rincipal), d(iffuse), f(undamental). Here are some figures showing the norm of the wave function for the lowest energy states of the hydrogen atom. The brighter the color, the larger the amplitude.
\begin{figure}[H]
 \begin{center}
  \includegraphics[width=0.5\textwidth]{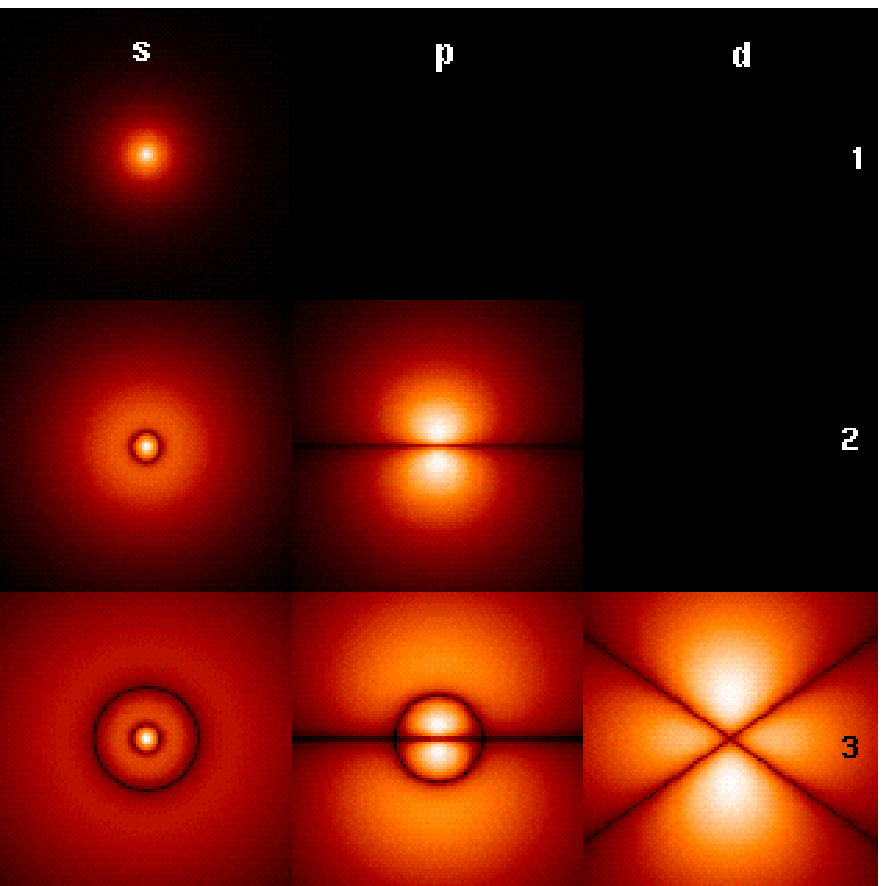}
  \label{fig:orbitals}
  \end{center}
\end{figure}
Some very nice things can be seen here. First, note that the higher the energy, the further away the wave function is located from the nucleus. Indeed, for $n=3$, the wave functions are more spread out (`larger') than for $n=2$ or $n=1$. This is the quantum analog of the classical expectation that for higher energies, the electron should be occupying a larger orbit. Let us be a bit more quantitative on the size of these orbitals. Using the fact that $L^\alpha_\beta$ is a polynomial of degree $\beta$, you can check that the leading term in the wave functions (\ref{eq:wavcomplete}) is given by
\be
\psi_{n, \ell, m} \sim r^{n-1} e^{- r/(na)}
\ee
where we have used $b=\frac{1}{na}$. The function on the right falls off quickly due to the exponential. The region over which it has an appreciable size, can be estimated by looking at the location of its maximum. There always is one, and you can check that it is given by
\be
\frac{\partial}{\partial r} (r^{n-1} e^{- r/(na)}) =0 
\quad \Leftrightarrow \quad r = n^2 a - na 
\ee
So we see that for larger $n$ the wave function is approximately limited to a region of size $r \sim a n^2$: 
\be
\textrm{state size} \quad \sim \quad n^2 a
\ee
The constant $a$ is called the \textbf{Bohr radius}. Its numerical value is 
\be
a = 0.529 \cdot 10^{-10} \,\,\textrm{m}
\ee
and this is indeed the approximate size of the hydrogen atom. (Others atoms are larger since they contain more electrons). So the size of the wave function grows quadratically with $n$. This explains to some extent the stunning success of Bohr's naive model. His mysterious assumption describes classical orbits, of which the radii just happen to coincide with the actual size of the quantum wave function.\footnote{Just combine (\ref{eq:bohrradeq1}) and (\ref{eq:bohrradeq2})  to get an expression for $r$, only involving $n$ and some constants, but not $v$. You get precisely the relation $r = n^2 a$.} On top of that, the corresponding classical energy to such an orbit just so happens to coincide with the right (quantum mechanical) value. For more complicated systems (atoms with more than one electron for example) there are no similar `smart guesses', and a quantum description is the only tool to get the right predictions. 

One other observation is that the solution we found, predicts that there are in general several states with the \textit{same} energy. (Since the energy only depends on $n$, not on $\ell$ or $m$.) This phenomenon is called \textbf{degeneracy} and finally is something that really contrasts with the Bohr description. It is possible to do very precise measurements and see the different states corresponding to the same energy. This again confirms that the quantum mechanical description really is the right one. 

Finally, some comments on the angular part of the wave function. You see that for increasing $\ell$ the spherical harmonics get more and more wobbly: higher and higher powers of $\sin \theta$ and $\cos \theta$ occur. This trend continues for greater $\ell$. Also the $\varphi$ dependence $e^{i m \varphi}$ becomes more and more oscillating. This is the quantum mechanical manifestation of angular momentum. The stronger a wave function varies along the angular coordinates, the more angular momentum it contains. This is a bit similar to the \textit{linear} momentum of a state. Since $P = - i \hbar \nabla$, the momentum of a state is proportional to its variation along the coordinate axes, its linear `wobbliness'. Of course, these statements are a bit shady: they only give a heuristic feel for the manifestation of angular and linear momentum in the language of wave functions. The right way to do this, is of course looking at the eigenvalues (or otherwise the expectation values) of all these operators. 

You can now say you truly understand what the hydrogen atom is. Historically, the fact that its structure can be fully derived from the Schr\"{o}dinger equation \textit{without} any extra assumptions was a very important sign that his equation (and quantum mechanics) are the right way to describe particles. 

\newpage
\section*{Exercises}

\begin{enumerate}
\item What is the wavelength a photon should have to excite a hydrogen atom from the third to the fourth level? 
\item Pretend an atom has a cubic shape, and that you can stack hydrogen atoms tightly next to each other. How many atoms then fit in a cubic centimeter? Estimate the number of sand particles in the Sahara desert. (Surface: $\sim 2500$ km $\times$ $4000$ km, depth $\sim 100$ m.) Would you say an atom is pretty small or very small? 
\item Using (\ref{eq:spherical laplacian}) and the fact that $L_z =- i \hbar \partial_\varphi$ in spherical coordinates, check for your favorite element of the table of spherical harmonics that it is an eigenvalue of $L^2$ and $L_z$, with the right eigenvalues.
\item Verify the transition of (\ref{eq:radial}) to (\ref{eq:ueq}).
\item If you bombard the electron with too much energy, it will be kicked away the hydrogen nucleus. Concretely, the states with high $n$ are only very weakly bound, and they are very close to an ionized state. How much energy do you need to kick an electron from $n=0$ to the `highest state' $n=\infty$?  If you impact a 20 eV photon on a hydrogen atom, how much kinetic energy will the (now free) electron have?
\item Challenge: consider a \textbf{hydrogen-like atom}: a nucleus with charge $+Z$ and a single electron around it. (For example, the case $Z=2$ corresponds to He$^+$, and $Z=1$ is just the hydrogen atom.) The potential is then Z times bigger. Skim through the computation of this chapter, and check which things change. Find the expression for the wave function. Hint: the functions occurring remain the same but everything (prefactors, function arguments) gets rescaled. If you do all well, you should in the end find that the energy levels of such a system are given by $E_n = - E_0 \frac{Z^2 }{n^2}$.
\item Consider a particle in a box - a true one, in three dimensions, with volume $L^3$. Write down the Hamiltonian: what is the kinetic part, what is the potential $V(x,y,z)$? Write down the time-independent Schr\"{o}dinger equation. To simplify things, use separation of variables: put $$\psi(x,y,z)=\psi_1(x)\psi_2(y)\psi_3(z)$$ Show that the equation breaks apart in three separate equations. Do you recognise these equations? What are the solutions? Write down the most general solution of the 3D system, and show that the energy levels are of the form $$\frac{\hbar^2 \pi^2}{2 m L^2}\cdot(n_1^2 + n_2^2 + n_3^2) $$ with $n_1$, $n_2$ and $n_3$ integers.
\end{enumerate}

%% file: H6.tex
\chapter{The commutator}
\subsection*{In this chapter...}
We will come back to the result we got for the hydrogen atom. We address some small issues, giving another view on the situation which (hopefully) extends your understanding.  A technical tool that will be important for this discussion is the \textit{commutator}. This concept has come up already several times, and it will re-appear even more often, so that is why it deserved this chapter's title. We start by improving our commutator computing skills, then (as promised) we spend some more time pondering about the result of the previous chapter, and we conclude by several small illustrating computations.

\newpage

\section[Commute along, cowboy]{\includegraphics[width=0.04\textwidth]{tool_c} Commute along, cowboy}
In the chapter on the measurement, we introduced the \textit{commutator} of two operators. It is defined as
\be
[A,B]=A B - B A.
\ee
If $A$ and $B$ were ordinary numbers, the above would obviously be zero. For operators however, the order in which you let them act on a function might matter; in that case the commutator is nonzero. Here in this section we briefly give some facts on commutators. Most of them are easy to check, just by writing out everything explicitly; it is a good exercise to do so. First of all, commutating is a linear operation:
\be
[a A + b B, C]= a [A, C] + b [B,C]
\ee
for complex numbers $a, b$, and operators $A, B, C$. Similarly
\be
[A, bB + cC] = b [A, B] + c [A,C]
\ee
The commutator of products is given by the \textit{`drag out sideways'}-rule (\includegraphics[width=0.03\textwidth]{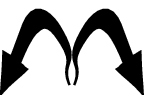}):
\be
[A,BC] = [A,B]C + B[A,C] \quad \textrm{and}  \quad [AB,C] = A [B,C] + [A,C]B.
\ee
Which can again be seen by just writing out both sides - carefully retaining the order of operators within in each term. We have also seen that
\be
[X,D_x]=-1
\ee
We showed this property before, but just to refresh your memory, a concrete mini-example illustrating it (for -say- the function $x^2$):
\be
[X,D_x] x^2 = X D_x\, x^2 - D_x\, X x^2 = X (2x) - D_x (x^3) = 2x^2 - 3 x^2 = -1(x^2).
\ee
Of course the same holds for other variables: $[Y,D_y]=[Z,D_z]=-1$. As a consequence  (using $P_i= - i\hbar D_i$)
\be
[X,P_x]=[Y,P_y]=[Z,P_z]=i\hbar
\ee
The above is sometimes called the \textbf{canonical commutation relation}. On the other hand, \textit{mixed} position-momentum commutators are zero:
\be
[X,P_y]=0
\ee
and similar for any other combination of two \textit{different} variables. To see this, just note that for any function $f(x,y,z)$,
\be
X D_y f = X ( \partial_y f) = x \partial_y f = \partial_y (x f) = D_y X f
\ee
and similar for any other combination of two \textit{different} variables. If the before last step is not immediately clear, just use the product rule for differentiation to see that
\be
\partial_y (x f) = (\partial_y x) f + x (\partial_y f)= 0+x \partial_y f.
\ee
We're done with the first part. Not all too harsh, right?

\section[Asking those questions...]{\includegraphics[width=0.04\textwidth]{once_c} Asking those questions...}

In the precious chapter, we uncovered the quantum structure of the hydrogen atom. You may have felt a bit like a passive audience though: most results fell out of the sky. In particular, the fact that we have to use \textit{three} labels to specify each state seems to have emerged `miraculously'. Of course, you may be satisfied with this, but if you are critical, there are several questions you can ask. First, notice that the third quantum number ($m$) describes the $z-$component of the angular momentum. Why this one? What is special about this direction? Didn't we start out with a perfectly spherically symmetric problem? Also, why did we find three quantum numbers, and not (say) four? How could we have seen this coming in advance? These are all quite good questions. Below, we will answer them in detail. Hopefully this adds some extra understanding to the matter.

\subsection{Z-supremacy?}
Why did we find the states of an electron in the hydrogen atom to be eigenstates of $L_z$, and not $L_x$ or $L_y$? Here is the explanation. Recall that we found that given the total angular momentum (labeled by $\ell$) there were \textit{several} solutions to the angular equation. So given some $\ell$, there are several solutions to
\be
L^2 F(\theta,\varphi)=\hbar^2 \ell(\ell+1) F(\theta,\varphi).
\label{eq:zsupre}
\ee
Now here is the key observation: the above equation is \textit{linear} - just like the Schr\"{o}dinger equation. So if you have two solutions $F_1(\theta,\varphi)$ and $F_2(\theta,\varphi)$ then any linear combination $a F_1(\theta,\varphi) +b F_2(\theta,\varphi)$ is a solution too. Thus, when we said there is more than one solution, we should actually have said: there are infinitely many solutions, and they form a vector space. This vector space can best be described by picking a basis - but which one you choose is up to you. In the previous chapter we picked a basis, namely functions $Y_\ell^m$, which (on top of $L^2$) were also eigenfunctions of the operator $L_z$, with eigenvalues $m$. Of course, it is possible to pick any other basis. In fact, it is equally possible to pick basis elements that are eigenfunctions of $L^2$ and $L_x$ (or of $L^2$ and $L_x$). In such a case, you would label your basis states by $\ell$ and their eigenvalue under $L_x$ (or $L_y$). So there really is nothing special about the z-direction. Using $L_z$ just happens to provide the simplest looking basis of the space of solutions of (\ref{eq:zsupre}). That is why people like to use that one. Originally, this comes from the fact that in the $(r,\theta,\varphi)$ system, $L_z$ has a very simple form, $-i\partial_\varphi$. The other two angular momentum operators look a bit more difficult, and so do their eigenfunctions - but nothing fundamental distinguishes them. As an illustration, we will actually perform the change of basis explicitly in the computation part of the chapter.  

\subsection*{Orthogonal states}
Here is another question that might have come to your mind. Are the states we found an \textit{orthogonal} basis? As emphasized in the chapter on observables, such a basis is very valuable. We have also shown that eigenstates of an observable with different eigenvalues are automatically orthogonal. Here, things look a bit different at first: since there are several states with the \textit{same} $n$.  
How does this work? First, let us be more clear on what we mean by orthonormal in this situation. Denoting by $|n\, \ell \,m\rangle $ the state corresponding to 
$\psi_{n\ell m}(x)$, we can express orthonormality by:
\be
\langle n\, \ell \, m | n' \,\ell' \, m'\rangle  = \delta_{n n'} \delta_{\ell \ell'} \delta_{m m'}.
\label{eq:orthogon}
\ee 
Let us think about this. States with different $n$ have a different energy. That assures that $\langle \psi_{n \ell m}|\psi_{n' \ell' m'}\rangle  =0$ whenever $n \neq n'$. But what if $n=n'$? Ah, remember that if $\ell$ is the eigenvalue under $L^2$. This is a Hermitian operator too, as it corresponds to the observable `angular momentum'. Hence, states with different eigenvalues $\ell$ under  $L^2$ are orthogonal too. And in the same fashion, eigenstates with a different $m$, have different eigenvalues under the (also hermitian) operator $L_z$, so that is sufficient to guarantee orthogonality too. In conclusion, the inner product $\langle \psi_{n \ell m}|\psi_{n' \ell' m'}\rangle  $ will be zero whenever one of the three quantum numbers $n$, $\ell$, or  $m$ are different. Stated differently: the above inner product is only nonzero when taking the inner product of a state with itself. If one also makes sure states are normalized ($\langle \psi_{n \ell m}|\psi_{n \ell m}\rangle =1 $) the equation (\ref{eq:orthogon}) is indeed satisfied.

\subsection{CoSCO}
Let's imagine what would happen if you were a computing beast which could solve differential equations simply by looking at them. Then, studying the hydrogen atom quantum mechanically would go a bit like this. You write down the Hamiltonian for an electron bound to a nucleus. You glance at that time independent Schr\"{o}dinger equation (the eigenfunction problem) and instantly solve it - without having to use silly tricks (splitting it into a radial and angular equation) which normal humans have to rely on: you just write down an infinite tower of solutions (which span the Hilbert space) labelled by their energy, and that's it.

But suddenly something bothers you: there are \textit{several} states belonging to each energy level. The entire set of solutions with the same energy actually forms a vector space, since the sum of eigenfunctions with the same eigenvalue is again an eigenfunction. Such a vector space is called a \textbf{subspace} of the total Hilbert space. This is not a problem per se, but you'd like to find a decent, \textit{orthogonal} basis of the Hilbert space - and picking a random basis of energy eigenstates -like you just did- is no good: states with the same energy eigenvalue are not necessarily orthogonal. 

So how do you find a basis that \textit{is} orthogonal then? You glance again (still in beast mode) and immediately realize that for every subspace, you can find a special set of states which are eigenstates of both $L^2$ and $L_z$. Picking these special states as your basis (elements of this basis are now labeled by their eigenvalue under $H$, $L^2$ and $L_z$) you have now obtained an orthogonal basis! Indeed, each two elements of your basis have a different eigenvalue with respect to at least one Hermitian operator ($H$ or $L^2$ or $L_z$) so they are guaranteed to be orthogonal. This finally gives you an orthonormal basis $|n\,\ell m\rangle$. (Being mere humans, we needed more work to obtain this basis.)

Since all states of the Hilbert space basis can be labeled by their eigenvalues under the observables $H$, $L^2$ and $L_z$, these are called a  \textbf{complete set of commuting observables}, or CoSCO for short. Let's clarify. The term `complete' means they specify every state uniquely: no two different states can be found with the same eigenvalues. Next (as suggested by the name) such a set of observables necessarily commutes. This is not too hard to show. Suppose one has two observables $A$ and $B$ and a basis of states $|i j\rangle $, which are eigen to $A$ with eigenvalues $a_i$ \textit{and} eigen to $B$ with eigenvalues $b_j$. So
\be
A|i j\rangle  = a_i |i j\rangle  \quad \textrm{and} \quad B|i j \rangle  = b_j |i j\rangle  \quad \forall i \, , \, \forall j
\ee
So $i$ specifies the $A$-subspace to which a state belongs, and $j$ the $B$-subspace. Then it follows that for a every state $|\psi\rangle =\sum_i \sum_j \alpha_{i j} |i j\rangle $ in the Hilbert space
\bea
A B |\psi\rangle  &=&A B \sum_i \sum_j \alpha_{ij} |i j \rangle  \nonumber\\
&=& A\sum_i \sum_j  b_j \alpha_{ij} |i j\rangle  \nonumber\\
&=& \sum_i \sum_j  a_i b_j \alpha_{ij} |i j\rangle  \nonumber\\
&=&\sum_i \sum_j  b_j a_i \alpha_{ij} |i j\rangle \nonumber\\
&=& B \sum_i \sum_j  a_i \alpha_{ij} |i j\rangle  \nonumber\\
&=& B A \alpha_{ij} |i j\rangle  = BA |\psi\rangle 
\eea
At several points the linearity of $A$ and $B$ was used. (This is OK because \textit{all} operators in quantum mechanics are linear.) So $AB |\psi\rangle  = BA|\psi\rangle $ for all states $|\psi\rangle $ in the Hilbert space, meaning
\be
AB = BA \quad \textrm{so}\quad [A,B]=0
\ee
Indeed, $A$ and $B$ \textit{must} commute. This explains the second C in CoSCO.

\subsection{Compatible observables}

The above combined with our finding of the states $|n\,\ell\, m\rangle $ implies that all the commutators between $H$, $L^2$ and $L_z$ have to be zero:
\be
[H,L^2]=[H,L_z]=[L^2,L_z]=0. 
\ee
But also a converse of the previous section is true: 
If the commutator $[A,B]$ of two observables is zero, then it is possible to find states that are eigenstates of $A$ and $B$ in the same time. 
Because of this property, commuting observables $A$ and $B$ are sometimes called \textbf{compatible observables}. 
But not any two observables are compatible. For example, we will see in the computation part that $[L_y,L_z]$ and $[L_x,L_z]$ are \textit{not} zero. This means that it is not possible to find a basis of states that are simultaneous eigenstates of more than one component of the angular momentum. So generally, if a state is an eigenstate of one of the components of the angular momentum, it can not be an eigenstate of any of the two other components. With respect to one of these two other components, it \textit{must} then be a superposition of several eigenstates, and measuring one of these other components will thus always yield a probabilistic outcome. This has very important consequences: in fact, it is the underlying reason for the so-called `uncertainty principle', which we will discuss in Chapter 8. Anyhow, we have now gotten the answer on the last question we posed at the beginning of this section. We now know that there are only 3 quantum numbers necessary (and not -say- 4) to describe states of the hydrogen atom because there are precisely three compatible observables for that system, no more and no less. Which set you take (f.e. $(H,L^2,L_z)$ or $(H,L^2,L_x)$ or ...) still has some freedom, but you always need three of them.

\section[Filling in the gaps]{\includegraphics[width=0.04\textwidth]{comput_c} Filling in the gaps}
In this section we fill in three gaps that we left open in the story part. First, we explicitly perform the basis change that we were talking about in the `z-supremacy'-section. Next, we give a baby-example of another problem where a CoSCO shows up. Finally, we compute the commutation relations between different components of the angular momentum to show that they are non-zero. This proves the different components of the angular momentum are non-compatible observables.

\subsection{Gap 1: Changing basis}
Let us denote the states $|2,1,-1\rangle $, $|2,1,0\rangle $ and $|2,1,1\rangle $ by the vectors
\be
\left( \begin{array}{c}
1  \\
0   \\
0\end{array} \right) \, , \quad
\left( \begin{array}{c}
0  \\
1   \\
0\end{array} \right) 
\quad \textrm{and}\quad
\left( \begin{array}{c}
0  \\
0   \\
1\end{array} \right) 
\ee
So the vector 
\be
\left( \begin{array}{c}
a  \\
b   \\
c\end{array} \right) 
\ee
is a shorthand for the state $a\, |2,1,-1\rangle  + \,b\,|2,1,0\rangle  + \,c\,|2,1,1\rangle $. The advantage of this symbolic notation is the following. Operators acting on these three states can be represented by a matrix acting on the corresponding vector. Indeed, the action of $L_z$ for example can be written as
\be
L_z= \hbar \left( \begin{array}{ccc}
1 & 0&0  \\
0 & 0 & 0 \\
0& 0 & -1\end{array} \right) 
\ee
So for example the equation $L_z |2,1,-1\rangle  = (-\hbar) |2,1,-1\rangle $ is represented in matrix form by
\be
\hbar\left( \begin{array}{ccc}
1 & 0&0  \\
0 & 0 & 0 \\
0& 0 & -1\end{array} \right) 
\left( \begin{array}{c}
0  \\
0   \\
1\end{array} \right) =
(-\hbar)\left( \begin{array}{c}
0  \\
0   \\
1\end{array} \right) 
\ee
Similarly, the operator $L^2$ acts on these states as 
\be
L^2= 2 \hbar^2 \left( \begin{array}{ccc}
1 & 0&0  \\
0 & 1 & 0 \\
0& 0 & 1\end{array} \right) 
\ee
because all three states are eigenstates of $L^2$ with eigenvalue $\hbar^2 \ell(\ell+1) = 2 \hbar^2$. Using the spherical coordinate expression for $L_x$:
\be
L_x= i \hbar \left( \sin \varphi \partial_\theta + \cot \theta \cos \varphi \partial_\varphi\right) 
\label{eq:Lx}
\ee 
and its explicit action on the three states under consideration, one can check that $L_x$ acts as the matrix
\be
L_x = \frac{\hbar}{\sqrt{2}} 
\left( \begin{array}{ccc}
0 & 1&0  \\
1 & 0 & 1 \\
0& 1 & 0\end{array} \right) 
\ee
Let us find the eigenvalues and eigenvectors of this matrix. First of all, the determinant  of $L_x - \lambda 1$ is $(-\lambda^3+\hbar^2\lambda)$, which has solutions $\lambda=-\hbar$, $\lambda=0$ and $\lambda=\hbar$. This means the eigenvalues of the matrix $L_x$ are $-\hbar$, $0$ and $\hbar$, just as for $L_z$. On the contrary, the eigenvectors are different. The eigenvector corresponding to $\lambda=-\hbar$ can be found by demanding
\be
 \frac{\hbar}{\sqrt{2}} 
\left( \begin{array}{ccc}
\sqrt{2} & 1&0  \\
1 & \sqrt{2} & 1 \\
0& 1 & \sqrt{2}\end{array} \right) \left( \begin{array}{c}
a   \\
b \\
c\end{array} \right)=0 
\ee
Which is solved by
\be
v_1=
\left( \begin{array}{c}
1  \\
-\sqrt{2}   \\
1\end{array} \right) 
\ee
or any multiple of that vector. Similarly, one can find the eigenvectors corresponding to the other two eigenvalues. They are
\be
v_2=\left( \begin{array}{c}
1  \\
0   \\
-1\end{array} \right) 
\quad \textrm{and}\quad
v_3=
\left( \begin{array}{c}
1  \\
\sqrt{2}   \\
1\end{array} \right) 
\ee
We are now very close to our original goal. Remember that we set out to find a basis of states that are eigenstates of the operator $L_x$ instead of $L_z$. By our above calculation, we have actually done so, in a very economic fashion. Indeed, the states corresponding to the vectors $v_1$, $v_2$ and $v_3$ are precisely eigenstates of $L_x$, with eigenvalues $-\hbar$, $0$ and $\hbar$. For convenience, we will denote these states by $|\psi_1\rangle $, $|\psi_2\rangle $ and $|\psi_3\rangle $. For example: $v_1$ corresponds to the state $|\psi_1\rangle  = 1 |2,1,-1\rangle  - \sqrt{2}| 2,1,0\rangle  + |2,1,1\rangle $, and one can check (although that is a bit tedious) that the corresponding wave function indeed is an eigenstate of $L_x$ with eigenvalue $-\hbar$. If we would switch to a notation where the vectors 
\be
\left( \begin{array}{c}
1  \\
0   \\
0\end{array} \right) 
\left( \begin{array}{c}
0  \\
1   \\
0\end{array} \right) 
\left( \begin{array}{c}
0  \\
0   \\
1\end{array} \right) 
\ee
would represent the states $|\psi_1\rangle $, $|\psi_2\rangle $ and $|\psi_3\rangle $, than $L_x$ would have precisely the form $L_z$ had before:
\be
\hbar\left( \begin{array}{ccc}
1 & 0&0  \\
0 & 0 & 0 \\
0& 0 & -1\end{array} \right) .
\ee
This confirms that one may just as well work with $L_x$ eigenstates by a simple change of basis. We can easily make this basis ortho\textit{normal} by getting the normalization right. We illustrate this with $|\psi_1\rangle $: redefine
\be 
|\psi_1\rangle  = N (1 |2,1,-1\rangle  - \sqrt{2}| 2,1,0\rangle  + |2,1,1\rangle )
\ee
 with $N$ a suitable constant. Demanding $\langle \psi_1|\psi_1\rangle =1$ gives
\bea
1&=&  ( \langle 2,1,-1| - \sqrt{2}\langle  2,1,0| + \langle 2,1,1|)\bar{N}\nonumber\\ 
&& \quad \quad \quad \quad \cdot N (1 |2,1,-1\rangle  - \sqrt{2}| 2,1,0\rangle  + |2,1,1\rangle ) \nonumber\\
&=& |N|^2 (1 + 2 +1)\nonumber\\
& \Rightarrow& |N|=\frac{1}{\sqrt{2}}\nonumber
\eea
So we could take for example $N = \frac{1}{\sqrt{2}}$ or $N e^{i \theta}$ with any value of $\theta$ - as we have seen before that they are all considered as a the \textit{same} physical state anyway. The very same thing can be done to normalize the other two vectors. 

Note: in the above example we have only changed basis from $L_z$ to $L_x$ eigenstates for the $n=2$ level; the general principle is hopefully clear.

\subsection{Gap 2: Another CoSCO}
Here, we do a baby example involving the use of a CoSCO. Consider the operator $D^2$. Its eigenfunctions with eigenvalue $- 1$ are given by $e^{i x}$  and $e^{- i x}$. Since these are two linearly different functions, they form an eigenspace of dimension two. More general, for each eigenvalue $-a^2$ there are two eigenfunctions $e^{iax}$ and $e^{-iax}$. This means that the label $a$ is not enough to specify which function we are talking about. Can we find a (Hermitian) operator that commutes with $D^2$? Yes, we can define the \textbf{reflection operator} $R$ as
\be
R: f(x) \rightarrow R f(x)= f(-x).
\ee
So $R$ maps a function to the function obtained by flipping around the vertical axis:
\begin{figure}[H]
 \begin{center}
  \includegraphics[width=0.6\textwidth]{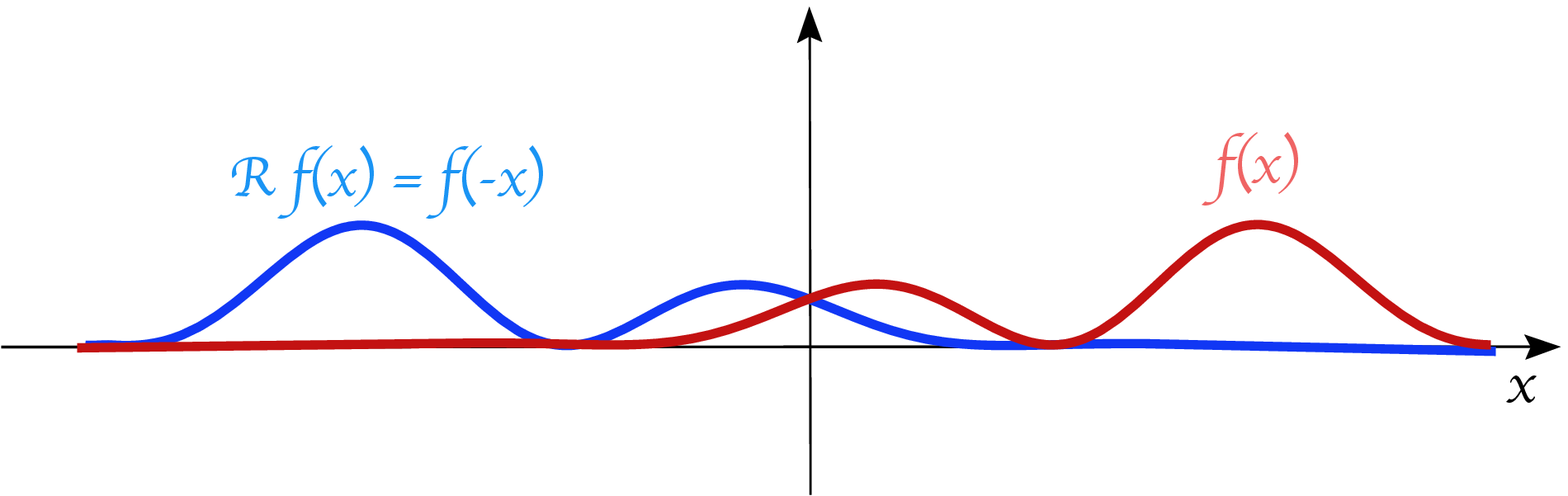}
  \end{center}
\end{figure}
One can check that this commutes with our other operator:
\be
[R,D^2]=0.
\ee
Here comes the nice thing: the eigenvalues of $R$ are $1$ and $-1$, and they split the degenerate eigenspaces of $D^2$ cleanly into two. Explicitly: \bea
R (e^{i a x} + e^{-i a x}) &=& e^{- i a x} + e^{i a x} \nonumber\\
 R (e^{i a x} - e^{-i a x}) &=& e^{- i a x} - e^{i a x} = - (e^{i a x} - e^{-i a x})
\eea
So the functions $ e^{- i a x} + e^{i a x} $ and $ e^{- i a x} - e^{i a x} $ provide a very nice basis: they are simultaneous eigenfunctions of $D^2$ and $R$, uniquely labeled by some $a$ and $\pm1$. That's all. Of course, the example has some shortcomings. We did not specify the total Hilbert space that we wanted to find a basis for. Also, the basis elements are is a bit special - they are not normalizable. This is somewhat unusual, but similar situations will be dealt with in the next chapter. The only point however was to show the CoSCO mechanism of using multiple labels to describe states uniquely. The obtained basis then \textbf{diagonalizes} all the operators of the CSCO. By `diagonalizing', we just mean `to form/pick a basis of eigenfunctions'  - just like in linear algebra. 

\subsection{Gap 3: The angular momentum commutation relations}
In the story part, we mentioned that the different components of the angular momentum do not commute, so that it is not possible to find simultaneous eigenstates. Here, we explicitly find these commutators. This also puts many of the formulae of first section into practice.
Let us start with $[L_x,L_y]$:
\be
[L_x,L_y]=[Y P_z - Z P_y, Z P_x - X P_z] 
\label{written out}
\ee
using linearity, one gets four commutators. One of them for example is $[Y P_z, Z P_x ]$. Now using the `drag out sideways' (\includegraphics[width=0.03\textwidth]{drag_out_sideways}) rule from above, we can rewrite that commutator as a sum of four terms:
\bea
[Y P_z, Z P_x ] &=&  Z [Y P_z, P_x ] + [Y P_z, Z ] P_x\nonumber\\
& =& Z Y [P_z, P_x ]+ Z [Y , P_x ] P_z+ Y[ P_z, Z  ]P_x+ [Y , Z ] P_z P_x
\eea
Phooh, this smells nasty. Luckily, the last expression is not as horrible as is looks. The only term involving a nonzero commutator is the $Y[P_z,Z] P_x$ term. So the entire above expression just equals $ - i \hbar Y P_x$. From this, we can conclude the following trick. Just look if there are a position and momentum operator of the same coordinate occurring in a commutator. Such terms are the only possible reason for a commutator to be nonzero. Another term of (\ref{written out}) is $[Z P_y, X P_z]$. Here, the only problem arises from the z-coordinate. Pulling out the two other operators, we get
\be
[Z P_y, X P_z] = P_y X [Z, P_z] = i \hbar X P_y
\ee
The two remaining terms of (\ref{written out}) have no possible problem: only mixed commutators, or position-position or momentum-momentum. So they are zero. In conclusion:
\be
[L_x,L_y]=i \hbar X P_y - i \hbar Y P_x = i \hbar L_z
\label{eq:ang mom 1}
\ee 
Analogously, one can find the two other commutators. Writing all results together, we have:
\be
[L_x,L_y] =  i \hbar L_z, \quad \quad  [L_y,L_z]= i \hbar L_x, \quad\quad [L_z,L_x] =   i \hbar L_y \nonumber\\
\label{eq:ang mom 2}
\ee
The above relations are called the \textbf{angular momentum commutation relations}. They show that the different components of the angular momentum aren't compatible observables. 
Computing them used everything we learned about commutators, so if you have reached/understood this page, you can rightfully call yourself a true commander of commutator craft. Good job!
\newpage
\section*{Exercises}
\begin{enumerate}
\item Explain in detail to someone (or yourself) what a CoSCO is. Can you reconstruct the reasoning?
\item Use the above angular momentum commutation relations and the `pull out sideways' rule to show that $[L^2,L_x]=0$.
\item Compute $[f(X),P]$ where $f$ is a function.
\item Show the \textbf{Baker-Campbell-Hausdorff formula}, which states that $e^{A}Be^{-A}=B+[A,B]+\frac{1}{2!}[A,[A,B]]+\frac{1}{3!}[A,[A,[A,B]]]+...$. 
(Hint 1:  the exponential of an operator is defined by the series expansion. Hint 2: The expression on the left can be seen as the function $F(s) = e^{sX}Ye^{-sX}$, evaluated at $s=1$. The value there can also be obtained by series expanding $F(s)$ around $0$. What relation is there between two consecutive coefficients of the series expansion?)
\item Find the states with $n=2$ which are eigenstates of $L_y$. 
\item Make up two operators and computer their commutator.
\item It was stated that two observables commute if and only if they can be simultaneously diagonalized (=admit a basis of states which are all eigenstates of both operators). What is the linear algebra analog of this statement? Can you prove this?
\end{enumerate}

%% file: H7.tex
\chapter{Position and momentum basis}

\subsection*{In this chapter...}
In the previous chapters, we have learned that a particle is not a point but an object smeared out over some region in space. This means the position is not sharply defined. In this chapter we will think a bit more careful about the \textit{momentum} of a particle. We will see that (just like position) the momentum of a particle is not sharply defined. In fact, we will see that a particle can be thought of as a packet spread out over different momenta, in the very same way that it is a packet spread out over different positions. An important mathematical tool will be the notion of Fourier transform. This will allow us to switch back and forth between the two descriptions, and give some more feel about the nature of the wave function. 
\newpage

\section[Fourier's trick]{\includegraphics[width=0.04\textwidth]{tool_c}  Fourier's trick}

As mentioned in the introduction, we will need a mathematical tool called \textbf{Fourier transform}. We introduce this tool here. If you have not learned about this before, be sure to read the following carefully. It is an important technique for physics and mathematics, and understanding it well will surely benefit you later on. If you have seen Fourier transforms before, just see the following section as a short recap. We work in two steps: first we study the baby brother of Fourier transform: \textbf{Fourier series}. After that, we are in a better position to understand the (closely related) topic of Fourier transforms. 

\subsection{Fourier series}
Imagine you have a function that is periodic, that is: it consists of a single piece, which is repeated over and over. For instance, it could look like this:
\begin{figure}[ht]
 \begin{center}
  \includegraphics[width=0.7\textwidth]{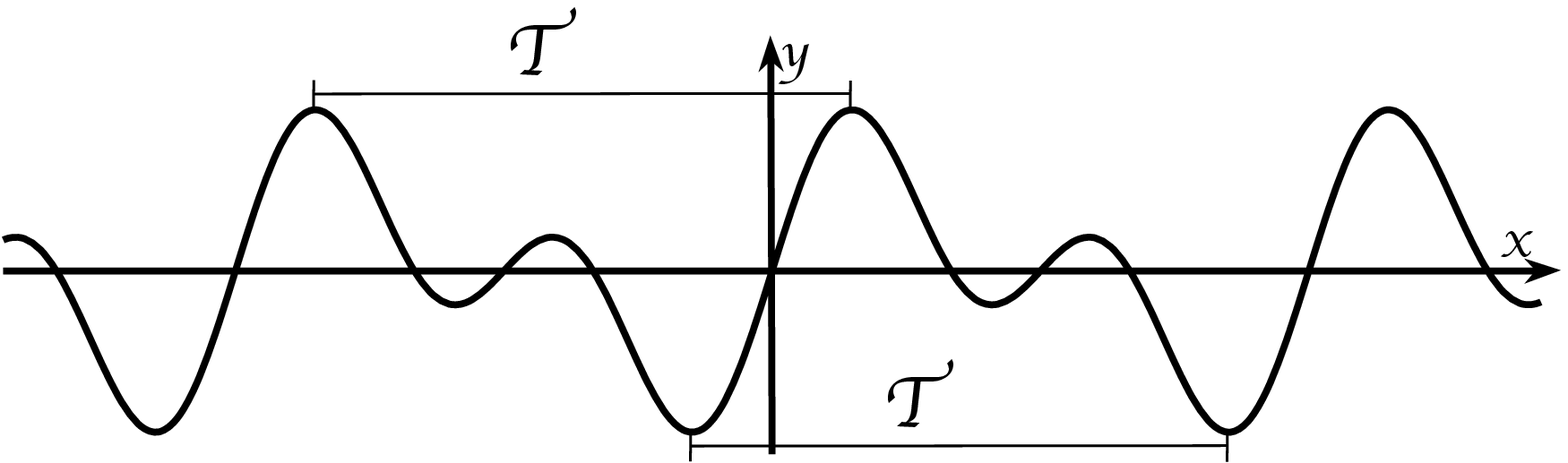}
  \label{fig:periode}
  \end{center}
\end{figure}

\begin{figure}[ft]
 \begin{center}
  \includegraphics[width=1.0\textwidth]{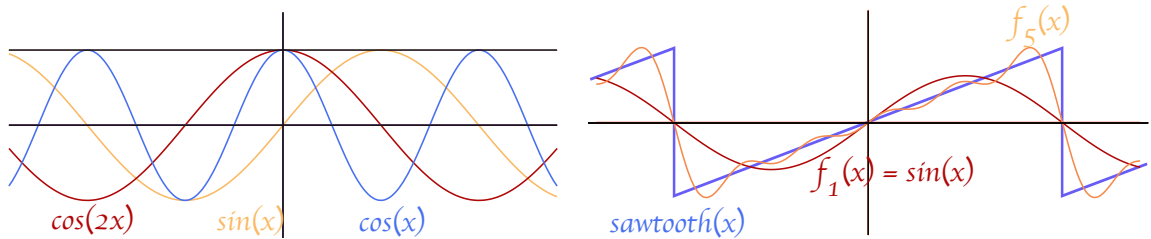}
  \caption{A Fourier series decomposes an arbitrary periodic function into harmonics. The left side shows some of the first harmonic functions involved in Fourier series. The right shows a function with the shape of a sawtooth. The line $f_1=\sin x$ includes only one term of the Fourier series. The function $f_5$ is a sum of more harmonic functions. Including even more terms, you can approximation the sawtooth with arbitrary precision.}
  \label{fig:fourier_series_c}
  \end{center}
\end{figure}

The length (in the x-direction) of the repeated piece is the \textbf{period} of the function.\footnote{For traveling waves, the period refers to the duration of the time cycle, and the wavelength refers to the periodic behavior in the spatial direction. Here, as we are considering a function of one variable only, you may call the length of a cycle either wavelength or period - the second one being slightly more popular.} For the function above, the period $T$ has been indicated on two places. Suppose that the period of a function is equal to $2 \pi$. In that case, you can capture all the information of the function effectively by giving its value on the interval $[-\pi,\pi]$. Indeed, the value of the function anywhere \textit{outside} the interval (say at some $x+ 2\pi k$, with $x\in [-\pi,+\pi]$ and $k$ an integer) is just given by
\be
f(x+ 2 \pi k) = f(x).
\ee
You know that there is a very important pair of functions functions that are of this kind (periodic with period $2 \pi$): the trigoniometric functions $\sin x$ and $\cos x$. In fact, all functions $\sin (n x)$ and $\cos (n x)$ (with $n$ a positive integer) are periodic with period $2 \pi$.\footnote{Actually, the functions $\sin (nx)$ and $\cos (nx)$ have period $2\pi/n$, but this implies they are also periodic over a distance of $2\pi$ - as this is just a multiple of their period.} The central question that lies at the heart of Fourier series, is the following: \begin{quote}Can we write the periodic function $f$ as a linear combination of the functions $\sin (n x)$ and $\cos (n x)$? 
\end{quote}
This question essentially asks how (and if) we can decompose an arbitrary periodic function into harmonic functions (sines and cosines). Because harmonic functions have some really nice properties, such a decomposition is obviously an interesting trick. It turns out that for almost every well behaved periodic function, this is indeed possible. An illustration of how this works is shown in Figure \ref{fig:fourier_series_c}. Let us try to see how this works in general. With some good faith, suppose such a decomposition is indeed possible. So assume that for every well-behaved $2\pi$-periodic function $f$ there exist numbers $a_i$ and $b_i$ (with $i\geq1$) and an $a_0$ such that
\be
f(x)=\frac{a_0}{2} + \sum_{n=1}^\infty \, [a_n \cos(nx) + b_n \sin(nx)]
\label{eq:decomp}
\ee
We have also included a constant function (first term), the last part is a sum over infinitely many harmonic functions. For this infinite sum to be well defined, we need the $a_n$ and $b_n$ to go to zero fast enough to ensure convergence. This is a mathematical problem which we will not touch upon here. A more important question one immediately comes up with is: given that this decomposition is possible, how do I \textit{find} the values of the $a_n$ and $b_n$? Here is a crucial observation: you can show that for all $n$ and all $m$
\bea
\int_{-\pi}^\pi \sin(nx) \cos(mx) dx &=&0 \label{eq:orthogser1}\\
\int_{-\pi}^\pi \sin(nx) \sin(mx) dx &=& \pi \delta_{nm} \label{eq:orthogser2}\\
\int_{-\pi}^\pi \cos(nx) \cos(mx) dx &=&\pi \delta_{nm}\label{eq:orthogser3}
\eea
where $\delta_{nm}$ is again the Kronecker delta:
\be
\delta_{nm} =  \left\{\begin{array} {cc}
1 &\textrm{if} \,\, n=m\\ 
0 &\textrm{if} \,\, n\neq m \end{array}\right.
\ee
In words: if one integrates a product of two different harmonic functions ($n\neq m$ or sine times cosine) the result will always be zero. 
Aha! Here is the miracle: if you multiply (\ref{eq:decomp}) by $\cos k x$ and integrate $x$ over $[-\pi,\pi]$, you get:
\be
a_k = \frac{1}{\pi}\int_{-\pi}^\pi f(x) \cos(kx)\, dx \quad \quad \forall k \ge 0
\ee
This expression also incorporates the expression for the constant term $a_0$. In a similar fashion you can show that
\be
b_k = \frac{1}{\pi}\int_{-\pi}^\pi f(x) \sin(kx)\, dx \quad \quad \forall k \ge 1
\ee
In conclusion: by performing simple integrals of the function $f$, we can find the constants $a_n$ and $b_n$. This gives a straightforward way to decompose any (well-behaved) periodic function into harmonics. Nice trick, Mr. Fourier!

\subsection*{Short story}

You might wonder about the possible uses of such a decomposition. If you ever have a course on differential equations, you are quite likely to learn the above trick, and see how it helps solving problems like heat evolution, the evolution of waves, etcetera. Another nice application are voice recognition and synthesizers. So either your favorite action movie scene (involving some fancy voice technology to open a safe) or your favorite music group might very well rely on Fourier transforms. As follows: every tone (of a voice or an instrument) typically has a frequency associated to it. However, a sound is seldom `pure', but consists of several superposed frequencies. These extra frequencies are sometimes called overtones. Their relative amplitude make the sound to what it is: they give it the typical total sound called \textbf{timbre}. It is precisely this what allows us to distinguish between different instruments and different voices: not just their pitch, but the \textit{combination} of different frequencies. Here, Fourier analysis comes into play. Imagine someone gives you a single sound signal (an amplitude as a function of time), and asks you to analyze it. This is not an easy task. However, if you use Fourier analysis (by f.e. some software that does the integrals/decomposition for you) you get a very nice overview of the frequencies that make up the signal. This allows you to very economically summarize the signal - as if you got it's fingerprint. This way, a synthesizer can store large numbers of different sounds, not relying on a sound database, but merely keeping some characteristic numbers for each one of them. For speaker recognition, the idea is similar: the composition of someone's voice can be analyzed and allows pretty well to uniquely characterize it, allowing for a rather secure identification method, just like a fingerprint does. Just to give a slight hint on the versatility of this technique.


\subsection{Fourier transform}
You may wonder: can make the above trick useful in the context of wave functions too? The answer is yes, but there are two important differences with the above. First, a wave function is not typically a periodic function. Second, a wave function is a \textit{complex} function while the above only dealt with real functions. This means we will have to adapt the above trick slightly, to a technique called \textbf{Fourier \textit{transform}}. We start out with the remark that the complex analogue of $\sin (kx)$ and $\cos (kx)$ is the function
\be
e^{i k x } \quad \textrm{with} \quad k \,\, \textrm{real}
\ee
So it is natural to go and try to decompose a complex function into several such $e^{ikx}$. Since in the context of a wave function no specific periodicity is required, $k$ can take on any real value. This suggests we need an \textit{integral} (not just a sum) over different $k$'s. More precise, we want to write
\be
f(x)=\frac{1}{\sqrt{2\pi}}\int_{-\infty}^{+\infty} dk \,\,g(k) \,e^{i k x}
\label{eq:FT}
\ee
Here, $g(k)$ is a complex function, the Fourier transform of $f$, and has the same role as the coefficients $a_n$ and $b_n$ before. The prefactor is just there for convenience.\footnote{Also, note that we have written the integration measure $dk$ before the integrand and not after it. In case you have not met this before: it is a different notation but has an identical meaning. If you are confused by it, just scratch away the $dk$ and write it at the end of the expression.}  The function $g(k)$ describes to what extent each of the waves $e^{ikx}$ is `present' in the shape of $f$. So if $f(x)$ has a piece that looks like a (complex) oscillation $e^{i k_0 x}$, then $g(k)$ will typically be large around this $k_0$. 
Just like with a Fourier series, we can obtain $g$ by performing an integral of $f$. It turns out that $g$ can be obtained as follows:
\be
g(k)=\frac{1}{\sqrt{2\pi}}\int_{-\infty}^{\infty} d x \,\, f(x) e^{-i k x}
\ee
Notice how the last two expressions are nicely symmetric (up to the minus sign). For this reason, the second one is called \textbf{inverse Fourier transform}. Let us try to show how the above definition of $g$ indeed is the correct one. In words, we want to show that the above definition of $g$ substituted into (\ref{eq:FT}) indeed gives $f$ as a result. If we plug in the definition, we get
\be
\frac{1}{\sqrt{2\pi}}\int_{-\infty}^{+\infty} d k \,\,\left[\frac{1}{\sqrt{2\pi}}\int_{-\infty}^{\infty} d x' \,\, f(x') e^{-i k x'}\right]\,e^{i k x}
\label{eq:plugg}
\ee
and we want to show that this indeed equals $f(x)$. (We have changed the integration variable to $x'$, since $x$ is already taken - its the argument of function $f$.)
\subsection*{Proof that (\ref{eq:plugg})$=f(x)$.}
\begin{figure}[ht]
 \begin{center}
  \includegraphics[width=0.35\textwidth]{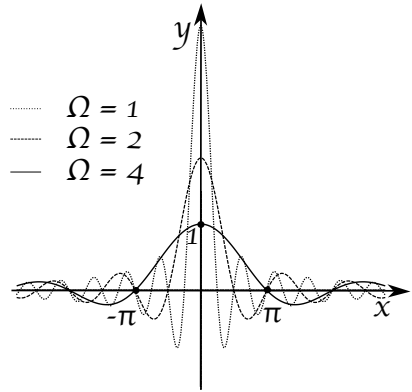}
  \caption{The sinc function $y = \frac{\sin(\Omega x)}{x}$, plotted for several values of $\Omega$.}
  \label{fig:sinc}
  \end{center}
\end{figure}

If we rearrange, and rewrite the range of integration as a limit, expression (\ref{eq:plugg}) becomes:
\be
\frac{1}{2\pi} \int_{-\infty}^\infty dx' \,\,f(x') \lim_{K \rightarrow \infty} \int_{-K}^{K} d k e^{i k (x-x')}
\label{eq:FTproof}
\ee
We have interchanged the integrals, which -in the context of physics- is no problem. (Mathematicians will tell you that this may not always be correct but we're not going to be that careful.) Now the integral on the right can be done explicitly (renaming $y=x-x'$):
\be
\int_{-K}^{K}\,\, d k  \, e^{i k y } = \left[\frac{1}{i k }e^{i k y}\right]_{k=-K}^{k=K}=2 \frac{\sin (K y) } {y}
\label{eq:14}
\ee
The function on the right side (not including the factor $2$) is called the \textbf{sinc function}. It's shape is shown in Figure \ref{fig:sinc}. There, you can see that the central peak is getting higher and more narrow with increasing $K$. The limit for large $K$ occurring in (\ref{eq:FTproof}) goes to ($\pi$ times) a very special function, the \textbf{Dirac delta function} (denoted $\delta$):
\be
\lim_{K \rightarrow \infty} \frac{\sin K y }{y}= \pi \delta(y)
\label{eq:15}
\ee
We will say more about this function in the next section, but we already give one of its properties, namely:
\be
\int_{-\infty}^\infty f(x) \delta(x)\,\, dx = f(0) 
\label{eq:propdel}
\ee
for every function $f$. This is a peculiar property, but as promised, the next section will explain this more carefully, so don't worry - just assume it is true for now. With this information, we can rewrite (\ref{eq:FTproof}) further. In terms of the new variable $y=x-x'$, the right hand side becomes
\be
\frac{1}{2\pi} \int_{-\infty}^\infty d y \,\,f(x-y) \lim_{\Omega\rightarrow \infty} 2 \frac{\sin \Omega y }{y}= \frac{1}{2\pi} \int_{-\infty}^\infty d y \,\,f(x-y) 2\pi \delta(y)
\ee
Using the special property of the delta-function, we can rewrite this last expression as
\be
\frac{1}{2\pi} \int_{-\infty}^\infty d y \,\,f(x-y)  \,\, 2 \pi \delta(y) = f(x-y)|_{y=0}=f(x)
\ee
Ah! This is precisely what we wanted. The above line shows that the definition of $g(k)$ is indeed the right one, and that it correctly gives the decomposition of $f(x)$ into a sum (integral) of oscillating functions $e^{i k x}$. This decomposition will turn out to have some nice applications in quantum mechanics, as we will see in the rest of this chapter. Before we go there, some more words on this mysterious delta function.

\subsection{Delta function}
Consider the following series of functions $\chi_n$ (with $n$ a positive integer):
\be
\chi_n(x)= 
 \left\{\begin{array} {cc}
n &\textrm{if} \,\, x \in [-\frac{1}{2n},\frac{1}{2n}] \\ 
0 &\textrm{everywhere else} \,\,\end{array}\right.
\ee
These functions have the following property: for large $n$ they are only nonzero within a very narrow interval. Despite this narrowing, the surface under each of these functions is always $1$, due to the fact that the height of the bump grows bigger and bigger (Figure \ref{fig:chi}). Explicitly:
\begin{figure}[H]
 \begin{center}
  \includegraphics[width=0.35\textwidth]{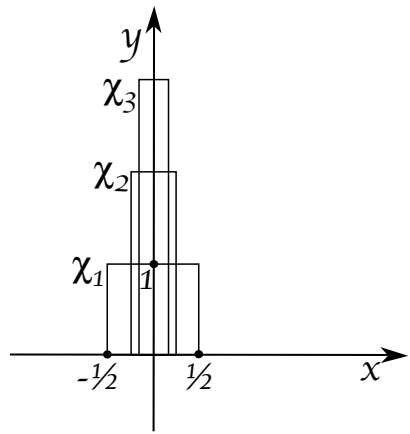}
  \caption{The functions $\chi_n$.}
  \label{fig:chi}
  \end{center}
\end{figure}
\be
\int_{-\infty}^{\infty} \chi_n(x) dx =1 \quad \forall \, n
\ee
Now here is an alternative (but equivalent) definition of the Dirac delta function we met above:
\be
\delta(x) = \lim_{n\rightarrow \infty} \chi_n(x).
\ee
Now if you like performing limits, you will immediately see that 
\be
\delta(x) =
 \left\{\begin{array} {cc}
\infty &\textrm{if} \,\, x=0 \\ 
0 &\textrm{everywhere else} \,\,\end{array}\right.
\ee
This is definitely not an ordinary function. In fact, the object $\delta$ is not a function, but a \textbf{distribution}. A distribution is an object that makes not so much sense on itself, only when it standing inside an integral. Indeed, although the above prescription looks a bit strange, the integral of $\delta$ is perfectly fine:
\be
\int_{-\infty}^\infty \delta(x) \,dx = \int_{-\infty}^\infty  \lim_{n\rightarrow \infty} \chi_n(x)\,\,dx =  \lim_{n\rightarrow \infty}  \int_{-\infty}^\infty  \chi_n(x)\,\,dx = \lim_{n\rightarrow \infty}1=1
\ee
Also, using the fact that for a continuous function $f$ 
\be
\lim_{\epsilon\rightarrow 0} \int_{x-\epsilon}^{x+\epsilon} f(y) dy = 2 \epsilon f(x)
\label{eq:cont func}
\ee
we get
\bea
\int_{-\infty}^\infty f(x) \delta(x) \,dx&=& \lim_{n\rightarrow\infty} \int_{-\infty}^\infty f(x) \chi_n(x)\,dx \\
&=& \lim_{n\rightarrow\infty} n \int_{-1/2n}^{1/2n} f(x)\,dx \\
&=& \lim_{n\rightarrow\infty} n \frac{1}{n} f(0)\\
&=& f(0).
\eea
This result is precisely the property we used in the section about Fourier series! In words: the central peak of $\delta(x)$ picks out the region around zero as the only contribution to the integral, giving $f(0)$ as a result. From this property, one can derive that
\be
\int_{-\infty}^\infty f(x) \delta(x-a)\,dx = f(a).
\ee
To see this, just rewrite this last expression in terms of a new integration variable $y=x-a$. Hopefully, you are not too much confused by the peculiar character of this object. Just keep in mind that the delta function is not an ordinary function, but a more general object: a distribution. As physicists are less careful about some things (like interchanging sums and integrals an limits in all the above) they tend to treat the object $\delta(x)$ just as a function, and even named it that way. You will get used to these properties later on.
\subsection*{Small remark}
A last remark: you may wonder whether the definition used here is really equivalent to the one from section on Fourier transforms. We will not really show this, but you may be convinced by looking back at Figure \ref{fig:sinc}. There too, we were dealing with an infinite series of functions (labeled by $K$ instead of $n$). There too, a central peak was getting higher and narrower, while maintaining the area under the graph constant. Hopefully you see that such a series of functions will converge to the same `object', the Dirac delta function. Just notice that all the above properties derived above only depended on the limiting behavior of the $\chi_n$, and not so much on their specific shape. So far the tools, let us get back to the world of quantum mechanics.

\section[Changing habits]{\includegraphics[width=0.04\textwidth]{once_c} Changing habits}
In the previous chapters, you learned that the states of a particle form a complex vector space, the Hilbert space. So far, we have studied three systems: a particle in a box, a particle in a harmonic potential, and the hydrogen atom. In each case we have found energy eigenstates as a natural basis (modulo soms complications that arose for the Hydrogen atom). Now let's try to change habits a bit, and see if we can also express states in terms of another basis. Here we will describe two other natural bases: the \textbf{momentum basis} and the \textbf{position basis}. Just to make things easier, we will work in one dimension - this avoids working with vectors - but all can be readily generalized to three dimensions.

\subsection{Momentum basis}
We have met the momentum operator already several times:
\be
P=-i \hbar D_x
\ee
You can easily check that this operator has eigenfunctions
\be
e^{i p x/\hbar}
\ee
with eigenvalue $p$. Now imagine someone gives you the wave function $\psi(x)$ of a particle, and asks you whether you can write that wave function as a sum of momentum eigenstates $e^{i p x/\hbar}$. You think about it for a moment, and then you realize: of course I can! This is precisely what a Fourier transform does. So yes, you can write as the decomposition
\be
\psi(x) = \frac{1}{\sqrt{2\pi\hbar} } \int_{-\infty}^\infty \,\,d p\,\, \psi(p) e^{i p x/\hbar}
\ee
where $\psi(p)$ is given by
\be
\psi(p)= \frac{1}{\sqrt{2\pi\hbar}} \int_{-\infty}^\infty \,\, dx\,\, \psi(x) e^{-i p x/\hbar}
\label{eq:momft}
\ee
Compared to the original expression for the Fourier transform, some extra $\hbar$'s have popped up in the prefactors. This is just a convention, and due to the fact that we decompose into $e^{ipx/\hbar}$ instead of the ordinary $e^{ikx}$. Furthermore, we have denoted the Fourier transform $\psi(p)$ by the same letter $\psi$ - in principle this is an ambiguous notation, but we will always write the argument explicitly to avoid confusion. So $\psi(p)$ and $\psi(x)$ are different functions. The Fourier transform $\psi(p)$ is called the \textbf{momentum space representation} of the wave function. For each $p$, it tells you to what extent the wave function looks like $e^{i\frac{p}{\hbar} x}$. To contrast, the wave function $\psi(x)$ which we worked with up to now, is called the \textbf{position representation}. The momentum representation (like with the position representation) is a function spread out over a certain region. So just like the fact that a wave function doesn't have a sharp position, it has a smeared out (complex) function describing its momentum. This means we have a very symmetric way to describe a particle: either by its wave function in position space $\psi(x)$, or by its wave function in momentum space $\psi(p)$.
\subsubsection*{Bra-ket notation}
We can also rewrite the above in bra-ket notation. First, we define the \textbf{momentum state} $|p\rangle $ by taking the corresponding wave function as follows:
\be
|p\rangle  \quad \leftrightarrow\quad \frac{e^{ipx/\hbar}}{\sqrt{2 \pi \hbar}}
\ee
With this, we can write the state $|\psi\rangle $ corresponding to the wave function $\psi(x)$ as follows:
\be
|\psi\rangle  = \int_{-\infty}^\infty \,\,dp\,\, \psi(p) |p\rangle 
\ee 
If you see the integral as an infinite sum, the right hand side is a linear combination of all the different $|p\rangle $. The $\psi(p)$ can thus be seen as the coefficients of all the $|p\rangle $ in this linear combination and is given explicitly by (\ref{eq:momft}). With this, we have reached the goal of this section: decomposing an arbitrary state $|\psi\rangle $ with respect to the basis of all momentum eigenstates $|p\rangle $. 
\subsubsection*{Measurements}
Beside giving the particle's momentum `components', the function $\psi(p)$ has another meaning. Given a state $|\psi\rangle $ and a device that measures the momentum, the chance to receive an outcome between momenta $p_0$ and $p_1$ is given by
\be
P(p_0< \textrm{(outcome of measurement)}< p_1)=\int_{p_0}^{p_1} |\psi(p)|^2 dp
\ee
So the object $|\psi(p)|^2$ acts as a \textbf{probability density}, just like $|\psi(x)|^2$. Yes, yes: it is all very symmetric between $x$ and $p$. Here is an example of a wave function $\psi(x)$  (real part shown) and its momentum representation $\psi(p)$:
\begin{figure}[H]
 \begin{center}
  \includegraphics[width=0.8\textwidth]{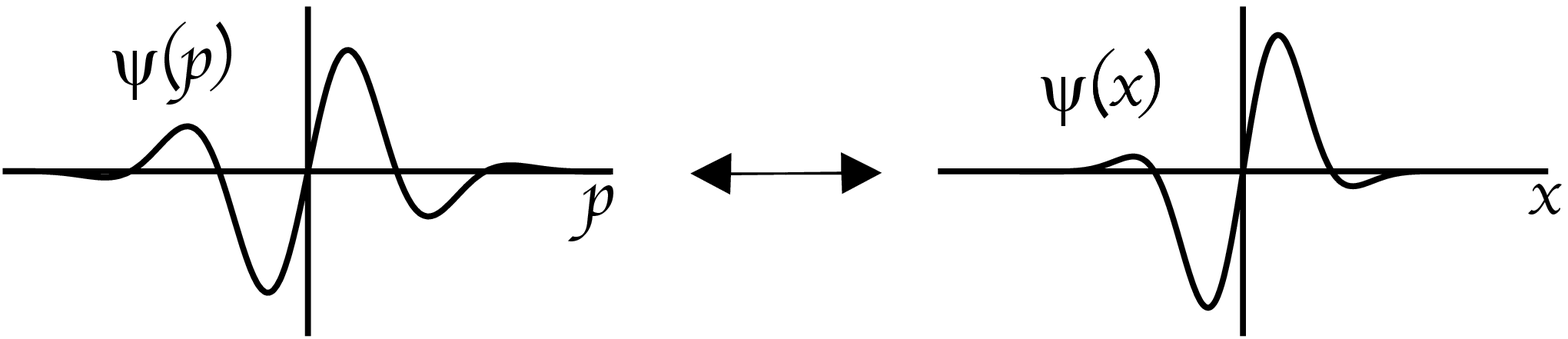}
   \end{center}
\end{figure}

\subsection{Position basis}
We now know there is something like a momentum basis. Let us now try to hunt for its counterpart, the position basis. First, think about this question more careful. The eigenstates we found for the momentum operator had a very clear physical meaning: they were infinite waves, with a sharp and well-defined momentum. Similarly, we are now looking for functions that can describe a wave function which is sharply localized, so that it has an exact position. We \textit{did} recently meet such `infinitely sharp' function: the delta function. Indeed, the set of functions
\be
f_{x_0}(x)=\delta(x-x_0)
\ee 
are all infinitly peaked. The peak of each $f_{x_0}$ is at $x_0$, as on that place the function value is $\delta(x_0-x_0)=\delta(0)=\infty$. So we suggest that the function $f_{x_0}$ describes a wave function of a state with exact (sharp) position $x_0$. We denote such a sharp position-state by $|x_0\rangle $, so
\be
\textrm{wave function} \, f_{x_0} \leftrightarrow \textrm{state}\,\, |x_0\rangle  \leftrightarrow \textrm{state sharply localized at position}\,\, x_0 
\ee
Let us do a check on this. For consistency, we should show that the position-states $|x_0\rangle $ are indeed eigenstates of the position operator $X$, just in the same way the momentum-states $|p\rangle $ are eigenstates of the momentum operator $P$. So we want to show:
\be
X |x_0\rangle  = x_0 |x_0\rangle  
\label{eq:prov1}
\ee
for all $x_0$. 
Using that $X$ acts as multiplication with the identity function (so $X: f(x) \rightarrow x \cdot f(x)$) the wave functions of both sides are
\be
x \, \delta(x-x_0) = x_0 \,\delta(x-x_0).
\label{eq:prov2}
\ee
As we have said before, $\delta$ is a distribution, and such an object makes better sense when standing inside an integral. Let us see what both sides of (\ref{eq:prov2}) give inside an integral with a random function $h(x)$:
\bea
\int_{-\infty}^{\infty}\,\,h(x) \left(x \, \delta(x-x_0) \right) \,dx&=& \int_{-\infty}^{\infty} \,\, (x h(x)) \, \delta(x-x_0) \,dx = x_0 \,h(x_0) \nonumber \\
\int_{-\infty}^{\infty} \,\,h(x) \left(x_0 \, \delta(x-x_0) \right)  \,dx&=& x_0 \int_{-\infty}^{\infty}\,\, h(x)  \, \delta(x-x_0)\, \,dx= x_0 \,h(x_0) \nonumber
\eea
The result is the same! As the above holds for any function $h(x)$, we have to conclude that the objects $x \delta(x-x_0)$ and $x_0 \delta(x-x_0)$ are indeed equal. This proves (\ref{eq:prov2}) and hence (\ref{eq:prov1}).

\subsubsection*{Bra-ket notation}
Can we now express any state as a linear combination of position states $|x_0\rangle $? By analogy with momentum eigenstates, you might guess that we can write an arbitrary state $|\psi\rangle $ as follows:
\be
|\psi\rangle = \int_{-\infty}^{\infty} \,dx_0\, \psi(x_0) |x_0\rangle .
\label{eq:prov3}
\ee
where $\psi(x_0)$ is the wave function, evaluated at $x_0$. This expression makes sense since the weight given to each position state $|x_0\rangle $ in the above `sum' is just $\psi(x_0)$: precisely the value of the wave function at position $x_0$. Better even, we can actually prove the above equation by showing that the state on the right side really has wave function $\psi(x)$. To see this, note that each $|x_0\rangle $ has wave function $\delta(x-x_0)$, so the total wave function of the right hand side is
\be
\int_{-\infty}^{\infty} \,dx_0\, \psi(x_0) \delta(x-x_0),
\ee
which just equals $\psi(x)$ by the standard property of the delta-function.
Expression (\ref{eq:prov3}) completes our second goal: writing out an arbitrary state $|\psi\rangle $ as a linear combination (integral) of position-basis elements. 
This means we now understand how to write an arbitrary state in several bases: 
\begin{itemize}
\item a basis of energy eigenstates (as in the first chapters), 
\item a basis of momentum eigenstates $|p\rangle $ (each with weight $\psi(p)$) 
\item or a basis of position eigenstates $|x\rangle $ (each with weight $\psi(x)$).
\end{itemize}
The last two give the momentum- and space representation we wanted to discover. Before we do some concrete examples, one last remark.

\subsection*{Caricature states}
We just got to know two new kinds of states: momentum states $|p\rangle $ and position states $|x\rangle $. (From now on, we drop the subscript $0$ on the position state, so by $|x\rangle $ we mean the position state located at position $x$.) Both are a bit peculiar: one is infinite in extend (an infinite wave) the other is infinitely sharp (a delta function). We already mentioned that real particles are never in such a state, but always a superposition (integral) of these, which results in a nice, finite and smooth wave function. So what are the states $|p\rangle $ and $|x\rangle $ then? The best way to describe them is as \textbf{caricature states}. They describe a particle, if you could squeeze it into a perfect state with exact momentum, or into a perfect state of exact position. A very good comparison are point masses in Newtonian mechanics. When dealing with problems of bodies moving and interacting, you were often told to `treat the bodies involved as point masses'. This is not a realistic description, but allows for simple calculations. A point mass does not exist, but the mathematics of such an object is much easier than that of a (physical, extended) object. In very much the same way, the states $|p\rangle $ and $|x\rangle $ do not `exist in nature', but they have some nice properties, like \textit{providing a clean basis} in which you can express a physical state $|\psi\rangle $. That is why we study them, nothing more, nothing less.

\section[Using the position and momentum representation]{\includegraphics[width=0.04\textwidth]{comput_c} Using the position and momentum representation}

\subsection{Example}
We will now use all the above in the simplest example we have met so far: the particle in a box. Recall that we found energy eigenstates labelled by $n=1,2,...$:
\be
\psi_n(x)=\sqrt{\frac{2}{L}} \sin \f{n \pi x}{L}
\ee
These satisfy $H \psi_n(x) = E_n \psi_n$, and their time evolution dictated by the SE is given by
\be
\psi_n(x,t)= e^{-i E_n t/\hbar}\psi_n(x)
\ee
The most general element in the Hilbert space is given by 
\be
\psi(x)=\sum_n a_n \psi_n(x)
\label{eq:energy}
\ee
and its time evolution is just given by $\psi(x,t)=\sum_n a_n \psi_n(x,t)$.

\subsubsection*{Position basis}
The last equation describes a general element $|\psi\rangle$ in the Hilbert space, decomposed in energy eigenstates. Here, we will try to express this very state $|\psi\rangle $ in the position- or momentum basis. Let us start with the easy one: the position-basis. First, note that every energy eigenstate $|\psi_n\rangle $ can be written as
\be
|\psi_n\rangle  = \int_{-\infty}^\infty \,dx\, \psi_n(x) |x\rangle  =  \int_{0}^L \,dx\, \left(\sqrt{\frac{2}{L}} \sin \f{n \pi x}{L}\right) |x\rangle 
\ee
The range of integration has changed in the last step: this is because the wave functions $\psi_n$ are actually zero outside $[0,L]$. Hence, for the state $|\psi\rangle $ itself, we have:
\bea
|\psi\rangle &=&\sum_n a_n |\psi_n\rangle  \\
&=& \sum_n a_n \left( \int_0^L \,dx\, \left(\sqrt{\frac{2}{L}} \sin \f{n \pi x}{L}\right) |x\rangle \right)\\
&=& \int_0^L \,dx\, \left(\sqrt{\frac{2}{L}} \sum_n a_n  \sin \f{n \pi x}{L}\right) |x\rangle 
\label{eq:position}
\eea
In the same way the $a_n$ are the coefficients of $|\psi\rangle $ with respect to the energy-eigenstate basis, the function $\left(\sqrt{\frac{2}{L}} \sum_n a_n  \sin \f{n \pi x}{L}\right)$ describes the `coefficients' of the same state when expressed in the position basis. In simple terms: this object is just the (position) wave function of $|\psi\rangle$. Can we also do the same for momentum? That is, can we also find a function $\psi(p)$ such that
\be
|\psi\rangle  = \int_{-\infty}^\infty \,dp\, \psi(p) |p\rangle 
\label{eq:tog}
\ee
Again we first try to do this for each $|\psi_n\rangle $ individually, so we will first look for functions $\psi_n(p)$ satisfying
\be
|\psi_n\rangle  = \int_{-\infty}^\infty \,dp\, \psi_n(p) |p\rangle 
\label{eq:indiv}
\ee
From the general formula (\ref{eq:momft}) for the Fourier transform, we get 
\bea
\psi_n(p)&=& \frac{1}{\sqrt{2 \pi \hbar}}\int_{-\infty}^\infty \,\,dx \,\, e^{-i p x /\hbar} \psi_n(x) \label{eq:FFT}\\
&=&  \frac{1}{\sqrt{2 \pi \hbar}} \int_0^L \,\,dx \,\, e^{-i p x /\hbar} \left(\sqrt{\frac{2}{L}} \sin \f{n \pi x}{L}\right) \nonumber\\
&=&  \sqrt{\frac{1}{\pi \hbar L}} \int_0^L \,\,dx \,\, e^{-i p x /\hbar} \left( \frac{e^{i\frac{n \pi x}{L}} - e^{-i\frac{n \pi x}{L}}}{2i}\right) \nonumber\\
&=& \frac{1}{2i} \sqrt{\frac{1}{\pi \hbar L}} \int_0^L \,\,dx \,\, \left[e^{i x \left(\frac{-p}{\hbar}+\frac{n \pi }{L} \right)}- e^{i x \left(\frac{-p}{\hbar}-\frac{n \pi }{L} \right)} \right]\nonumber\\
&=& \frac{1}{2i} \sqrt{\frac{1}{\pi \hbar L}} \left[\frac{1}{i c_1}e^{i x c_1} - \frac{1}{i c_2}e^{i x c_2} \right]_{x=0}^{x=L} \label{eq:psi_p}
\eea
In the last step we have defined $c_1=\left(\frac{-p}{\hbar}+\frac{n \pi }{L} \right)$ and $c_2=\left(\frac{-p}{\hbar}-\frac{n \pi }{L} \right)$. With these expressions, you can write out the end result and simplify it a little bit, but we don't do this effort here. Just note that the end result depends on $p$ - via the objects $c_1$ and $c_2$. We have just done our first Fourier transform! Congratulations. 
With this, the state $|\psi\rangle $ can then be written as follows:
\bea
|\psi\rangle  &=& \sum_n a_n |\psi_n\rangle  \nonumber\\
&=& \sum_n a_n \left(\int_{-\infty}^\infty dp \,\, \psi_n(p) |p\rangle  \right)\nonumber\\
&=& \int_{-\infty}^\infty dp \left(  \sum_n a_n \psi_n(p)\right) |p\rangle  
\label{eq:momentum}
\eea
where the object between brackets -with $\psi_n(p)$ given in (\ref{eq:psi_p})- is just one function of $p$.
This is the finish line! We have succeeded in writing $|\psi\rangle $ not only in energy representation (\ref{eq:energy}) but also in the position basis (\ref{eq:position}) and momentum basis (\ref{eq:momentum}). If you managed to follow this last derivation, you have probably understood most of this chapter so far.

\subsection{Projection, inner product and completeness}
As a last application of what we have learned in this chapter, we give the so-called \textbf{completeness relations}. These are very useful equations, and provide a nice mnemonic for the formula of Fourier transform. First, recall that in a vector space we could take the \textit{projection} of a vector on a basis element:
\be
\langle \vec{e}_i, \vec{v} \rangle  = v_i
\ee
If the basis vectors $e_i$ are orthonormal, we can get back the original vector $\vec{v}$ by summing up all its projections along the basis vectors:
\be
\vec{v} = \sum_i e_i v_i = \sum_i \vec{e}_i \langle \vec{e}_i, \vec{v}\rangle 
\ee
Or if we write the right hand side with a Hermitian conjugate:\footnote{In the context of ordinary linear algebra, the Hermitian adjoint is denoted by an asterisk $*$ instead of a $\dagger$.}
\be
\vec{v} = \sum_i \vec{e}_i (\vec{e}_i ^* \vec{v}) =\left( \sum_i \vec{e}_i \vec{e}_i ^*\right) \vec{v}
\ee
In the last step, we used associativity of matrix multiplication. This is true for all $\vec{v}$, which implies
\be
\sum_i \vec{e}_i \vec{e_i}^*   = 1 \label{eq:compv2} 
\ee
where $1$ stands for the identity matrix. For example, in a two dimensional vector space with orthonormal basis $e_1 = \left(\begin{array}{c} 1 \\ 0 \end{array}\right) $ and $e_2 =  \left(\begin{array}{c} 0 \\ 1 \end{array}\right) $ we have
\be
\sum_i \vec{e}_i \vec{e_i}^*   = \left(\begin{array}{c} 1 \\ 0 \end{array}\right) \left(\begin{array}{cc} 1 & 0 \end{array}\right) +  \left(\begin{array}{c} 0 \\ 1 \end{array}\right) \left(\begin{array}{cc} 0 & 1 \end{array}\right)  = \left(\begin{array}{cc} 1& 0 \\ 0&1 \end{array}\right). 
\ee
Equation (\ref{eq:compv2}) is the completeness relation (for vector spaces). The name is derived from the fact that such a relation only holds for a complete (and orthonormal) basis. Again, it is possible to extend this to the context of quantum mechanics. First, we give some results that you can show yourself (see exercises): 
\bea
\langle x|\psi\rangle  &=& \psi(x)\label{eq:psix} \label{eq:eerste_comp}\\
\langle p|\psi\rangle  &=& \psi(p)\label{eq:psip} \\
\langle x|p\rangle &=&\frac{e^{i p x}}{\sqrt{2\pi\hbar}}\label{eq:psixp}\\
\langle x_1|x_2\rangle &=&\delta(x_1-x_2)\\
\langle p_1|p_2\rangle &=&\delta(p_1-p_2) \label{eq:laatste_comp}
\eea
Especially the last two equations are very suggestive: the states $|x\rangle $ and $|p\rangle $ are `orthonormal': only then with a Dirac delta function instead of a Kronecker delta. This suggests maybe here too, we have something like a completeness relation. Indeed, it turns out that
\bea
\int_{-\infty}^\infty dx\, |x\rangle \langle x| &=& 1\label{eq:compx}\\ 
\int_{-\infty}^\infty dp\, |p\rangle \langle p| &=& 1.\label{eq:compp}
\eea
Where $1$ is the identity operator (sends every state to itself). These are the completeness relations for quantum mechanics. You might be a bit puzzled if the objects on the left side are really operators. The object $\int_{-\infty}^\infty dx |x\rangle \langle x|$ for example is to be read as the command: starting with a state, take the inner product with $\langle x|$. Then, multiply this (complex) number by the corresponding state $|x\rangle $ and then sum over all $x$. This operation indeed sends a state to a state. Hopefully, you see the full parallel with (\ref{eq:compv2}). Here are nice consequences of the above equations. First:
\bea
|\psi\rangle  &=&1 |\psi\rangle \\
&=& \left(\int_{-\infty}^\infty dx \,|x\rangle \langle x|\right) |\psi\rangle  \\
& =& \int_{-\infty}^\infty dx \, |x\rangle \langle x|\psi\rangle  \\
&=& \int_{-\infty}^\infty dx \, |x\rangle  \psi(x) \\
&=& \int_{-\infty}^\infty dx \, \psi(x) |x\rangle .
\eea
Ah! If you compare the first and last part, you see that we precisely get the decomposition of $|\psi\rangle $ into position states. That's a really short way to get there. Another example:
\bea
\psi(x) &=& \langle x|\psi\rangle  \\
&=& \langle x|1|\psi\rangle \\
&=& \langle x|\left( \int_{-\infty}^\infty dp \, |p\rangle \langle p|\right) |\psi\rangle  \\
&=& \int_{-\infty}^\infty dp\,  \langle x|p\rangle \langle p|\psi\rangle \\
&=&  \frac{1}{\sqrt{2\pi\hbar}}\int_{-\infty}^\infty dp\,\, e^{i p x/\hbar}\,\,\psi(p)
\eea
In the before last step, when pulling out the integral to the left, we have used linearity of the inner product. Once again comparing first- and last part, we see that we have fully rediscovered the Fourier transform, out of the hat! This ultrashort `proof' of the Fourier transform formula is a very clear sign of the power of the bra-ket notation. You may even start to like it one day... . 

\subsection*{Conclusion}
In this chapter, you have learned to view a wave packet in a different way: as a superposition of different infinite waves (momentum-eigenstates), each with some weight. This should caution you a bit when speaking about a particle's momentum - just like with position the particle's state comprises an extended region of momenta. 
Of course, you can still speak of the \textit{expectation value} of both, these are well-defined. 

When you think back of the de Broglie relations, you may be a bit confused: there, the particle's momentum, energy, wavelength occur. Actually, it is best to see the variables over there as expectation values, for a state which is sufficiently centered around these values.

\newpage

\section*{Exercises}
\begin{enumerate}
\item Show (\ref{eq:eerste_comp}) - (\ref{eq:laatste_comp}).
\item Argue that the probability density of the momentum and position representation are actually mere applications of the Postulate on the measurement, with the only difference that sums are replaced by integrals, and probabilities by probability densities.
\item Check that (\ref{eq:orthogser1}) is true (should not be much work). Can you find an easy way to show (\ref{eq:orthogser2}) for the cases $n \neq m$? (Hint: the states $\psi_n$ of the particle in a box have different eigenvalues under $H$. What does this imply? Can you relate this to what we want to show here?) To show the case $n=m$, you may want to use $\sin^2{m x}= \frac{1- \cos 2 m x}{2}$. Use the result to check that we properly normalized the states of the particle in the box by putting the prefactor $\sqrt{2/L}$.
\item Consider the function $f(x)$, which has period $2\pi$, has $f(x)=-1$ if $-\pi<x<0$ and $f(x)=1$ if $0<x<\pi$. This function is called the \textbf{square wave}. What does this function look like: a sine or a cosine function? So which component of the series do you expect to be largest? Now go ahead and compute its Fourier series. (You best split the integrals in two parts. Which Fourier components are zero?)
\item Consider the wave function $\psi_n(x)=x$ which equals $\sqrt{n}$ on $\frac{-1}{2n}<x<\frac{1}{2n}$ and zero elsewhere. What is the Fourier transform of $\psi_n(x)$? Which function do you recognize? 
\end{enumerate}

%% file: H8.tex
\chapter{The uncertainty principle}

\subsection*{In this chapter...}
In the chapter on commutators we showed that compatible observables (observables admitting a basis elements, the elements of which are eigenstates to \textit{both} operators) always commute. In this chapter, we spend some time on a logical consequence of this: if operators don't commute, it is impossible to find such a basis of common eigenstates. A quantitative consequence is the so-called uncertainty principle. This is an important principle, with a very high degree of quantum flavor to it: it will help you to get a better feel for the nature of wave functions. 

In the tool part, we show an important example of a Fourier transform which already suggests the uncertainty principle. In the story, we explain in more detail what the principle is, where it comes from. We then give a proof and conclude with several consequences/applications.

\newpage

\section[Fourier transform of Gaussians]{\includegraphics[width=0.04\textwidth]{tool_c} Fourier transform of Gaussians}
As promised in the intro, the tool part consists of an extra example on Fourier transforms. Two other small things we will review are the notion of variance, and the Cauchy-Schwarz inequality. Ready? Go!

\subsection{Variance}
In statistics a function $\rho(x)$ is called a \textbf{probability density function} if for each $x$, $\rho(x)$ is a measure for the chance to have $x$ as an outcome. The usage of this term in quantum mechanics is just a special case of this. For the total probability to be one, we need:
\be
\int_{-\infty}^\infty \rho(x) \,dx=1
\ee
One then defines the expectation value of $x$, as follows:
\be
\langle x\rangle =\int_{-\infty}^\infty x \rho(x) \,dx.
\ee
For example, if $\rho(x)$ is the chance to win an amount $x$ at the lottery, then $\langle x\rangle $ will give the amount you win on average per game. (Which is always lower than the cost of you ticket, alas.) Equally interesting, one may ask how much your winnings will typically be \textit{away} from the average winnings. A quantity expressing this, is given by
\be
\langle (x-\langle x\rangle )^2\rangle  = \int_{-\infty}^\infty (x-\langle x\rangle )^2 \rho(x) \,dx.
\ee
This quantity indeed says something about the typical deviation: it gives the expectation of the squared deviation from average. This quantity is the \textbf{variance} $\textrm{Var}(x)$, and its square root is called the \textbf{uncertainty} $\Delta x$:
\be
\langle (x-\langle x\rangle )^2\rangle  = \textrm{Var}(x) = (\Delta x)^2
\ee
Of course you see that the above is what we have been using in the context of quantum mechanics. Just write the probability density $|\psi(x)|^2$ instead of the probability density function $\rho$. The quantity $\inft \, |\psi|^2 x \, dx\,$ then nicely gives the expectation value of the position:
\be
\langle X\rangle  = \langle \psi|X|\psi\rangle = \int_{-\infty}^\infty \psi^*(x) X \psi(x) \,dx = \int_{-\infty}^\infty x |\psi(x)|^2 \,dx.
\ee
Similarly the uncertainty of the position is given by
\be
\Delta X = \sqrt{\langle (X-\langle X\rangle )^2\rangle }
\ee
with 
\be
\langle (X-\langle X\rangle )^2\rangle = \int_{-\infty}^\infty (x - \langle X\rangle )^2 |\psi(x)|^2 \,dx.
\ee
More general, for \textit{any} observable $O$, we can define the expectation value $\langle O\rangle  = \langle \psi|O|\psi\rangle $ (see chapter on observables), and the corresponding uncertainty is
\be
\Delta O = \sqrt{\langle (O-\langle O\rangle )^2\rangle }
\ee
Although the name sounds somewhat mystical, there is nothing deep about the notion of uncertainty: it just expresses how far possible measurement outcomes of the observable $O$ on a given state $|\psi\rangle $ are lying apart. For example: a wave packet that is rather localized will have a small uncertainty $\Delta X$, a packet which is more spread out will have a much larger position uncertainty. Last remark: although the last two expressions don't seem to contain the state $|\psi\rangle $, they really \textit{do} depend on it as the expectation values $\langle \rangle $ are to be taken between $\langle \psi|$ and $|\psi\rangle $. To avoid confusion, some people write subscripts:
\be
\langle \Delta O\rangle _\psi = \sqrt{\langle (O-\langle O\rangle _\psi)^2\rangle _\psi}
\ee

\subsection{Fourier transform of Gaussians}
A very important function in math and physics, is the Gaussian distribution. It is given by
\be
f(x) = \frac{1}{\sqrt{2 \pi \sigma^2}} e^{- \frac{x^2}{2 \sigma^2}} 
\ee
and has variance $\textrm{Var}(x)=\sigma^2$. Also, using the property
\be
\inft e^{- \alpha x^2} \, dx= \sqrt{\frac{\pi}{\alpha}}
\label{eq:gaussint}
\ee
it is easy to show that $f$ is a properly normalized probability density function.  
\begin{figure}[H]
 \begin{center}
  \includegraphics[width=0.6\textwidth]{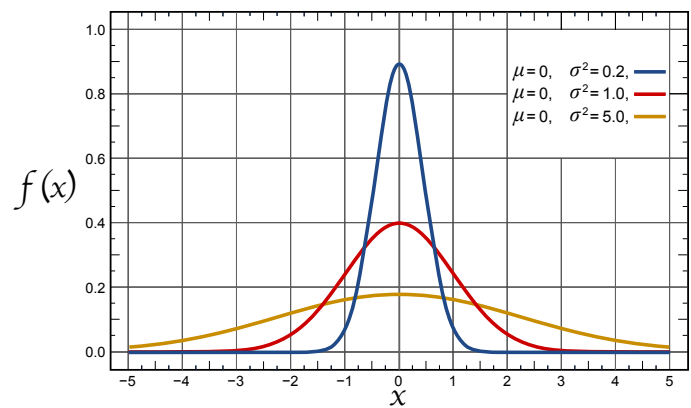}
  \caption{The Gaussian distribution, for different values of the variance}
  \label{fig:gaussian}
  \end{center}
\end{figure}
The fact that $f$ is normalized, means that its square root 
\be
\phi(x) = \frac{1}{(2 \pi \sigma^2)^{1/4}} e^{- \frac{x^2}{4 \sigma^2}} 
\label{eq:gauss}
\ee 
is a normalized wave function: $\inft |\phi(x)|^2 \, dx =1$. 
Now we ask the question: what is the Fourier transform of this function? In other words: what is the momentum representation $\phi(p)$ of the state $|\phi\rangle $? We know that
\be
\phi(p) = \frac{1}{\sqrt{2\pi \hbar} } \inft \, dx\,  \phi (x) e^{- i p x/\hbar}  
\ee
so we get 
\be
\phi(p) = \frac{1}{\sqrt{2\pi \hbar} } \frac{1}{(2 \pi \sigma^2)^{1/4}} \inft \, dx\,  e^{- \frac{x^2}{4 \sigma^2}+i p x/\hbar}  
\label{eq:tussenstap}
\ee
At first sight, performing the integral is not so obvious. However, there is a smart way to recast it into a form that allows to use (\ref{eq:gaussint}). This is done by completing the square standing in the exponent, as follows. Go to a new variable $x'=x- i \sigma p /\hbar$. Then
\be
x'^2= \left(x- \frac{2 i \sigma^2 p}{\hbar}\right)^2 = x^2  - 4  i x \sigma^2 \frac{p}{\hbar}  - 4 \sigma^4 \frac{p^2}{\hbar^2}
\ee  
and 
\be
\frac{- x'^2}{4\sigma^2} = \frac{- x^2}{4\sigma^2}  +  i x  \frac{p}{ \hbar}  +  \sigma^2\frac{ p^2}{ \hbar^2}.
\ee
Exponentiating, we get
\be
e^{\frac{- x'^2}{4\sigma^2}} = e^{\frac{- x^2}{4\sigma^2} } e^{i x  \frac{p}{ \hbar}} e^{\sigma^2\frac{ p^2}{ \hbar^2}}
\ee
which is the integrand of (\ref{eq:tussenstap}) up to the last (x-independent) factor. This means that in terms of $x'$ and $dx'(=dx)$,
\bea
\phi(p) &=& \frac{1}{\sqrt{2\pi \hbar} } \frac{1}{(2 \pi \sigma^2)^{1/4}} \inft \, dx'\,  e^{- \frac{x'^2}{4 \sigma^2}}   e^{-\sigma^2\frac{ p^2}{ \hbar^2}}\\
&=& \frac{1}{\sqrt{2\pi \hbar} } \frac{1}{(2 \pi \sigma^2)^{1/4}} e^{-\sigma^2\frac{ p^2}{ \hbar^2}} \inft \, dx'\,  e^{- \frac{x'^2}{4 \sigma^2}} \\
&=& \frac{1}{\sqrt{2\pi \hbar} } \frac{1}{(2 \pi \sigma^2)^{1/4}} e^{-\sigma^2\frac{ p^2}{ \hbar^2}} \cdot \sqrt{4 \sigma^2 \pi}
\eea
where in the last step we have used (\ref{eq:gaussint}). Rearranging, and defining $\sigma' = \frac{\hbar}{2 \sigma}$, we finally get
\be
\phi(p)= \frac{1}{(2 \pi \sigma'^2)^{1/4}} e^{-\frac{p^2}{4 \sigma'^2}}. 
\ee
Comparing to (\ref{eq:gauss}), you see that this distribution is again Gaussian, but now with variance $\sigma'^2$ instead of $\sigma^2$. This result is shown in the next figure:
\begin{figure}[H]
 \begin{center}
  \includegraphics[width=0.8\textwidth]{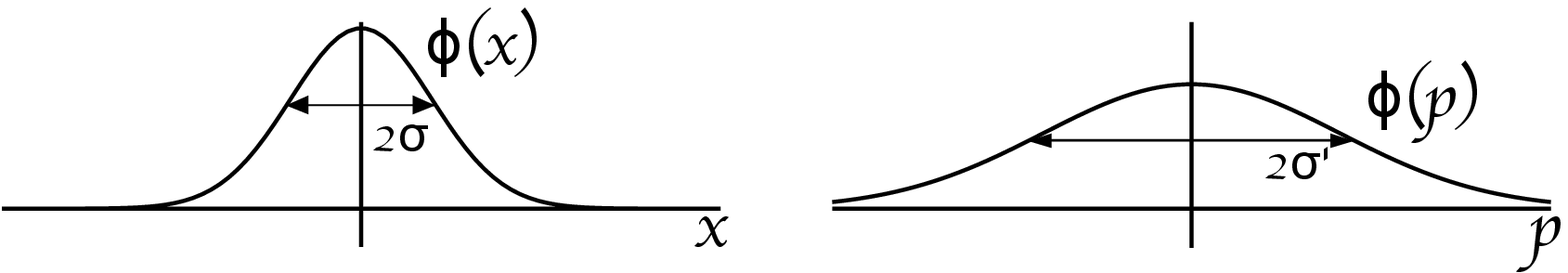}
  \end{center}
\end{figure}
This is quite remarkable: if a Gaussian wave packet is sharply peaked, its momentum representation will be very spread out, and vice versa. This is already a first glimpse of the uncertainty principle which we are about to meet. Before we go there, another small recap which might be useful:

\subsection{Cauchy-Schwarz inequality}
You probably remember the famous Cauchy-Schwarz inequality from a course on linear algebra. It states that for two vectors $\vec{x}$ and $\vec{y}$, 
\be
|\vec{x}\cdot\vec{y}|\leq \|\vec{x}\| \,.\, \|\vec{y}\|
\ee
In words: the magnitude of the inner product of two vectors is always smaller than the product of their lengths.
The reason for this inequality is very simple. For vectors, the inner product is just given by $\vec{x}\cdot\vec{y} =  \|\vec{x}\| \,.\, \|\vec{y}\| \,. \, \cos \theta $ with $\theta$ the angle between the two vectors. 
Taking absolute values on both sides, you then get
\be
|\vec{x}\cdot\vec{y}| = \|\vec{x}\| \,.\, \|\vec{y}\| \,. \, |\cos \theta\, | \leq \|\vec{x}\| \,.\, \|\vec{y}\|
\ee
This inequality can also be expressed in terms of inner products only. Writing $\langle \vec{x},\vec{y}\rangle $ instead of $\vec{x}\cdot\vec{y}$ and using $\|\vec{x}\|  = \sqrt{\langle \vec{x},\vec{x}\rangle }$, we get
\be
|\langle \vec{x},\vec{y}\rangle |^2\,\, \leq \,\,\langle \vec{x},\vec{x}\rangle \,\langle \vec{y},\vec{y}\rangle .
\ee
It turns out that this inequality is not only true for ordinary vector spaces $\mathbb{R}^n$, but for \textit{any} space with an inner product. This means that also in the context of the Hilbert space
\be
|\langle \chi|\psi\rangle |^2 \,\,\leq\,\, \langle \chi|\chi\rangle  \, \langle \psi|\psi\rangle 
\ee
for \textit{all} states $|\chi\rangle $ and $|\psi\rangle $. We will use this form of the Cauchy-Schwarz inequality later on in this chapter. Time for the story part... 

\newpage
\section[The uncertainty principle]{\includegraphics[width=0.04\textwidth]{once_c} The uncertainty principle}
\subsection{Commuting observables, revisited}
In the chapter on the commutator, we learned that if two observables are compatible (=there exists a basis of states $|\alpha_i,\beta_j\rangle $, labeled by $i$ and $j$, that are simultaneous eigenstates to both $A$ and $B$) then these operators commute: $[A,B]=0$: 
\bea
A|\alpha_i,\beta_j\rangle  = \alpha_i |\alpha_i,\beta_j\rangle  &\textrm{and}&  B|\alpha_i,\beta_j\rangle  = \beta_j |\alpha_i,\beta_j\rangle \nonumber\\
\Rightarrow \quad &[A,B]=0&
\eea
So in the case of commuting observables, a particle can have a definite value for both of them at the same time: this is indeed true for all the states $|\alpha_i,\beta_j\rangle $. (Of course, it does not \textit{have} to be this way: if the particle is in a superposition  - say a state proportional to $|\alpha_i,\beta_j\rangle  + |\alpha_{i'},\beta_{j'}\rangle $ then it isn't an eigenstate of any of the two observables. But the fact that you can find an entire basis of states which \textit{are} simultaneously eigen is quite special.) 

Now the question we want to address in this chapter is the following: what happens if $[A,B]\neq0$? Well, by the negation of the above expression, if two operators do not commute, they are incompatible and you can not find a basis of simultaneous eigenstates. So simultaneous eigenstates will be rather rare in this case. In practice, there are often no such states for non-commuting observables.
So in such a situation we expect that an eigenstate $|\alpha\rangle $ of $A$ can (generally) only be a sum of different eigenstates of $B$:
\be
|\alpha\rangle  = \sum_i  |\beta_i\rangle   \quad \textrm{with} \quad B |\beta_i\rangle  = \beta_i |\beta_i\rangle 
\ee
So for incompatible observables, if a state has a definite value with respect to one of them, it usually does not have a definite value with respect to the other one. 

To make this more concrete, let us think of two observables that do not commute. Ah! Position and momentum are such a pair. Recall that
\be
[X,P] = i \hbar
\ee
From the above, we expect that eigenstates of $X$ will not be eigenstates of $P$. After the previous chapter, we know both sets of eigenspace very well. The first are the position eigenstates $|x\rangle $ (delta functions) and the second are the momentum eigenstates $|p\rangle $ (infinite waves). For sure, these are two very different sets of functions: they do not have any overlap. There is not a single state that is both a position eigenstate and a momentum eigenstate. 

\subsection{The Gaussian distribution}
So the above suggests that for every state, \textit{at least one of the observables $X$ and $P$ has to be `spread out'}. They can not both have a definite value. Let us now look back at the computation we did in the first section. We found that for a Gaussian wave function 
\be
\phi(x) = \frac{1}{(2 \pi \sigma^2)^{1/4}} e^{- \frac{x^2}{4 \sigma^2}} ,
\ee
the momentum representation is also Gaussian
\be
\phi(p) = \frac{1}{(2 \pi \sigma'^2)^{1/4}} e^{- \frac{p^2}{4 \sigma'^2}} 
\ee
and the variances $\sigma^2$ and $\sigma'^2$ are related by
\be
\sigma'^2 = \frac{\hbar^2}{4 \sigma^2}.
\ee
Remember that the square root of the variance is the uncertainty, so $\Delta X = \sigma$ and $\Delta P = \sigma'$. In terms of these, the above relation becomes 
\be
\Delta X \Delta P = \frac{\hbar}{2}
\label{eq:gausseq}
\ee

\subsection{The uncertainty principle of Heisenberg}
The relation (\ref{eq:gausseq}) is a very striking result. It implies that a narrow distribution in position space necessarily has a broad distribution in momentum space. It turns out that this phenomenon also occurs for many other cases - not only for Gaussian wave packets. Every time, the product $\Delta X \Delta P$ is of the order of $\hbar$. 
Actually, the Gaussian wave packets have a bit of a special role: they realize the \textit{lowest} value for $\Delta X \Delta P$. For other wave functions, the product of the uncertainties tends to be bigger. This behavior has leads to the following inequality:
\be
\Delta X \Delta P \geq \frac{\hbar}{2}
\ee
for every state $|\psi\rangle $. This relation is called \textbf{the uncertainty principle} and is due to  the physicist Werner Heisenberg. It is the central result of this chapter. We will actually prove it in the next part. Before we go there, some small remarks:
\bit
\item 
Note that $\Delta X$ and $\Delta P$ are quantities that depend on the specific state of the particle. This is because the expectation values $\langle \rangle $ are always to be taken with respect to some state. The uncertainty principle states that there is a universal relation between those two quantities which is satisfied for \textit{every physical state}.
\item This relation has an immediate consequence for measurements. Remember that measuring a quantity collapses a state into a narrow region around it. For example, measuring the position collapses a state in a very narrow region centered around the value you read of from the apparatus. Ideally, this is a delta-function (exact position eigenstate) but in real (imperfect) measuring devices this is just a `very very narrow function'. For such a wave function, the uncertainty $\Delta X$ will be small. The uncertainty principle then implies that $\Delta P$ is very large. This means that a consecutive measurement of momentum can yield a very broad range of results. This means it is very hard to predict the outcome of this second measurement. Of course, a similar thing occurs when you first measure the momentum and directly after that the position.
\item The last lines explained that there are \textit{consequences} of the uncertainty principle for the process of measurement. However, it is important to bear in mind that the inequality also has a meaning besides the context of measurement. It is not just a limitation of a researcher to measure particular quantities of a system, but it is a statement about the nature of the wave function. Actually, it can be seen as a \textit{mathematical} statement, relating the variances of a function and its Fourier transform. That's why we can \textit{prove} it. 
\eit
\begin{figure}[ht]
 \begin{center}
  \includegraphics[width=0.7\textwidth]{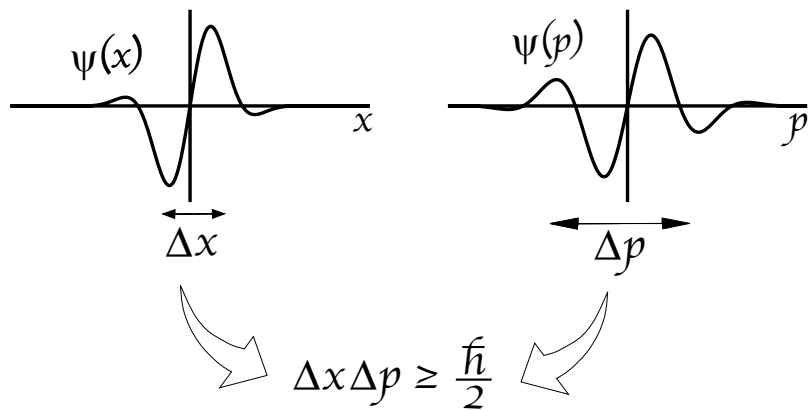}
  \caption{The uncertainty principle: a relation between the variances of the position and momentum representation of the wave function.}
  \end{center}
\end{figure}

\section[Proof and applications]{\includegraphics[width=0.04\textwidth]{comput_c} Proof and applications}

\subsection{Proof}
Consider two operators $A$ and $B$, and a state $|\psi\rangle $. Then
\be
\| A |\psi\rangle  \|^2 \|B |\psi\rangle \|^2 = \langle \psi|A^\dagger A|\psi\rangle \langle \psi|B^\dagger B|\psi\rangle  \ge |\langle \psi|A)(B|\psi\rangle |^2
\label{eq:firststep}
\ee
In the first step we used the definition of the norm, and the fact that $(A|\psi\rangle )^\dag = \langle \psi| A^\dag$ and similarly $(B|\psi\rangle )^\dag = \langle \psi| B^\dag$. The last step is nothing but the good old Cauchy-Schwarz inequality, reviewed earlier in this chapter, applied on the states $A|\psi\rangle $ and $B|\psi\rangle $. First, look at the left hand side of (\ref{eq:firststep}). If we restrict $A$ and $B$ to be  \textit{observables}, they have to be hermitean, so $A^\dag=A$ and $B^\dag = B$. This means the left hand side is equal to 
\be
\| A |\psi\rangle  \|^2 \|B |\psi\rangle \|^2 = \langle \psi|A A|\psi\rangle \langle \psi|B B|\psi\rangle  = \langle A^2\rangle _\psi \langle B^2\rangle _\psi.
\ee
Now the right hand side of (\ref{eq:firststep}). This is the norm squared of the complex number $\langle \psi|A B|\psi\rangle $. Since the norm of a complex number is alway greater than the square of its imaginary part, we have
\be
|\langle \psi|AB|\psi\rangle  |^2 \ge {\left\vert{1\over 2i} \langle \psi|AB - BA |\psi\rangle \right\vert}^2
\ee  
If we now plug in the above two equations in (\ref{eq:firststep}), we get:
\be
\langle  A^2 \rangle_\psi  \langle  B^2 \rangle_\psi  \ge {1\over 4} |\langle  [A,B]\rangle_\psi  |^2
\ee
which is already very close to the general uncertainty relation. For the last step, note that the above equation is true for all operators $A$, $B$, and all states $|\psi\rangle $. In particular, it will also be true for the operators $\tilde{A} = A - \langle A \rangle_\psi$ and $\tilde{B} = B - \langle B \rangle_\psi$. Since $\langle \tilde{A}^2\rangle  = (\Delta A)^2$ , $\langle \tilde{B}^2\rangle  = (\Delta B)^2$ and $[\tilde{A},\tilde{B}] = [A,B]$, application of the above equation to $\tilde{A}$ and $\tilde{B}$ gives
\be
(\Delta A)_\psi^2 (\Delta B)_\psi^2 \ge {1\over 4} |\langle  [A,B]\rangle_\psi  |^2
\ee
or, upon taking square roots, and dropping subscripts
\be
\Delta A\Delta B \ge {1\over 2} |\langle  [A,B]\rangle  |
\ee
This is the general form of the uncertainty relation. It holds for any pair of observables $A$ and $B$. For the operators $X$ and $P$ in particular, you can check that it yields the result
\be
\Delta X \Delta P \ge {\hbar\over 2}
\ee
Also, note that (as we suggested above) there may be \textit{some} simultaneous eigenstates states of noncommuting observables. Indeed, this is possible if this particular state is an eigenstate with value 0 of the operator $[A,B]$. In that case, the uncertainty principle imposes no conditions on the uncertainties of $A$ and $B$ for that state.

\subsection{Applications}
\subsubsection*{Application: uncertainty principle for other observables}
Another meaningful uncertainty relation can be stated in three dimensions for any pair of the angular momentum operators. For example the uncertainties of $J_x$ and $J_y$ are related by:
\be
\Delta J_x \Delta J_y \ge {1\over 2} |\langle  [J_x,J_y ]\rangle  | =  {1\over 2} |J_z |
\ee
Another variant is obtained as following. The observables $X$ and $P$ are in many ways related. This is why they are sometimes called `conjugate quantities'. Another pair of quantities that display such a `paired' behavior in many situations are time and energy. This inspired some physicists to write 
\be
\Delta E \Delta t \ge {\hbar\over 2}
\ee
This relation is a bit more subtle though, as the quantity time ($t$) is not an observable, it is a \textit{parameter}. Hence, the above equation does not make immediate sense. However, in the context of particle physics and decay processes it turns out that a similar relation does actually hold, in the following way. If a system decays from one state to another, then the typical time after which this decay occurs and the uncertainty of the energy released are indeed related by the above inequality. To understand this better: an unstable state typically does not have a very sharply defined energy, but rather a small range of energies. This means the energy released (energy of the initial unstable state minus the final energy) is not always the same, and has a small uncertainty ($\Delta E$), called the \textbf{natural linewidth}. It turns out that the typical time of existence of an unstable state (denoted $\Delta t$) and the linewidth ($\Delta E$) are indeed related by the above formula. This means the more unstable a state is (if it decays faster) the bigger the linewidth is. Conclusion: the above uncertainty relation has some experimental meaning, but you should be a bit careful with the meaning of $\Delta E$ and $\Delta t$ there. (Note that it is \textit{not} just another application of the general uncertainty relation: time ($t$) is not an observable, it's a parameter!)

\subsubsection*{Application: exotic atoms.}
With the uncertainty principle in our hands, we can understand the structure of atoms a bit better. Think again of the hydrogen system. The Hamiltonian was
\be
H = - V(r)   + \frac{P^2}{2m} = -\frac{ke^2}{r} - \frac{\hbar^2 }{2m}\Delta
\ee
Imagine a wave packet needs to be put in the lowest energy state possible. You know the outcome of this problem, but let us ignore this for a moment. If a packet wants to have a low potential energy (first term of the Hamiltonian), it needs to be well localized around the nucleus (around $r=0$). Such a packet will have a rather small $\Delta X$. However, from the uncertainty principle, we know this implies $\Delta P$ is large. This means that -when viewed in the momentum perspective- the wave packet necessarily contains components of large momentum $P$. (Otherwise it can't have a large momentum variance.) This in turn is disadvantageous for the total energy, as it blows up the kinetic part $P^2/2m$. It is precisely a balancing between the two extremes that gives an atom a finite size: the optimal way in between.
A question one can ask, is the following: what if the electron was heavier? What would this change in the story? Well. As the mass of a particle is standing in the denominator of the kinetic term, a larger mass means the relative cost of kinetic energy is getting lower - at least if one fixes the total momentum.\footnote{This is just the same statement as the fact that in classical mechanics, the kinetic energy $p^2/2m$ decreases with mass - if one fixes the momentum. Of course this is something very different from keeping the \textit{velocity} fixed while decreasing the mass. In that case, the kinetic energy $mv^2/2$ decreases, as expected.} 
So taking the same wave function (hence the same $P^2$) but for a particle with a higher mass $m$ will have less kinetic energy $P^2/2m$. In such a case, the particle might just as well sit a bit closer to the nucleus, lowering its potential energy, while not driving up his kinetic energy too much. Based on this reasoning, one expects a `heavier electron' (if such existed) to sit closer to the nucleus, and hence to be better bound. Recalling the expression for the Bohr radius and the binding energy of the lowest energy state ($n=1$), 
\begin{figure}[t]
 \begin{center}
  \includegraphics[width=0.55\textwidth]{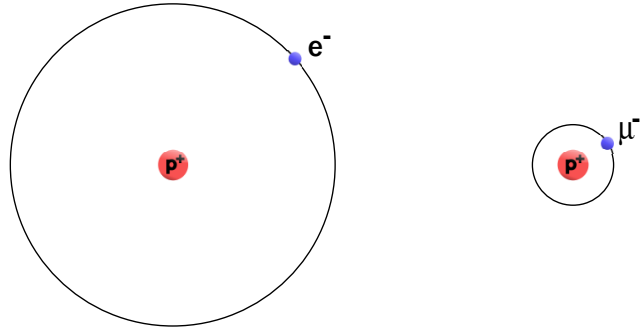}
  \caption{A muonic hydrogen atom is almost identical to an ordinary hydrogen atom. The only difference is that the electron is replaced by it's heavier brother, the muon. This particle sits in much smaller wave functions, and is much stronger bound to the nucleus.}
  \end{center}
\end{figure}
\bea
E(n=1)&=& - E_0 \frac{1}{1^2}= - E_0 = - \frac{m_e k^2 e^4}{2  \hbar^2} \\
a&=&\frac{\hbar^2}{m_e k e^2}
\eea
we see that the binding energy is indeed proportional to $m_e$, and the Bohr radius $a$ is inversely proportional to it. This confirms our reasoning.

You may wonder if all this philosophizing is of any use. After all, the electron mass is just a fixed number, there is no way of changing it. Well, here is a reason why we would actually like to think about it. Apart from the particles you already know (electron, proton, neutron and photon) there are quite some other, more rare particles. They are generally short lived (getting created in some very special processes like particle collisions and nuclear decay) and rapidly decay to the more familiar particles. An example is the \textit{muon}. It lives for about 2 microseconds, weighs 200 times as much as the electron, but has all other characteristics identical to the electron. It is actually possible to bind such a muon to a nucleus (during its short lifetime) forming a so-called \textit{muonic atom}. The energy-levels and atomic size are quite well described by the above two formula, if one at least puts in a mass 200 times the electron mass. This confirms both the above reasoning, and that the muon nicely obeys the Schr\"{o}dinger equation, like all other particles!
 
\subsubsection*{Application: what the universe is made of.}

\begin{figure}[t]
 \begin{center}
  \includegraphics[width=0.8\textwidth]{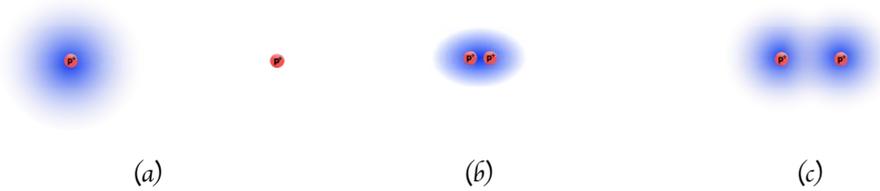}
  \caption{The $H_2^+$ molecule. (a): Two nuclei far apart (no lowering of the energy). (b): Two nuclei close together (strong repulsion). (c): Intermediate distance: lowest energy state (bound system).}
  \label{fig:H2}
  \end{center}
\end{figure}

With a reasoning similar to the one above, we can also understand the nature of chemical bonds a bit better. Try to think of the simplest molecule: $H_2^+$. This consists of two hydrogen nuclei, and one electron `circling around' and binding the system together. How does the electron manage to keep the system together, and how does the system get its size? If you look at Figure \ref{fig:H2}
you see that there are three players in the game. First, there is the attraction of the electron to both nuclei. Second, there is the mutual repulsion between the two nuclei. Last, there is the uncertainty principle, or better: the electron wants to minimize its kinetic energy. If the two nuclei are too close together, their mutual repulsion will drive them apart. If they are far away of each other, the electron might just as well sit on one of the nuclei only, and the system breaks apart. However, if the two nuclei are at an intermediate distance, something interesting happens. On the first place, the electron can orbit around both nuclei, getting a low potential energy. On the second place, as the wave function of the electron is now larger, $\Delta X$ is big, and the uncertainty principle allows a very small $\Delta P$. In other words: the electron can take on a very narrow range of low momenta, thus having only a very small kinetic energy. By `taking on low momenta', we mean that $\psi(p)$ is only nonzero for very small $p$, which necessarily corresponds to small $\Delta p$. This system obviously has a lower energy than any other option, meaning it forms a genuine bound state. This is what keeps the molecule together. The very same effects are responsible for the structure of any other molecule, even if more electrons and more nuclei are involved. A rule of thumb is that electrons like to occupy a lot of space, taking on low momenta, while combining this with a wave function close enough to the nuclei. This is precisely what a chemical bond is. These bonds make molecules to what they are, and hence give structure to what most of the (known) universe is made of. 

\newpage
\section*{Exercises}
\begin{enumerate}
\item Explain to someone what the uncertainty principle is. What are the implications for measurements? How is it related to commutativity?
\item By using linearity, show that $\textrm{Var}(x)=\langle X^2 \rangle - \langle X \rangle^2$. Since the variance is a positive quantity (it's the expectation value of a square) this implies $\langle X^2 \rangle > \langle X \rangle^2$. Check this property for the wave function $\psi(x)$ which equals one on the interval $[-1/2,1/2]$ and equals zero outside that interval. 
\item For a particle in a box, $L$ gives a vague idea of the position uncertainty. How large do you estimate the momentum uncertainty to be then? What is the (classical) energy of a particle with a momentum lying somewhere around that value? Compare this to the expression for the energy level that we found. (Just the form, not the exact prefactors - this exercise is only a vague estimate.)
\item A team experimentalists can prepare atoms in an excited state. These atoms decay to a stable state, with a lifetime of $10^{-8}$ s. What is the uncertainty of the energy of the photons produced in their experiment?
\item In the last exercise of the previous chapter, you computed the Fourier transform of the rectangular wave functions $\psi_n(x)$. (You should have gotten something proportional to a sinc function.) Estimate roughly the variance of both $\psi_n$ and its transform (i.e. give a number characterizing the magnitude of this variance) and verify that again $\Delta X \Delta P \sim \hbar$. 
\item Compute $\Delta X$ and $\Delta P$ for the first excited level of the harmonic oscillator. (Hint: express $X$ and $P$ in terms of creation and annihilation operators. Be careful with prefactors like in (\ref{eq:lowermat}) and (\ref{eq:othermat})!) Do these satisfy the uncertainty principle? What about the ground state? (Look at its wave function; it has a special form!)
\end{enumerate}

%% file: H9.tex
\chapter{Reflection and transmission of particles}
\subsection*{In this chapter...}

The chapters of this Part 3 will study different kinds of systems then what we have met so far. If you think about it, all the examples we have done considered particles which were \textit{trapped} in a potential. In order of appearance: a box-potential, an oscillator-potential, and the Coulomb potential of the hydrogen atom. If a particle is not trapped/bound in a potential, we call it a \textit{free particle}.  This chapter will precisely deal with such free particles. We will see that -just like trapped particles- they have some special and unexpected properties. One of them is the phenomenon of tunnelling, where a particle can move trough a `wall' (potential barrier) which according to classical mechanics would be impenetrable. 

In the first part, we prepare ourselves a bit: we study the SE in potentials which consist of several constant regions and we introduce the notion of probability current. Then, we explain the phenomenon of tunnelling and why it is relevant in nature. Finally, we compute the probability for a particle to actually tunnel though a potential barrier.

\newpage

\section[Particles in constant potentials]{\includegraphics[width=0.04\textwidth]{tool_c} Particles in constant potentials}

For this chapter, we will study particles in potentials, which are composed of several regions in which the potential is constant. To understand the behavior of such systems, we need to solve the SE in each part separately, and then glue together the solutions. Here, we will tackle the first part of that problem: solving the SE in a region where the potential is constant. 

So imagine we a particle, with total energy $E$, traveling through a region where the potential is constant: $V(x)=V$. Because the particle has energy $E$, we have
\be
H |\psi\rangle= E \, |\psi\rangle
\ee 
with $\psi(x)$ the wave function of the state $|\psi\rangle$. This means that
\be
H \psi(x)=-\frac{\hbar^2}{2 m} \frac{\partial^2 \psi (x) }{\partial x^2} + V \psi(x) = E \psi(x).
\ee
so
\be
\frac{\partial^2 \psi (x) }{\partial x^2}  = - \frac{ 2m (E - V)}{\hbar^2} \psi(x).
\label{eq:cte pot}
\ee
There are two possibilities. Either the value of the potential is lower than the total energy of the particle, or it is higher. We treat those cases separately.
\subsubsection*{Case 1: $V < E$}
In this case, $E-V$ is positive, and we define a real and positive parameter $k$:
\be
k^2 =  \frac{ 2m (E - V)}{\hbar^2} 
\ee
With this definition equation (\ref{eq:cte pot}) becomes:
\be
\frac{\partial^2 \psi (x) }{\partial x^2}  = - k^2 \psi(x)
\ee
This is a differential equation that we have met before. We can write the solution as a sum of a sine and a cosine, or in terms of complex exponentials. Here we will choose for the latter, so we write the general solution of the above equation as 
\be
\psi(x)= A e^{i k x} + B e^{-ikx}
\ee 
with $A$ and $B$ complex numbers. (If you use $e^{ix} = \cos x + i \sin x$, the above expression can be turned into a sum of a sine and a cosine again, be it with different coefficients.) This means the full, time dependent solution is given by:
\be
\psi(x,t) = \psi(x) e^{- i E t/\hbar} = A e^{i k x - i \omega t} + B e^{-ikx - i \omega t}
\ee
with $\omega = E/\hbar$. This is just the sum of two waves, one traveling to the left, and the other traveling to the right. Using 
\be
kx - \omega t = c \Rightarrow x = (c/k) + (\omega/k) t
\ee
we see that the wave of amplitude $A$ is traveling to the right, with speed $v = \omega/k$, and the wave with amplitude $B$ is traveling to the left, with speed $v = - \omega/k$.

The interpretation of this should be quite clear: if a particle has enough energy to be in a certain region (so $E>V$), then the solutions are just waves. By superposition of such waves, we can build wave packets traveling around in such a region.

\subsubsection*{Case 2: $V > E$}
Here, it is $V-E$ which is positive, so we define a real and positive parameter $\mu$ as follows:
\be
\mu^2 =    \frac{ 2m (V - E)}{\hbar^2} .
\ee
Equation (\ref{eq:cte pot}) now becomes:
\be
\frac{\partial^2 \psi (x) }{\partial x^2}  = \mu^2 \psi(x)
\ee
The solutions are now: 
\be
\psi(x)= A e^{ \mu x} + B e^{-\mu x}
\ee 
with $A$ and $B$ again complex numbers. These solutions are not ordinary waves anymore, but represent wave functions that are either damped or blowing up for large $x$. This means something special is happening - we will say more about this later. Anyhow, you might have expected something special to happen: if in a certain region $V>E$, it means that the particle does not have enough energy to be at that place - according to classical mechanics. So it is not strange that we do not find any ordinary wave solutions here anymore. This behavior will be explained later in this chapter. 

\subsection{Probability current}
Another thing we will use in this chapter is the notion of \textit{probability current}. Typically, in a system where there is a density $\rho$ in play (energy density, particle density, ...) people like to write an equation of the following form: 
\be
\frac{\partial \rho}{\partial t} = - \vec{\nabla} \cdot \vec{j} = - (\partial_x j_x + \partial_y j_y + \partial_z j_z)
\ee
This is called the \textbf{continuity equation}. In just one dimension, it becomes:
\be
\frac{\partial \rho}{\partial t} =  - \partial_x j_x 
\ee
This can easily be shown to imply that the total quantity $Q=\inft \rho(x)\, dx$ is conserved. By conserved, we mean that it does not grow or shrink in the course of time: $\partial_t Q=0$. (Showing this needs the assumption $j_x(+\infty)=j_x(-\infty)=0$.) The object $j(x)$ is then called the \textbf{conserved current} associated to the density $\rho$. For example: if $\rho_E(x)$ represents an energy density, and the \textit{total energy} $\inft \rho_E(x)\,dx$ of a system is conserved (does not grow or shrink with time) then there will typically be a conserved current $j_E(x)$ such that together with $\rho_E$ it satisfies the continuity equation.
Similarly, if $\rho_p(x)$ is a particle density, and the total number of particles $\inft \rho_p(x) \,dx$ is conserved, then there typically is some conserved current $j_p(x)$ in play. 
In quantum mechanics, there is a continuity equation as well. Recall that $|\psi(x)|^2$ has the meaning of a probability density, and since the total probability $\inft dx\, \rho (x) = \inft dx |\psi(x)|^2 $ is constant (it equals one for any time $t$) it is conserved. This means we can probably set up a continuity equation for the probability density. Let us try to find it. By the product rule of derivatives,
\be
\frac{\partial}{\partial t} |\psi|^2  = \frac{\partial \psi^*}{\partial t} \psi + \frac{\partial \psi}{\partial t} \psi^*
\label{eq:contpsi}
\ee
To proceed, let us consider a freely moving particle. This means we consider $V(x)=0$. In that case, the SE tells us that
\be
\frac{\partial \psi}{\partial t} = \frac{i \hbar}{2m} \Delta \psi
\ee
and by taking the complex conjugate
\be
\frac{\partial \psi^*}{\partial t} = \frac{-i \hbar}{2m} \Delta \psi^*.
\ee
If this step is not obvious, try to write out $\psi$ as a sum of its real and imaginary part, write out the derivatives, and then apply complex conjugation to arrive at the above expression. Plugging these two equations into (\ref{eq:contpsi}), we get
\bea
\frac{\partial}{\partial t} |\psi|^2  &=&\frac{i \hbar}{2m} (\psi^* \Delta \psi - \psi \Delta \psi^*)\\
&=&\frac{i \hbar}{2m} \vec{\nabla} (\psi^* \vec{\nabla} \psi - \psi \vec{\nabla} \psi^*)
\eea
If the second step is not clear, write out the derivatives in components, like $\Delta \psi = \sum_i \partial^2_i \psi$ or $ \vec{\nabla} (\vec{v}) = \sum_i \partial_i v_i$. Ah! The above equation is a continuity equation indeed, for $\rho=|\psi|^2$ and 
\be
\vec{j} = \frac{i \hbar}{2m} (\psi^* \vec{\nabla} \psi - \psi \vec{\nabla} \psi^*) = \frac{\hbar}{m} \textrm{Im} (\psi^* \vec{\nabla} \psi)
\ee
The interpretation of $\vec{j}$ is then as follows: if a wave packet is moving from one place to the other, the probability is \textit{flowing} from this one region to the other. The size of the flow at each position $x$ is precisely given by $\vec{j}(x)$.

\section[Tunnelling and nuclear decay]{\includegraphics[width=0.04\textwidth]{once_c} Tunnelling and nuclear decay}
In this story, we will fresh up our knowledge about the nucleus. You know it consists of protons and neutrons. You also know that these are bound together very tightly. You may wonder how this is possible: after all, the protons are all positively charged, so they repel each other. The secret (which you may or may not have heard of) is the so-called strong force. This is a force between particles, just like electromagnetism and gravitation, 
but with some special features. First of all: it has a very short range. As you know, the influence of gravitation and electromagnetism can work on very large distances. Arbitrary large, in fact. Both forces have a $1/r^2$ behavior - which gets small for large $r$ but never vanishes. However, for the strong force, this is not true. It can only work between particles that are very close together, at a distance of about $10^{-15}$m, approximately the size of the nucleus of an atom. Another special feature is that this force is always attractive, and only felt by protons and neutrons. That means electrons do not feel this force at all. This is a bit similar to the fact that the electric force is not felt by the neutron (because it is neutral). Lastly, this force is very strong - or better: within its small range, it dominates the electromagnetic force by far. This in turn is a bit similar to the fact that the electromagnetic force between individual particles is much much larger than their mutual gravitational force - that's why we can safely ignore gravitation when dealing with small systems. To summarize: 

\begin{center}
\begin{tabular}{|c|c|c|c|}
\hline
Force: & Range: & Strength: & Affected particles:\\
\hline\hline
Strong & $\sim 10^{-15}$m & very strong & p$^+$, n$^0$\\
\hline
Electromagnetic & $\infty$& intermediate & charged particles \\
\hline
Gravitational & $\infty$ & very weak 
& \textit{all} particles \\
\hline
\end{tabular}
\end{center}

With this information, you can understand better why a nucleus sticks together. If neutrons and protons are very very close together, they are within the range of eachs others attractive strong force. This force is much larger than the electrostatic repulsion between the protons, so the nucleons stick together tightly and well-bound. This way, they form a nucleus, and the size of that nucleus is of the same order as the range of the strong force.

This sounds very cosy, but it turns out that if there are too many nucleons sitting together, the nucleus gets crowded, and some of the nucleons might want to get out. Dividing the nucleons over two nuclei might be energetically more advantageous than packing them all together in a single crowded one. This is precisely what is known as radioactivity. Typically, elements with too many nucleons - such as uranium, for example - have the tendency to fall apart after a while.\footnote{Also some smaller nuclei can be unstable, but this is for other reasons.} By falling apart, they release some energy: the energy of the initial state minus the energy of the final state. So in a large bunch of atoms of an unstable kind, there are all the time decays occuring. Each such decay releases a bit of energy. It is this ongoing energy emission - in the form of radiation - what signals that a chemical element is radioactive.

\begin{figure}[t]
 \begin{center}
  \includegraphics[width=0.6\textwidth]{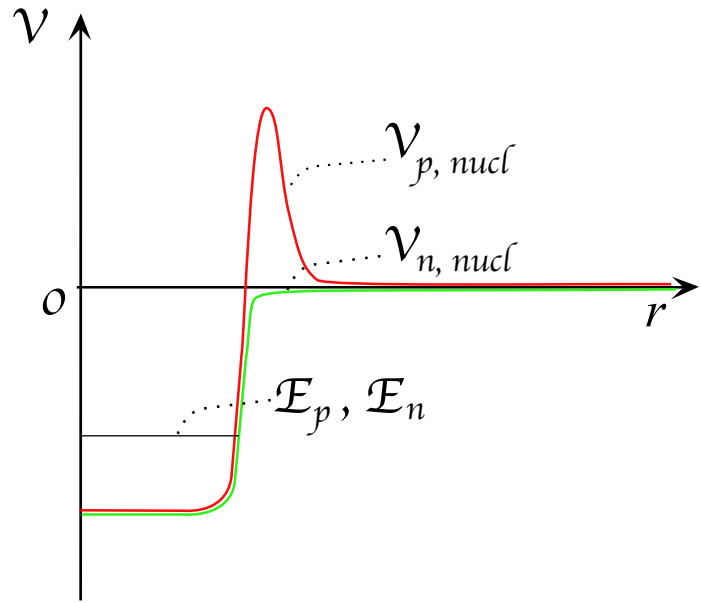}
  \caption{The total nuclear potential for a proton and a neutron. Both feel the attractive strong force. This creates the deep pit around the origin. On top of that, the proton experiences the repulsive coulomb potential. For both particles the typical energy lies below the height of the barrier, meaning they can not escape.}
  \label{fig:nucpot}
  \end{center}
\end{figure}

Hopefully you now have a better understanding of what the nucleus is. However, there is one logical gap in the explanation. We have said that the strong force (within its range) is always bigger than the electrostatic repulsion. So it seems that once the protons and neutrons are locked up in a nucleus, they can never get out. Not even when the end product has a lower energy and thus is favorable. From this we would incorrectly conclude that there are no unstable nuclei. How can we understand this better? To answer this question, look at Figure \ref{fig:nucpot}. There, the total potential (of the electrostatic force + the strong force) is shown. If a nucleon was a point particle, it would just tend to move hence and forth in the potential well. Every time it reaches the boundary of the nucleus, it slows down, stops and moves back. 
So in this description, a single particle (or a cluster of them) can never escape, and we can not explain radioactivity. 

\begin{figure}[ht]
 \begin{center}
  \includegraphics[width=1.0\textwidth]{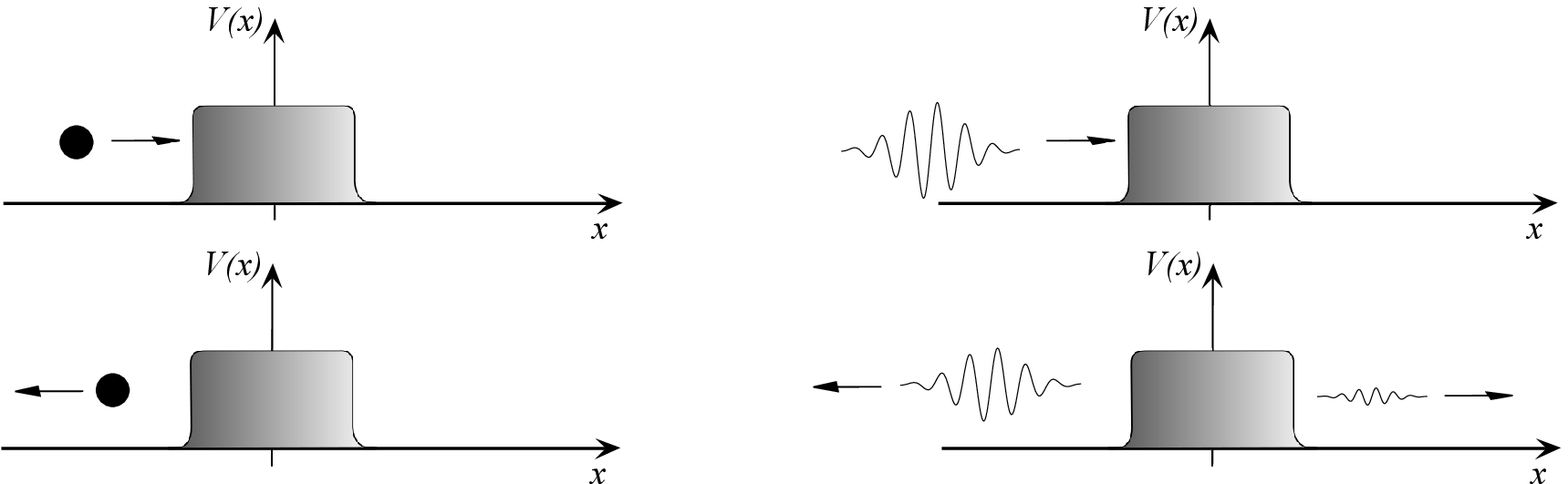}
  \label{fig:clqubarr}
  \caption{When a classical (point) particle hits a potential higher than its energy, it is reflected. When a wave function hits a barrier, there is always a part which is transmitted. This implies there is a nonzero chance for the particle to go through the barrier.}
  \end{center}
\end{figure}

However, quantum mechanics tells us this description is wrong. We know that the protons and neutrons (just like electrons) are each described by a wave function, which has a nonzero size. 
Now the result of such a wave packet hitting a potential barrier is very interesting. There is \textit{always} a small part of the wave function that crosses the barrier, \textit{even} if the energy of the wave packet is much lower than the height of the potential barrier. This is the so-called \textbf{tunnelling effect}. It is shown in Figure \ref{fig:clqubarr}. We will study this effect quantitatively at the end of this chapter, and give an expression for the amount of the wave packet that is let through the barrier. For now, knowing this effect is enough to understand the puzzle here. Indeed, the tunnel effect implies there is always a small chance for a a particle to cross a barrier, and hence for one or more nucleons of an unstable atom to leave the nucleus. This means nuclei can decay (at least: if the end result has a lower energy) even if the barrier to escape the nucleus is classically impenetrable.

This quantum effect was first studied by George Gamow, who used the tunnel effect to describe the process of \textbf{alpha decay}. In alpha decay, two neutrons and two protons leave the nucleus at once. These four particles are exactly the same as a helium nucleus, and such a cluster is also called an `alpha particle'.  Using the tunnelling probability formula (which we will derive shortly) Gamow managed to explain a relation between the emitted energy in alpha decay process and the decay time. The decay time is the typical time an unstable atom lives before decaying and depends on the specific isotope. This relation matches very well with experiments, confirming the quantum mechanical computation.

Before we go to the calculations, two small remarks on the explanation above. First, you may wonder why we can describe the nucleus as a point when we are dealing with electrons in an atom. Indeed, in the hydrogen case we just took the nucleus to be a point charge. The above explanation suggests we should describe all particles in the nucleus by their wave function, not as a single point. This would obviously make the problem of studying electrons in an atom much harder. What saves the day is of course the \textit{strength} of the strong force, which squeezes the wave functions of nucleons together very tightly. This makes the nucleus look extremely small - much smaller than the atom itself. A popular comparison is that is the nucleus sitting in the atom is as small as an ant sitting in a large football stadium.\footnote{The size of the total atom is of course determined by the approximate size of the cloud of electrons.} That is why we can safely treat it as a single point when we are describing the electrons. A second pitfall we want to prevent is a too literal interpretation of Figure \ref{fig:clqubarr} in the context of the nucleus. Just like the wave functions in a box, the wave functions of nucleons stretch out over the entire nucleus. They are not just small wave packets bouncing around, so Figure \ref{fig:clqubarr} gives a bit a simplistic image of this situation. For cases where a free particle hits a potential barrier, this figure \textit{is} precisely what happens, and it will be this situation which we will describe below. So the two situations (free particle hitting a potential barrier and an alpha particle leaving the nucleus) both involve the same tunnelling effect - although the situations are not strictly identical. Got it?

\section[Particles and barriers]{\includegraphics[width=0.04\textwidth]{comput_c} Particles and barriers}

We will study two problems in this part. The first one considers a particle, traveling freely, until at some point in space it encounters a sudden potential step. We will ask what happens to the particle. If it travels onwards (into the region where $V$ is higher), we say the particle is transmitted. When it is sent back to where it came from, we say that the particle has been reflected. We want to find out the relative probabilities for those two options. The other  problem deals with a potential barrier: a narrow but high bump in the potential. Classically, the particle will be reflected by the bump if it is higher than the particles' energy, but according to quantum mechanics, there will always be a nonzero chance for the particle to boldly cross the barrier. This is precisely the tunnel effect we were talking about before.

\subsection{The potential step}
This describes a particle traveling in a potential that has a \textit{step} at some place, say at $x=0$:
\be
V(x) = \left\{ \begin{array}{c} 0 \quad \,\textrm{for} \quad x<0\nonumber\\
 V  \quad \textrm{for} \quad x\geq0\nonumber \end{array}\right.
\ee
Also, let us take the energy $E$ of the particle to be larger than $V$. We then know from the first part of the chapter, that the solution of the SE in each part of the potential is given by ordinary waves:
\be
\psi(x) =\left\{ \begin{array}{c}\psi_1(x) = A_1 e^{i k_1 x} + B_1 e^{-ik_1 x} \quad \,\textrm{for} \quad x<0\nonumber\\
\psi_2(x) =  A_2 e^{i k_2 x} + B_2 e^{-ik_2 x} \quad \textrm{for} \quad x\geq0\end{array}\right.\label{eq:two waves}
\ee
with $k_1 = \sqrt{ 2m E }/\hbar$ and  $k_2 = \sqrt{ 2m (E - V)}/\hbar$. The time evolution of this solution is just given by multiplying the above wave functions by $e^{-iEt}$. To have a healthy solution we need to glue together the part $\psi_1$ on the left to the part $\psi_2$ on the right. In particular, to have a descent wave function we need the function $\psi(x)$ to be \textit{continuous} at $x=0$. Also, to ensure the wave function does not have a strange kink, we will demand the derivative of $\psi$ to be continuous at $x=0$ as well. These requirements give the following system of equations:
\bea
\psi_1(0)=\psi_2(0) &\Rightarrow& A_1 + B_1 = A_2 +B_2\\
\psi_1'(0)=\psi_2'(0) &\Rightarrow& k_1 (A_1 - B_1) = k_2 (A_2-B_2)
\label{eq:stelsel}
\eea
Now some more physical input. The solution (\ref{eq:two waves}) contains four parts: two waves in the region $x<0$ (one going to the left and one to the right), and similarly two waves in the region $x>0$. We want to describe a particle coming from the left, falling on the potential step, and check how much of the initial wave is reflected and how much is transmitted. This means that in the region $x>0$ we only want to consider waves that are going to the right. Hence we will put
\begin{figure}[t]
 \begin{center}
  \includegraphics[width=0.6\textwidth]{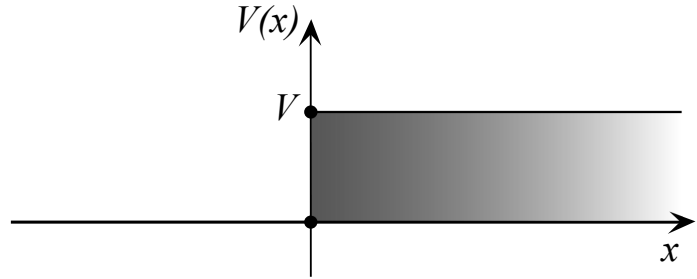}
  \caption{The potential step. When a particle with energy $E>V$ hits this barrier, there is a chance $R$ to be reflected and a chance $T$ to be transmitted. This contrasts with the description of particles as points. There a particle will \textit{always} be transmitted when $E>V$.}
  \end{center}
\end{figure}
\be
B_2=0.
\ee
Also, we only want to study the ratios of the transmitted and reflected waves compared to the incoming wave. This means we can just normalize with respect to the incoming wave, and put the amplitude of the incoming wave to one:
\be
A_1=1.
\ee
The conditions (\ref{eq:stelsel}) then become
\be
1+B_1 = A_2 \quad \textrm{and} \quad k_1(1-B_1) = k_2 A_2 \nonumber
\ee
which has as solutions
\be
A_2 =  \frac{2 k_1}{k_1+k_2}, \quad  B_1 =\frac{k_1-k_2}{k_1+k_2}.
\ee

\subsubsection{Wave packets}
It should be clear that the coefficients $A_2$ and $B_1$ are closely related to the probabilities for a particle to be transmitted or reflected by the barrier. If $A_2$ is large, a large portion of an incident wave will be going into the region $x>0$, which can be interpreted as a large chance for the particle to be transmitted. 

However, real particles are never an infinite wave, but a limited packet. (Recall that we described an infinite wave as mere caricature of a wave packet of which the momentum is sharply peaked around some value.) But we know from the chapter on Fourier transforms that such a packet can nevertheless be described by a \textit{superposition} of infinite waves. So we only need a small step to convert the above results into a statement about particles. Say we start out with a wave packet $\psi(x)$ that is initially located in some region on the side $x<0$, and that is is traveling to the right. In momentum representation, we can write
\be
\psi(x)=\frac{1}{\sqrt{2\pi}} \inft dk_1\, e^{ik_1 x} \psi(k_1).
\label{eq:pakketje}
\ee
In this last equation, we the waves $e^{ikx}$ in terms of the wavenumber, and not in terms of the momentum $p = \hbar k$. This means we do not have to include any extra factors $\hbar$ anywhere: we can just use the original expression for a Fourier transform which we met in the first section of Chapter 8. The normalization condition phrased in terms of $\psi(k_1)$ is
\be
\inft dk_1 |\psi(k_1)|^2 =1
\label{eq:norma}
\ee
The previous section showed that a wave $e^{i k_1 x}$ which falls on the potential step is broken up in two parts: 
\be
e^{ik_1 x} \rightarrow B_1 e^{-i k_1 x} + A_2 e^{ i k_2 x}
\ee
(We suppress the time evolution for elegance.) This is true for  all $k$, so for the wave packet (\ref{eq:pakketje}) in its entirety will break up into
\be
\quad\quad \rightarrow\frac{1}{\sqrt{2\pi}} \inft dk_1\, B_1 e^{ - ik_1 x} \psi(k_1) +\frac{1}{\sqrt{2\pi}} \inft dk_1\, A_2 e^{i k_2 x} \psi(k_1)
\label{eq:entirety}
\ee
What we want to do, is to find the amplitudes of the two parts. First, let us consider the reflected part. Its total amplitude (=the norm squared) will be interpreted as the \textit{chance} for reflection:
\be
R=\inft dk_1\, |B_1|^2 |\psi(k_1)|^2.
\label{eq:refl}
\ee
Similarly the total amplitude of the transmitted part can be interpreted as the chance for the particle to go over the potential step. We compute the values of $R$ and $T$ below.
\subsubsection{Finding $R$ and $T$}
We start with the reflected part. To do the integral (\ref{eq:refl}) note that $B_1$ depends on $k_1$, so it is not just a constant. However, things simplify if we assume that the initial wave function $\psi(k_1)$ is sharply peaked around a single value (say $\bar{k}_1$) and zero at other places. Then the relevant range of integration in the above integral becomes a very small region around $\bar{k}_1$, and we can take $B_1$ to be approximately constant: namely, just its value when evaluated for $\bar{k}_1$. Using this, and the normalization (\ref{eq:norma}), we get
\be
R = |B_1|^2 \inft dk_1\,  |\psi(k_1)|^2= |B_1|^2
\ee
As said, it is understood that $B_1$ is evaluated at the peak value $\bar{k}_1$ of the incident wave function.
Now let us consider the transmitted part, the last term in (\ref{eq:entirety}). Let us denote that wave packet by $\psi_T(x)$. Because of the relation between $k_1$ and $k_2$
\be
k_1^2-k_2^2= \frac{2 E m}{\hbar^2} - \frac{2 (E-U) m }{\hbar^2} = \frac{2 U m }{\hbar^2}
\label{eq:k1k2}
\ee
we can change the integration variable to $k_2$, and get
\be
\psi_T(x)=\inft dk_1\, A_2 e^{i k_2 x} \psi(k_1) =\inft dk_2 \,  \frac{\partial k_1}{\partial k_2}\, A_2 e^{i k_2 x} \psi(k_1(k_2)) 
\label{eq:psi2golf}
\ee
In the object on the right hand side, the integration variable ($k_2$) is the same object as what occurs in the exponentials - unlike the integral on the left.
So the momentum representation $\psi_T(k_2)$ is to be read off from the integral on the right. Using (\ref{eq:k1k2}) we can see that $\frac{\partial k_1}{\partial k_2} = \frac{k_2}{k_1}$ and we find 
\be
\psi_T(k_2)=\frac{k_2}{k_1}\, A_2 \,\psi(k_1)
\ee
where $k_1$ depends on $k_2$ via (\ref{eq:k1k2}). With this, the total amplitude of the transmitted part becomes
\be
T=\inft dk_2\, \frac{k_2^2}{k_1^2} |A_2|^2 |\psi(k_1)|^2 = \inft dk_1  \frac{\partial k_2}{\partial k_1} \frac{k_2^2}{k_1^2} |A_2|^2 |\psi(k_1)|^2  
\ee
Now again we use the fact that $\psi(k_1)$ is sharply peaked, so we can approximate the objects inside the integral by using in the value $\bar{k}_1$ and the corresponding $\bar{k}_2$:
\be
T=  \frac{ \bar{k}_1}{\bar{k}_2} \frac{\bar{k}_2^2}{\bar{k}_1^2} |A_2|^2 \inft dk_1  |\psi(k_1)|^2   = \frac{\bar{k}_2}{\bar{k}_1} |A_2|^2.
\ee
Where $A_2$ is to be evaluated at $\bar{k}_1$ and $\bar{k}_2$. In conclusion, the chances for our incident particle to be reflected or transmitted, are given by:
\be
R =  |B_1|^2 = \left(\frac{\bar{k}_1-\bar{k}_2}{\bar{k}_1+\bar{k}_2}\right)^2 \quad \textrm{and} \quad T = \frac{\bar{k}_2}{\bar{k}_1} |A_2|^2 =  \frac{4 \bar{k}_1\bar{k}_2}{(\bar{k}_1+\bar{k}_2)^2}.
\ee
Getting here was a bit nasty, but as a rather non-trivial check, you can verify that these probabilities nicely add up to one:
\be
R+T=1.
\ee
Two more remarks to wrap this all up. First: you may wonder why we changed integration variables twice when deriving $T$. Once from $k_1$ to $k_2$ and once from $k_2$ to $k_1$. Is this really necessary? Couldn't we just have left $k_1$ there all the time? The answer is no. We had to switch to $k_2$ to correctly read off the wave function $\psi_T(k_2)$, since the waves in the integrand were expressed in terms of $e^{ik_2 x}$ and not $e^{i k_1 x}$. The second change of variables was there to pull out the integral $\inft dk_1  |\psi(k_1)|^2=1$ again, like we did in the reflection part.  A second comment is about time dependence. We have at no point written any of the $e^{-i E t/\hbar}$ factors. Is that ok? Well, since we were considering waves with different $k$, these also have different energies $E$. If the wave function $\psi(k)$ is sharply peaked around one $k$, the energies of the different components are very comparable, and the time dependence really is just an irrelevant factor $e^{-i E t/\hbar}$ showing up everywhere. So no, it does not change much if we include this factor or not, but yes, to be correct we should have. This would introduce some phases that are present in the outgoing wave packets, but which of course do not change the norms, and hence not the outcome. The main goal of this derivation was not to include all the details, but we did manage to find expressions for $R$ and $T$. And as we anticipated, they are just proportional to $B_1$ and $A_2$.

\subsubsection{Probability currents}
The above construction of wave packets has two faces. On one hand, it makes clear what is happening: a wave packet gets split and the amplitude of both parts precisely give the probability for transmission or reflection to occur. On the other hand, it is a bit a long way round to get the end result. Is there a shorter way to obtain $T$ and $R$? You bet. The tool we can use is the probability current we met in the beginning of this chapter. The current on the left side ($x<0$) and right side ($x>0$) are:
\bea
j_1(x) &=& \frac{\hbar}{m} \textrm{Im}\left(\psi_1(x)^*\frac{\partial \psi_1(x)}{\partial x}\right) = \frac{\hbar k_1}{m} (1- |B_1|^2)\\
j_2(x) &=& \frac{\hbar}{m} \textrm{Im}\left(\psi_2(x)^*\frac{\partial \psi_2(x)}{\partial x}\right) = \frac{\hbar k_2}{m} |A_2|^2
\eea
The current $j_1$ consists of two parts. The positive piece $j_1^+(x)= \frac{\hbar k_1}{m}$ corresponds to a probability current moving to the right. This is the incoming wave. The negative piece  $j_1^-=\frac{\hbar k_1}{m} |B_1|^2$ is the reflected part. The current $j_2$ is just the probability current corresponding to the transmitted wave. Relative to the incoming current $j_1^+$, the transmitted and reflected currents are
\bea
T &=& \frac{j_2}{j_1^+}= \frac{k_2}{k_1} |A_2|^2=\frac{4 k_1k_2}{(k_1+k_2)^2}.\\
R&=& \frac{j_1^-}{j_1^+}=|B_1|^2=\left( \frac{k_1-k_2}{k_1+k_2}\right)^2
\eea
Which confirms our previous results via a nice shortcut. Nevertheless, the interpretation of $T$ and $R$ remains most clear in the wave packet view. We just bare in mind that infinite waves are not fully physical (not normalizable) - but the wave functions constructed from them are.

A last comment on the result of the computation. Note that (unlike what you would guess from classical mechanics) there is \textit{always} some nonzero chance for reflection. So even if a particle can easily cross a potential step, there is a (small) chance for it to be bounced right back. There are several experimental situations where particles meet a small potential step, typically when particles are traveling from one type of material into another one. Even if this step is small enough for all particles to pass (given their energy) there are some which are reflected - in accordance to the above expression. 
\subsection{The potential barrier}
We now consider a potential barrier:
\be
V(x) =\left\{\begin{array}{c} V \quad \textrm{for}\quad -a<x<a\nonumber\\
0 \quad \textrm{everywhere else} \end{array}\right.
\ee
and a particle coming in from the left. The solution of the SE in the three regions is:
\be
\psi(x)= \left\{\begin{array}{ll}\psi_1= A_1 e^{ikx} + B_1 e^{-ikx} & \quad \textrm{for}\quad x<-a\\
\psi_2= A_2 e^{-\mu x } + B_2 e^{\mu x} &\quad \textrm{for}\quad -a<x<a\\
\psi_3 = A_3 e^{ikx}& \quad \textrm{for}\quad x>a \end{array} \right.
\ee
with
\be
k= \frac{\sqrt{2 m E}}{\hbar}  \quad \mu = \frac{\sqrt{2 m (V - E)}}{\hbar}
\ee
Demanding continuity at $x=-a$ and $x=a$ amounts to
\bea
\psi_1(-a) = \psi_2(-a) &\Rightarrow& A_1 e^{-ika} + B_1 e^{ika} = A_2 e^{\mu a} + B_2 e^{-\mu a} \\
\psi'_1(-a) = \psi'_2(-a) &\Rightarrow& i k  A_1 e^{-i k a} - i k B_1 e^{ika} = -\mu A_2 e^{\mu a} + \mu B_2 e^{-\mu a} \\
\psi_2(a) = \psi_3(a) &\Rightarrow& A_2 e^{-\mu a} + B_2 e^{\mu a} = A_3 e^{i k a}\\
\psi'_2(a) = \psi'_3(a) &\Rightarrow& -\mu A_2 e^{-\mu a} + \mu B_2 e^{\mu a} = i k A_3 e^{ika}
\eea
This is a homogeneous system of 4 equations with 5 unknown parameters $A_1$, $B_1$, $A_2$, $B_2$ and $A_3$. We will put $A_1=1$, just like we did for the potential step. This gives 4 equations and 4 unknown parameters. Solving for $A_3$ yields
\be
|A_3|^2 = \frac{(2 k / \mu)^2}{(1-k^2/\mu^2)^2 \sinh^2(2\mu a)+ (2k/\mu)^2\cosh^2 (2 \mu a)}
\label{eq:A3}
\ee
The currents on the left and right are now
\bea
j_1 &=& \frac{\hbar k}{m} (1- |B_1|^2) = j_1^+ - j_1^- \\
j_3 &=& \frac{\hbar k}{m} |A_3|^2
\eea
so we see that $|A_3|^2$ ($= j_1^+ / j_3$) is the transmission probability. Striking enough, even if $E>V$ (which is implicitly expressed by the reality of  $\mu$) there is a nonzero chance for the particle to cross the barrier. In the limit $\mu a >> 1$ (broad barrier, large energy deficit) the transmission probability (\ref{eq:A3}) becomes
\be
T = |A_3|^2=4 e^{- 4 \mu a} \left(\frac{k\mu}{k^2 + \mu^2}\right)^2=4 e^{- 4 \mu a} \frac{E}{V}\left(1-\frac{E}{V}\right)
\ee
so the transmission probability decreases exponentially with the barrier width $a$. The incoming, reflected, and transmitted probability currents are shown on the figure below:
\begin{figure}[H]
 \begin{center}
  \includegraphics[width=0.6\textwidth]{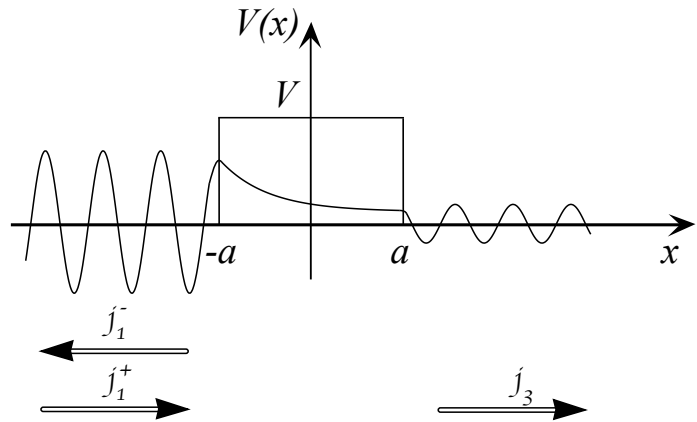}
  \caption{The incoming, reflected, and transmitted probability currents at a potential barrier. The amplitude of the transmitted current gives the probability for a particle to tunnel through the barrier - even if this is not allowed classically.}
  \end{center}
\end{figure}

\newpage
\subsection{Application: Scanning Tunnelling Microscope}
A nice application of the tunnel effect is the Scanning Tunnelling Microscope. This is an extremely high-resolution microscope, which relies on the tunnelling effect. The device consists of a sharp `scanning tip', which is put at an elevated electrical potential with respect to the sample one wishes to study. By bringing the tip close to the sample, electrons will tunnel through the empty space between the tip and the sample - even when the two are not touching. This way, a current arises. The size of the current tells something about the distance between the tip and the sample. This way, one can `feel' the shape of the sample, at a distance. (This distance is very small: the tip has to be extremely close the sample to realize a nonzero tunnelling current.) A modern STM has such a good resolution, that it can easily distinguish single atoms on the sample surface. Pretty impressive! 
\begin{figure}[H]
 \begin{center}
  \includegraphics[width=0.7\textwidth]{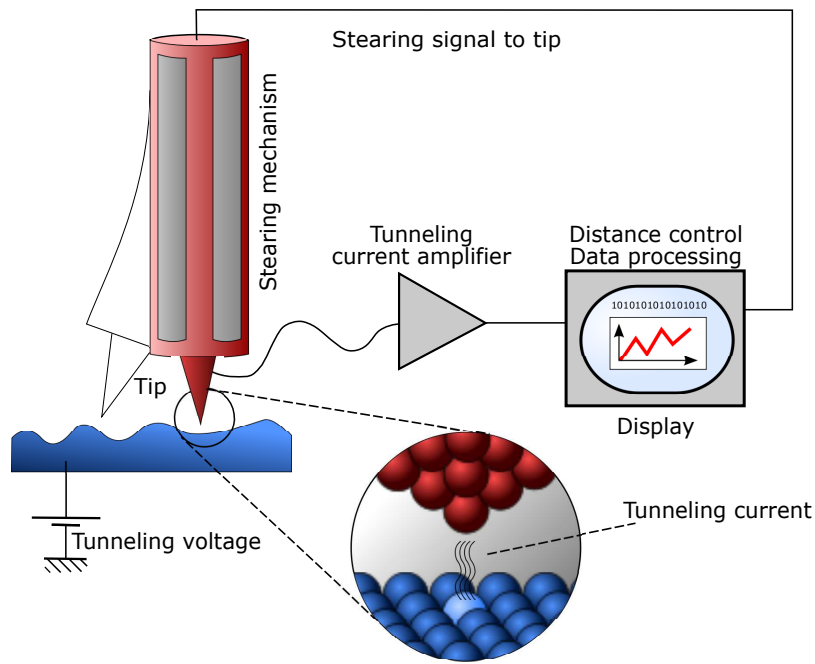}
  \caption{A Scanning tunnelling microscope. The amplitude of the tunnelling current allows to scan the surface in an extremely sensitive way. This allows to see objects as small as individual atoms.}
  \label{fig:STM}
  \end{center}
\end{figure}

\subsection{Application: It's warm outside!}
You probably know that our beloved sun is a very large fusion reactor. At its extremely high temperature, hydrogen nuclei fuse to helium nuclei. These are more strongly bound, so there is an energy gain in the process. This energy is radiated away and eventually reached us on earth. If you use what you have learned in this chapter, you may see a problem. The very barrier which prevents particles from the nucleus to escape is also there when two nuclei try to merge. So classically, it is very hard (almost impossible) for two hydrogen nuclei to fuse into a helium nucleus. Ah - this means that once again tunnelling plays a crucial role. Indeed, the tunnelling process gives rise a very small but nonzero fusion probability, just enough to make the sun into the bright shiny object we see in the sky. So the next time you and your friends are enjoying a sunny day, make sure to mention that none of this would be possible without quantum tunnelling. Caution: for the well-being of you social life you might better not say things like that out loud. Well, there still is the sweet inner joy of insight, right?

\newpage
\section*{Exercises}
\begin{enumerate}
\item An electron with energy $6$ eV hits a $10^{-10}$ m long barrier of $8$ eV high. What is the chance for the electron to go through it?
\item The same question, but now the barrier is $0.01$ m long. Do you have to be afraid that your cup of tea spontaneously empties itself by tunnelling of particles (or molecules) through the sides?
\item A proton with energy $2$ keV suddenly travels into a region where the electric potential is $10$ V higher. What would happen if a proton was a charged point particle? Using quantum mechanics, what is the chance for the proton to be reflected?
\item Prove that the charge corresponding to a conserved current is indeed conserved in time. Can you do the three dimensional case? What expression would you take for $Q$? Can you show it's conserved?
\item Study a situation where a particle (coming from the left) \textit{drops} of a potential step. So $V(x)=V$ on the left, and $V=0$ on the right. Solve the SE, write boundary conditions and find $R$ and $T$ using probability conditions. Compare the result to what you expect classically. 
\item Challenge: try to solve the potential well: a particle stuck in a potential which has $V(x)=-V$ for $-L<x<L$ and zero elsewhere. Can you recover the particle in a box by taking the appropriate limit?
\end{enumerate}

%% file: H10.tex
\chapter{Spin}
\subsection*{In this chapter...}
By now, you have become quite experienced in dealing with particles in potentials. As a consequence, you also know how to deal with electric fields. Just write down the corresponding potential (such that $-\vec{\nabla} V= F =q \vec{E}$), solve the Schr\"{o}dinger equation of a particle moving in this $V$, \textit{et voila}. In fact, this is precisely what we did in the chapter on the hydrogen atom. For a random electric field and the corresponding potential, solving the SE may be a hard problem, but you can always do this in principle. You or your computer just need to be good enough at solving differential equations.

That is very nice, but it leaves an important question unanswered. What about \textit{magnetic} fields? You know that there is no such thing as `the magnetic potential' so the influence of magnetic fields can not be summarized by giving a potential $V$. 
So what about those situations, in which a particle is immersed in a magnetic field? It is clear that there will be some new input here. Indeed, we will meet a new and important concept: \textbf{spin}. This is a property of particles which not only determines their behavior in magnetic fields, but also in other situations. Reason enough to have a closer look!

\newpage

\section[Tools]{\includegraphics[width=0.04\textwidth]{tool_c} Tools}

\subsection{Magnetic moment}
As you very well know, a free electric charge will start to move when immersed in an electric field. If the electric field is created by two distant source charges of opposite sign, the free charged particle will move towards the source charge with opposite sign. Something nice happens when you attach two small opposite charges to each other to form an \textit{electric dipole} still immersed in the same electric field. They will each pull towards one side, and stay put by force balance. Hence, the only thing that happens is the alignment of the electric dipole with the electric field. Now let us compare this with magnetism. In that case there are no monopoles, that is: there are no isolated magnetic charges. But of course there are magnetic dipoles - which can be thought of as a little magnet. Just like the electric dipoles, they tend to align with an external field, which is magnetic in this case. 
So the energy for a magnetic dipole to be aligned with an external field is lower than when it is anti-aligned. If we denote the magnetic moment (=strength of the dipole) by $\vec{M}$, and the external field by $\vec{B}$, the energy $W$ of the dipole is
\be
W = - \vec{M}\cdot\vec{B}
\ee
This is indeed lowest when $\vec{M}\parallel \vec{B}$, so that alignment is preferred. The situation is shown in the figure below:
\begin{figure}[H]
 \begin{center}
  \includegraphics[width=0.9\textwidth]{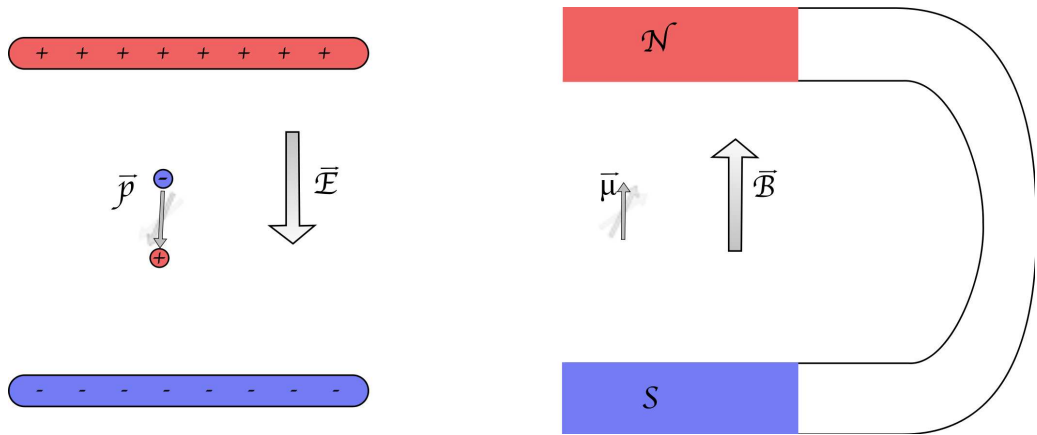}
  \caption{An electric dipole aligns with the external E-field. Similarly, a magnetic dipole aligns with the external B-field. Notice that the electric dipole is conventionally defined as a vector pointing \textit{from the negative to the positive} particle - not to be confused with the definition of the electric field, which points from positive to negative.}
  \end{center}
\end{figure}

\subsection{Larmor precession}
We just refreshed the well-known fact that a magnetic dipole tends to align with an external magnetic field. This of course can only happen if the dipole can dispose of the energy released by aligning. Otherwise, it will typically perform an oscillating movement, not able to find its equilibrium value because of conservation of energy. Similarly, a ball can only roll to the bottom of a pit and \textit{stay there} because of the energy lost while rolling down. If there was no friction, the ball would be doomed to roll in and out constantly. In the following, we will show that a magnetic field indeed can keep a dipole in an oscillatory motion, or more precise: a \textit{rotation}. This phenomenon is called \textbf{Larmor precession}.

\begin{figure}[t]
 \begin{center}
  \includegraphics[width=0.6\textwidth]{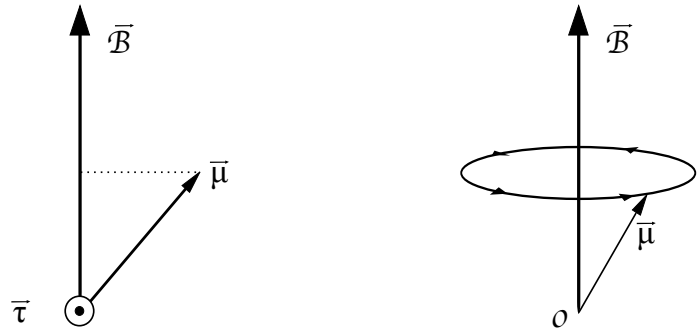}
  \caption{Left: direction of the torque, exerted on a magnetic dipole in an external field. Both $\vec{B}$ and $\vec{\mu}$ are in the plane of the paper, while $\vec{\tau}$ is pointing directly towards you. Right: the circular movement arising from the torque: Larmor precession. This movement is quite similar to that of a planet in its orbit. In that case the \textit{linear momentum} is constantly changing due to the \textit{force} acting. For Larmor precession the \textit{angular momentum} of the dipole is constantly changing due to the \textit{torque} acting.}
    \label{fig:larmor}
  \end{center}
\end{figure}

To be concrete, let us consider a small electric loop current - like a charge running in a circle. As you know from electromagnetism, the circle movement of a charge creates a magnetic field: a dipole. If $\vec{S}$ is the angular momentum of the circling charge, the induced magnetic moment $\vec{M}$ is proportional to it:
\be
\vec{M} = \gamma \vec{S} 
\ee
The proportionality constant depends on the details of the loop current (which we do not consider here) and is called the \textbf{gyromagnetic ratio}. As we just refreshed, a magnetic dipole feels a force trying to align it with the external B-field. This can be expressed as a torque, trying to turn the dipole in the direction of $\vec{B}$:
\be
\vec{\tau}  = \vec{M} \times \vec{B}
\ee
The direction of the torque is shown in Figure \ref{fig:larmor}. Just like a force causes the momentum to change, the torque causes the angular momentum to change:
\be
\vec{\tau} = \frac{d}{dt} (\vec{r} \times \vec{p}) = \frac{d}{dt} \vec{S}
\ee 
Where $\vec{S}$ as before denotes the angular momentum $\vec{r}\times\vec{p}$ of the orbiting charge. Combining the last three expressions, we get
\be
\frac{d}{dt} \vec{S} = \gamma \vec{S} \times \vec{B}.
\ee
Taking the magnetic field directed along the z-axis, we get $\vec{B} = (0,0,B)$ with $B$ the magnitude of the field. Hence, the components of the above equation are:
\bea
\frac{d}{dt} S_x &=& \gamma (S_y B_z - S_z B_y) =\,  \gamma S_y B\\
\frac{d}{dt} S_y &=& \gamma (S_z B_x - S_x B_z) = \, -\gamma S_z B\\
\frac{d}{dt} S_z &=& \gamma (S_x B_y - S_y B_x)  = 0
\eea
The last equation implies $S_z$ is constant in time. The first two combine to $\frac{d^2 S_x}{dt^2} = -\gamma^2 B^2 S_x$, which has solutions of the form
\be
S_x = A \cos (\omega t - \phi )
\ee
with $\omega = |\gamma| B$. This number is known as the \textbf{Larmor frequency}. 
The total dynamics is thus given by
\be
\vec{S} = (A \cos (\omega t - \phi ), \pm A \sin (\omega t - \phi ), S_z)
\ee 
with the sign of $S_y$ determined by the sign of $ \gamma$. The total motion is shown in Figure \ref{fig:larmor}.
 
\section[Spin]{\includegraphics[width=0.04\textwidth]{once_c} Spin}
\subsection{Zeeman effect}
\begin{figure}[H]
 \begin{center}
  \includegraphics[width=0.65\textwidth]{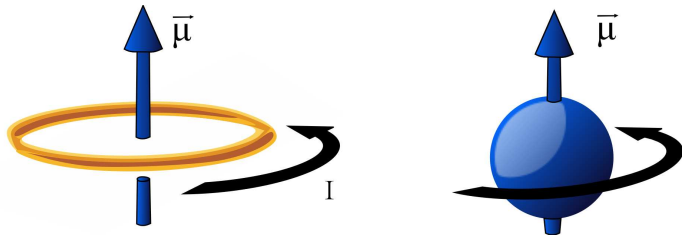}
  \caption{A current in a loop creates a magnetic moment. Similarly, a rotating charged sphere would do so. This gives a \textit{pictorial} idea on how the intrinsic magnetic moment of an electron might arise. Be careful: this picture is too simplistic. We will give a correct and more detailed description below.}
  \label{fig:spincurrent}
  \end{center}
\end{figure}

In the previous example we considered a magnetic dipole arising from an orbiting charge. Where could we naturally meet such a system? Precisely, the electron in an atom! Indeed, it is charged and every state with $\ell\geq 1$ has nonzero angular momentum: 
\be
L^2 |n,\ell,m\rangle  = \hbar^2 \ell (\ell+1) |n,\ell,m\rangle 
\ee
From this you might infer all states with $\ell\geq 1$ have a magnetic dipole moment. For a classical magnetic dipole that is caused by an angular moment, the motion is just the Larmor precession described above. But what happens according to the full-fledged quantum mechanical description? First, write
\be
\vec{M} = \gamma \vec{L}
\ee
where $\gamma$ is the gyromagnetic ratio relating the angular momentum and the magnetic moment of the `orbiting' electron. This makes $\vec{M}$ is the quantum mechanical \textit{operator} corresponding to the magnetic moment. Moreover, 
$
\vec{M}^\dag = \gamma^* \vec{L}^\dag = \gamma \vec{L} = \vec{M}
$
so that $M$ is Hermitian -  meaning it is an observable. As we noted before, the \textit{classical} energy of a dipole in a B-field is given by $W= -\vec{M}\cdot\vec{B}$. This suggests that for the quantum description, we have to add to our Hamiltonian a piece:
\be
H_L = - \vec{M} \cdot \vec{B} = - \gamma \vec{L} \cdot \vec{B}
\ee
So that the total Hamiltonian for an atomic electron in a B-field $(0,0,B)$ becomes
\be
H = H_0 + H_L =\left( -\frac{ \hbar^2}{2m} \Delta +V\right) - \gamma B L_z \cdot 
\ee 
The electron states $|n\,\ell\, m\rangle $ will also be eigenstates of this new Hamiltonian, but with a \textit{different} energy. Indeed: 
\begin{figure}[t]
 \begin{center}
  \includegraphics[width=0.7\textwidth]{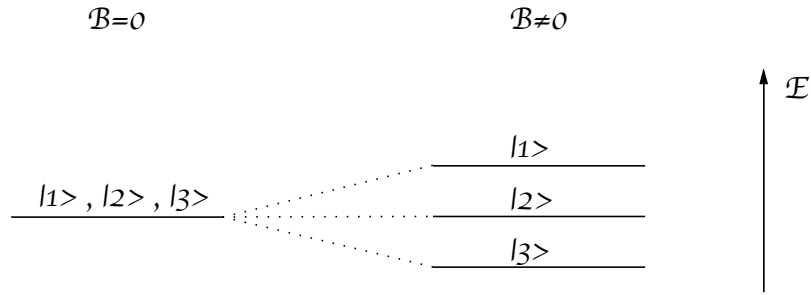}
  \caption{Under the presence of a B-field, the degenerate energy levels of an electron are split into several distinct levels. This confirms that the electron's angular momentum creates a magnetic moment that couples to the external B-field. However, this is not enough to explain the observed splitting. It seems that there is an extra contribution. This is due to the intrinsic angular momentum of the electron (the spin) and the corresponding intrinsic magnetic moment.}
  \label{fig:zeeman}
  \end{center}
\end{figure}
\be
H | n, \ell,m\rangle  = (H_0 + H_L) | n, \ell,m\rangle  = (E_n - \gamma B \hbar m)  | n, \ell,m\rangle  
\label{eq:split}
\ee
This is a very interesting result. Without the B-field, all states with the same main quantum number $n$ have the same energy $E_n$. However, in the presence of an external magnetic field the total energy $E=E_n - \gamma B \hbar m$ also depends on $m$. Hence, the energy degeneracy is partly lifted: it now depends on $n$ and also on $m$.
Originally, it was Zeeman who discovered this shift of the energy levels of atoms in the presence of a B-field. The Zeeman effect for hydrogen partly agrees with the above reasoning: indeed, spectral lines split up due to the magnetic field. That's nice, but there is one big problem: the splitting is not \textit{just} $ - \gamma B \hbar m$. It seems as if on top of the angular momentum of the electron(s), there is an \textit{extra} contribution to the magnetic moment, not due to the angular momentum of the wave function. After a long history of guesses and other experiments, physicists found a resolution to this paradox: the electron must have an \textbf{intrinsic angular moment} and a corresponding \textbf{intrinsic magnetic moment}. This property is called the \textbf{spin} of the electron. A way to visualize this spin, is shown in Figure \ref{fig:spincurrent}. Imagine an electron would be a small charged sphere, rapidly rotating around one axis. Then it would have an angular moment, and a magnetic moment due to the rotating charge of the sphere. This picture is not really correct (an electron obviously is \textit{not} just a charged sphere) but it gives a nice classical idea on how such an intrinsic angular momentum may arise. 

\subsection{Stern-Gerlach experiment}
Another important experiment in the discovery of the electron spin, was that of Stern and Gerlach. They prepared vaporized silver atoms in a furnace, and fired these into a region with an inhomogeneous magnetic field: 
\begin{figure}[t]
 \begin{center}
  \includegraphics[width=0.8\textwidth]{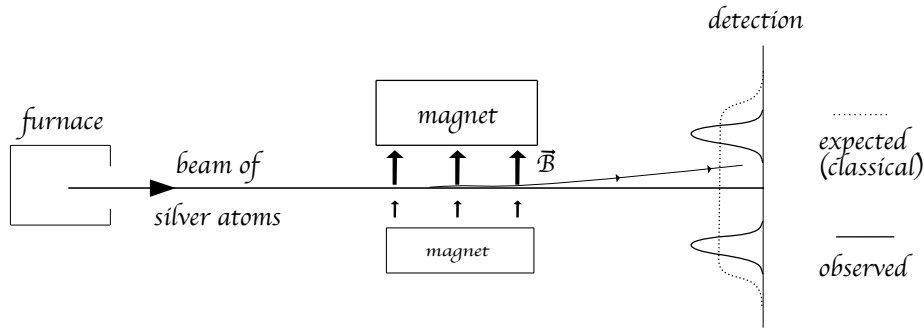}
  \caption{The Stern-Gerlach experiment. The B-field is directed along the z-axis, and also increases in strength in that direction: $\partial_z B(z)> 0$. On the right, the result of the experiment is shown: detection occurs at two central bright spots. This contrasts with the classical expectation, which predicts detection along a wide and smeared out region.}
  \label{fig:SG}
  \end{center}
\end{figure}
\be
\vec{B} = (0,0,B(z)) \quad \textrm{with} \quad \partial_z B(z) >  0
\ee
The setup is shown in Figure \ref{fig:SG}. If the atom has magnetic moment $\vec{M}$, the force exerted on it by passing through the field is given by
\be
\vec{F} = - \vec{\nabla} W = \vec{\nabla} (\vec{M}\cdot \vec{B}) 
\ee
The last expression only has a nonzero z-component, so the force acting on the atom is given by
\be
F_z= M_z \partial_z B(z).
\ee
There are several remarks in place. First, the force is only due to the \textit{gradient} of the magnetic field. Also note it is proportional to the z-component of the magnetic moment only. From the part on Larmor precession, we know that the effect of a B-field on the magnetic moment is rather simple: precession occurs in the x- and y- direction, but $M_z$ stays constant. In conclusion: the atom is deflected by the field, in the direction of the force $F_z\propto M_z$. If $M_z> 0$, the atoms will be sent upwards, for $M_z< 0$ they bend off to lower $z$. Since the magnetic moment of the atoms can be oriented in any direction, we expect $M_z$ to take on a continuous region of values. This would imply the particles hit the wall on the right of the detector along a continuous region. This is shown with the dashed line (`classical expectation') on the right in Figure \ref{fig:SG}.

\subsubsection*{The outcome}
As shown in the figure, the outcome of the experiment is rather different. Instead of particles impacting along a broad region of the screen, they only hit two well-localized spots. The location of the spots correspond to the following magnetic moments:
\be
M_z = \pm \gamma_e \frac{\hbar}{2}
\label{eq:stern}
\ee
With $\gamma_e$ is a numerical constant. This is a pretty shocking result if you are a die-hard fan of classical physics. Just by sending a beam of ordinary silver atoms through some magnetic field, it splits in two very sharp lines! Obviously, it's hard to explain this with classical physics.

Can we understand this result with the spin picture? The answer is yes. First, one can show that the magnetic moment of a silver atom is due to only \textit{one} of its electrons. All other electrons pair up in states with opposite spin, and also the total angular momentum of all the electrons in silver is zero. This will be explained better in next chapter, just buy for now that the experiment really measures the intrinsic magnetic moment of one electron only. If we define the constant $\gamma_e$ to be the electron's (intrinsic) gyromagnetic ratio, equation (\ref{eq:stern}) readily implies that
\be
S_z = \pm \frac{\hbar}{2}
\label{eq:spinstern}
\ee
Hence, from the Stern-Gerlach experiment, we draw two conclusions
\begin{itemize}
\item The \textbf{size} of the intrinsic angular momentum of the electron is equal to $\hbar/2$.
\item The \textbf{direction} of the spin is very restrained: it can only point upwards or downwards- nothing in between.  
\end{itemize}
Let us put this in a quantum mechanical style. The electron (ignoring the spatial dependence of the wave function) can be in two states: \textbf{spin up} and \textbf{spin down}, which we denote as follows
\be
\textrm{spin up} \quad \leftrightarrow   \left(\begin{array}{c}
    1 \\ 
    0 \\ 
  \end{array}\right) \quad \textrm{} \quad \textrm{spin down} \quad \leftrightarrow   \left(\begin{array}{c}
    0 \\ 
    1 \\ 
  \end{array}\right)
\ee
We also define the operator corresponding to the z-component of the intrinsic angular momentum (the \textbf{spin operator}) to be
\be
S_z = \frac{\hbar}{2}\left(\begin{array}{cc}
    1 &0\\ 
    0 &-1\\ 
  \end{array}\right)
\ee
This operator acts on the spin up and spin down states by ordinary matrix multiplication. Denoting the spin up and spin down state by $|\uparrow\rangle $ and $|\downarrow\rangle $, we have
\be
S |\uparrow\rangle  = \frac{\hbar}{2}\left(\begin{array}{cc}
    1 &0\\ 
    0 &-1\\ 
  \end{array}\right) 
 \left(\begin{array}{c}
    1 \\ 
    0 \\ 
  \end{array}\right)
=\frac{\hbar}{2}  \left(\begin{array}{c}
    1 \\ 
    0 \\ 
  \end{array}\right) = \frac{\hbar}{2} |\uparrow\rangle 
\ee
and similarly
\be
S |\downarrow\rangle  = \frac{\hbar}{2}\left(\begin{array}{cc}
    1 &0\\ 
    0 &-1\\ 
  \end{array}\right) 
 \left(\begin{array}{c}
    0 \\ 
    1 \\ 
  \end{array}\right)
=-\frac{\hbar}{2}  \left(\begin{array}{c}
    0 \\ 
    1 \\ 
  \end{array}\right) = -\frac{\hbar}{2} |\uparrow\rangle 
\ee
From this we conclude that the spin up state $|\uparrow\rangle $ has eigenvalue $\hbar/2$ under the operator $S_z$, and the spin down state $|\downarrow\rangle $ has eigenvalue $-\hbar/2$. This corresponds precisely to the two spots on the detection screen. When the electron is in the up state, it is deflected upwards by the Stern-Gerlach apparatus. When it is in the down state, it will hit the spot lower on the screen.

\begin{figure}[H]
 \begin{center}
  \includegraphics[width=0.5\textwidth]{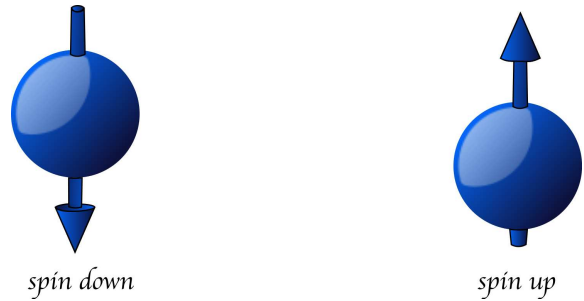}
  \caption{Again relying on the naive depiction of an electron as a rotating sphere, we can see the spin up and spin down states as two different rotations. Rotating in one direction gives spin up, rotating in the opposite direction gives spin down. Again, we stress this is a simplistic visualization of the spin.}
  \end{center}
\end{figure}

\subsubsection*{Superpositions}
According to the rules of quantum mechanics, all linear combinations of these two should be physical states too. So any state of the form
\be
\alpha \left(\begin{array}{c}
    1 \\ 
    0 \\ 
  \end{array}\right) + \beta \left(\begin{array}{c}
    0 \\ 
    1 \\ 
  \end{array}\right) = 
\left(\begin{array}{c}
    \alpha \\ 
    \beta \\ 
  \end{array}\right)
\ee
is a good state too, for any complex numbers $\alpha$ and $\beta$. (If you require normalization, you need $|\alpha|^2 + |\beta|^2 =1$.) What happens to such a state when it enters the apparatus? As we have mentioned several times by now, the Schr\"{o}dinger equation (for whatever system of Hamiltonian) is \textit{linear}. Hence, if a state is the sum of two parts, each part will evolve in its own way - as if the other was not there. From this we infer that the $\alpha  \left(\begin{array}{c}
    1 \\ 
    0 \\ 
  \end{array}\right)$ part will travel upwards, and the $\beta  \left(\begin{array}{c}
    0 \\ 
    1 \\ 
 \end{array}\right)$ will fly downwards. Hence the wave function will \textbf{split} in two parts. The total amplitudes of the two parts of the wave function are given by $|\alpha|^2$ and $|\beta|^2$. You know that the measuring device (etecting screen) will then report impact on the upper spot with probability $|\alpha|^2$ , and impact on the lower spot with probability $|\beta|^2$. So actually, we should say that an electron can be: \\
\,\,\,\,\,\quad  $\circ$ in the up state\\
\,\,\,\,\,\quad  $\circ$ in the down state\\
\,\,\,\,\,\quad  $\circ$ in a superposition\\
However, for any of these possibilities there will only be one place at which detection occurs, so it \textit{always} looks as if the electron was strictly up or down, even if the state entering the apparatus was a superposition. So just by adding the notion of spin to our quantum theory, we can successfully explain the stunning experiment of Stern and Gerlach.

\subsubsection{Playing with spins}
\begin{figure}[t]
 \begin{center}
  \includegraphics[width=1.0\textwidth]{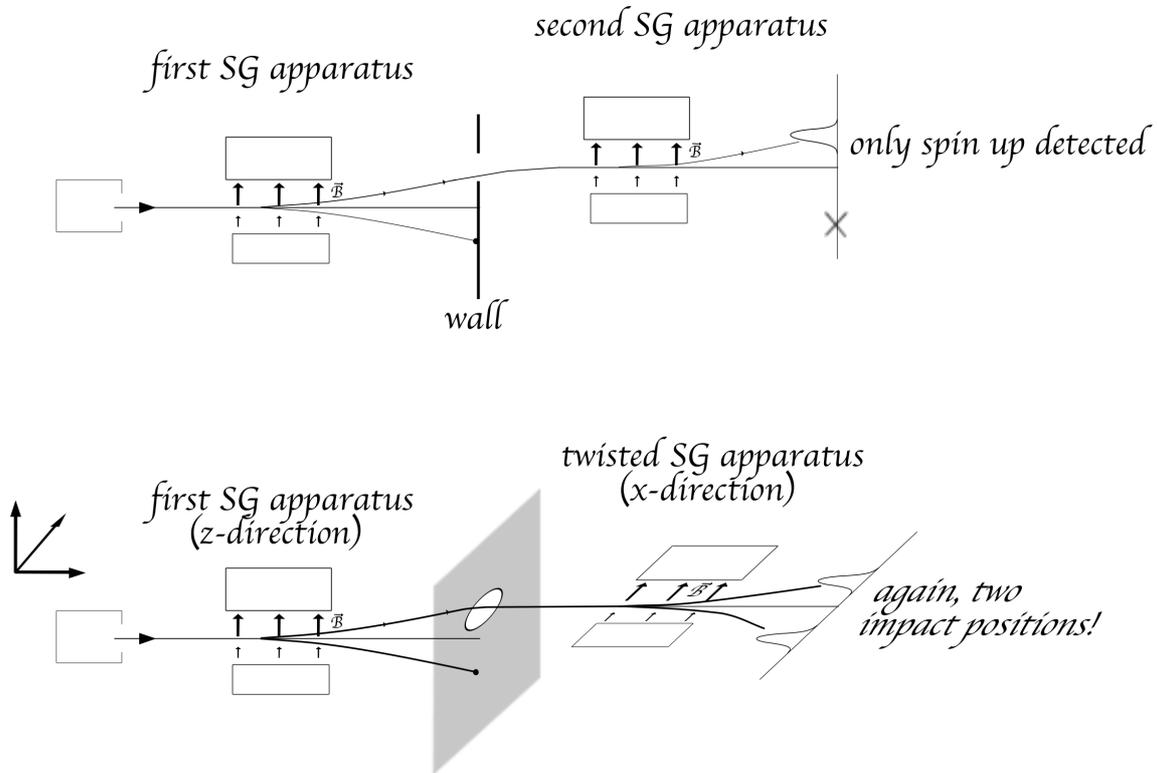}
  \caption{Playing with different SG apparatuses. Top: drilling a hole at the spin-up spot and placing another appartus after that. The outcome is as expected: only spin-ups are detected in the second device. Bottom: when placing a second, but rotated device, things become more interesting. Again, two spots emerge, one at $S_x = \hbar/2$ and one at $S_x = - \hbar/2$. This means that a state with spin up along the z-axis does not have a definite value for $S_x$.}
  \label{fig:SGs}
  \end{center}
\end{figure}

This clarifies the z-component of the spin, but what about the other components, $x$ and $y$? Well, Stern-Gerlach apparatuses are like Lego pieces: you can put them together in many ways, and have lots of fun. Consider drilling a hole in the detecting screen, at the spin up spot, and behind it placing another Stern-Gerlach apparatus, as in Figure \ref{fig:SGs}. The outcome is not so surprising: the screen after the second apparatus will only detect spin ups. Indeed: the hole served as some kind of a \textit{filter} - letting through only the $|\uparrow\rangle $ states. Hence, at a second splitting, there will only be particles going upwards, as confirmed by experiment. Now do something special. Keep the first one, but twist the second apparatus over an angle of $90^\circ$, so that the B-field is in the x-direction. That way it selects states by $S_x$, the $x$-component of their intrinsic magnetic moment. The outcome is surprising: again two spots are seen on the final screen: one with magnetic moment $S_x = \frac{\hbar}{2}$, and one spot for $S_x = - \frac{\hbar}{2}$. How can this be? Somehow, the state $ \left(\begin{array}{c}
    1\\ 
    0 \\ 
 \end{array}\right)$ of definite $S_z$ does \textit{not} have a definite $S_x$. It is a superposition of an spin up and spin down when viewed in the x-direction. A similar result holds when studying $S_y$. 
Well, let us now look at it from the quantum side. Both $S_x$ and $S_y$ are -like $S_z$ - physical quantities: they can be measured. They have to correspond to \textit{operators} acting on the spin state. So, just like $S_z$ they have to be represented by a two-by-two matrix acting on the states $\left(\begin{array}{c}
    \alpha\\ 
    \beta \\ 
 \end{array}\right)$. So we need
\be
S_x = \left(\begin{array}{cc}
    a_x &b_x\\ 
    c_x & d_x \\ 
 \end{array}\right) \quad \textrm{and } \quad 
S_y =\left(\begin{array}{cc}
    a_y & b_y \\ 
    c_y & d_y \\ 
 \end{array}\right)
\ee
for complex numbers $a_x$ , ... , $d_y$. How can we guess the form of $S_x$ and $S_y$? Because the operators $S_x$ and $S_y$ are observables, they have to be Hermitian. Hence the corresponding matrices need to be so too. So we need
\be
\left(\begin{array}{cc}
    a & b \\ 
    c & d \\ 
 \end{array}\right) = \left(\begin{array}{cc}
    a^* & c^* \\ 
    b^* & d^* \\ 
 \end{array}\right).
\ee
This is a very restricting property. You can easily check that (up to linear combinations) there are only 4 Hermitian 2-by-2 matrices:
\be
\left(\begin{array}{cc}
    1 & 0 \\ 
    0 & 1 \\ 
 \end{array}\right) , 
\quad
\left(\begin{array}{cc}
    0 & 1 \\ 
    1 & 0 \\ 
 \end{array}\right), \quad
\left(\begin{array}{cc}
    0 & -i\\ 
    i & 0 \\ 
 \end{array}\right),
\quad
\left(\begin{array}{cc}
    1 & 0 \\ 
    0 & -1 \\ 
 \end{array}\right). 
\ee
The first one is trivial: the identity matrix. This not really an interesting operator. The last one you should also recognize: it's our friend $S_z$ modulo the factor $\hbar/2$ which has been stripped of. A fond guesser might then conjecture that
\be
S_x =\frac{\hbar}{2}\left(\begin{array}{cc}
    0 & 1 \\ 
    1 & 0 \\ 
 \end{array}\right)\quad \textrm{and} \quad
S_y =\frac{\hbar}{2}\left(\begin{array}{cc}
    0 & -i \\ 
    i & 0 \\ 
 \end{array}\right)
\ee
Well, this actually is a good guess. To convince you a bit, we will use the above guess to explain the outcome in Figure \ref{fig:SGs}. On the first place, 
\be
S_x \left(\begin{array}{c}
    1  \\ 
    0  \\ 
 \end{array}\right)=\frac{\hbar}{2} \left(\begin{array}{cc}
    0 & 1 \\ 
    1 & 0 \\ 
 \end{array}\right) \left(\begin{array}{c}
    1 \\ 
    0  \\ 
 \end{array}\right) = \frac{\hbar}{2}\left(\begin{array}{c}
    0 \\ 
    1  \\ 
 \end{array}\right)
\ee
so clearly $\left(\begin{array}{c}
    1  \\ 
    0  \\ 
 \end{array}\right)$ is \textit{not} and state with definite $S_x$. However, it can be broken in two parts
\be
\left(\begin{array}{c}
    1  \\ 
    0  \\ 
 \end{array}\right) = 
\left(\begin{array}{c}
    1/2  \\ 
    1/2  \\ 
 \end{array}\right) +
\left(\begin{array}{c}
    1/2  \\ 
    -1/2 \\ 
 \end{array}\right)
\ee
And these parts \textit{are} eigenstates of $S_x$:
\bea
S_x \left(\begin{array}{c}
    1/2  \\ 
    1/2  \\ 
 \end{array}\right)&=&
\frac{\hbar}{2}\left(\begin{array}{c}
    1/2 \\ 
    1/2  \\ 
 \end{array}\right)\nonumber\\
S_x \left(\begin{array}{c}
    1/2  \\ 
    -1/2  \\ 
 \end{array}\right)&=&
-\frac{\hbar}{2}\left(\begin{array}{c}
    1/2 \\ 
    -1/2  \\ 
 \end{array}\right)
\eea
This explains the outcome of the `twisted' Stern-Gerlach experiment. The first part selects the state to be in $\left(\begin{array}{c}
    1 \\ 
    0  \\ 
 \end{array}\right)$. This is an eigenstate of $S_z$, but not of $S_x$. Hence, when passing through the second B-field, the state gets split again.  The part $\left(\begin{array}{c} 
    1/2 \\ 
    1/2  \\ 
 \end{array}\right)$ had positive $S_x$, hence feels a force $F_x > 0$ and goes to positive $x$. The part $\left(\begin{array}{c}
    -1/2 \\ 
    1/2  \\ 
 \end{array}\right)$ has negative $S_x$, so is subject to $F_x < 0$ and goes the other way. Similar reasonings hold when considering $S_z$ and $S_y$ or $S_x$ and $S_y$: if an electron is in a definite state with respect to one of the three operators $S_x$, $S_y$ of $S_z$, it will be a superposition of the up and down states along any other direction.

\subsection{The Zeeman effect, ctd.}
We can now see why Zeeman's experiment gave a `strange' outcome. Consider a free electron (so not bound in an atom). Put it in a magnetic field: $(0,0,B)$. The Hamiltonian of the electron is:
\be
H = - \vec{B} \cdot \vec{M} = - \gamma_e  \vec{B} \cdot \vec{S} = - \gamma_e  B S_z
\ee
Hence, the spin-up and the spin-down states are energy eigenstates:
\be
H |\uparrow\rangle  = - \gamma_e B \frac{\hbar}{2}|\uparrow\rangle  \quad H |\downarrow\rangle  =  \gamma_e B \frac{\hbar}{2} |\downarrow\rangle 
\ee
Their energy difference is given by:
\be
\Delta E =  \gamma_e B \frac{\hbar}{2}  - \left(- \gamma_e B \frac{\hbar}{2} \right) = \gamma_e B \hbar
\ee
So in conclusion, when you want to know the energy of an electron in an atom, not only do you need to consider the energy shift due to its angular momentum (proportional to the quantum number $m$) you also need to specify its spin state (up or down) to get the right expression. If you carefully account for both of those, your prediction very tightly coincides with Zeeman's measurements. 

This concludes the story part. You now understand the outcome of both the Zeeman and Stern-Gerlach experiment thanks to the concept of spin. Of course, you may not be fully comfortable with it yet. To get more grip on it, we will work out some more details and concrete examples in the last part of this chapter.

\section[Working with spin]{\includegraphics[width=0.04\textwidth]{comput_c} Working with spin}

\subsection{Bra's and ket's}
A nice thing about spin, is that realizes the bra and ket formalism in a very simple way. For a general spin ket $|s\rangle  = \alpha |\uparrow\rangle  + \beta |\downarrow\rangle $, we define its bra as follows:
\be
|s\rangle  =  \left( \begin{array}{c}
    \alpha \\ 
    \beta \\ 
  \end{array}\right) \quad \quad \Rightarrow\quad  \quad\langle s| = \left(\begin{array}{cc}
    \alpha^* &\beta^* \\
  \end{array}\right)
\ee
The inner product of states is nothing but the inner product of vectors:
\be
\langle s_1|s_2\rangle  =  \left(\begin{array}{cc}
    a & b \\
  \end{array}\right) \left( \begin{array}{c}
    c \\ 
    d \\ 
  \end{array}\right) = ac + bd
\ee
In particular, the square of the norm of a spin state is given by 
\be
\|s_1\|^2 = \langle s_1|s_1\rangle  =   \left(\begin{array}{cc}
    \alpha^* &\beta^* \\
  \end{array}\right)  \left( \begin{array}{c}
    \alpha \\ 
    \beta \\ 
  \end{array}\right) = |\alpha |^2 + |\beta |^2
 \ee
This clarifies our remark in the previous section: a spin state $\left( \begin{array}{c}
    \alpha \\ 
    \beta \\ 
  \end{array}\right) $ is normalized if and only if $|\alpha |^2 + |\beta |^2=1$.
This is a nice moment to write down the spin up and down states along the $x$ and $y$ axes. Denoting them by $|\pm\rangle _x$ and $|\pm\rangle _y$, we have
\be
|\pm\rangle _x = \left(  \begin{array}{c}
    1/\sqrt{2} \\ 
    \pm 1/\sqrt{2} \\ 
  \end{array}\right)
\quad \textrm{and} \quad |\pm\rangle _y = \left(  \begin{array}{c}
    1/\sqrt{2} \\ 
    \pm i/\sqrt{2} \\ 
  \end{array}\right)
\ee
You can verify their normalisation and that
\be
S_x |\pm\rangle _x = \pm \frac{\hbar}{2}|\pm\rangle _x \quad \textrm{and} \quad S_y |\pm\rangle _y = \pm \frac{\hbar}{2}|\pm\rangle _y.
\ee
Also, it is nice to check explicitly that the inner product between the two eigenstates of any $S_i$ is zero. This has to be so, since these pairs are eigenstates (eigenvectors) of a Hermitian operator (matrix) with different eigenvalues.

\subsection{Angular momentum}
If you are very (very) critical, you may remark that the Zeeman and Stern-Gerlach experiment do not strictly prove the existence of an electron's angular momentum. They clearly show an electron has an intrinsic magnetic dipole moment - how else could it get an extra energy term $\vec{M} \cdot \vec{B}$ in the presence of a magnetic field? From this observation then, we just \textit{assumed} the underlying reason for this dipole is an intrinsic angular momentum - which we called \textit{spin}. This is a very natural thing to do, but how do we show more directly that this \textit{spin} really is a form of angular momentum? Or more general, if one has three operators $(O_x, O_y, O_z)$ what makes one conclude that these give the angular momentum of a quantum system? Classically, an object is an angular momentum if it gives an object's `amount of rotation' but in quantum mechanics we can not use this pictorial definition. It might be good at this point to go back quite a bit, to the point where we computed the commutation relations of the angular momentum operators $\vec{L}=(L_x,L_y,L_z)$. We found that
\be
[L_x,L_y] = i \hbar L_z \quad \textrm{and cyclic}
\ee
By `cyclic' we mean that you can shift all three variables $x\rightarrow y\rightarrow z$ to get the other two commutators. 
These commutation relations are actually quite special. They are so special, that whatever set of operators satisfying similar commutation relations is \textit{defined} to be a \textbf{generalised angular momentum}. This gives a nice and strict definition of angular momentum in quantum mechanics. Do the spin operators $\vec{S} = (S_x,S_y,S_z)$ satisfy the above commutation relations? You bet: with the matrix representation it is easy to show that
\be
[S_x,S_y] = i \hbar S_z  \quad \textrm{and cyclic}
\ee
Hence, we can comfortably state that spin really is a (generalized) angular momentum, just because of the structure of the commutators of its three components. 

In analogy with the total angular momentum operator $L^2$, we can also write out the operator $S^2 $:
\be
S^2 = \vec{S} \vec{S} = \sum_i S_i S_i= \hbar^2 \frac{3}{4}   \left(\begin{array}{cc}
    1 & 0 \\ 
    0 & 1 \\ 
  \end{array}  \right) =  \hbar^2  \frac{1}{2}\left(\frac{1}{2}+1\right)  \left(\begin{array}{cc}
    1 & 0 \\ 
    0 & 1 \\ 
  \end{array}  \right)
\ee
So every spin state $|s\rangle $ is an eigenvector of $S^2$, with eigenvalue $\hbar^2  \frac{1}{2}\left(\frac{1}{2}+1\right)$:
\be
S^2 |s\rangle  = \hbar^2  \frac{1}{2}\left(\frac{1}{2}+1\right) |s\rangle .
\ee
The reason to write $3/4$ in a funny way, is to make clear the (formal) resemblance to the atomic energy states, where we found 
\be
L^2 |n , \ell, m\rangle  = \hbar^2\ell(\ell +1) |n,\ell,m\rangle . 
\ee
So just like we labeled $|n,\ell,m\rangle $ with an \textit{orbital} angular momentum quantum number $\ell$, spin states can be interpreted as carrying an \textit{intrinsic} angular momentum quantum number $1/2$. Because this value is the same for all spin states (all spin states are eigenstates of $S^2$ with the same eigenvalue) this is an intrinsic property of the particle itself. People say that the electron is a \textbf{spin-$\frac{1}{2}$ particle}.

\subsection{Quantum Larmor precession}
Ah, we have arrived at our final computation. We start out with a small question. The operators $S_x$, $S_y$, and $S_z$ describe the spin of a particle along three main axes. What about all other directions? In spherical coordinates, every direction can be described by a unit vector $\vec{u} = (\sin \theta \cos \varphi ,\, \sin \theta \sin \varphi ,\, \cos \theta)$. Hence it is natural to define the operator of the \textbf{spin along direction $\vec{u}$} as:
\be
S_u = \vec{u}\cdot \vec{S} = \frac{\hbar}{2}  \left(   \begin{array}{cc}
    \cos \theta & \sin \theta e^{- i \varphi } \\ 
   \sin \theta e^{i \varphi } & -\cos \theta \\ 
  \end{array}\right).
  \label{eq:spinu}
\ee
A trivial example: the x-direction is given by angles $\theta =\frac{\pi}{2}$, $\varphi =0$. This corresponds to the unit vector $\vec{u}_x =(1,0,0)$, and the associated operator written out above is nothing but $S_x$.
For a general direction $\vec{u}$ and the corresponding operator $S_u$, the spin eigenstates along that direction are given by
\be
|+\rangle _u = \left(   \begin{array}{c}
  e^{- i \varphi /2}   \cos (\theta/2)  \\ 
 e^{ i \varphi /2}    \sin (\theta/2) \\ 
  \end{array}\right)\quad \textrm{and}\quad |-\rangle _u = \left(   \begin{array}{c}
    -e^{- i \varphi /2} \sin (\theta/2  ) \\ 
   e^{i \varphi /2} \cos (\theta/2 )\\ 
  \end{array}\right)
\ee
Again, a trivial check: for the z-direction, $\theta=\varphi =0$, and the above expressions give $|+\rangle _z = \left(  \begin{array}{c}
    1 \\ 
    0 \\ 
  \end{array}\right)$ and $|-\rangle _z= \left(  \begin{array}{c}
    0 \\ 
    1\\ 
  \end{array} \right)$ as should be.
We have now all necessary tools to describe the \textit{quantum mechanical} version of Larmor precession. Let us take a state, which is spin up along a direction $\vec{u}$:
\bea
|\psi(t=0)\rangle &=&|+\rangle _u\\
 &=& \left(   \begin{array}{c}
  e^{- i \varphi /2}   \cos (\theta/2)  \\ 
 e^{ i \varphi /2}    \sin (\theta/2) \\ 
  \end{array}\right) \\
 &=& e^{- i \varphi /2}   \cos (\theta/2)|+\rangle _z + e^{ i \varphi /2}    \sin (\theta/2)  |-\rangle _z
\eea
We now want to study how this state evolves through time, given that we put a magnetic field $(0,0,B)$. Well, earlier we found that the magnetic field gives an energy $E^+=  \gamma_e B \frac{\hbar}{2}$ to the up state and $E^-=- \gamma_e B \frac{\hbar}{2}$ for the down state. The Schr\"{o}dinger equation $ i \partial_t |\psi\rangle  = H |\psi\rangle $ then tells us that the time evolution of the above state is given by
\be
|\psi(t)\rangle = e^{- i \varphi /2}   \cos (\theta/2) e^{-iE^+ t /\hbar}|+\rangle _z + e^{ i \varphi /2}    \sin (\theta/2) e^{-i E^- t/ \hbar} |-\rangle _z
\ee
From this, we can compute the \textit{expectation value} of the different spin components:
\bea
\langle S_x\rangle _\psi (t) &=& \langle \psi(t) |S_x |\psi(t) \rangle  = \frac{\hbar}{2} \sin \theta \cos(\varphi + \omega_0 t)\\
\langle S_y\rangle _\psi (t) &=& \langle \psi(t) |S_y |\psi(t) \rangle  = \frac{\hbar}{2} \sin \theta \sin(\varphi + \omega_0 t)\\
\langle S_x\rangle _\psi (t) &=& \langle \psi(t) |S_z |\psi(t) \rangle  = \frac{\hbar}{2} \cos \theta
\label{eq:qmlar}
\eea
Where $\omega_0 =  \frac{2 E^+}{\hbar} = \gamma_e B$, which is nothing but the classical Larmor frequency. The above result is very nice: if you look back, you see that the \textit{expectation values} of the spin components have the same time dependence as the components of the \textit{classical} counterpart. 
Such similarities show up quite often in quantum mechanics. Recall for example how the (semi-)classical radius of the electron orbit (the Bohr radius) was quite good an indicator for the size of the correct quantum wave function. In general, classical intuition can often `predict' the evolution expectation values of systems, and orders of magnitude. But this is only a guidance. Equally often, the classical intuition is of no good at all - think of the tunneling effect for example. 

Also with spin, one should be careful. The above result on Larmor precession may seems so natural, that one would almost think that the naive picture of spin (an electron rotating around an axis) captures the situation accurately.  Although this would make the concept of spin more easy to grasp, this idea really is incorrect. Spin is a bit more tricky than that. Take a look at Figure \ref{fig:point}. Say there is an electron with spin up along axis $z$. You may be tempted to represent this as an arrow pointing upwards along $z$. (And such is done quite often.) This is not bad, but some students tend to infer from this drawing that the components of the spin along axis $x$ are zero. So they might write
\be
S_x |+\rangle _y = 0. \quad \quad \textrm{(WRONG)}
\ee
This is very wrong: we have seen in this chapter that the spin state - expressed in the basis of the $x$ direction will be a \textit{superposition}:
\begin{figure}[t]
 \begin{center}
  \includegraphics[width=1.0\textwidth]{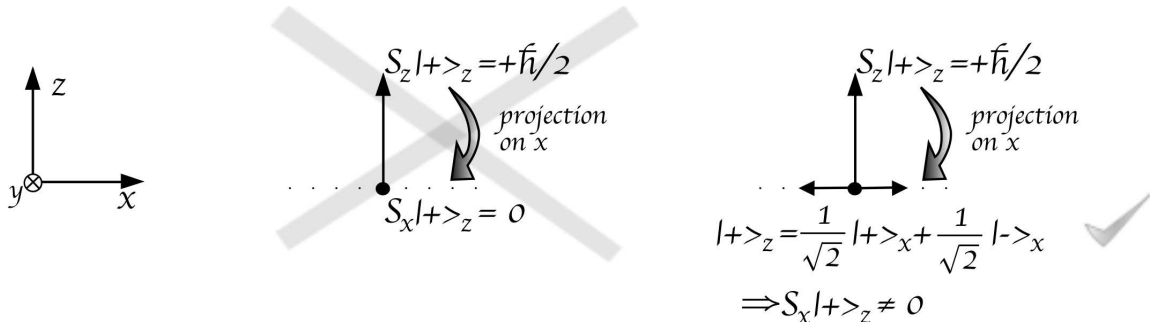}
  \caption{The spin of a particle does not point. Even though the expectation values define a vector (which does point) one would draw several incorrect conclusions from treating spin as a vector. The most obvious pitfall is concluding that $S_x|+\rangle _z `='0$ by projecting the spin on the x-axis: this is of course wrong. The right decomposition is shown on the right. Clearly this is not the way ordinary vectors work.}
  \label{fig:point}
  \end{center}
\end{figure}
\be
|+\rangle _z =\frac{1}{\sqrt{2}} |+\rangle _x\, +\frac{1}{\sqrt{2}} |-\rangle _x
\ee
so that
\be
S_x |+\rangle _z = \frac{\hbar}{2}  \frac{1}{\sqrt{2}} |+\rangle _x\, +\left(-\frac{\hbar}{2}\right) \frac{1}{\sqrt{2}}  |-\rangle _x \neq 0
\ee
So if you really want to use arrows, make sure you understand the right side of Figure \ref{fig:point}. But even then, you might get into trouble when trying to draw \textit{complex} spin components - so there probably all intuition breaks down. The safest thing is to remember the following (not so catchy) catchphrase: 
\begin{quote}
The spin of a particle is represented by two (possibly complex) numbers. You can decompose this as a linear combination of a spin-up and a spin-down state, with respect to any direction. In particular, if the state is up (or down) along a direction $\vec{u}$, then it is eigen under the operator $S_u$ with eigenvalue $ \hbar/2$ (or $- \hbar/2$). However, this does \textit{not} mean this spin points up (or down) along $\vec{u}$. \textit{Spin does not point in any direction}; it is a state - not a vector. 
\end{quote}
Especially the last sentence is important. You might refute this: don't the expectation values of $\vec{S}$ define a vector for each state  - like in the equation (\ref{eq:qmlar})? This thing points, right? Well, that is very true, but then recall that a state is more than just it's expectation value. In a similar way, we can associate to each wave function a well-defined number $\langle x\rangle $, but this does not yet mean a particle is a point. 

So this makes spin a bit subtle to understand. There is a pretty common saying that there is nothing we \textit{understand} in life - we just get \textit{used} to things. That is probably true here too - and don't worry, you will get used to the idea of spin at some point.

\subsection*{Application: Sunspots}
Due to the high temperature, many of the atoms in the solar atmosphere are excited or even (partly) ionized. As a result, there is an abundant absorbing/emitting of photons by the electrons kicked up/falling down from one level to another. This is visible in the light that reaches: the spectrum has so-called \textbf{spectral lines} (less intensity at specific frequencies). It is possible to identify which lines are due to which elements and hence deduce the composition of the sun without actually having to go there. (Historically, this is how we got to know the sun is composed of mainly hydrogen and helium. Before it was thought the sun had a composition similar to earth.) However, if you look at those spectral lines carefully, you can see that the degenerate lines (coming from transitions between degenerate energy levels) are actually split! With what you have learned, you would guess that the sun has a magnetic field, and that the splitting you are seeing is actually the Zeeman effect. This is indeed true. By doing precise measurements, you can observe that this magnetic field varies along the solar surface. Maybe you have once heard of the small but numerous dark spots on the sun (not so imaginatively called `sunspots'). It turns out the magnetic field is significantly larger at these places. This very crucial: the darkness of the spots turns out to be caused by the magnetic field, which influences the dynamics of the gas and plasma around it - in a way that the temperature is lowered which in turn makes that region shine less bright. The exact dynamics of the magnetic field of the sun and the formation of sunspots is a vast field of research; and the underlying experimental observations would definitely not be possible without (an understanding of) the Zeeman effect!

\newpage
\section*{Exercises}
\begin{enumerate}
\item Explain to yourself (without looking at the text) how the Zeeman and Stern-Gerlach experiments work and why the lead to the idea of spin.
\item In the chapter on the uncertainty principle, you learned that non-commuting observables are incompatible. How is this related to spin and the `twisted' version of the Stern-Gerlach experiment (Figure \ref{fig:SGs})? 
\item  Check that the commutation relations of the spin operators satisfy the angular momentum commutation relations. Are these still satisfied if you absorb an extra factor - say multiply all three matrices by 2? (Now look again at the transition (\ref{eq:stern})$\rightarrow$($\ref{eq:spinstern}$). Could you also have taken $\gamma_e$ only half as large, and $S_z=\pm\hbar$?)
\item Check the completeness relations for the basis $\{|+\rangle _u,|-\rangle _u\}$
\item Find the eigenvalues and eigenvectors of the matrix (\ref{eq:spinu}). Do you find the vectors stated in the text?
\item A typical value of the B-field in a solar spot is 0.15 Tesla. Ignoring angular momentum, what is the difference between the two spin states of an electron of an atom near that spot? (The size of the gyromagnetic ratio of the electron spin is $\gamma_e=1.76\cdot10^{11}$ s$^{-1}$T$^{-1}$.) What is the frequency of the emitted photon when the electron goes from one spin state to another?
\item Check the following property of the Pauli matrices: $\sigma_a \sigma_b = \delta_{ab} \cdot I + i \sum_c \varepsilon_{abc} \sigma_c$ where $I$ is the $2\times2$ identity matrix. With this, prove $ (\vec{a} \cdot \vec{\sigma})(\vec{b} \cdot \vec{\sigma}) = (\vec{a} \cdot \vec{b}) \, I + i \vec{\sigma} \cdot ( \vec{a} \times \vec{b} ) $.
\item It is natural to guess that the \textit{total angular momentum} of an electron is given by $\vec{J} = \vec{L} + \vec{S}$: just the sum of its orbital and intrinsic angular momentum. Because $\vec{J}$ works on the wave function, and $\vec{S}$ on the spin state of the electron, $[\vec{S},\vec{L}]=0$. 
Show that this implies that $\vec{J}$ is indeed a angular momentum operator (i.e. that it satisfies the right commutation relations).
\item Given a set of operators $J_x$, $J_y$, $J_z$ which satisfy the angular momentum commutation relations, define the following two operators:
$$J_+ = J_x + iJ_y \quad \textrm{and}\quad J_- = J_x - iJ_y$$. Show that $\left[J_z,J_\pm\right] = \pm\hbar J_\pm$. Show also $J_z J_\pm|m\rangle = \hbar\left(m \pm 1\right)J_\pm| m\rangle$ if $m$ is an eigenstate of $J_z$ with eigenvalue $\hbar m$. Interpret this result. (Hint: think back of the situation with the creation and annihilation operators for the harmonic oscillator.) 
\end{enumerate}

%% file: H11.tex
\chapter{Many particles}

\subsection*{In this chapter\dots }
We are about to explore a topic left untouched so far: describing the quantum behaviour of \textit{collections} of particles, instead of just a single one. An important concept governing the dynamics of such systems is the notion of \textit{identical particles}. This will lead to a division of particles in two types: bosons and fermions. These concepts go a long way understanding the structure of larger systems. It will soon become clear that they have very direct consequences on the touch and feel of our own (macroscopic) world.
\newpage
\section[Product and sum of vector spaces]{\includegraphics[width=0.04\textwidth]{tool_c} Product and sum of vector spaces}

We will give a short review on the \textbf{direct sum} and \textbf{direct product} of vector spaces. You might have seen these before, but for what follows, it's useful to have it all fresh in your mind. 

As an ode to those golden times of high school math, we will illustrate the two concepts with an example involving marbles - oh yes. Imagine you have a collection of marbles. They are of three kinds: yellow, blue and red. You can summarize the content of your collection with three numbers: $(n_y,n_b,n_r)$. These quantities give the total number of yellow, blue and red marbles you have. The space of all possible $(n_y,n_b,n_r)$ forms a vector space. (We ignore the issue of these numbers being integer or not.) For example: $(1,2,2)$ and $(2,4,1)$ give two possible marble collections, but also their sum $(3,6,3)$ describes a possible collection. The same for multiples. Imagine you are even \textit{more} lucky - and not only have marbles, but also three empty jars. Woah, full of enthusiasm you put your marbles in the three jars. That sure is some good fun, but a sudden problem strikes you. How can I now describe my collection, now that it is in the jars? Here is a first way to do so. On top of specifying $(n_y,n_b,n_r)$, you can give another vector, which gives the total weight of marbles in each jar: $(w_{1}, w_{2}, w_{3})$. These numbers too form a vector space, by very similar arguments. 
Better even, your collection (and how it is put in the jars) can be compactly summarised by one long vector:
\be
(n_y,n_b,n_r, w_1,w_2,w_3).
\ee
This just glues the two pieces of information together in one single object. That is possible because the two pieces of information are completely independent: the number of marbles per color and the weight of the marbles in each jar have nothing to do with each other. Making a new vector space in this fashion, is called the \textbf{direct sum} of vector spaces. 

Being satisfied with this answer, you are just about to go and play with the marbles and jars, but another issue strikes you. You could also describe the marbles in the jars in another way. Why not say how much marbles there are in each jar, \textit{per color}? 
So something like:
\be
\left(  \begin{array}{ccc}
    w_{1,b} & w_{1,y} & w_{1,r} \\ 
    w_{2,b}& w_{2,y} & w_{2,r}\\ 
    w_{3,b} & w_{3,y} & w_{3,r} \\ 
  \end{array}\right)
\ee
with for example $w_{2,y}$ the total weight of yellow marbles in jar 2. In this alternative description, we get \textit{more} information on the system. This is because we are not keeping track of the weights in jars and the types of marbles \textit{separately} anymore: we give mixed information on how many of each type is in each jar. This new vectorspace (in the form of  3-by-3 matrices) is called the \textbf{direct product} of the original two vector spaces.

What you should remember from this somewhat pathetical example is the following. Every pair of vector spaces can be combined into a new (larger) one. There are two ways to do this. The first (direct sum) glues vectors together. It means that you keep track of both vectors (marble types and jar weights) in a rather trivial way. These two bits of information need not have any relation at all. The second possibility is taking the direct product. In that case, you keep \textit{mixed} information (here: how much of each of the marble type is in each of the jars). This construction is suited for situation in which you want to keep track of a finer sort of information.

In a slightly more mathematical language: let there be given two vector spaces $V$ and $W$ of dimension $n$ and $m$, with bases $v_1,\dots  v_n$ and $w_1,\dots w_m$. The direct sum of these vector spaces is another vector space, denoted $V \oplus W$, with basis
\be
\{v_1,\dots ,v_n, w_1,\dots ,w_m\}
\ee
Obviously, this vector space has dimension $n+m$. Next to this, we can also construct the direct product, denoted $V \otimes W$. This space has as a formal basis:
\be
\left\{\begin{array}{cccc}
v_1 w_1, & v_1 w_2 &  \dots & v_1 w_m \\
\vdots   & \vdots & \vdots  &\vdots \\
v_n w_1, & v_n w_2, & \dots & v_n w_m
\end{array}\right\}
\ee
and has dimension $n m$.

\section[Many particles]{\includegraphics[width=0.04\textwidth]{once_c} Many particles}
\begin{figure}[t]
 \begin{center}
  \includegraphics[width=0.8\textwidth]{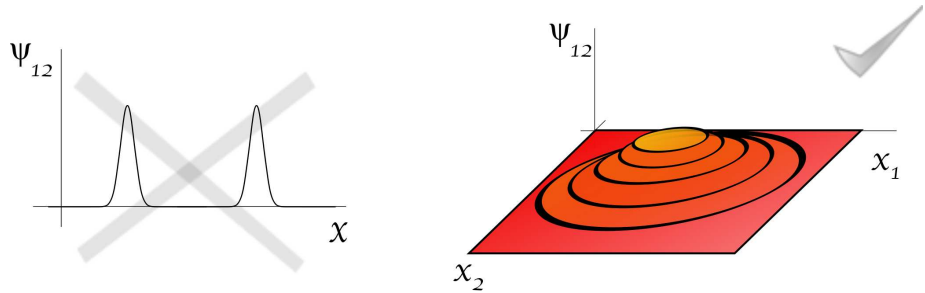}
  \caption{Left: if the wavefunction $\psi_{12}$ of two particles would be just the sum of the wavefunctions, such a state could not be distinguished from a single particle whose wavefunction happens to consist of to lumps. Right: if we describe the two-particle system with a function $\psi_{12}=\psi(x_1,x_2)$ (real part shown) this problem does not occur. So this gives the right description.}
  \label{fig:colwav}
  \end{center}
\end{figure}

So here we start our quest to understand systems with more than one particle. For concreteness, consider two particles. How can we describe those two together? Since both of them are wavepackets ($\psi_1$ and $\psi_2$), we could guess that the total system should be described by the sum of the two wavefunctions $\psi_1+\psi_2$. Something like this is suggested in Figure \ref{fig:colwav}. However, such a description quickly runs into trouble: how could we then distinguish between those \textit{two} particles and just \textit{one} particle of which the wavefunction just happens to consist of two lumps? We need something else. A good starting point is the following question: what if we measure their positions? In that case we get a concrete outcome like $(x_1,x_2)$. Here $x_1$ is the measured position of one particle, and $x_2$ the detected position of the other. From our experience with single particle systems, we can guess that such an outcome is not certain, but only has a \textit{probability} associated to it. From this we guess there must be a function 
\be
\psi(x_1,x_2)
\ee
such that $|\psi(x_1,x_2)|^2$ is the probability density to measure the particles' positions to be $(x_1,x_2)$. This turns out to be the right description. So when we correctly want to describe two particles at once, we should not consider two separate wavefunctions, but one \textbf{collective wavefunction}. As an illustration: the probability to measure $a_1<x_1<b_1$ and simultaneously $a_2<x_2<b_2$ is given by
\be
P( a_1<x_1<b_1\,\,\& \,\,a_2<x_2<b_2) = \int_{x_1=a_1}^{x_1=b_1} \int_{x_2=a_2}^{x_2=b_2} |\psi(x_1,x_2)|^2 \,dx_1\,dx_2
\ee
Hence, the normalisation of a collective wavefunction takes the form 
\be
\inft \inft |\psi(x_1,x_2)|^2 \,dx_1\,dx_2 =1
\ee
Because a collective wavefunction has more arguments (here two: $x_1$ and $x_2$) it is somewhat harder to visualise than a single wavefunction. An attempt at drawing a particular $\psi(x_1,x_2)$ is made in Figure \ref{fig:colwav} on the right. 

\subsection{Direct product}
We have just seen that the wavefunction of a system of two particles keeps track of both particles \textit{at once}, not just the conjunction of two individual wavefunctions. If you got the marble story, you might smell a direct product coming up here. Say that a basis of the Hilbert space of particle one is given by $|\chi_1\rangle, ... ,  |\chi_n\rangle$, and a basis for the states of the second particle by $|\phi_1\rangle, ... , |\phi_m\rangle$. Then the combined system of the two particles has as basis
\be
|\chi_i\rangle |\phi_j\rangle \quad \textrm{with} \quad 1\leq i\leq n, \,\, 1\leq j \leq m.
\ee
The state $|\chi_i\rangle |\phi_j\rangle$ is understood as a situation in which the first particle is in state $|\chi_i\rangle$ and the second is in $|\phi_j\rangle$. The inner product between two collective states is defined in the following natural way:
\be
(\langle\chi_i| \langle \phi_j| ) (|\chi_{i'}\rangle |\phi_{j'}\rangle ) = \langle \chi_i|\chi_{i'}\rangle \langle \phi_j|\phi_{j'}\rangle
\ee
But this is not enough: we need the combined states to form a Hilbert space. So we also want to allow linear combinations of the above basis of states. 
This means the most general state of the two-particle system can be written as
\be
|\psi\rangle = \sum_{ij} c_{ij} |\chi_i\rangle |\phi_j\rangle
\label{eq:sysdecomp}
\ee
(By $\sum_{ij}$ we mean $\sum_{i=1}^{n} \sum_{j=1}^m$ of course.) For this $|\psi\rangle$, the chance to measure the first particle to be in state $|\psi_i\rangle$ and the second in $|\chi_j\rangle$ is given by
\be
P(1: |\chi_i\rangle \,\,\&\,\, 2: |\phi_j\rangle) = |c_{ij}|^2
\ee
Hence $\psi$ is normalised if 
\be
\sum_{ij} |c_{ij}|^2 =1.
\ee
We can now better understand the collective wavefunction $\psi(x_1,x_2)$ from above. A (continuous) basis we can take for any particle is the position basis $|x\rangle$. Hence, we can take as a basis for the collective states all products $|x_1\rangle|x_2\rangle$. These are wavefunctions sharply peaked at one specific $x_1$ and $x_2$. So just like
\be
|\psi\rangle = \int dx\,\, \psi(x) |x\rangle 
\ee 
for a single particle, we have
\be
|\psi\rangle = \int \int dx_1\,\, dx_2\,\, \psi(x_1,x_2) |x_1\rangle |x_2\rangle
\label{eq:colwavposrep}
\ee
for two particles. So the collective wavefunction we were talking about before, is just the position representation of the combined system: 
\be
\psi(x_1,x_2)=(\langle x_1|\langle x_2|) |\psi\rangle
\ee
This means we can also obtain (\ref{eq:colwavposrep}) by inserting two completeness relations:
\bea
|\psi\rangle  &=& \left(\int\, dx_1 \,|x_1\rangle \langle x_1| \int dx_2 \, |x_2\rangle \langle x_2|\right) |\psi\rangle \nonumber\\
&=& \int \int dx_1\,\, dx_2\,\, \psi(x_1,x_2) |x_1\rangle |x_2\rangle
\eea
Before moving on, we want to point out a very classical pitfall. Although this is so for all basis states, not every state can be written in the product form $|\psi\rangle = |\chi\rangle|\phi\rangle$. We illustrate this subtlety with an example. Consider two particles, each of which can be in states $|a\rangle$ and $|b\rangle$. For example, having both particles in state $a$ is written as
\be
|a\rangle_1|a\rangle_2
\ee
with the subscript denoting we are talking about particle 1 and 2 respectively. In the same way, the system can be in state 
\be
|a\rangle_1|b\rangle_2
\ee
The system could also be in a (normalised) superposition of the above two: 
\be
\frac{1}{\sqrt{2}}|a\rangle_1|a\rangle_2 + \frac{1}{\sqrt{2}} |a\rangle_1 |b\rangle_2 = |a\rangle_1\left(\frac{1}{\sqrt{2}} |a\rangle_2 + \frac{1}{\sqrt{2}} |b\rangle_2 \right)
\ee
This is a product form again! Indeed, if we rename the state $(|a\rangle+|b\rangle)/\sqrt{2} $ as $|c\rangle$, the above state is just $|a\rangle_1|c\rangle_2$.  In contrast, you can show that the state
\be
\frac{1}{\sqrt{2}}|a\rangle_1|a\rangle_2 +\frac{1}{\sqrt{2}} |b\rangle_1|b\rangle_2 
\ee
can \textit{never} be rewritten as a product $|d\rangle_1|e\rangle_2$ - whatever definition of $d$ and $e$ you would try. So if someone says you a system of two particles is in a state $|\psi\rangle$ you can \textit{not} in general decompose this state as $|\psi\rangle = |\chi\rangle|\phi\rangle$. The only decomposition that is always possible is (\ref{eq:sysdecomp}).

\subsection{Identical particles}

In the previous section, we did not at all specify which two particles or systems we were talking about. They could have been anything. Let us be more specific here, and say that the two objects under consideration are two particles of the \textit{same species}: so two protons, or two electrons, two muons - whatever.  You may or may not have thought about this before, but there is a very special thing about particles. They don't have hair and they don't wear clothes, so if you see two particles of the same species, they are really \textbf{identical}. To illustrate this, imagine studying two electrons. The chance to detect electron 1 around some $x$ and electron 2 around $x'$ is proportional to the probability density:
\be
P(1:\,x\, \, \& \,2: \,x') \propto |\psi(x,x')|^2
\ee
In a similar fashion, you would say that the chance to find electron 1 around \textit{$x'$} and electron 2 around \textit{$x$} to be like
\be
P(1:x'\, \,\& \,2: x) \propto |\psi(x',x)|^2
\ee
This could a priori be two different probabilities. However, if you think more critical, you may remark that identical particles neccesarily are \textbf{indistinguishable}. If you detect one electron at site $x$ and another at $x'$, how would you say which ones they are? The names that you gave them (electron 1 and electron 2) are not written on them, so if you do measurements, you can never tell which detection corresponded to which electron. So purely on the basis of consistency, we need to have
\be
|\psi(x,x')|^2 =  |\psi(x',x)|^2  \quad \textrm{for identical particles}.
\ee
Indeed, both quantities describe one and the same measurement outcome (an electron found at $x$ and an electron found at $x'$) so they have to yield equal probabilities. So we conlude
\be
\psi(x,x') = \eta\, \psi(x',x)
\label{eq:antisy}
\ee
with $\eta$ a complex number of modulus one. In fact, by interchanging twice, we get the same state again, so $\eta^2=1$ implying $\eta =\pm1$. Inspired by this result, we define the \textbf{exchange operator} $P$. Take again two identical particles. 
If particle 1 is in state $|\chi\rangle$ and particle 2 in state $|\phi\rangle$, then $P$ acts as follows:
\be
P |\chi\rangle_1 |\phi\rangle_2 = |\phi\rangle_1 |\chi\rangle_2
\ee
That is, $P$ puts particle 1 in the state of particle 2 and vice versa. Obviously, $P^2 =1$: if you interchange the states of the particles twice the state is again as before. From $P^2 =1$, it follows that $P$ can only have eigenvalues $\pm 1$. (Indeed, if $P |\psi\rangle = \lambda |\psi\rangle$, and $P^2=1$ then  $|\psi\rangle = P^2 |\psi\rangle = \lambda^2 |\psi\rangle$ so $\lambda^2 = 1$.) If the eigenvalue under $P$ of the pair of identical particles is $-1$, we call them \textbf{fermions}, and if their eigenvalue is $+1$, we call the particles \textbf{bosons}. So
\bea
|\chi\rangle_1 |\phi\rangle_2 &=& |\phi\rangle_1 |\chi\rangle_2 \quad \textrm{(bosons)} \\ 
|\chi\rangle_1 |\phi\rangle_2 &=& - |\phi\rangle_1 |\chi\rangle_2 \quad \textrm{(fermions)}
\eea
This is a very fundamental classifications of elementary particles. Electrons, neutrons and protons are all fermions. Photons are bosons.  As you know, there are many other particles, and all of these are either bosons or fermions. Let us consider fermions as an example. The two-particle wave function is:
\be
|\psi\rangle = \int \int dx\, dx'\,\, \psi(x,x') |x\rangle_1|x'\rangle_2
\ee
and since $ |x\rangle_1|x'\rangle_2 = -  |x'\rangle_1|x\rangle_2$ only the antisymmetric part of $\psi$ contributes to the integral. So without any loss we can keep only the antisymmetric part of $\psi(x,x')$:
\be
\psi(x,x') = - \psi(x',x)
\ee
This way, we see that $\eta=-1$ for fermions. In a similar fashion, $\eta=1$ for bosons and $\psi(x,x')$ is symmetric in that case. So for either bosons or fermions, the chance to detect particle 1 here and particle 2 there is equal to the chance to detect particle 2 here and particle 1 there - as should be for indistinguishable particles.
\begin{figure}[H]
 \begin{center}
  \includegraphics[width=0.7\textwidth]{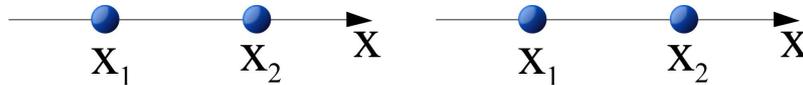}
 \caption{Two particles: one is at location $x_1$ and the other is at location $x_2$. Because the particles are identical, this can not be distinguished from a situation where we interchange the particles' positions. This has important consequences for the collective wavefunction: it should either be symmetric or anti-symmetric in terms of the position variables. }
  \end{center}
\end{figure}

\subsection{The Pauli exclusion principle}

So far we have only considered two particles. We will not consider more than two particles in a general way. This is a bit sloppy - expressions with many indices and that sort of visual foltering- and it gives few extra extra insight: all rules are just a quite straightforward extrapolation of the above conclusions. Taking direct products, bosons vs. fermions, minus signs popping up, that kind of things. So we skip the general treatment here. 

A special case we \textit{will} treat is a system of $N$ fermions, all of the same kind. If every fermion separately can be in states $|\psi_i\rangle$, then the Hilbert space of the entire system is spanned by
\be
|\psi_{i_1}\rangle|\psi_{i_2}\rangle ... |\psi_{i_N}\rangle
\label{eq:span}
\ee
The above state describes a situation where fermion 1 is in state $|\psi_{i_1}\rangle$, fermion 2 in $|\psi_{i_2}\rangle$, etcetera. Now by the definition of a fermion, exchanging the states of two fermions is minus the original state. So for example, interchanging the states of fermion $1$ and $3$, we get: 
\be
|\psi_{i_1}\rangle|\psi_{i_2}\rangle|\psi_{i_3}\rangle ... |\psi_{i_N}\rangle =  - |\psi_{i_3}\rangle|\psi_{i_2}\rangle|\psi_{i_1}\rangle ... |\psi_{i_N}\rangle
\ee
Up to the sign, the right hand side is to be read as: fermion 1 is in state $i_3$ and fermion 3 in state ${i_1}$. Once more, note that this rule is not so strange: both above states yield the \textit{same} observations (you'd measure indistinguishable electrons in states $i_1,...,i_N$) so they have to be identical up to a phase factor, which by definition is -1 for fermions.\\
From the above equation, we conclude that there is some redundancy in the states (\ref{eq:span}). Any two combinations that are a permutation of the individual states have to be proportional to each other. So many of them are just the copies of another up to a possible sign.

But there is a more striking remark. Suppose the system is a state like (\ref{eq:span}) but now we put fermions 1 and 2 in the same state, so $i_1=i_2$. Hence the total system is $|\psi_{i_1}\rangle|\psi_{i_1}\rangle|\psi_{i_3}\rangle ... |\psi_{i_N}\rangle$. If we blindly interchange the states of fermions 1 and 2 (even though they are the same) the above rule tells us we get a minus sign:
\be
|\psi_{i_1}\rangle|\psi_{i_1}\rangle|\psi_{i_3}\rangle ... |\psi_{i_N}\rangle = - |\psi_{i_1}\rangle|\psi_{i_1}\rangle|\psi_{i_3}\rangle ... |\psi_{i_N}\rangle
\ee
This is a bit of a funny equation: an object can only equal its opposite if it is zero. So a state with $i_1=i_2$ can not exist! In general, we conclude: 
\begin{quote}
\textbf{It is impossible for two fermions (of the same species) to be in the same quantum state.}
\end{quote}
This is a very profound statement, with far-reaching consequences (we will meet some later on in this chapter). It was first formulated by Pauli, and is called \textbf{the exclusion principle}. 

In the unlikely case you mix up the meanings of bosons and fermions, you can use a trick. A fermion is \textit{firm} (it `repels' other fermions) while a boson likes to get \textit{close on} (it likes to sit together with other bosons). OK, there are better sounding rhymes in the world but well. Time to use all these new goodies for some applications.

\section[Applications]{\includegraphics[width=0.04\textwidth]{comput_c} Applications}
With the ideas and techniques presented in the theory part, you can say a very great deal about larger physical systems. By `large' we mean: an atom with two or more electrons, a molecule, or even an large piece of metal. The Pauli exclusion principle is the final piece of the quantum mechanical jigsaw that we have been putting together in this course. Below are some nice applications.

\subsection{The atomic structure}
\begin{figure}[t]
 \begin{center}
  \includegraphics[width=0.85\textwidth]{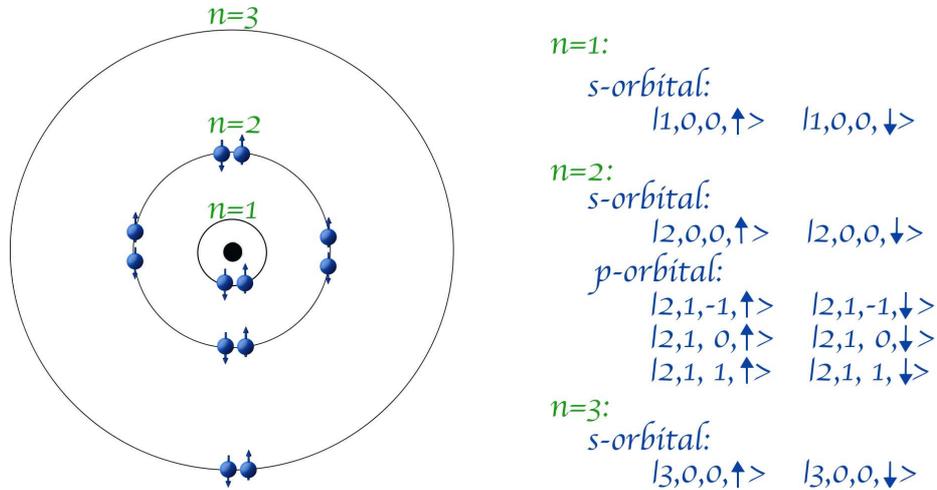}
  \caption{The atomic structure of magnesium. The twelve electrons sit in the lowest possible energy states. On the left these are pictorially shown as Bohr orbits. The correct description involving states is shown on the right. States with the same $n$ are called a \textbf{shell}, and states with the same $n$ and $\ell$ comprise a \textbf{subshell}. So for magnesium two shells ($n=1,2$) are completely filled and one shell ($n=3$) is partly filled.}
  \label{fig:exclusion}
  \end{center}
\end{figure}

You know that the energy eigenstates of a single electron in an atom are given by $|n\,\ell\, m\rangle$. In the absence of a magnetic field, the energy depends on $n$ only. 
If an electron is in a high energy state $n>1$, it will typically fall down after a while, until it has reached $n=1$. With each transition, the energy difference is released in the form of radiation. So a hydrogen atom at rest will typically be in the $n=1$ state, the ground state. Now what about larger atoms? These have more electrons in them, and the nuclear charge is now $Z\neq1$ but it turns out you can still find states of the form $|n,\ell,m\rangle$. A bold guess would be that in such systems then, all electrons would sit in the lowest energy state. So you would write
\be
|\psi\rangle = |1\,0\,0\rangle_1 ... |1\,0\,0\rangle_N  \quad \textrm{(WRONG)}
\ee 
In words: the system is given by all ($N$) electrons sitting in the lowest energy state $|1\,0\,0\rangle$. However, this is not what Pauli taught us to do. No two electrons can sit in the same state, so the above configuration is just not possible. If no two electrons can sit in the same state, what is the best we can do (energy-wise)? Ah, just filling up the n lowest energy states of course. Also, we know that electrons have a spin degree of freedom. Hence, in each energy level, there can be a spin-up electron and a spin-down. As an example of this, consider the Lithium atom. Two electrons can sit in the $|1\,0\,0\rangle$ state, if at least they have opposite spins - this way they are in a different state, and the exclusion principle poses no problem. The third electron however is the odd one out: we can not do anything but putting it in a higher energy level. So the ground state of the 3-electron system of Lithium is given by
\be
|\psi\rangle_{Li} = |1\,0\,0\uparrow\rangle_1 |1\,0\,0\downarrow\rangle_2 |2\,0\,0\,s\rangle_3
\ee
Here, we have just put the spin states as an extra parameter in the list of quantum numbers. Note that the spin state $s$ of the third electron can be whatever it likes to be - any superposition of up and down is allowed. Also, the spin up and down (of the first two electrons) can be with respect to any direction - as long as they are opposites, they describe states without overlap. In principle, the above notation is fine, but one has to keep in mind that it is equals to $-|1\,0\,0\downarrow\rangle_1 |1\,0\,0\uparrow\rangle_2 |2\,0\,0 \, s\rangle_3$ because this just interchanges states of electrons 1 and 2. To prevent any confusion, people like to write states explicity in an anti-symmetrized form. Here, this amounts to summing up all six permutations, with the right signs, and each with an individual weight $1/\sqrt{6}$:
\bea
|\psi\rangle_{Li}  &=& \frac{1}{\sqrt{6}} |1\,0\,0\uparrow\rangle |1\,0\,0 \downarrow\rangle |2\,0\,0\,s\rangle  -  \frac{1}{\sqrt{6}} |1\,0\,0 \uparrow\rangle  |2\,0\,0\,s\rangle|1\,0\,0\downarrow\rangle \nonumber\\
&+&  \frac{1}{\sqrt{6}} |1\, 1\, 0 \downarrow\rangle |2\,0\,0\,s\rangle |1\,0\,0\uparrow\rangle  -  \frac{1}{\sqrt{6}} |1\,0\,0\downarrow\rangle |1\,0\,0\uparrow\rangle |2\,0\,0\,s\rangle  \nonumber\\
&+&  \frac{1}{\sqrt{6}} |2\,0\,0\,s\rangle  |1\,0\,0\uparrow\rangle |1\,0\,0\downarrow\rangle -  \frac{1}{\sqrt{6}}  |2\,0\,0\,s\rangle |1\,0\,0\downarrow\rangle |1\,0\,0 \uparrow\rangle \nonumber
\eea
In principle, this is just the same state as the one up higher, but it is explicitly anti-symmetrised, and hence we can freely drop all subscripts denoting the particle numbers, these are not relevant anymore in the above notation. In a similar fashion, we can construct the states of the electrons in arbitrary large atoms. This `stacking' of electrons in energy levels explains a very wide range of chemical properties of elements. As shown in Figure \ref{fig:exclusion} it can describe the configuration of electrons around larger atoms. These kind of diagrams can in turn be used to explain which elements can be bound to which others, and so on. So combining the exclusion principle with our earlier results opens up the vast world of chemistry. That's quite pretty... . 

Of course, there are some things one should bear in mind when considering larger atoms. There are several effects that \textit{shift} the energy levels of the states. These make the energy of a state depend on \textit{all} quantum numbers, not just $n$. So the most economic way of filling an atom's orbitals becomes a bit more subtle. You may remember some rules from chemistry that tell you in which order to fill the orbitals. These rules are precisely a consequence of the energy shifts we are talking about. Here we mention two main effects (in order of importance) determining the shifts and splittings of the energy levels:
\subsubsection*{Effect 1: Coulomb repulsion}
\begin{figure}[t]
 \begin{center}
  \includegraphics[width=0.8\textwidth]{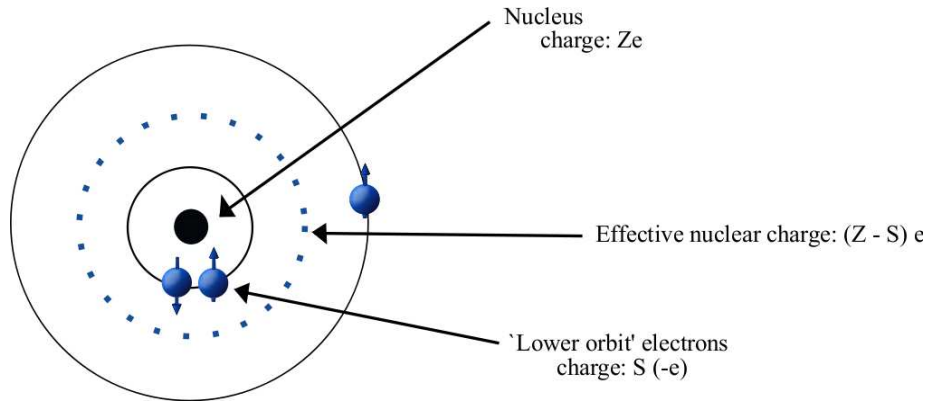}
  \caption{The shield effect, as experienced by the outer electron in lithium. The electrons in the lowest energy states cover up the nucleus charge $+Z e$. So for the outer electron, the effective nuclear charge is $(Z-S)e$, where $S$ is the number of shielding electrons. Here $Z=3$ and $S=2$. Note that $S$ depends on the particular electron we are considering.}
  \label{fig:shield}
  \end{center}
\end{figure}

In larger atoms, the Coulomb force is not acting between the nucleus and the electrons, but also amongst each pair of electrons. This is an extra energy contribution, and hence one needs to put the corresponding term into the Hamiltonian. Physically, what is happening, is that the electrons in low energy states (small $n$ and small $\ell$) are covering up a part of the charge of the nucleus. So electrons in more extended wavefunctions (larger $n$ or larger $\ell$) see a lower \textbf{effective nuclear charge}. An approximate equation expressing this, is
\be
Z_{\textrm{eff}} = Z - S
\ee
with $Z_{\textrm{eff}}$ the nuclear charge, seen by an electron that has $S$ electrons below itself. This is called the \textbf{shield effect}. It is shown in Figure \ref{fig:shield}. Hence, the binding energies of electrons in high orbits are lower than one would expect. Indeed, a lower (effective) nuclear charge means that they are less attracted to the nucleus, and hence less well bound. Also, for states with the same $n$ but a different $\ell$ will not have the same energy any more. A larger $\ell$ corresponds to a more extended wavefunction (recall from your chemistry class the dumbbell shape p-orbital versus the spherical s-orbital) and will be more strongly experience the shield effect.

\subsubsection*{Effect 2: Spin-orbit interactions}
Recall that the angular momentum $\vec{L}$ of a state creates a magnetic moment. For example, we found a Zeeman effect, even when not including spin. But there are also magnetic moments due to the spin $\vec{S}$ of the electrons. Because any pair of magnetic moments tend to align, there is an extra term in the Hamiltonian of the form $- \vec{L} \vec{S}$, which is called the \textit{spin-orbit coupling}.\footnote{The word `coupling' is a very popular term to indicate that two quantities influence each other. In the context of quantum mechanics, it means that the there is a term in the Hamiltonian involving the two quantities. This means that the energy depends on the relative values of those quantities, so they `feel' each other, they are not independent degrees of freedom.} Actually, even in the case of just one electron, this term is present. The effect of this term is less drastic than the shield effect, but for precise results, one should include it. The structure of atomic energy spectra obtained when including this effect is called the \textbf{fine structure}. Typically, it will involve small shifts of energy levels (and splits the degeneracy with respect to the spin) which can be observed in precise experiments.

\subsubsection{Consequences}
There are several consequences of the above effects. From the \textit{physics side}, we see that the structure of electronic states is more involved than just antisymmetrising the single-electron results. Most important, the energy levels turn out not to depend on $n$ only. In the shield effect, the angular momentum plays a role, and upon including spin-orbit coupling also the spin state becomes relevant. In the end, few is left of the degeneracy that we might have expected. Actually this is not only an effect on larger atoms: spin-orbit coupling also takes place in ordinary hydrogen, so even in that situation the fine structure is non-trivial: there is a very small splitting of the energy states of equal $n$.
We stress again that the Coulomb interaction between the electrons is the largest effect. Since this is absent in the single electron case, the result there is only subject to the spin-orbit interaction. 
So only if one really does precise measurements, the spectral lines of hydrogen are split. But to a very good accuracy, the energy levels are given by the $n$-dependent result we found in chapter 6. The size of the splitting is about a thousand times smaller than the energy differences between states with different $n$.

From the \textit{physicist's side} this means that much more work needs to be done to describe many-electron atoms properly. In practice, one will have to rely on numerical techniques, such as the \textbf{Hartree-Fock} methods. But to apply those techniques, one really needs to know what one is doing. So there is also quite some theoretical input. In conclusion, the study of atoms is a rich field. If you want to know more about it, be sure to take a class on \textbf{atomic physics} once. With this course here, you sure have enough background to embark it. 

\begin{figure}[t]
 \begin{center}
  \includegraphics[width=0.6\textwidth]{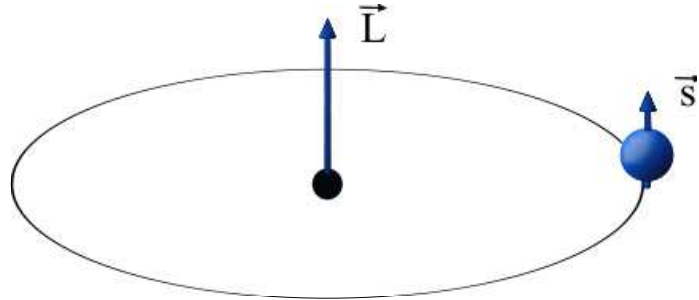}
  \caption{Classical depiction of the spin-orbit coupling. Both the orbital angular momentum $\vec{L}$ of the electron and the intrinsic angular momentum $\vec{S}$ lead to a magnetic moment. These two magnetic moments tend to align. So a parallel state as shown here has lower energy than when the two are anti-parallel. This means states with the same $\ell$ but different spin have different energies. With precise measurements these (very small) energy differences can be measured.}
  \end{center}
\end{figure}

\subsection{Application: Lasers}
As you know, electrons in high energy states tend to fall back after a while, emitting a energy: a photon. For this reason, energy levels that are not the ground state are called \textbf{excited states}. The energy of the emitted photon from a transition between states is given by
\be
\Delta E = E_i - E_f
\ee
where $E_i$ and $E_f$ stand for the energy levels of the initial and final state of the electron. Such a process is called \textbf{spontaneous emission}. There is however a trick to make this spontanteous decay process happen earlier. If you send a photon onto an atom that is excited, \textit{and} this photon has exactly energy $\Delta E$, the decay process will occur right away. That is: \textit{two photons} will instantly leave the atom, both of energy $\Delta E$. This phenomenon is called \textbf{stimulated emission}. The process of stimulated emission was predicted a long time ago, by Einstein. This process was soon verified some years after, and soon raised the interest of many scientists for practical applications. What if you would be able to put $N$ atoms all in the same excited state. If one of them decays, the escaping photon would trigger all others to decay as well. This \textit{cascade process} would almost instantly release $N$ photons, all of the same energy $\Delta E$. The result would be: an intense flash of light, consisting of photons that are all of identical energy (= identical wavelength). Repeating the process, you can get a continuous beam of very `pure' light. Since the photons all have the same phase and frequency, such a source is called \textbf{coherent and monochromous}. This invention got the name LASER: Light Amplification by Stimulated Emission of Radiation. Since the realisation of the first lasers, it has become an important player in eye surgery, metal cutting industry, measuring devices, printers, CD and DVD readers, and so on. Why do we mention this here? Well, to stress that this important invention would never work if photons were fermions. For each individual flash, the photons exiting the laser are all in the same state: same energy, same position, etcetera. This would be impossible under the asiocial behaviour of fermions. Only bosons like to sit together in the same state, fermions just can't. So the next time you watch your favourite high definition action movie DVD, be sure to thank those photons in the reading beam for being sociable bosons, otherwise we might still be working with those charming but impractical black tapes of our parent's generation! 

\begin{figure}[H]
 \begin{center}
  \includegraphics[width=0.6\textwidth]{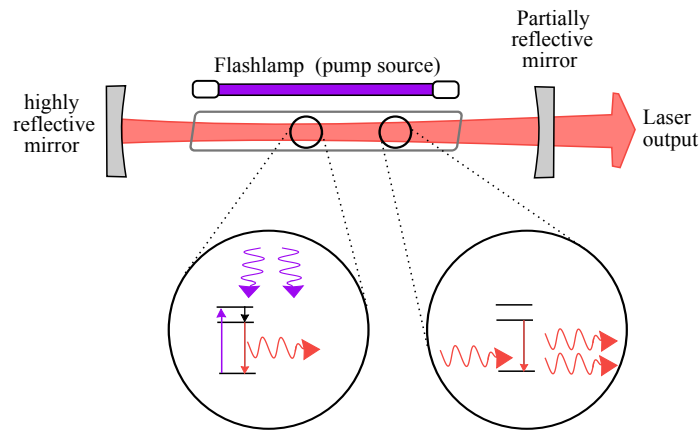}
  \caption{A possible design of a laser. A flash lamp (purple light) lifts the atoms below it to an excited energy level. The decay may consist of several steps: typically first a small step (no visible light) and then a larger one - for example emitting red light. By stimulated emission, each photon rips out an entire cascade of identical photons of all the excited atoms met on the way, giving a very monochromous and coherent light beam.} 
  \label{fig:laser}
  \end{center}
\end{figure}

\newpage
\section*{Exercises}
\begin{enumerate}
\item Without looking, try to reconstruct as much as you can of the reasoning which leads to the distinction between bosons and fermions, and their defining property.
\item In a many-particle system, the Hamiltonian will depend on the positions and momenta of all the particles. Also, the Hilbert space will be bigger. However, the Schr\"{o}dinger equation still holds. Convince yourself that the postulates of Quantum mechanics still apply despite these generalizations.
\item In the text, the inner product of the product space was defined, based on the inner product of the single particle spaces. How would you have defined it if we had taken the direct sum of the spaces?
\item Someone is arguing with you whether quantum mechanics is different from classical mechanics. He goes: `Ok, particles are tiny pieces of wave. But there are waves in classical physics too'. Explain to him that (as just one possible counter argument) the quantum physics of many particles is \textit{not} a classical theory of waves. (Hint: look back at Figure \ref{fig:colwav}.) In some sense, the conceptual shift to multi-particle quantum mechanics departs much further from our intuition than the step we made to describe a single particle - luckily at least the mathematical framework stays the same in this step.
\item Symmetrize the state $|a\rangle |b\rangle |c\rangle |d\rangle$ (four bosons), respecting normalization. You may assume that $|a\rangle$, $|b\rangle$, $|c\rangle$, $|d\rangle$ are all properly normalized. 
\item Consider two fermions, both in a harmonic potential. The Hamiltonian is then $$H= H_1+H_2= -\f{\hbar^2}{2 m}D_{x_1}^2+\f{m\omega^2}{2}X_1^2 -\f{\hbar^2}{2 m}D_{x_2}^2+\f{m\omega^2}{2}X_2^2.$$ This Hamiltonian acts on states of the form $\psi(x_1,x_2)$. If we denote by $|n\rangle_1$ the n-th harmonic oscillator state of the first particle (which is described by wave function with argument $x_1$) then $H_1|n\rangle_1=E_n |n\rangle_1$. Similarly, $H_2|n\rangle_2=E_n |n\rangle_2$ for the second particle. Show that the state $|n\rangle_1|m\rangle_2$ is an eigenstate of $H$. What is the corresponding energy? What is the time evolution of this state? What is the anti-symmetrised version? Which states are forbidden by Pauli's principle? The solution to the combined system is so easy here because the Hamiltonian doesn't impose any interaction between the particles: the potential only depends on the particle's positions, not their \textit{relative} distance.
\item Consider two particles, but now with Hamiltonian: $$H=-\f{\hbar^2}{2 m}D_{x_1}^2+\f{m\omega^2}{2}(X_1-a_1)^2 -\f{\hbar^2}{2 m}D_{x_2}^2+\f{m\omega^2}{2}(X_2-a_2)^2 + \frac{K}{|(X_1-X_2)|} $$ with $a_1$, $a_2$ and $K$ constants. What is the classical meaning of such a Hamiltonian? Hint: interpret the three potential terms. For which molecule might this Hamiltonian be a model? Here, the two particles are coupled: they feel each other as the energy depends on their relative position, so the solutions aren't simply products of the single-particle solutions anymore - so you don't have to solve the system here.
\item In the exercises of Chapter 5, you computed the spectrum of particles in a three dimensional box. Here, we will look for the Fermi energy of such a system, when $N$ (noninteracting) fermions are in such a potential with volume $V$. To find the Fermi energy, note that at $T=0$, the $N$ lowest energy states are occupied. Each state is characterized by a vector $\vec{n}=(n_x,n_y,n_z)$, and the energy of a state is essentially a constant times the length of this vector. So roughly speaking, at zero temperature, all states which are filled have their $\vec{n}$ inside a sphere with radius $R$. (At least, the region where all components $n_x$, $n_y$, $n_z$ are positive - one eight of a sphere that is.) By computing the sphere volume, estimate the number of states lying inside the sphere:$N(R)$. Invert this to find the radius corresponding to $N$ fermions at zero temperature. The energy of a state on this sphere is then the Fermi energy of the system. Prove that $$E_F = \frac{\hbar^2}{2m} \left( \frac{3 \pi^2 N}{V} \right)^{2/3}$$.
\end{enumerate}

%% file: H12.tex
\chapter{Metals, insulators and all that's in between}
\subsection*{In this chapter\dots }
In this last chapter, we will move up one step in scale, and derive properties of metals (but also insulators and semiconductors) from the microscopical quantum laws. The understanding of these systems has an impact on our daily life which can hardly be overestimated. An obvious example is the development of computers. These would still be limited to constructions of rotating wheels or vacuum tubes if not at some point the transistor 
was invented - a simple but versatile application of the physics of semiconducting materials. Again, quantum mechanics is the gateway to truly understand the underlying physics.

\newpage 
\section[Particle statistics]{\includegraphics[width=0.04\textwidth]{tool_c} Particle statistics}

When describing large systems, one typically relies on statistical approaches. This is not only a useful technique, it is often the only way out. For example, a container of gas can in principle be described by the motion of all individual molecules, but this is so complex that it is not feasible in practice.
What you do then, is describe how much of the molecules would on average have this or that momentum, be on this or that position, etcetera. You make statistical statements. This is the domain of Statistical Mechanics - on which may have had a course already. If you haven't, don't worry: we'll briefly review the concepts we need in this context.

\subsection{Population numbers}
A large collection of particles or molecules is typically described by a \textit{temperature}. The notion of temperature is intimately connected to statistical properties. For example, if a (monoatomic, ideal) gas is at temperature $T$, the \textit{average} energy per particle is given by
\be
\langle E \rangle    = \frac{3}{2} k T
\ee
with $k$ the \textbf{Boltzmann constant}. Not all individual particles have energy $\frac{3}{2} k T$ though: on the contrary, many particles will have a much larger or much lower energy. Again, the brackets (not to be confused with the quantum mechanical expectation value) denote that the above is only an \textit{average} over the entire collection of particles.

Luckily, the above can be refined a bit. Suppose that the states that can be taken by the individual particles form a discrete set, with corresponding energies $E_i$. Then the \textbf{Boltzmann distribution} states that the number $n_i$ of particles in energy level $E_i$ is given by:
\be
n_i = \frac{1}{Z} e^{-\frac{ E_i }{ kT}}
\label{eq:boltz}
\ee
The numbers $n_i$ are also called \textbf{occupation numbers}. This law should be understood as follows. If a the energy $E_i$ of state $i$ is big, then the exponential will be very small, so the state is only occupied by a few particles. States with lower $E_i$ have a larger occupation number $n_i$. The effect of high temperature is to make states of large energy more easily accessible, since the exponential suppression becomes less harsh. Again, the above expression is a statistical one: it expresses how many particles will on average be in $E_i$ if the system as a whole is at temperature $T$. In particular, $n_i$ does not have to be an integer. The proportionality constant $Z$ has to be chosen such that $\sum_i n_i$ equals to total number of particles. This constant can be ignored if one only considers ratio's of population numbers:
\be
\frac{n_i}{n_j} = \frac{e^{- E_i/kT}}{e^{-E_j/kT}} = e^{-\frac{E_i - E_j}{kT}}
\ee

\begin{figure}[t]
 \begin{center}
  \includegraphics[width=0.6\textwidth]{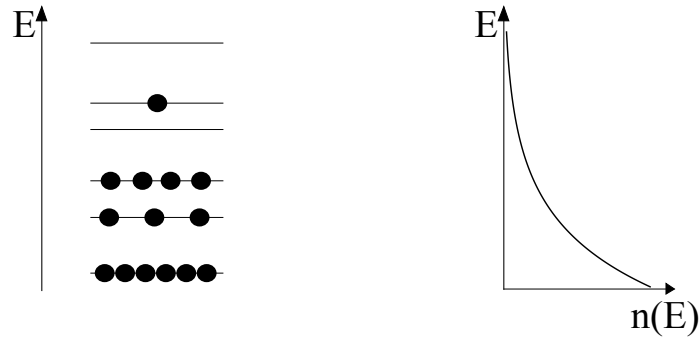}
  \caption{On the left a system at temperature $T$ with several energy levels and some particles is shown. The specific arrangement is not determined by   the temperature: the only thing we can talk about is the \textit{average} occupation number of each level of this system. This is given by a function $n(E)$: the Botzmann distribution. Here, we have inverse-plotted the function to highlight the similarity to the figure on the left. }
  \end{center}
\end{figure}

\subsubsection*{Bosons and fermions}
From the previous chapter, we know that particles can not always be in one state together. This is only the case for bosons. A fermionic state can either be occupied or empty, nothing else. A more careful analysis for those cases shows that for bosons and fermions the Boltzmann distribution has to be replaced by:
\be
n_i = \frac{1}{e^{\frac{ E_i - \mu }{ kT}} \pm1}
\label{eq:BEFD}
\ee
with $\mu$ is a constant, explained below. The lower sign (minus) is to be used for bosons. Such a distribution of particles over states is called \textbf{Bose-Einstein-statistics}.  
The upper sign (plus) is to be used for fermions. Such a distribution is called \textbf{Fermi-Dirac statistics}. Clearly, in that case all $n_i$ are smaller than one, as should be for fermions.
 
The number $\mu$ in the above expression is again a normalization constant - although it has now been absorbed into the exponential. It is to be taken such that the sum of all occupation numbers equals the number of particles of the system. It is called the \textbf{chemical potential}. It is interesting to see what happens at low temperatures. First, consider bosons. Denote the lowest energy level by $E_0$. Of course, we need $\mu<E_0$, to ensure that all occupation numbers are positive. If you now let $T\rightarrow 0$ and also $\mu \rightarrow E_0$, then the number of particles in the lowest energy state ($n_0$) becomes large (the denominator becomes $1-1$ since $\mu \rightarrow E_0$) while all others ($n_1$, $n_1$, ... ) go to zero (the denominator becomes $\infty-1$ because $T\rightarrow 0$). This means all particles settle in the lowest energy state. This phenomenon is  called \textbf{Bose-Einstein condensation}. This interesting phenomenon is only possible because bosons can all sit in the same state. Now, consider fermions. You can check that in the limit $T \rightarrow 0$, the occupation numbers $n_i\rightarrow 1$ for all energy levels with $E_i<\mu$ and that $n_i\rightarrow 0$ for all $E_i>\mu$. This is not so strange: at low temperatures, the fermions seek the lowest energy configuration: stacking all lowest energy levels with one particle each is the best you can do.
\subsubsection{Low occupation limit}
You may ask yourself what the use of (\ref{eq:boltz}) is if we have (\ref{eq:BEFD}). After all, every particle (even an entire atom or molecule!) is either a boson or a fermion. Why don't we see that a gas consists of bosons or fermions? From the above distributions you might infer that both cases behave drastically different. The answer is that for most gases the average occupation number of a state is extremely low. Not because there are few particles - on the contrary - but there is an even more overwhelming amount of states. Hence, all $n_i$ will be very small. This can only be due to the fact that $e^{(E_i-\mu)/kT}$ is very large, so for all these systems the $\pm 1$ in the numerator becomes negligible, and (\ref{eq:BEFD}) neatly reduces to (\ref{eq:boltz}). Phrased differently: only if there are very few states available ($\mu$ high, i.e.: comparable to particle energies) or if the temperature is very low (small $e^{E_i/kT}$) the difference between bosons and fermions becomes visible. For typical gas systems neither is the case, so the gas is well described by the ordinary Boltzmann distribution. So far the Stat-Mech recap, time for the story...

\section[Metals, insulators and all that's in between]{\includegraphics[width=0.04\textwidth]{once_c} Metals, insulators and all that's in between}
\begin{figure}[t]
 \begin{center}
  \includegraphics[width=0.5\textwidth]{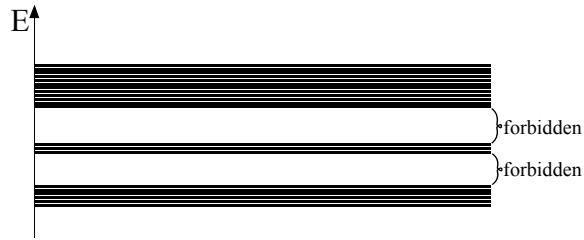}
  \caption{The typical allowed energy levels for a crystal. Several continuous energy regions can be filled with electrons: these are the energy bands. In between them, there are no states available: these are the gaps.}
  \label{fig:bands}
  \end{center}
\end{figure}

In this course, we have met two types of systems: \textit{bound particles} trapped in a potential, and \textit{free particles} which can move around freely - possibly meeting some obstacles. In the first case, the spectrum typically consists of a discrete set of states that can be taken. In sharp contrast, the spectrum of a free particle is continuous: by adjusting its momentum one can in principle give it any energy. Here, we meet a third type of systems: crystals. By a `crystal' we mean a solid in which the constituting atoms are arranged in a periodic and ordered way (a lattice). Electrons in such an environment are surely bound: they can not just walk out of the crystal. However, they are not strictly attached to one single atom: within the grid, they can freely move around. In some sense this puts crystals somewhere in between free and bound systems. And this is also reflected in the spectrum. For a crystal, the energy states that can be occupied by the electrons consist of several continuous regions, \textbf{energy bands}, which are separated from each other by regions without accessible states: \textbf{band gaps}. An example is shown in Figure \ref{fig:bands}. There, electrons can sit in energies in three bands. The bands are separated by two gaps. There are no states of which the energy lies in this gap regions, so no electron can occupy a state with that energy - they are `forbidden' energy zones.

\subsubsection*{Level filling}
We just explained that every crystal structure has energy bands and band gaps associated to it. The placement of these regions on the energy spectrum depend on several things: the atoms making up the crystal (their atomic numbers) the spatial arrangement of the atoms (the lattice type) and possibly also impurities. It are these data that decide where the energy bands and band gaps lie. But the \textit{location} of these energy regions is not the end of the story. An important property is how far these energy levels are \textit{filled up} by the electrons present in the crystal. Once again, we note that the exclusion principle prevents any two electrons from sitting in the same energy level. Hence, the best electrons can do is filling up all the lowest energy states available, not putting any two in the same state. Pictorially speaking, the electron filling defines a surface: the highest energy level that has to be occupied, given that all levels below have already been filled. This surface is called the \textbf{Fermi surface}, and the corresponding energy is the \textbf{Fermi energy}.\footnote{Combining with the tool part, you may remark that a low temperatures these should tend to the chemical potential.} Now we have all necessary ingredients 
to classify into three large categories: conductors, insulators and semiconductors.  We review them in that order.

\subsection{Conductors}

\begin{figure}[H]
  \begin{center}
  \includegraphics[width=1.0\textwidth]{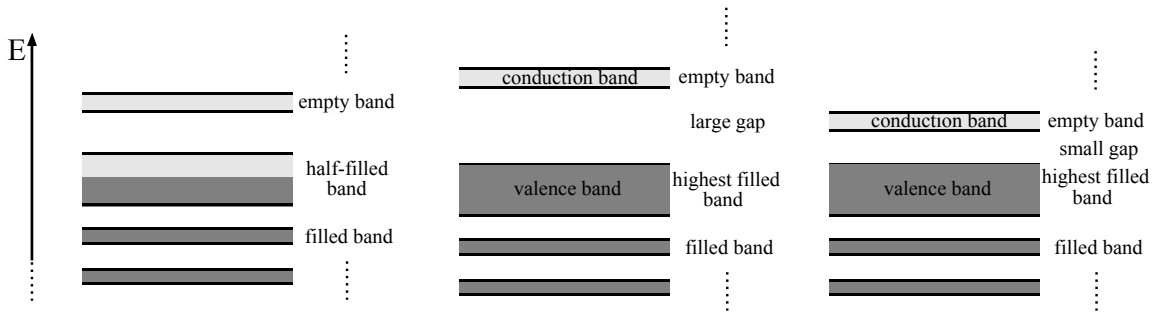}
  \caption{The different types of crystals according to their electrical properties. Left: a conductor has a broad band, in the middle of which lies the Fermi surface. This leaves a lot of available states open right above the occupied states. Such a situation easily conducts an electric current. Middle: when a large gap separates the (filled) valence band from the (empty) conduction band, electrons can hardly overcome that gap. This means few or electrons will participate in the current: the material is an insulator. }
  \label{fig:between}
\end{center}
\end{figure}

Conductors are materials that easily produce an electric current when an external potential is applied. What is the mechanism behind this current? As follows: the electrical field gives some electrons an extra energy, they are accelerated, and lifted up to a slightly higher energy level. These electrons of a slightly higher energy tend to have a momentum direction determined by the electric field. Hence they give rise to a macroscopic current. However, it is not always possible for an electron to take on a slightly higher energy level. To do so, there needs to be a vacant state available, right above the original state of the electron. Clearly, the electrons occupying low energy states can not do this: all states above them are already taken. So these electrons do not take part in the process of conduction. For electrons near the Fermi surface, the situation is different. If the Fermi surface lies somewhere inside an energy band then the electrons there can jump up to higher (unoccupied) energy states. In this scenario, the material can conduct a current very well: it is a conductor. 
This case is shown on the left in Figure \ref{fig:between}.

\subsection{Insulators}
We just explained that if the Fermi surface lies somewhere inside an energy band, the electrons near that surface can jump up in energy and conduct a current. Another possibility that can occur, is that the Fermi surface lies at the top of an energy band. This energy band is then completely filled, and is called the \textbf{valence band}. In such a case, the lowest vacant energy states lie much higher: at the bottom of the next energy band called the \textbf{conduction band}. So electrons need to jump up over the entire band gap to reach those states. If the band gap is wide, the applied potential is typically insufficient for a significant number of electrons to cross the band. Hence, the current will consist of only very few (or even none) of the electrons. Such a material does not conduct well, or even not at all: it is an insulator. 

\subsection{Semiconductors}
The third category of materials is a bit of an in-between. The Fermi surface again lies at the top of an energy level, but now the valence band and the conduction band are much closer together. The energy gap to be crossed is then much smaller. Here, the role of temperature becomes very crucial. In the beginning of this chapter we met an expression for the occupation of states for a collection of Fermions at some temperature $T$, which can be rewritten as follows:
\be
n_i = \frac{1}{e^{(E_i-\mu)/kT }+1}
\ee
The shape of the distribution at several temperatures is shown in Figure \ref{fig:FDT}.

Back to the band gap of a semiconductor. Since the occupation numbers are low in the conduction band, the statistics can be safely approximated by the Boltzmann distribution. Indeed: $(E-\mu) >>  kT $ there, whence 
\be
\frac{1}{e^{(E-\mu)/kT}+1} \sim e^{-(E-\mu)/kT}
\ee 
Denoting the size of the energy gap by $\Delta E$, we find that the occupation number $n_{cond}$ of the conduction band is proportional to:
\begin{figure}[t]
 \begin{center}
  \includegraphics[width=0.5\textwidth]{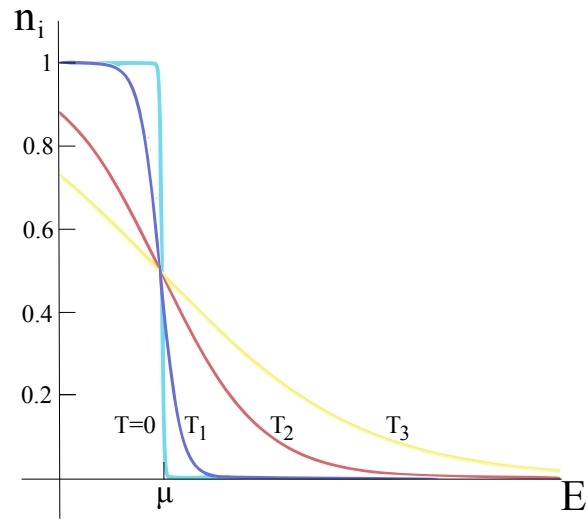}
  \caption{The Fermi-Dirac distribution for different temperatures. At zero temperature, the system takes on the lowest possible energy state. This means all states below some $\mu$ are filled and all higher states are empty. When raising the temperature, thermal excitations start to show up: some electrons migrate to higher energies. The distribution smoothly transitions to a more spread out distribution, the tail of which has a $e^{-E_i/kT}$ shape.}
  \label{fig:FDT}
  \end{center}
\end{figure}
\be
n_{cond.} \sim e^{-\Delta E/ kT}
\ee
since $(E-\mu) \approx \Delta E$ for low-lying energy levels in that band. This factor strongly depends on the temperature, so we conclude that the conductivity of the semiconductor is varies strongly with temperature. Also, note that for each electron in the conduction band, a \textbf{hole} is left in the valence band. In the background of the lattice, this hole looks like a positive charge, and hence it can take part in the conduction too.

Besides the temperature, the conductivity of a semiconductor is strongly influenced by impurities. That is because some impurities \textit{donate} freely moving electrons (or holes). Since only very few electrons of the semiconductor take part in the conduction, the addition of a small amount of impurities may easily donate an equally large or much larger amount of available electrons. So adding impurities allows to manipulate the properties of semiconductors strongly. This is the underlying reason of the widespread use of semiconductors in electronics. In particular, simple semiconductor based applications like the transistor have revolutionized our world and this has even lead some people to state it is the single most important invention of the 20$^{th}$ century. Again, the physics underlying this invention would not have been understood without the advent of quantum mechanics. 
Time for an example...

\section[A simple model]{\includegraphics[width=0.04\textwidth]{comput_c} A simple model}
\begin{figure}[t]
 \begin{center}
  \includegraphics[width=0.75\textwidth]{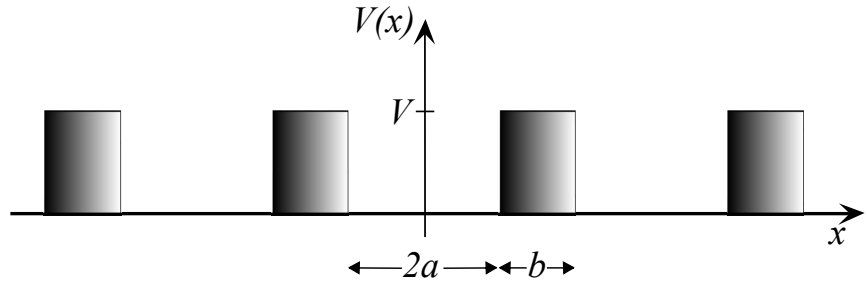}
  \caption{A periodic potential. The value is alternating between $V(x)=0$ and $V(x) = V$. The width of the peaks is $b$ and they are separated by a distance $2a$.}
  \label{fig:perpot}
  \end{center}
\end{figure}

Striking enough, we can obtain a (semi-realistic) example of a system containing energy bands and gaps without all too much effort, by the so-called \textbf{Kronig-Penney model}. As we explained, a crystal is a solid material in which the atoms are arranged in a periodic and ordered fashion. By consequence, the Coulomb potential felt by electrons is periodic as well. We now mimic such a periodic potential by $V(x)$ as shown in Figure \ref{fig:perpot}. The potential is zero over a distance $2a$, peaks at some $V(x)=V$ over a distance $b$, and this is repeated. The period of the potential is given by $c = 2a +b$:
\be
V(x+c)=V(x)
\ee
 We now introduce a translation operator $T$, which is shifts any function horizontally over a distance $c$: 
\be
T f(x) = f(x+c)
\ee
For example $T V(x) = V(x+c) = V(x)$. Using this and the fact that $\frac{\partial}{\partial x} = \frac{\partial}{\partial (x+c)}$ (which follows from the chain rule) we get
\bea
T H \psi(x) &=& T \left(-\frac{\hbar^2}{2m} \frac{\partial^2}{\partial x^2}+ V(x) \right) \psi(x) \\
&=& \left(-\frac{\hbar^2}{2m} \frac{\partial^2}{\partial (x+c)^2}+ V(x+c) \right) \psi(x+c) \\
&=& \left(-\frac{\hbar^2}{2m} \frac{\partial^2}{\partial x^2}+ V(x) \right) \psi(x+c)\\
&=& H T \psi(x)
\eea
So $[H,T]=0$. This means that if $|\psi\rangle   $ is an eigenstate of $H$ with eigenvalue $E$, the state $T |\psi\rangle   $ will be so too, and with the same eigenvalue: 
\be
H (T |\psi\rangle   ) = T H |\psi\rangle    = T E |\psi\rangle    = E (T|\psi\rangle   ).
\ee
If we assume that there is only one state for each energy, $|\psi\rangle   $ and $T|\psi\rangle   $ must be the same physical states. So their wave functions $\psi(x)$ and $\psi(x+c)$ must be equal up to a phase factor:
\be
\psi(x) = e^{i\alpha} \,\,T\psi(x) = e^{i\alpha} \psi(x+c)
\ee
The equation $\psi(x) = e^{i\alpha} \psi(x+c)$ is very restrictive. It can be shown that it can only be satisfied for functions of the following form:
\be
\psi(x) = e^{i K x} u(x)
\label{eq:bloch}
\ee
with $u(x)$ a periodic function. Wave functions of this kind are called \textbf{Bloch waves}. They are not necessary periodic over a distance $c$, since the `rotating' part $e^{i K x}$ can have any periodicity (not necessary $c$). Also note that the above equation implies that
\be
\psi(x+c)=e^{i K c} \psi(x).
\label{eq:bloch2}
\ee
Now, let us put a particle of energy $E$ in the potential. For concreteness, take $E<V$, the other case works out in a similar fashion. Defining
\be
\kappa^2 = \frac{2m}{\hbar^2} (V- E) \quad \textrm{and} \quad k^2 = \frac{2m}{\hbar^2} E 
\ee
the solution in region 1 (see Figure \ref{fig:perpot}) is given by:
\bea
\psi_1 (x) &=&  A_1 e^{\kappa x} + B_1 e^{-\kappa x} \quad \textrm{for}\quad -b-a<x<-a\\
&=&A_2 e^{ikx} + B_2 e^{-ikx} \quad \textrm{for}\quad -a<x<a
\eea
From the above and (\ref{eq:bloch2}) it follows that the solution in region 2 looks like:
\bea
\psi_2 (x) &=& e^{iKc}\left[ A_1 e^{\kappa (x-c)} + B_1 e^{-\kappa (x-c)}\right]  \quad \textrm{for}\quad a<x<a+b\\
&=&e^{iKc}\left[ A_2 e^{ik(x-c)} + B_2 e^{-ik(x-c)}\right] \quad \textrm{for}\quad a+b<x<a+c
\eea
Demanding continuity of the wave function and its derivative at the points $x=-a$ and $x=a$ gives the following four constraints:
\bea
A_1 e^{-\kappa a} + B_1 e^{\kappa a}  &=& A_2 e^{-ika} + B_2 e^{ika}\\
\kappa  A_1 e^{-\kappa a} -\kappa B_1 e^{\kappa a}  &=& ik A_2 e^{-ika} -ik B_2 e^{ika}\\
A_2 e^{ika} + B_2 e^{-ika}&=& e^{iKc}\left[A_1 e^{\kappa(a-c)}+ B_1 e^{-\kappa(a-c)}\right]\\
i k \left[A_2 e^{ika} - B_2 e^{-ika}\right]&=& e^{iKc}\kappa\left[A_1 e^{\kappa(a-c)}- B_1 e^{-\kappa(a-c)}\right]
\eea
This is a homogenous system of four equations in the unknowns $A_1$, $B_1$, $A_2$, $B_2$. The condition to have solutions is thus
\be
\left|  \begin{array}{cccc}
e^{-\kappa a}  & e^{\kappa a}  & -e^{-ika} & -e^{ika}\\
\kappa   e^{-\kappa a} &-\kappa  e^{\kappa a} &-ik e^{-ika} & ik e^{ika}\\
 - e^{iKc} e^{\kappa(a-c)} & -e^{iKc} e^{-\kappa(a-c)}& e^{ika} & e^{-ika} \\
 - \kappa e^{iKc} e^{\kappa(a-c)}& \kappa e^{iKc} e^{-\kappa(a-c)} & i k e^{ika} &- ik e^{-ika} 
  \end{array}\right|=0
\ee
To evaluate the determinant, we simplify it a bit. First divide away the factors $e^{-\kappa a}$, $e^{\kappa a}$, $- e^{-ik a}$ and $- e^{ika}$ from the four rows, and use $2a-c = - b$ to simplify some exponents. The first row then looks like $(1,1,1,1)$. Then put (column 1+ column 2)/2 and (column 1 - column 2)/2 instead of the original columns 1 and 2, and similarly for the columns 3 and 4. Taking away a last factor $i$ from column 4, we get: 
\be
\left|  \begin{array}{cccc}
1  & 0  &  1 &0\\
0  & \kappa   & 0&  k \\
 - e^{iKc}\cosh (\kappa b ) & e^{iKc}\sinh (\kappa b)&- \cos (2 ka) & -  \sin 2 k a \\
 \kappa e^{iKc}\sinh(\kappa b)& - \kappa e^{iKc}\cosh{ \kappa b} &k \sin(2 ka) &  -k \cos( 2 k a) 
  \end{array}\right|=0
\ee
A lot of zeros have popped up! By expanding this as a sum of cofactors of the elements of first row, the above is just a sum of two 3-by-3 determinants. These are not too hard: they each lead to four terms. Collecting terms and simplifying expressions like $\sin^2 \theta + \cos^2 \theta =1$ and $\cosh^2 x -\sinh^2 x =1$, the whole thing becomes less and less ugly. The end result reads:
\be
\cos (2 k a) \cosh (\kappa b ) + \frac{\kappa^2 - k^2}{2 \kappa k } \sin(2 k a)\sinh (\kappa b) = \cos (Kc)
\label{eq:endres}
\ee
So to recap: for the system to have smooth solutions at energy $E$, the above equation needs to be satisfied. Since the right hand side lies between -1 and 1, the left hand side must be so to in order to allow solutions. In conclusion, we need
\be
-1\leq f(E)\leq 1,
\ee
with
\begin{figure}[t]
 \begin{center}
  \includegraphics[width=0.7\textwidth]{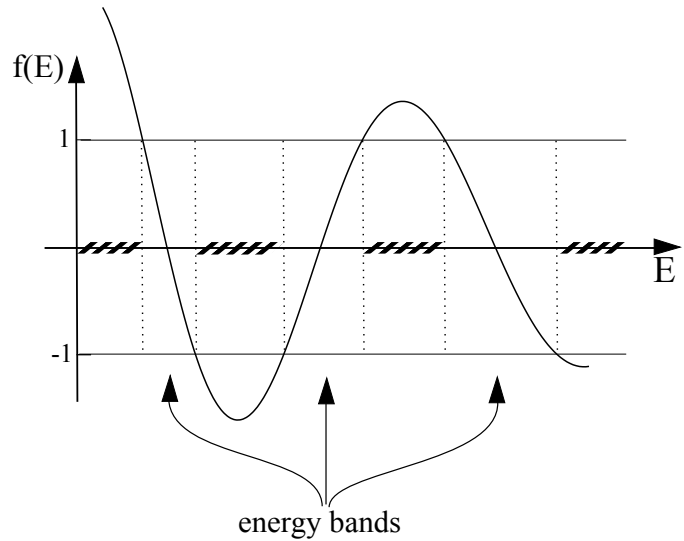}
  \caption{The function $f(E)$ for a specific value of $a$ and $b$. There are no states with energy $E$ in the regions where $|f(E)|>  1$. So these zones are gaps, as indicated on the figure. In the regions where $|f(E)|<1$ there \textit{are} energy states available, so these constitute the energy bands.}
  \label{fig:kronig}
  \end{center}
\end{figure}
\be
f(E) = \cos (2 k a) \cosh (\kappa b ) + \frac{\kappa^2 - k^2}{2 \kappa k } \sin(2 k a)\sinh (\kappa b) 
\ee
The function $f(E)$ is shown in Figure \ref{fig:kronig} for some arbitrary values of the parameters $a$, $b$ and $V$. It is immediately clear that 
there are genuine energy bands and energy gaps. This behavior occurs rather generic: the parameters $a$, $b$ and $V$ need not to be finely tuned to arrive at pictures similar to the Figure \ref{fig:kronig}. So even for a very simplistic potential we obtain a semi-realistic energy spectrum. One can upgrade this model by taking more Coulomb-like potentials (of which the charge is partly shielded by the other electrons) by making the model three-dimensional and by including imperfections of the crystal. So with the very same principles, one can obtain even more realistic results. 

\newpage
\section*{Exercises}
\begin{figure}[H]
 \begin{center}
  \includegraphics[width=0.7\textwidth]{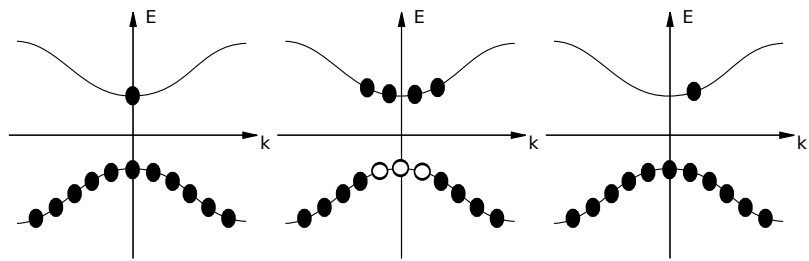}
  \caption{An N-type semiconductor in different physical situations.}
  \label{fig:Ntype}
  \end{center}
\end{figure}

\begin{enumerate}
\item Explain that the electrons crystals are somewhere in between bound state and free. How is this reflected in the energy spectrum?
\item It is sometimes stated that a conductor is a limiting case of a semiconductor, where the valence and conduction band are right on top of each other, without gap. 
Explain. 
\item The surface of the sun is at about 5800K. For every hydrogen atom in the 2s state, how many atoms can you find in the 1s state? The Boltzmann constant is $k=8.62\cdot10^{-5}$ eV/K$=1,38\cdot10^{-23}$ J/K.
\item Redo the periodic potential, but now for the case $E>V$. (There is a shortcut to obtain this result: if $E>V$, then $\kappa$ becomes imaginary. Putting $\kappa=i k_2$ and using $\sinh(ix)=i \sin x$ and $\cosh (ix)=\cos x$ you should be able to find the analog of (\ref{eq:endres}) in terms of $k$ and $k_2$. This should correspond to what you found in the `brute' way.) 
\item For the Kronig-Penney model, we found the energy spectrum. Recall that the states are also characterized by their momentum $k$. (Just like for a free particle, the energy is a function of the wave number.) So actually, the spectrum can also be depicted as lines in a $E-k$ diagram, instead of just bands on an energy axis. Look at Figure (\ref{fig:Ntype}): the set of allowed states are lines in this diagram. The dots denote states (actually, they form a continuous spectrum, but well) and a white dot denotes an empty state. How do you see the energy gap on the pictures? The material shown is an N-type semiconductor: a semiconductor doped by a material which donates some extra electrons to the lattice. The occupation of states is shown for the same material, in three different situations. Explain which physical situations you see on the three different pictures. In the situation where a net current is flowing (due to an external E-field) in which direction is it going?
\end{enumerate}

%% file: H13.tex
\chapter*{Outro}
\addcontentsline{toc}{chapter}{Outro}

That's it! This is the end of the course. You can now say that you know quite a bit about quantum mechanics. When you think back, you may realize that quantum mechanics is not just a funny and strange theory invented by even stranger physicists. It is an accurate and comprehensible theory about nature on its smallest scale. Moreover, it is the beating heart of the reality as we see it. Indeed, atoms, molecules, crystalline solids, radiation: all these crucial ingredients of nature are very strongly determined by the quantum mechanical laws that govern their behavior. Also, the role of quantum mechanics in technological applications is much bigger than most people think. Lasers, microprocessors, precision thermometers and STM microscopes: none of these inventions could have been possible without understanding the laws quantum mechanics. Hopefully you have a more broad vision on this now, but far more important: I hope you have enjoyed this textbook. 

%% file: H14.tex
\chapter*{Bonus track: philosophical note}
\addcontentsline{toc}{chapter}{Bonus track: philosophical note}

\section{Quantum mystery and God's dice} 

\begin{wrapfigure}{l}{0.3\textwidth}
  \begin{center}
    \includegraphics[width=0.25\textwidth]{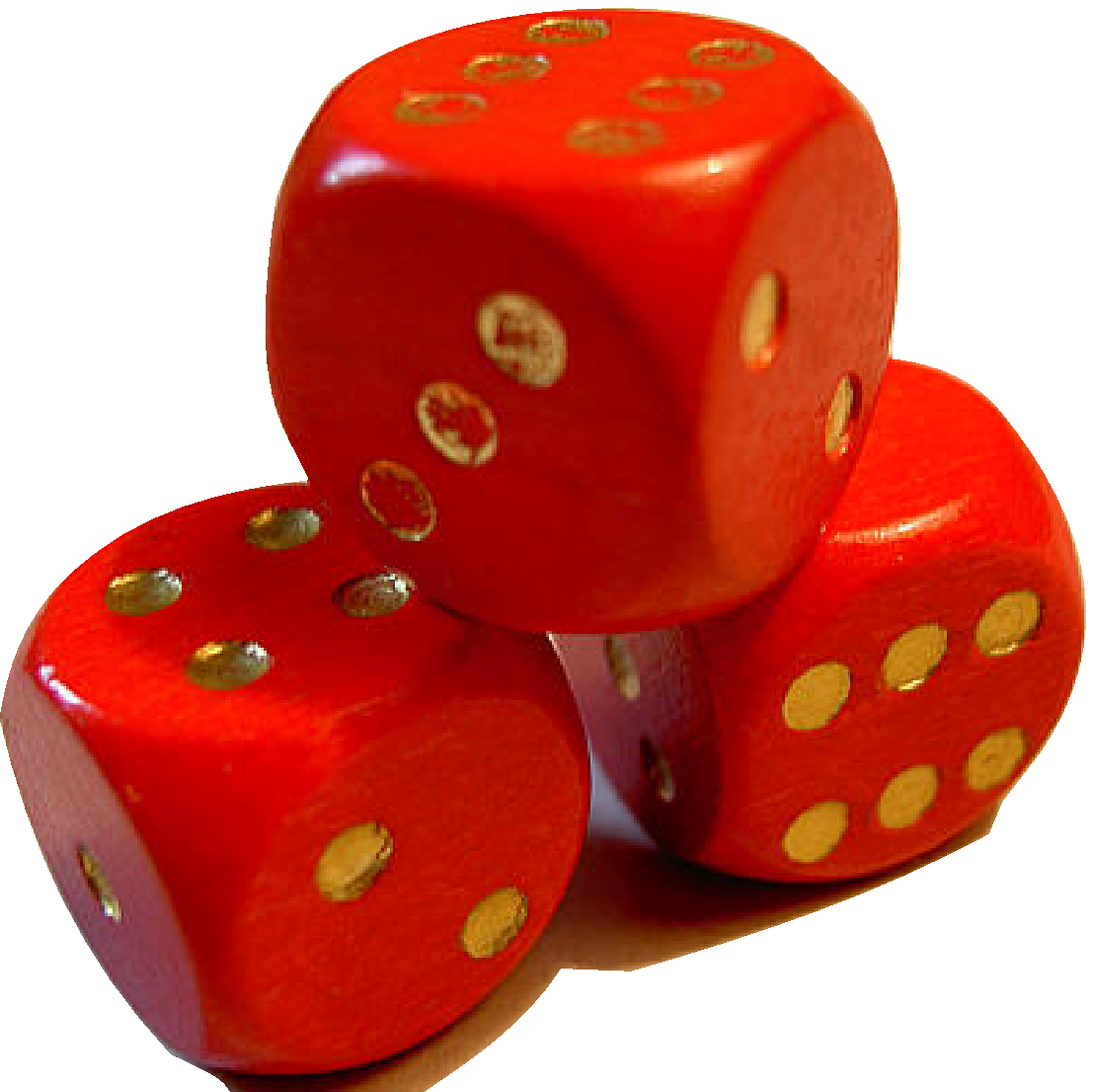}
    \label{fig:dice}
  \end{center}
\end{wrapfigure}
 
In the outro, emphasis was put on the practical value of quantum mechanics in quite some technological applications. This is a fact which seems to be a bit ignored by popular culture. On the contrary, to a large public, quantum mechanics has a somewhat mysterious or even transcendental status. This despite the fact that the theory came to life at the very beginning of the 20th century - the absolute prehistory on the scientific and technological timescale. So where does this esoteric status of quantum mechanics come from? An important element in this quantum-mystery is the widespread idea that quantum mechanics is not a deterministic theory. This contrasts to more or less every other theory in physics, and would surely be an uttermost mysterious conclusion of one of the most refined theories of nature. 

Let us think about this more carefully. Indeterministic ... where did we see \textit{that} word popping up in this course? For sure there are no `unknown' parameters in the Schr\"{o}dinger equation. Whether we are talking about one particle, or a bunch of them, or even an entire lump of matter, the rules of quantum mechanics give a very clear \textit{and} deterministic prescription on how this system will evolve in time. If you know the system's Hamiltonian, you can solve the time evolution, and say what it will look like at any later moment in time. So in that aspect, quantum mechanics is not even slightly different from -say- the Newtonian description of point particles. It is only a bit more complex, and a lot more accurate. 

`Ah, but what about the measurement', you say. We did learn indeed that measuring the position of a particle for example yields a more-or-less unpredictable outcome. The answer will typically be some value lying inside the approximate location of the wave function, but we can never predict what the exact outcome will be. Back then, we summarized this by the statement that a measuring device gives a `reasonable, but random answer'. It is this aspect of the theory which people refer to when they say that quantum mechanics is indeterministic. And it is a puzzling feature indeed, which has caused a lot of confusion over the decades. A famous quote of Einstein in this context is: `God does not play dice'. This summarizes his and many others' dissatisfaction with the issue.

So when considering the issue of determinism, it seems like quantum mechanics shows two faces. On one hand, every system (small or big) evolves perfectly deterministic by the Schr\"{o}dinger equation, on the other hand, the act of measurement introduces a random outcome, and collapses the system onto the state that is was measured to be in. So measuring devices seem to play a very special role here. Of course this is a bit paradoxical. If the laws of quantum mechanics are correct (and there are very good reasons to believe this) also measuring devices have to behave according to those laws. So they too, need to behave deterministic. What is going on here?

In principle, the issue is almost of mathematical clarity. Either measuring devices evolve according to the deterministic laws of quantum mechanics, or they have a special status that allows them to give fundamentally indeterministic answers. The problem is of course that studying the behavior of a measuring device on the basis of quantum mechanics is a horrible problem: it is practically impossible to do any realistic computation for such a large and complex system. There are some attempts to do so, but not with all too great success - and it will probably stay like that for a while.

So the problem of deciding whether or not quantum mechanics is a deterministic theory is essentially one of computational complexity. Actually, there is quite a lot to this phrase of Einstein. What he refers to is of course that it seems unnatural for God\footnote{a term that he used as a metaphor for the most fundamental laws of nature} to include `random' events in the evolution of any system - as if he was throwing dice for every measurement. Funny enough, when you think about dice, you may remark that these (depite being our most universal symbol for randomness) do not behave random either. They look like they do, but you know they just obey Newton's laws of mechanics. Let us illustrate this with a discussion between Newton and a gambler.

\begin{quote}
Gambler: Hey Newton, do you have any fancy theory to predict the outcome of my dice? If you have, I'll learn physics right away!\\
Newton: Well, actually, I have. They fall, bounce and twist just according to my laws of mechanics.\\
Gambler: Oh, really? Can you predict what I will get right now? \textit{(Throws dice.)} Aha, a double six! did you see that one coming?\\
Newton: Not really. In principle, if you would give me all data, the momentum you gave the dice, their shape and that of the surface they land on, I could predict the outcome. But even with all the right data, it's a very nasty computation. Probably even harder than computing orbits of planets.\\
Gambler: \textit{(disappointed)} So actually, you can not say anything interesting about it. \\
Newton: Well, yes. There are probability laws.
These will tell you that you are much more likely to throw a total of seven than a total of twelve, which you got just now. These laws are of great value, even though they are not able to predict any individual outcome.\\
Gambler: Ah, I see. So I better take a course on statistics then?\\
Newton: If you want to be a better gambler, yes. Good luck with it!\\
Gambler: Thanks, Sir Newton, good luck with your laws! 
\end{quote}
In a very similar way, you may believe that measuring devices too behave deterministic.\footnote{This similarity was very likely \textit{not} intended by Einstein's phrase, but the possible analogy between the underlying mechanisms is rather striking.} Only in the very end, the dazzlingly complex behavior gives rise to a relatively simple probabilistic law: $P\sim|\psi|^2$. But how could such a simple law be the end result of such a huge complexity? Well, after all a measuring device is a very special system, designed to do a very special job. In the same way, a dice is precisely designed to give a nice equal chance distribution resulting from its dynamics - despite the fact that it was probably not designed with the active purpose to `develop a solid object, of which the dynamical and elastic characteristics are such that the final stage after a throw on a smooth surface ends up in an almost perfectly homogeneous probability distribution over the possible outcomes'. In the same way, the simplicity of the probabilistic law of measurements in quantum mechanics may be the result of `accidental' smart design of the instrument.\footnote{No intelligent design pun intended.} 

\section{A jungle of interpretations}
So far a partly demystifying view on the whole subject. However, don't think there is a consensus on this matter though. There is a jungle of possible interpretations of the physics of the measurement in quantum mechanics. Just check out `interpretation of quantum mechanics' on any search engine, and this jungle opens up. There are roughly three categories of interpretations:
\begin{itemize}
\item Indeterministic theories. These take the process of measurement as an indication that quantum mechanics is intrinsically non-deterministic. This leads to very deep questions on what reality is, what an observer is and what connects them. Probably the most hard and (sorry for the word) confusing interpretation, these theories seem to dominate the view on the matter presented to the large public. Without doubt the origin of the widespread mystery surrounding quantum mechanics.
\item Indifferent theories. These do not even pose the question whether quantum mechanics is deterministic or not, and categorize the issue as irrelevant and philosophical. Here, the arguments (quite rightful) are that quantum mechanics is a perfectly healthy theory. It is consistent, allows to describe systems and the (be it probabilistic) outcomes of measurements. Where this probabilistic law comes from, is not a question the physicist should ask him- or herself. A quote supporting this: `Shut up and calculate.' If you are not fond of philosophy, this might be your favorite viewpoint.  
\item Deterministic theories. These assume that the process of measurement is governed by the same quantum laws that describe other systems. Some arguments for this view were presented above. Conceptually, this is probably the least confusing option. (Unless you are more dogmatic and like `indifferent theories' better.) However, the question on how this precisely works (i.e. how the probability rule emerges exactly) is a very hard one, and remains largely unanswered. 
\end{itemize}

As said, each of the categories comprises numerous theories and variations, and probably equally many of them do not fall strictly into one of the three types outlined here. Maybe you will develop a view of your own one day. Feel free to do so, the interpretation of the measurement of quantum mechanics is one of the (many) open problems of physics.

%% file: H15.tex
\chapter*{Figure credits}
\addcontentsline{toc}{chapter}{Figure credits}

\subsubsection*{Thanks} 
In creating this work, I have benefitted from the free software, media and knowledge provided by the Latex team, the Inkscape team and the Wikimedia foundation. I also want to thank my colleges and friends for their moral support. Many of the images were based on existing ones. Here is a list of attributions. The images presented in this document are under the Creative Commons Attribution-Share Alike 3.0 Unported license, unless required differently by the license of their originals.

\begin{center}
	\begin{tabular}{ | l | l | p{5cm} | l |}
    \hline   
      Fig. & Original work & Author/user name & License \\ \hline \hline
\ref{cat} &  Doubleslitexperiment &  Koantum, Trutz Behn & by-sa\\ \hline 
\ref{cat} &  Blocked\_cat &  Binnette & 	GFDL/by-sa\\ \hline 
\ref{complex_number} & Complex\_number & Karol Ossowski, MesserWoland & GFDL/by-sa\\ \hline 
\ref{max} & Onde\_electromagnetique & SuperManu &  GFDL/by-sa\\ \hline 
\ref{doble} & Double-slit\_schematic & Peter Suppenhuhn, Trutz Behn & public domain\\ \hline 
\ref{well_well}& Infinite\_potential\_well & Benjamin D. Esham for the Wikimedia Commons & public domain\\ \hline 
\ref{fig:pib} & Particle\_in\_a\_box\_wavefunctions & Papa November & by-sa\\ \hline 
%
 \ref{fig:SHO} & HarmOsziFunktionen & AllenMcC.  & by-sa\\ \hline 
\ref{fig:bohr_zelf}& Bohr-atom-PAR &PAR, JabberWok & GFDL\\ \hline 
%
\ref{fig:orbitals} & HAtomOrbitals & FlorianMarquard & GFDL/by-sa\\ \hline 
%
\ref{fig:gaussian} & Normal Distribution PDF & Inductiveload  & public domain\\ \hline 
\ref{fig:STM} & Rastertunnelmikroskop-schema & Michael Schmid, Grzegorz Pietrzak &  by-sa\\ \hline
\ref{fig:zeeman} & Zeeman\_effect & Bogdan   & by-sa\\ \hline
\ref{fig:spincurrent}&  Gloriole  & Nae'blis & GFDL\\ \hline
\ref{fig:laser} & Lasercons  & DrBob & GFDL/by-sa\\ \hline
\ref{fig:FDT} & FD\_e\_mu  & Unc.hbar	 & public domain\\ \hline
(dice) & Wuerfel\_rot & W.J.Pilsak & by-sa\\ \hline
 \hline
    \end{tabular}
\end{center}